\documentclass[sotoncolour]{uosthesis}      

\graphicspath{{Figures/}}   
\usepackage{attrib}            
\hypersetup{colorlinks=true}   
\newcommand{\BibTeX}{{\rm B\kern-.05em{\sc i\kern-.025em b}\kern-.08em T\kern-.1667em\lower.7ex\hbox{E}\kern-.125emX}}

\newcounter{address}
\newcounter{alg}

\usepackage{bm}
\usepackage{tikz}
\usepackage{amsmath}
\usepackage{amssymb}
\usepackage{amsfonts}
\usepackage{graphicx}
\usepackage[style=ddmmyy]{datetime2}
\usepackage[normalem]{ulem}

\usepackage[english]{babel}

\usepackage[sorting=none,style=numeric-comp]{biblatex}
\numberwithin{equation}{chapter}
\counterwithout{figure}{chapter}
\usepackage{csquotes}
\usepackage{color}

\usepackage{color}

\newcommand{\stkout}[1]{\ifmmode\text{\sout{\ensuremath{#1}}}\else\sout{#1}\fi} 

\department  {School of Mathematical Sciences}
\DEPARTMENT  {\MakeUppercase{\deptname}}
\group       {[Group name]}
\GROUP       {\MakeUppercase{\groupname}}
\faculty     {Faculty of Social Sciences}
\FACULTY     {\MakeUppercase{\facname}}
\title      {Superconducting phases of strongly-interacting matter in large magnetic fields}
\authors    {Geraint Wyn Evans} 
\addresses  {\groupname\\\deptname\\\univname}
\date       {June 21, 2023}

\qualifications{MSci}
\orcidid{0000-0001-6106-4567}
\subject    {}

\begin{document}
\pagenumbering{gobble} 
\copyrightDeclaration{} 
\raggedright                  
\frontmatter
\maketitle

\begin{abstract}
    \footnotesize{

       The large magnetic fields of neutron stars and produced in heavy-ion collisions motivate investigation into the response of strongly-interacting matter to extreme magnetic forces beyond just theoretical interest. Furthermore, the varying temperature $T$, baryon chemical potential $\mu_B$, and aforementioned magnetic fields $B$, of these systems leads to questions concerning the phase structure of Quantum Chromodynamics (QCD) at large. At low temperatures, superconducting phases become a possible candidate for the ground state in the $\mu_B$-$B$ plane. This thesis investigates two scenarios where these phases emerge at $T=0$, with an emphasis on the type-II regime.
       
       The first scenario concerns type-II superconductivity at large $\mu_B$. At asymptotically high baryon density the ground state of QCD is a colour superconductor where gluonic fields can experience a Meissner effect. In the two-flavour pairing (2SC) and colour-flavour locked (CFL) colour-superconducting phases, a small admixture of the photon with a gluon is also expelled which means an applied external magnetic field will experience a slight Meissner effect. Therefore, these phases act as very weak electromagnetic superconductors. Previous works have shown that with massless quarks the type-II 2SC phase is preferred in a certain parameter region where the magnetic defects are domain walls. We introduce corrections for a finite strange quark mass in a Ginzburg-Landau approach, and find that the domain wall defects are replaced by a cascade of multi-winding flux tubes, among other changes to the phase diagram. Due to the emergence of a second colour-superconducting condensate emerging in the core, the magnetic flux is confined into ``rings'' where both condensates are depleted, forming pipe-like structures.

       In the second scenario, $\mu_B$ is low enough such that the presence of nucleons is not yet energetically favourable. It was previously shown that, using Chiral Perturbation Theory and incorporating the chiral anomaly via a Wess-Zumino-Witten term, the ground state in this region above a certain critical magnetic field is a Chiral Soliton Lattice (CSL) of neutral pions - an inhomogeneous phase consisting of a series of domain walls. It was further shown that the CSL becomes unstable to charged pion fluctuations at an even higher second critical field. We argue this instability corresponds to a second order phase transition to a type-II superconducting vortex lattice phase and construct this phase in the chiral limit. We find the type-II vortex lattice phase is preferred and has a non-zero, inhomogeneous baryon number density, leading to a two-dimensional crystalline structure.

       Preceding the presentation of these works, the $\mu_B$-$B$ plane of the QCD phase diagram is reviewed such that the results can be placed in the wider context of this plane.
       }
\end{abstract}

\tableofcontents
\authorshipdeclaration{
\newline
\newline\fullcite{Evans:2020uui}
\newline
\newline\fullcite{Evans:2022hwr}
}
\acknowledgements{
First, I would like to thank Andreas Schmitt for his patience, guidance, honesty and support over the course of my PhD. His door is almost always open, ready for my unexpected visits and questions. I have learnt much from him and I could not have asked for a better supervisor.

Thank you to the Neutron Star group for being so welcoming. I thank Nils Andersson for his advice and making the group meetings so entertaining each week. Special thanks also goes to Thomas Celora and Savvas Pitsinigkos for their friendship and the experiences we shared in conferences and workshops. 

From the wider postgraduate researcher community in Mathematical Sciences there are too many people I am grateful to have met to name here. Special thanks goes to Ysanne Pritchard for co-organising Maths \& Mingle with me. I found running the seminars a very enjoyable and rewarding experience and they could not have happened without her hard work. Thank you to the current and past denizens of office 54/8035 for providing endless distraction, especially Briony Eldridge for her company and conversation over the past months.

Furthermore, thank you to Sebastian Chenery, Philip Wells, Dawid Drelinkiewicz and Jack Mitchell for being good friends and housemates over the years. 

I would also like to thank Monica Motoc, Robert Leigh, James Smedley, Benjamin Doyle, Matthew Mettson and Benjamin Hopper for their continued friendship post our undergraduate degrees and my close friends from back home Harry Powell, Joshua White and Evan Rees. Thank you all for your support and spirit-raising company.

Finally, thank you to my family. Thank you Anti Sheila for interesting discussions on physics and keeping me driven. Thank you Mam, Dad and Siwan whom I love and am deeply grateful for their unwavering support over not just my PhD, but my life in general. Diolch, a caru chi gyd.
}
\mainmatter
\chapter{Introduction}
\label{chpt:Intro}
\section{The strong and electromagnetic interactions}
\label{Intro:sec:Strong}

The discovery of quantum mechanics and special relativity in the early 20\textsuperscript{th} century marked a shift in our understanding of the natural world. The combination of these two pillars of modern physics lead to Quantum Field Theory (QFT) and the much celebrated Quantum Electrodynamics (QED) \cite{Tomonaga:1946zz,Schwinger:1948iu,Schwinger:1948yk,Feynman:1949zx,Dyson:1949bp,Feynman:1950ir}. This described the fundamental interaction of electromagnetism at the subatomic scale. Following the discovery of QED, Quantum Chromodynamics (QCD) was proposed to describe the strong interaction. Alongside the weak interaction, which was shown to merge with electromagnetism at high energies \cite{Glashow:1959wxa, Glashow:1961tr,Weinberg:1967tq}, these two theories form the cornerstones of the Standard Model and high energy physics today. While they are similar in many ways, there are key differences which makes QCD more challenging to use. Aside from being non-abelian (as opposed to abelian like QED) the application of perturbation theory in QCD is limited. Whereas one can safely expand QED in powers of the coupling constant $e$ for a wide range of energy scales that are of interest (e.g. in particle colliders), the coupling constant of the strong interaction $g$, changes significantly with the energy scale. It is only at high energies when $g$ is sufficiently small that one can use perturbative techniques, or perturbative QCD (pQCD) to make reliable predictions. This feature of the theory is called asymptotic freedom \cite{Gross:1973id,Politzer:1973fx}. Thus, we can use QCD at high energies but are limited to other theoretical tools at lower energy scales, such as numerical lattice simulations and effective theories like Chiral Perturbation Theory (ChPT) \cite{Weinberg:1978kz,Gasser:1983yg}.

It is the equilibrium of the electromagnetic force of QED and the residual strong force of QCD which is responsible for the structure of everyday baryonic matter. Together, these forces sculpt the nuclei within atoms - the long range repulsive Coulomb force between protons is balanced with the short range attractive nuclear force between nucleons. Typically, one must go out of one's way to manufacture situations where protons and neutrons are the degrees of freedom rather than the atoms and nuclei they constitute. Even more extreme conditions are required for the degrees of freedom to be something other than atoms or nucleons, where matter would be formed of deconfined quarks or mesons for example. However, there is a naturally occurring system where such conditions for this so-called nuclear or quark matter may be realised. The incredibly high densities reached in the core of a neutron star (NS) could result in nucleonic matter emerging in the form of neutron superfluidity and proton superconductivity \cite{MIGDAL1959655,Bailin:1983bm,Chamel:2017wwp}. It is hypothesised that even deeper in the core, the chemical potential is high enough to support exotic states such as deconfined quarks, strange/hyperonic matter and colour superconductivity (see reviews \cite{Weber:2004kj,Alford:2007xm,Schmitt:2010pn}). The required conditions for exotic phases can also be achieved in heavy-ion collision (HIC) experiments, where quarks are deconfined, forming a Quark-Gluon Plasma (QGP) \cite{BRAHMS:2004adc,PHENIX:2004vcz,PHOBOS:2004zne,STAR:2005gfr,Rafelski:2015cxa}. Both of these examples not only provide a way of probing the QCD phase diagram but possess/produce large magnetic fields \cite{1991ApJ...383..745L, Kharzeev:2007jp,Ferrer:2010wz,Potekhin:2011eb,PhysRevC.85.039802, STAR:2015wza}, and therefore motivate investigation into the response of strongly interacting matter to electromagnetic fields beyond pure theoretical curiosity. These considerations warrant a wider exploration than the typically considered baryon chemical potential $\mu_B$ and temperature plane $T$ plane of the QCD phase diagram. One can imagine extending the diagram along a third axis for the magnetic field $B$ and ask what is the phase structure in the $B$-$T$ and $\mu_B$-$B$ planes? 

The $B$-$T$ plane can be investigated using lattice QCD at low $\mu_B$, which is uninhibited by the numerical sign problem \cite{deForcrand:2009zkb} that plague these calculations at finite $\mu_B$.
 This has allowed for a wide theoretical study of this plane, see reviews Refs.\,\cite{Endrodi2013QCDPD,Philipsen:2016lr,Guenther:2020jwe}. Experimentally, one expects that HICs \cite{Kharzeev:2007jp, Kharzeev:2015znc,STAR:2015wza,Li:2020dwr} and measurements of primordial magnetic fields from the early universe \cite{Vachaspati:1991nm,Grasso:2000wj} can provide insight. In the $\mu_B$-$B$ plane there is a significant lack of input from first-principle calculations at low temperatures, with lattice calculations hindered due to the sign problem. Many of the results rely on effective theories and models. 
 
 Experimental input is also limited, but data from future multi-messenger observations of compact stars and future colliders that could operate at lower temperatures \cite{Schmitt:2016pre,Ablyazimov:2017rf,Kekelidze:2017ual,Xiaohong:2018weu,Hachiya:2020bjg} are promising. Due to these considerations, this plane has been left comparatively unexplored. This makes it an interesting area to probe with many theoretical questions to address. It leads one to wonder about what possible phases and transitions we would expect in the $\mu_B$-$B$ plane at $T=0$.
 
\section{Superconductivity from the strong interaction}
\label{Intro:sec:SC}
While high energy physics was in the midst of making breakthroughs at the smallest scales, condensed matter physics was experiencing its own revelations. The late 1950s (about a decade after the discovery of QED) saw great theoretical advancements in our understanding of superconductivity, a phase where quantum effects are manifest at macroscopic scales. Perhaps the most notable development was the microscopic Bardeen-Cooper-Schrieffer (BCS) theory in 1957 \cite{Bardeen:1957kj}. It gave accurate predictions for many superconducting parameters \cite{Bardeen:1957mv} and introduced the concept of Cooper pairs. Cooper pairs constitute the condensate responsible for the Meissner effect characteristic of superconductors, whereby magnetic flux is expelled from a superconductor's interior below a certain critical temperature $T_c$ and critical magnetic field(s). The phenomenological Ginzburg-Landau (GL) theory \cite{Landau:1950lwq} first introduced in 1950 could also be derived from BCS theory \cite{Gorkov:1959}. By writing the free energy as an expansion near $T_c$ in the order parameter (which is often interpreted as the wavefunction of the condensate), one can more easily deal with spatially inhomogeneous systems, like the vortex lattice phase in type-II superconductors \cite{Abrikosov:1956sx, PhysRev.133.A1226}.

Superconductivity is a phenomenon which is sensitive to changes in the magnetic field. One can therefore ask whether superconducting phases emerge in the QCD phase diagram and have interesting phase structure as we vary $B$. A simple argument in favour of the appearance of these phases in the QCD phase diagram comes from considering the criterion for Cooper pairing. In essence, according to BCS theory, we require charged fermions with an attractive interaction between them to form Cooper pairs. All known quarks (one of the fundamental degrees of freedom of QCD) possess electric charge as well as the colour charge associated with the strong interaction. As a result, many of the hadrons that they constitute also have an overall electric charge. It follows that quark and nuclear matter could have superconducting states with the attractive interaction being supplied by the strong force in some way, shape, or form. This is comparatively simpler than the attractive force between electrons in a conventional superconductor, which is usually attributed to the electron-phonon scattering, requiring an underlying lattice of ions to be present. 

The large number of charged particle species that interact through the strong and residual strong force also implies a rich phase structure. 
In conventional superconductivity, Cooper pairing occurs between one particle species i.e.\ the electrons. When we consider strongly-interacting matter, Cooper pairing can potentially occur between many different charged particle species. This is not confined to pairing within the same particle species either. Cross species pairing is also possible for instance in colour superconductivity (CSC) where pairing is possible between quarks of different flavours.
All this leads to the possibility of not only a greater variety of superconducting phases but also transitions between them based on the number of charged particle species available for pairing alone. Add the consideration of type-I/type-II superconductivity and Cooper pair spin (see, for example, spin $1$ CSC, Ref.\,\cite{Schmitt:2004hg} and references therein) and the possible phase structure becomes very rich indeed\footnote{Analogous phases are present in electronic superconductors also i.e.\ type-I/type-II superconducting phases and phases where the Cooper pairs have total spin greater than $0$ e.g.\ d-wave superconductivity.}. The particle number of each charged particle species is controlled by its own chemical potential. In principle, there is a potential $\mu_f$ for each quark flavour $f$. These potentials become linked via weak interaction processes and imposed global charge neutrality such that we can consider a single quark chemical potential $\mu_q$ for all flavours. From analogous arguments involving weak interactions we define $\mu_B$. We can relate $\mu_q$ to $\mu_B$ by considering all quarks carry $1/3$ baryon number charge, such that $\mu_q=\mu_B/3$. Therefore, we can study the phase structure of both nuclear and quark matter by varying $\mu_B$ and might expect to move between different types of superconductivity as we vary $\mu_B$. Furthermore, the GL parameter which determines whether we are in the type-I or type-II regime for a superconductor can also depend on $\mu_B$. This makes the $\mu_B$-$B$ plane an interesting plane to study from the perspective of superconductivity, and is the over-arching motivation behind this thesis. There is also the option of using the isospin chemical potential $\mu_I$ for both quark and nuclear matter if we wish to investigate the effects of imbalance between the number density of up and down quarks or protons and neutrons. Phase transitions due to varying $\mu_I$ will not be considered in this thesis.

We will also work at $T=0$ throughout. Of course, superconductivity is also sensitive to temperature. While we would not expect it to appear in the $B$-$T$ plane where $\mu_B=0$ (with the possible exception of vacuum superconductivity \cite{Chernodub:2010qx,Chernodub:2011gs}), superconducting phase transitions due to temperature could be of interest when considering the wider three-dimensional structure of the $\mu_B$-$B$-$T$ phase diagram. Indeed, proton superconductivity and colour-superconducting phases are often included in the $\mu_B$-$T$ plane at large $\mu_B$ and could be present in NSs. Currently, outside of NS observations, exploring the temperature transitions of these superconducting phases is difficult experimentally. While $T=0$ is a good approximation for these stars, simulations (e.g.\ Ref.\,\cite{Prakash:2021wpz}) indicate $T\sim 10-100\,\rm{MeV}$ in NS mergers. The critical temperatures of proton superconductivity and CSC is within this range at NS densities \cite{Wambach:1992ik,Rajagopal:2000wf} and so future observations of these events could give insight into their phase structure. With plans to perform more HICs at lower temperatures, it should go without saying that extending the calculations to finite-temperature should be done in future for more realistic and widely applicable results. 

\section{Structure of the thesis, units and conversions}
\label{Intro:sec:structure}

This thesis compiles work done at two extremes of the $\mu_B$-$B$ plane at $T=0$  regarding superconducting phases. The first can be found in Chapter \ref{chpt:Project1}, which concerns dense, three-flavour colour-superconducting matter and its phase structure due to an externally applied magnetic field. Usually, one can treat the quarks as massless at asymptotically high $\mu_B$, but in NSs the strange quark mass $m_s$ becomes comparable in magnitude to $\mu_q$.
We incorporate corrections for $m_s$ and analyse the effect on the phase diagram. The majority of this chapter is published work from \cite{Evans:2020uui}. Where appropriate, details have either been added or removed for the purposes of the thesis. It follows almost the same structure and contains the same figures. Sections (\ref{Project1:sec:setup}), (\ref{Project1:sec:results}), (\ref{Project1:subsec:Hc1}), (\ref{Project1:subsec:Hc2}) and Appendix \ref{app:Asymptotic} have been taken from the paper with some small editions for continuity, language and formatting purposes (including brief introductions for some sections). Sections that are not taken from Ref.\,\cite{Evans:2020uui} are very similar to the corresponding sections in this reference.

Work at the other extreme can be found in Chapter \ref{chpt:Project2}. As shown in \cite{Brauner:2016pko}, when the chiral anomaly is included in ChPT via a Wess-Zumino-Witten (WZW) term, the ground state of QCD above a certain critical field is a Chiral Soliton Lattice (CSL) of neutral pions. Furthermore, this phase is unstable to charged pion fluctuations above a second, larger critical field. Through comparing this instability to that of a conventional type-II superconductor, we argue that the system should transition into a superconducting, charged pion vortex lattice phase. We construct this phase in the massless limit, where the baryon number density is inhomogeneous and is constrained into ``baryon tubes'', resembling a two-dimensional crystal. This work was published in Ref.\,\cite{Evans:2022hwr} and both sections of this chapter (once again, with minor changes for this thesis) are taken from it, including all figures and Appendices \ref{app:Anomalous} and \ref{app:Compute}. Also, large parts of Sec.\,\ref{Background:subsec:GL} and \ref{Background:subsec:Hc2} are taken from Appendix A in Ref.\,\cite{Evans:2022hwr}.

The preceding Chapter \ref{chpt:Background} reviews useful background for the aforementioned chapters. It is split into two halves. The first summarises the current state of the QCD phase diagram in the $\mu_B$-$T$ and $B$-$T$ planes at $B=0$ and $\mu_B=0$ respectively. Here is also provided a review of the literature in the $\mu_B$-$B$ plane at $T=0$ and discussion about how a schematic diagram of this plane may look. The second is devoted to reviewing conventional superconductivity in the GL regime including determining the critical fields of both type-I and type-II superconductors. We also outline the derivation of the Abrikosov vortex lattice in detail. Chapter \ref{chpt:Outlook} gives an overall summary, discussion and outlook for possible extensions, improvements and future applications of our work. 

Our convention for the metric is $g^{\mu\nu}=\rm{diag}\left(1,-1,-1,-1\right)$. Throughout the thesis we use natural units $\hbar=c=k_{B}=1$ and Heaviside-Lorentz units for electromagnetism, in which the electric charge $e=\sqrt{4\pi\alpha}$ where $\alpha$ is the fine-structure constant. In these units, useful unit conversions are $1\,\text{fm}^{-1}=(\hbar^{(\text{SI})}c^{(\text{SI})}/e^{(\text{SI})})10^9\,\text{MeV}\simeq 197.3\,\text{MeV}$ and $1\,\text{K}=(k_B^{(\text{SI})}/e^{(\text{SI})})10^{-6}\,\text{MeV}\simeq 8.617\times10^{-11}\,\text{MeV}$ where the SI superscript denotes the magnitude of the constants in SI units. To convert the magnetic field into Gaussian units, we note $1\, \text{G} = 1\, \text{g}^{1/2}\text {cm}^{-1/2}\text{s}^{-1}$ and thus  $1\, \text{G} = \left(\hbar^{(\text{SI})}c^{(\text{SI})}\right)^{3/2}/[\left(e^{(\text{SI})}\right)^2\sqrt{10}]\, \text{eV}^2\equiv \beta\, \text{eV}^2$ in natural units, where $\beta\simeq 0.06925$. With $\sqrt{4\pi} \, B_{{\rm HL}} = B_{{\rm G}}$, where $B_{{\rm HL}}$ and $B_{{\rm G}}$ are the magnetic fields in the Heaviside-Lorentz and the Gaussian system of electromagnetism respectively, we conclude that $1\, {\rm eV}^2$ in natural Heaviside-Lorentz units corresponds to $\sqrt{4\pi}/\beta\, {\rm G} \simeq 51.189\, {\rm G}$ in the Gaussian system. 

\chapter{Background}
\label{chpt:Background}
\section{The QCD Phase Diagram}
\label{Background:sec:QCD}
Phase diagrams encapsulate our knowledge of how matter and its properties change under different conditions. It shows which phase of matter is energetically preferred at thermodynamic equilibrium as a function of chosen macroscopic, thermodynamic variables. Bounding each phase on the diagram are phase transition which are typically classified as first order, second order, or a crossover. The Ehrenfest classification distinguishes different phase transitions by divergences in the derivatives (of any order) of the free energy with respect to some thermodynamic variable. An $n$-th order phase transition means the $n$-th derivative of the free energy diverges. A crossover is an infinite-order phase transition i.e.\ there are no discontinuities in the derivatives  of the free energy. Note that higher order phase transitions than first and second order are possible within this classification but these will not be relevant for our discussion. Another way of distinguishing these transitions is by which symmetries are obeyed in which phase. While first and second-order transitions involve a change in symmetry of the system, a crossover does not. The change in a specific symmetry can be measured by an associated order parameter which is finite in one phase but not in another, indicating when the symmetry is broken or restored.

The classic example of a phase diagram is the temperature-pressure phase diagram for the compound H\textsubscript{2}O, marking at which temperatures and pressures transitions occur between its solid (ice), liquid (water) and gas (water vapour) phases. Phase diagrams are not limited however to compounds and materials, nor to solid, liquid or gaseous phases either. There are conditions under which the electromagnetic forces, which govern the everyday world of molecules and atoms, can be overcome. Plasma phases, for instance, are formed at temperatures high enough for electrons to become unbound, the kinetic energy imparted from thermal collisions allowing them to escape the electromagnetic potential of nuclei. At even greater extremes, it is hypothesised that the nuclear forces that hold nuclei together and even the strong force that confines quarks to nucleons can be overcome to form new states of matter. These states of matter would then belong in the QCD phase diagram, which would contain our knowledge of matter under the most extreme conditions currently imaginable. 

Given that there are a number of different variables one can consider, in principle the full QCD phase diagram is a multi-dimensional plot. Clearly, it is not feasible to visualise the whole diagram. While the entire diagram is of interest from the perspective of QCD being a fundamental theory of nature, more focus is given to the parameters and planes which pertain to additional scientific questions. The QGP phase is thought to have been the ground state of matter in the early stages of the Universe after the Big Bang \cite{Boyanovsky:2006bf} where temperatures and densities would have been extremely high. Investigating this primordial matter is one of the motivations behind HIC experiments which also involve large magnetic fields and angular momenta. Furthermore, the ground state of matter at the centre of NSs is still unknown. High baryon densities, magnetic fields and angular motion are conditions which are also expected in compact stars as well as high temperatures if one includes mergers. These physical scenarios motivate study of matter under these extreme conditions which feature on the QCD phase diagram. Therefore, the parameters most commonly considered are temperature ($T$), particle density via baryon chemical potential ($\mu_B$) and isospin chemical potential ($\mu_I$), magnetic field ($B$), and angular velocity ($\nu$).

It should be noted that even the most experimentally accessible plane is largely schematic, meaning that the structure of the QCD phase diagram is mainly a conjecture at present. Furthermore, many constructs are considered qualitative, as some lines have been extrapolated beyond the regions of validity of the methods used to calculate them. This chapter summarises the current progress on the phase structure in three planes from the parameter space spanned by $T$, $B$ and $\mu_B$. With the exception of Sec.\,(\ref{Background:subsec:exp}), each section explores the two-dimensional phase diagram of two of these variables with the remaining kept at zero. Beginning from the most well known to the least, we will discuss the $\mu_B$-$T$ plane at $B=0$, $B$-$T$ plane at $\mu_B=0$ and $\mu_B$-$B$ plane at $T=0$ in Sec.\,(\ref{Background:subsec:muTPlane}), (\ref{Background:subsec:BTPlane}) and (\ref{Background:subsec:muBPlane}) respectively. Due to the relative lack of experimental input, the focus will be mainly on theoretical results from lattice QCD and other numerical and analytical calculations. However, we do provide some experimental and observational discussion, including data from colliders and multi-messenger astrophysics, primarily in Sec.\,(\ref{Background:subsec:exp}).

Before we discuss these planes, there are a few concepts which will be relevant to more than one which are worth mentioning before we continue. A key prediction of QCD is that the strong force between quarks becomes large at long distances/low energy, but is small at short distances/high energy. To demonstrate this, we can look at the potential energy $V(r)$ of a quark in the presence of an anti-quark at distance $r$ away. At large distances,
\begin{equation} V(r) \propto r \,,\end{equation}
i.e.\ it grows linearly with $r$, which means the more the quark and anti-quark are separated, the more energy is required to continue to separate them. To move the quark an infinite distance away, i.e.\ for it to be a  free particle, would require an infinite amount of energy. Due to this, quarks are considered to be bound at large distances. In the opposite limit, the potential has the form
\begin{equation}    V(r) \sim  -\frac{\alpha_s}{r}\,,\end{equation}
where $\alpha_s\propto g^2$ is the dimensionless strong coupling of QCD. As written, it looks similar to the attractive potential of the Coulomb force. However, $\alpha_s$ also depends on the energy scale and therefore on $r$. At very small $r$, $\alpha_s\sim-1/\ln{r}$, weakening the potential. It is therefore implied that quarks become non-interacting at small distances and making the quarks essentially non-interacting. This is a consequence of asymptotic freedom. From all of this we gather that quarks cannot be separated from one another and are ``confined'' as the constituents of hadrons at low energies, and they are non-interacting and ``deconfined'' from hadrons at high energies. Therefore, it is expected that there should be a ``deconfinement transition'' present in the QCD phase diagram. For $T$ (with all other thermodynamic variables set to zero), the transition is expected when $T\sim \Lambda_{\rm{QCD}}\sim 100\,\rm{MeV}$  where $\Lambda_{\rm{QCD}}$ is the strong coupling scale, the characteristic energy scale of QCD. Currently, there is no known order parameter which perfectly describes this transition. The Polyakov Loop can be used as an approximate order parameter since it can be related to the free energy of a single quark $F_q$ by 
\begin{equation}
    \langle \Omega(x)\rangle = e^{-F_q/T}\,,
    \label{PolyakovLoop}
\end{equation}
where $\langle \Omega (x) \rangle$ is the expectation value of the Polyakov loop  $\Omega (x)$ \cite{Ukawa:1995tc}. As mentioned, it takes an infinite amount of energy to separate quarks when they are confined. Thus, for a quark to exist in isolation (i.e.\ to be a free particle) in the confined phase it would need an infinite amount of energy. This implies $F_q=\infty$ in the confined phase and so $\langle \Omega(x)\rangle=0$. It becomes non-zero in the deconfined phase when quarks become non-interacting and $F_q$ is finite. As an order parameter, $\langle \Omega (x) \rangle$ indicates whether the system is invariant to transformation by the centre of the  $SU(3)$ colour group of QCD which is isomorphic to the $\mathbb{Z}_3$ group. In the confined phase, $\Omega(x)$ is invariant under transformation by elements of the group $\mathbb{Z}_3$ and it is therefore a symmetry of the system. In the deconfined phase, this symmetry is spontaneously broken and $\langle \Omega (x) \rangle\neq 0$. Unfortunately, relation \eqref{PolyakovLoop} only holds in the infinite quark mass limit i.e.\ for static quarks. This means it only fits the role of order parameter for the deconfinement transition exactly in pure gauge theory (or ``glue'' where only gluons interact). However, it can still be useful for insight into where the deconfinement transition approximately occurs in the QCD phase diagram.

A similar concept is the chiral transition or chiral restoration. With the two lightest quark flavours\footnote{One can in principle work with more flavours but the following approximation works best with only the two lightest, as will be alluded to.}, $u$ and $d$, QCD has an approximate global symmetry, $SU(2)_L \times SU(2)_R$  called chiral symmetry, where $SU(N_f)_L$ and $SU(N_f)_R$ are the special unitary group of transformations on left-handed and right-handed spinors ($\psi$) for the number of flavours $N_f$ respectively. With massive quarks, it is explicitly broken. Since the up ($u$) and down ($d$) quarks have comparatively small masses at the $\mathrm{GeV}$ scale, we can approximate them as massless at these scales. In this limit, the symmetry becomes exact and is now spontaneously broken by the QCD vacuum. This is referred to as chiral symmetry breaking, which follows the symmetry breaking pattern
\begin{equation} 
SU(2)_L \times SU(2)_R \rightarrow SU(2)_{V}\,,
\label{SSBchiral}
\end{equation}
where $SU(N_f)_{V}$ is the diagonal subgroup of $SU(N_f)_L \times SU(N_f)_R$. Associated with this broken symmetry is the chiral condensate $\langle\bar{\psi}\psi\rangle$. Chiral symmetry breaking is responsible for the existence of the pseudo-scalar mesons (like the pions) and for the masses of baryons being much higher than the sum of the bare masses of their constituent quarks. We can think of quarks as having an effective mass much larger than their bare mass when they are constituents of hadrons due to chiral symmetry breaking. When we reach higher energies, the vacuum that spontaneously breaks the symmetry is no longer relevant and chiral symmetry would be restored. The effective mass of the quarks then drops to their bare masses (or in the approximation, massless). This can be identified with a phase transition where the chiral condensate goes to zero. For this reason, $\langle\bar{\psi}\psi\rangle$ is often used as an order parameter for this transition. However, just as with the Polyakov loop, it is only an approximate order parameter. Since the chiral symmetry is not exact in the physical case, the condensate does not go to zero exactly at the transition. It is only an order parameter in the massless quark or chiral limit (the opposite limit for which the Poylakov loop is an order parameter). The chiral condensate is not the only way chiral symmetry can be broken. This can also be achieved by diquark condensates $\langle \psi\psi\rangle$ of some colour superconductors which we will encounter later.

While confinement is the restriction of quarks to being only the constituents of hadrons at low energies, chiral symmetry breaking is responsible for the masses of these hadrons at these low energies. It can be tempting to assert that deconfinement and chiral restoration are related or even that one necessitates the other \cite{Banks:1979yr}. Indeed, in the planes we shall discuss the transitions are often shown or conjectured to occur in very close proximity. An aside, when one takes the limit $N_c\rightarrow\infty$ where $N_c$ is the number of colours, there is a phase which could differentiate the deconfinement and chiral restoration transitions, existing in between them. This is a nuclear matter phase where the pressure is proportional to $N_c$ \cite{McLerran:2007qj}. Proportionality of the pressure to $N_c$ is usually attributed to quark matter phases, which is unusual given that the quarks are confined and therefore the phase (everywhere) is colour neutral. For this reason, it is sometimes called Quarkyonic Matter (see Ref. \cite{McLerran:2020rnw} for an introduction to this phase). It is proposed that the phase consists of a Fermi sea of quarks with a thin shell of baryon states at the surface, such that only the baryon states are accessible but the bulk properties, like the pressure, are determined primarily by the quarks. While difficult to picture physically, the coexistence of the quark and baryon states does leave possibility for matter to be chirally restored but still confined. Whether chiral symmetry remains broken in this phase is ambiguous (see discussions in Refs.\,\cite{McLerran:2008ua,Kojo:2011cn,Glozman:2011yz} and references therein). However, it remains to be seen whether any of the results at $N_c=\infty$ are relevant to the QCD phase diagram where $N_c=3$. For this reason, we shall not discuss this phase in much further detail but note that it is occasionally included in phase diagrams as an intermediate phase between the deconfinement and chiral transition.

\subsection{The $\mu_B$-$T$ plane at $B=0$}
\label{Background:subsec:muTPlane}

\begin{figure}
    \centering
    \includegraphics[width=0.75\textwidth]{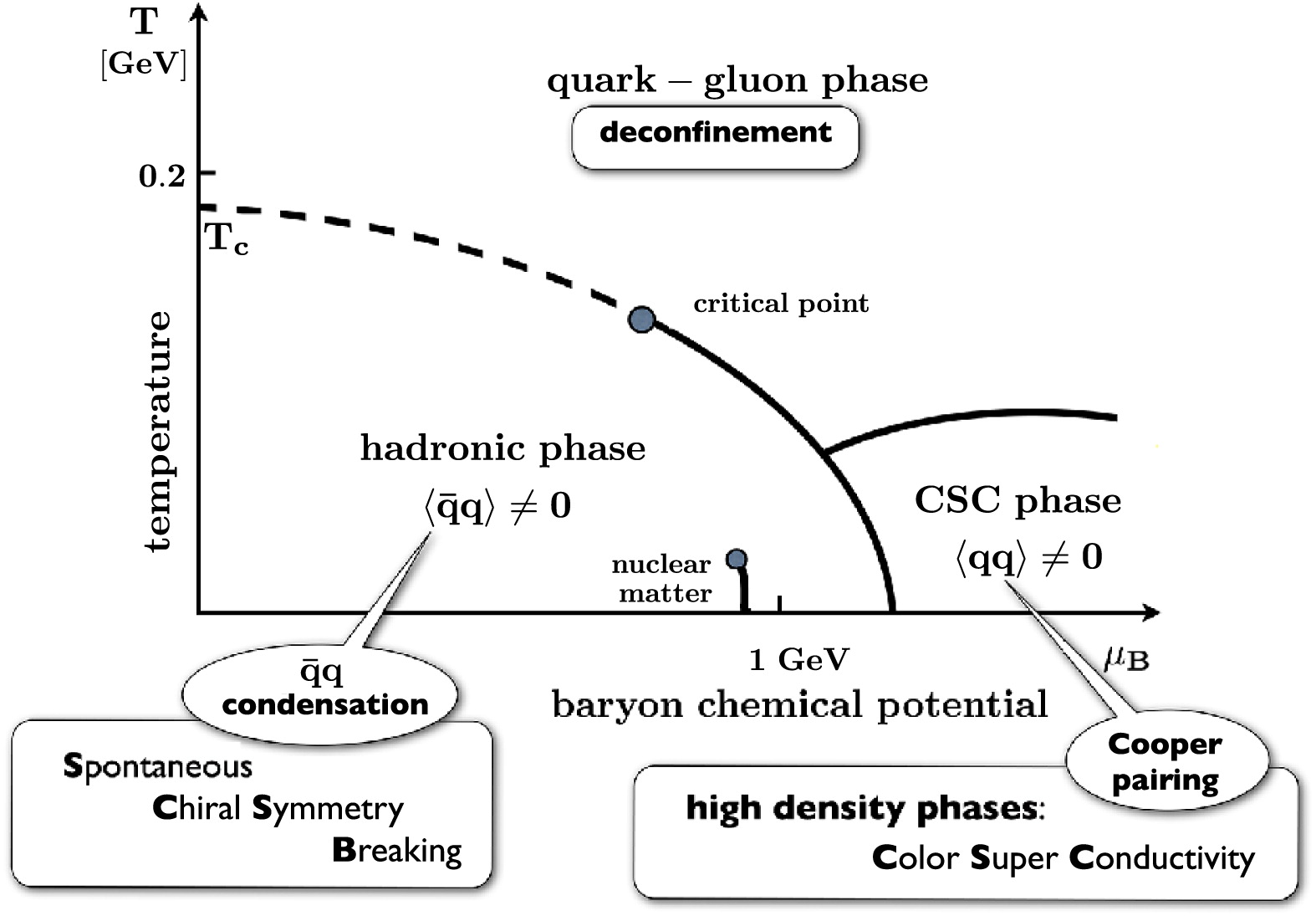}
    \caption{A schematic $\mu_B$-$T$ plane at $B=0$. Solid lines indicate first-order phase transitions, dotted lines denote crossovers. Here, $\langle\bar{q}q\rangle$ and $\langle qq\rangle$ are the chiral and diquark condensates ($\langle\bar{\psi}\psi\rangle$ and $\langle \psi\psi\rangle$ respectively in the text) associated with chiral symmetry breaking and colour-superconductivity respectively. In the text, we discuss the pseudo-critical temperature $T_{pc}$ rather than $T_c$. Taken from \cite{Kawagoe:2012iky}.}
    \label{fig:TmuPlane}
\end{figure}

By far the most active research is in the $\mu_B$-$T$ plane (see Fig.\,\ref{fig:TmuPlane}). Indeed, it is often referred to as \textit{the} QCD phase diagram. This is due to it overlapping the conditions found in NSs and their mergers and the early Universe, which makes it relevant to outstanding questions in high-energy physics, astrophysics and cosmology. The effort to answer these questions translates to greater experimental and theoretical interest. Observational data from gravitational and electromagnetic waves produced by NSs and NS mergers (NS-NS and NS-black hole) and collider experiments provide verification and constraints for theoretical predictions as well as insight where theoretical models are less successful. As a result, this plane boasts a large number of (possible) features. 

To give an overview, the plane at $B=0$ can be split into roughly three regions, corresponding to the confined phases, the QGP phase and low temperature, high density phases where we expect CSC to emerge. The confined phase is the natural place to start. As the name suggests, it comprises hadronic matter and is bounded by a deconfinement/chiral transition, taking up the lower left-hand region of Fig.\,\ref{fig:TmuPlane}. Matter in this phase consists of the vacuum, mesons and baryons. At low $\mu_B$ and $T$ we are unlikely to find any hadrons as thermal excitation is exponentially suppressed, and at $T=0$ we expect no excitations at all. Instead we have only the QCD vacuum, where quark and gluon condensates are present \cite{Savvidy:1977as,Shifman:1978bx,Ioffe:2005ym}, including the chiral condensate already discussed in the context of chiral symmetry breaking. 

Mesons are insensitive to $\mu_B$, and therefore become more abundant when we increase $T$. The upper left area of this region can thus be described as a meson gas. The lightest mesons are the pions with mass $m_{\pi}\approx 140\,\rm{MeV}$,  such that this gas will (mainly) be formed of pions. The lightest baryons on the other hand are the nucleons with mass $m_N = 939 \,\rm{MeV}$, thus it is only near and above this value their presence is expected. At low temperatures, this manifests as a liquid-gas phase transition which can be seen in Fig.\,\ref{fig:TmuPlane}. The transition line extends from the point\footnote{As a point of interest, matter in the everyday world in which we live is located at this point (room temperature is $\approx 300\,\rm{K}\sim 10^{-10}\,\rm{MeV}\approx 0\,\rm{MeV}$) in a mixed phase of liquid nuclear matter and the vacuum.} $\mu_B = 923 \,\rm{MeV}$, $T=0$ ending in a critical point beyond which it is a crossover. We expect the critical point to be located at $T\approx 16$-$20\,\rm{MeV}$ from both theoretical calculations and results from experiments (see reviews \cite{Chomaz2002TheNL,Baldo_2012,BORDERIE201982} and references therein). Note the transition at $T=0$ is not exactly at $\mu_B=m_N$ because of the binding energy of nucleons.

The shape of the liquid-gas phase transition is echoed by the deconfinement/chiral transition that encloses the hadronic part of the plane. It too is expected to be first-order up to some critical point where it becomes a crossover. Our best knowledge of this curve is of the crossover portion i.e. at low $\mu_B$ where we can use lattice calculations - numerical simulations of QCD where the path integral is evaluated on a discretised spacetime grid or lattice. Elsewhere where $\mu_B$ is appreciably finite, they are plagued by the infamous sign problem. Briefly, in lattice QCD one usually deals with a partition function $\mathcal{Z}$ of the form
\begin{equation}
    \mathcal{Z}=\int \mathcal{D}\phi\, \mathrm{det}[M(\mu_B)] e^{-S_\phi[\phi]}\,,
\end{equation}
where $\mathcal{D}\phi$ is the sum over all path configurations of the gluon field $\phi$ and $S_\phi$ is the remaining part of the action pertaining to the gluons. The quark part of the path integral has been done analytically, leading to a factor of the determinant of the operator $M$. The sign problem occurs because including a real-valued $\mu_B$ leads to $\mathrm{det}[M(\mu_B)]$ being complex, which means it cannot be evaluated via the usual numerical method of Monte Carlo importance sampling used in lattice simulations. For more on this issue (and approaches to circumvent it) see review \cite{deForcrand:2009zkb}.

Lattice calculations predict that the transition from hadronic matter to QGP is a crossover. This is shown in Fig.\,1 of Ref.\,\cite{Aoki:2006we} by a peak in the chiral susceptibility. These peaks do not scale with the system volume, which indicates that it is a crossover. In this case the order parameter is the chiral condensate and so the transition is a chiral transition. As mentioned, this order parameter is not exact as the chiral symmetry of QCD is not. The associated temperature where the chiral condensate almost vanishes is then usually referred to as the pseudo-critical temperature, $T_{pc}$. Due to the non-exactness of this transition, one can measure $T_{pc}$ in variety of ways. In the fairly recent Ref.\,\cite{HotQCD:2018pds} (see Fig.\,2 and 4), they computed five different critical temperatures from different criteria and took the average to give $T_{pc}=156.5 \pm 1.5 \,\rm{MeV}$ at $\mu_B=0$. They also extended their calculations by Taylor expansion up to $\mu_B =300 \,\rm{MeV} $ and found that $T_{pc}\sim 150 \,\rm{MeV}$ over the range. Simulations of the Polyakov loop seem to be in agreement (see \cite{Borsanyi:2010bp} and Fig.\,7), with $T_{pc}$ expected in the region of $150\,\rm{MeV}$ where the expectation value begins appreciably changing.

\begin{figure}
    \centering
    \includegraphics[width=0.5\textwidth]{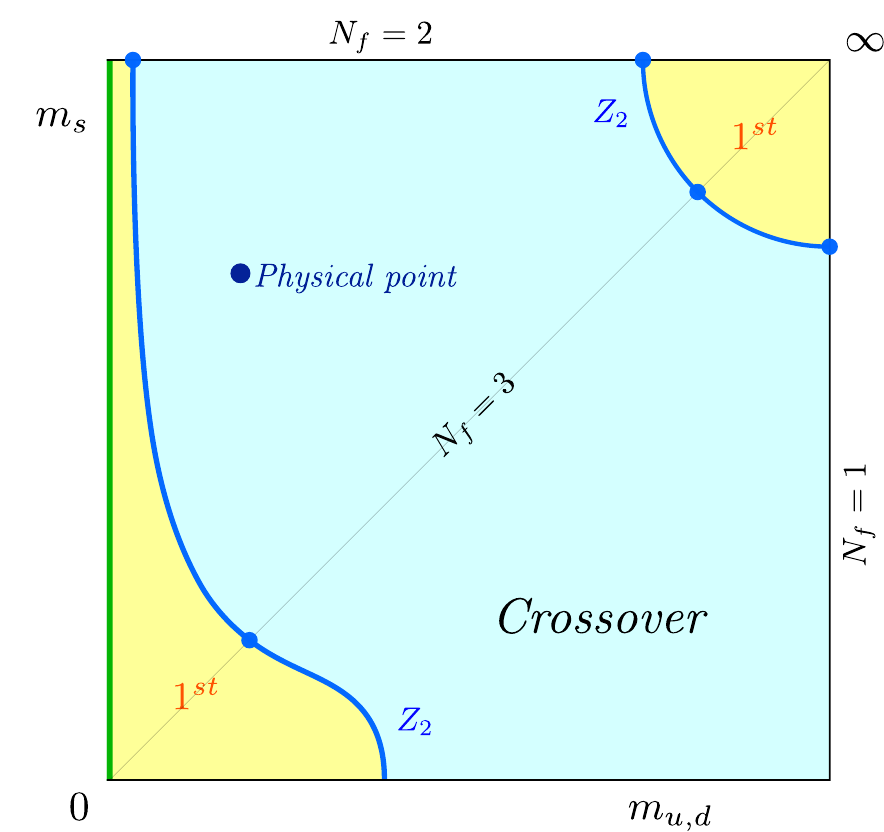}
    \caption{Columbia plot determined by lattice calculations. On the horizontal axis is the mass of the $u$ and $d$ quark, and the strange quark mass on the vertical axis. Plotted are the critical quark masses over which the order of the chiral transition changes, separating the plane into regions where it is a crossover or first-order. $Z_2$ ($\mathbb{Z}_2$) denotes the universality class of the critical masses. Shown also is how the order changes with $N_f$ quark flavours but with equal quark masses. Taken from Ref.\,\cite{Cuteri:2017gci}.}
    \label{fig:Columbia}
\end{figure}

All the results discussed were done for ``physical'' quark masses. In the best case this means using the bare quark mass values measured by experiment $m_f$,  for each of the three lightest quark flavour $f=u,d,s$. The bare masses of the remaining three quarks are too large to be relevant at this energy scale. In addition to the $u$ and $d$ quarks mentioned earlier, we now have the strange (s) quark flavour. It is not always feasible to match the physical case however, and often one approximates $m_u=m_d\equiv m_{u,d}\lesssim m_s/10$ i.e.\  the up and down quarks are taken to be equal to a set mass $m_{u,d}$ which is an order of magnitude smaller than the separate strange quark mass. This approximation is not far from the physical case when we consider that $m_{u}/T_{pc}\approx\ m_{d}/T_{pc}\sim 10^{-2}$. These considerations are important, since the order of the phase transition may depend on the quark masses we choose. This may not be unexpected given that the breaking/restoring of the symmetries which our order parameters measure are only exact in certain limits of the quark masses. Such a dependence is usually demonstrated via a Columbia plot \cite{Brown:1990by}, of which Fig.\,\ref{fig:Columbia} is one. The deconfinement and chiral transition become first-order in the limits $m_f\rightarrow\infty$ \cite{Gottlieb:1985ug,Boyd:1996bx} and $m_f\rightarrow 0$ \cite{Pisarski:1983ms,Gavai:1987dk, Kogut:1987ah} respectively. It is only in between these limits they both become crossovers. (For an idea about the effect of the quark masses on the wider phase diagram, see Ref.\,\cite{Rajagopal:1999cp}.)

Using lattice QCD we have obtained much insight into the deconfinement/chiral transition at low $\mu_B$. In Fig.\,\ref{fig:TmuPlane}, this transition is shown to exist all the way down to $T=0$ and $\mu_B>1\,\rm{GeV}$. Thus far we have only discussed results pertaining to the transition from lattice QCD, omitting results from effective theories and models. Given that lattice simulations are predictions of QCD, these results have been given precedence over less fundamental approaches. Several reviews referenced at the end of this section discuss the results from effective theories and models for those interested. However, at larger $\mu_B$, the sign problem rules out using the lattice, and we must rely on effective theories and models to gain insight into the remaining pieces of the  phase transition. The next best thing to the QFT is an Effective Field Theory. For QCD at low energies, this is ChPT (more on this in Chapter\,\ref{chpt:Project2}). It is therefore the obvious choice for studying the confined phases. Its usefulness is limited when discussing the chiral transition however. As it is built around chiral symmetry breaking, it becomes less and less valid as we approach chiral restoration. This translates into including more and more higher order terms to the leading order momentum expansion (which breaks down at $\sim 1\,\rm{GeV}$) to yield relevant results. See the recent review \cite{GomezNicola:2020yhm} for more on the role of effective theories regarding the transition. 

\begin{figure}
    \centering
    \includegraphics[width=0.75\textwidth]{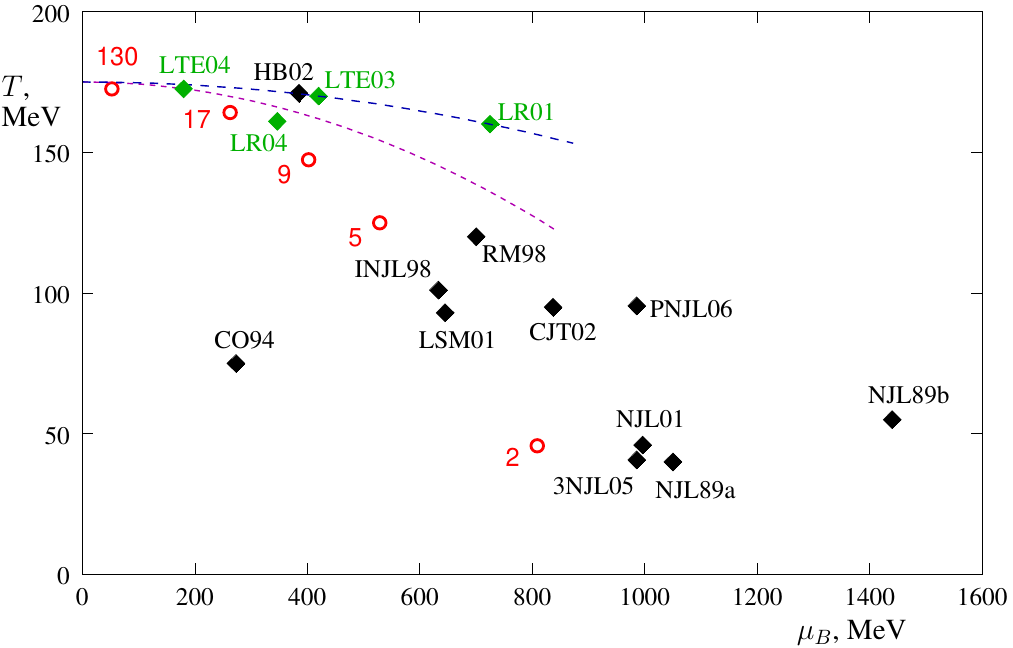}
    \caption{Location of the critical point in the $\mu_B$-$T$ plane at $B=0$ according to different models, lattice QCD and HICs. The black dots are model predictions, including some referenced in the text. Green points and dashed lines are predictions and extrapolations respectively from lattice QCD. The red circles are the location of chemical freeze-out points from HICs. Taken from Ref.\,\cite{Stephanov:2006zvm}, see there for more information and a full list of references for each data point and curve in the plot. }
    \label{fig:ModelFirstOrder}
\end{figure}

This leaves theoretical models as our main source of knowledge for the remainder of the phase transition from the theoretical side. In the opposite limit near $T=0$, many different approaches indicate the chiral transition is first-order \cite{Stephanov:2004xs}. This was first conjectured in Ref.\,\cite{Pisarski:1983ms} using a linear sigma model \cite{Gell-Mann:1960mvl} and several other models have been used to demonstrate this since, including Nambu-Jona-Lasinio (NJL) \cite{Nambu:1961tp} models \cite{Asakawa:1989bq, Berges:1998rc,Scavenius:2000qd}, Non-linear sigma  models \cite{Scavenius:2000qd}, Random Matrix models \cite{Halasz:1998qr} and Composite Operator Formalisms \cite{Barducci:1993bh} to name a few. (Note that the references given are not exhaustive. See Refs.\,\cite{Stephanov:2004xs,Stephanov:2006zvm} for better lists.) Some predict that this is a transition to colour-superconducting or other high density phases as shown in Fig.\,\ref{fig:TmuPlane} rather than the QGP phase like other sketches of the $\mu_B$-$T$ plane. In between the $\mu_B\approx 0$ crossover region predicted by lattice QCD and the $T\approx 0$ first-order transition predicted by models the nature of the chiral transition must therefore change, indicating the existence of a critical point, as shown in Fig.\,\ref{fig:TmuPlane}. It is therefore a similar situation to that of the liquid-gas phase transition. The question now becomes where is this critical point located along the transition line, which can be tackled from either end of the curve. While models generally agree that the transition is first-order at low temperatures, their predictions for the position of the critical point varies. This is well demonstrated by the (slightly dated) Fig.\,\ref{fig:ModelFirstOrder} from Ref.\,\cite{Stephanov:2006zvm}. Also, this figure includes data from HICs and estimated phase transition curves from lattice QCD (like in Ref.\,\cite{HotQCD:2018pds}),  
which have recently provided more accurate constraints on its position. To obtain these curves requires working around the sign problem (again see review in Ref.\,\cite{deForcrand:2009zkb}). There are a few ways of doing this. Examples include re-weighting, Taylor expanding around $\mu_B\approx 0$, and extrapolating from imaginary $\mu_B$. A review of lattice constraints on the chiral phase transitions is given in Ref.\,\cite{Philipsen:2021qji}, with focus on constraining the location of the critical point using the two latter methods. Both involve expansions in $\mu_B/T$ and one can calculate expansions of the pressure in this parameter. It is implied that the critical point can be treated as an upper bound on the radius of convergence of these expansions since the phase transition becomes non-analytic at first-order i.e.\ it is no longer a crossover. Therefore, one can calculate the radius of convergence to check whether there is a critical point in the vicinity. From this method and the lattice data from different sources, Ref.\,\cite{Philipsen:2021qji} estimates the bound on the critical point to be $\mu_B\gtrsim 3T$. If we take the critical point to be at $T\geq 125\,\rm{MeV}$ this would put it in the region $\mu_B \gtrsim 375 \,\rm{MeV}$. This agrees with most of the theoretical models referenced. It should be noted however that estimates for the critical point are far from conclusive \cite{Schmidt:2017bjt}. 

Exploring the deconfinement transition and critical point location at finite $\mu_B$ has been less well explored as static quarks are a less accurate approximation than massless quarks. Nevertheless, there have been some attempts from models and the lattice (see Ref.\,\cite{Fischer:2013eca,Fukushima:2007fc} for examples). A particularly interesting model that bares mentioning is the Polyakov loop NJL (PNJL) model \cite{Meisinger:1995ih, Fukushima:2003fw, Megias:2004hj} which incorporates both the chiral condensate and Polyakov loop in one framework, enabling simultaneous analysis of the chiral and deconfinement transitions. There is another aspect of these transitions which is not shown in Fig.\,\ref{fig:TmuPlane}, which is the quark-hadron continuity. It is possible that at low temperatures that there is no transition between nuclear matter and deconfined quark matter, resulting in another critical point. Our brief discussion of this topic is best left until after the discussion of the deconfined phases.

Beyond the deconfinement and chiral transition, the phase diagram is broken up into the QGP and CSC regions. Taking up the high temperature region of the $\mu_B$-$T$ plane of Fig.\,\ref{fig:TmuPlane} is the QGP phase \footnote{Another point of interest, according to the Big Bang theory, matter in the early Universe is located in the upper left corner of the $\mu_B$-$T$ plane. Therefore, all matter in some of the earliest moments of the Universe was in the QGP phase (see Ref.\,\cite{Boyanovsky:2006bf} and references therein for more on this topic).}. Sometimes it is referred to as a quark ``soup'', where individual quarks and gluons are the degrees of freedom. It is a plasma in the sense that there can be a net colour charge locally unlike the confined phase. To probe the properties of this phase, lattice QCD can still be used for a segment of this region. In fact, because the aforementioned methods used to extend lattice calculations to finite $\mu_B$ rely on $\mu_B/T<1$, the region gets larger as we move to higher $T$ and $\mu_B$. Essentially, one can draw a diagonal along $T=\mu_B$ in Fig.\,\ref{fig:TmuPlane} and the upper half region would in principle be accessible by lattice calculations where they could at least be useful qualitatively \cite{Philipsen:2021qji}. When we go to $\mu_B/T> 1$ however, we must rely on other tools. Thankfully, the weak coupling approximation becomes valid away from the deconfinement transition due to asymptotic freedom, which means we can perform analytical calculations to make predictions using pQCD in thermal field theory. 

Introducing pQCD into thermal field theory is not entirely straightforward. In a direct, naive application, infrared (IR) divergences (low energy/long distance contributions to momentum integrals causing divergences) appear due to contributions from interactions between quarks where only a small amount of momentum (low energy gluons) are exchanged e.g.\ small angle scattering. These need to be appropriately dealt with if thermal pQCD is to be used and is achieved through a process known as resummation. The two common schemes for this process are Hard Thermal Loop perturbation theory (HTLpt) and Dimensional Reduction (DR) \cite{Ghiglieri:2020dpq}. In addition, there are two formalisms one can adopt; real-time and imaginary time. Both are equivalent but are better suited in different contexts. The former can be used for out of equilibrium calculations and is therefore better for observables in HIC experiments. In the latter, thermal equilibrium is assumed from the beginning which makes it easier to use for determining bulk thermodynamic properties. Two important thermodynamic properties - the pressure and trace anomaly - derived in this formalism show good agreement with lattice QCD at $\mu_B=0$ all the way down to $T=200\,\rm{MeV}$. There is even agreement with lattice results at finite $\mu_B$ down to $T=300\,\rm{MeV}$ which lends to the possibility that thermal pCQD can be extended to all but the lowest temperatures \cite{Kurkela:2016was}. This is promising, especially since thermal pQCD can in principle then be used to explore the extremes of the $\mu_B$-$T$ plane at $B=0$ i.e. everything but the hadronic phases. Calculations have thus far ignored the possibility of pairing and therefore CSC. As such, thermal pQCD calculations of this kind do not tell us much about the locations and orders of phase transitions in this plane. For more details on pQCD in thermal field theory and the aforementioned results, see Ref.\,\cite{Ghiglieri:2020dpq}. 

At low $T$ and high $\mu_B$ we expect to enter the CSC region displayed in the bottom right corner in Fig.\,\ref{fig:TmuPlane}. Weak coupling still applies for asymptotically large $\mu_B$ and we can continue to use HTLpt and DR if we so desired to analyse deconfined quark matter in these conditions, with quarks and gluons as the only degrees of freedom. However, in this region the possibility of Cooper pairing must be taken into account which is a non-perturbative phenomenon, making the application of pQCD less straightforward. Quarks are expected to pair and form diquark condensates in this regime. These need not be electrically neutral nor colour neutral and could in general be both superfluid and superconducting. The Meissner effect could then apply to chromo-magnetic fields leading to CSC, pioneered in Refs.\,\cite{Barrois:1977xd,Frautschi:1978rz,Bailin:1983bm}. 

In the CSC region, we yet again consider the three lightest quarks. Multiple different quark flavours leads to different Cooper pairing patterns which means there are several possible colour-superconducting phases. This makes the structure of this region of the plane rich but also difficult to determine due to the large number of possible transitions that could take place between different colour-superconducting phases as we vary $\mu_B$ and $T$. What can be said with a high degree of confidence is that the colour-flavour locked (CFL) \cite{Alford:1998mk} phase is the ground state of QCD at (very) high $\mu_B$. At these high densities, $\mu_B$ is much larger than the effective masses of the quarks $M_f$, which typically depend on $\mu_B$. (Deconfined quarks might imply that we can use the bare quark masses $m_f$, but since there are still interactions i.e. Cooper pairing, we should technically still discuss the effective mass $M_f$.) The strange quark mass being the largest, this translates to $M_s\ll\mu_B$, such that the $u$,$d$ and $s$ quarks can be treated as massless and are only differentiated by their flavour and charge. The CFL phase has the largest condensation energy and thus the lowest free energy. This can be appreciated due to it retaining a high degree of symmetry. Colour-flavour locked pairing  spontaneously breaks the symmetry of massless, three-flavour quark matter according to the pattern
\begin{equation}
    SU(3)_c \times SU(3)_L \times SU(3)_R \times U(1)_B \rightarrow SU(3)_{c+L+R} \times \mathbb{Z}_2\,.
    \label{SSBCFL}
\end{equation}
Here, the global baryon number symmetry of QCD represented by the unitary group $U(1)_B$ is broken, as are the colour (represented by $SU(3)_c$) and flavour (represented by $SU(3)_L\times SU(3)_R$) symmetries. However, the phase is still invariant under certain simultaneous rotations in colour and flavour space, leaving the global $SU(3)_{c+L+R}$ intact. This gives the phase its name, as the colour and flavour rotations are ``locked'' together. Note, that the breaking of the individual $L$ and $R$ groups is the same as in pattern \eqref{SSBchiral}. The CFL thus also breaks chiral symmetry. It should also be noted that the $U(1)_Q$ symmetry associated with electromagnetism hidden in the $SU(3)_L\times SU(3)_R$ becomes modified to $U(1)_{\tilde{Q}}$ in the $SU(3)_{c+L+R}$. The modified symmetry manifests as a mixing between the photon and gluons and corresponds to an invariance to a simultaneous electric and colour charge rotation. As a result, seven gluons and one photon-gluon admixture are Meissner screened, while the remaining photon-gluon is unaffected. Therefore, it is a colour and electromagnetic superconductor in principle, though the very small mixing angle ($e>>g$ at weak-coupling) between the photon and gluons means that the expulsion of magnetic fields is very weak. It is usually not classified as an electromagnetic superconductor for this reason (rather, it is a transparent insulator \cite{Manuel:2001mx}). The CFL phase is also a superfluid due to the spontaneously broken baryon number symmetry. In its place is the $\mathbb{Z}_2$ group structure under which the phase is invariant under transformations of the quark fields of the kind $\psi\rightarrow -\psi$. 

The critical temperature $T_{c}$ of the CFL phase can be calculated in weak-coupling, yielding
\begin{equation}
    \frac{T_c}{\Delta}= 2^{1/3}\frac{e^{\gamma}}{\pi} \,,
    \label{TcCFL}
\end{equation}
where $\Delta$ is the zero-temperature pairing gap of the CFL phase and $\gamma \simeq 0.577$ is the Euler-Mascheroni constant. This only differs from the standard BCS result \cite{Bardeen:1957mv} by a factor of $2^{1/3}$ \cite{Schafer:1999fe,Brown:1999yd,Brown:1999aq,Pisarski:1999bf,Pisarski:1999tv,Schmitt:2002sc}. All we need is a value for $\Delta$, which behaves like $\Delta/\mu_q\propto \exp{[-1/g(\mu_q)]}$ (where $g(\mu_q)$ is the coupling as a function of $\mu_q$) at asymptotically large $\mu_B$ and can be calculated in this limit. We can therefore sketch this transition line in Fig.\,\ref{fig:TmuPlane} at large $\mu_B$ between CFL and the QGP. Calculation of $\Delta$ is done in QCD in the weak-coupling regime. The IR divergences of the momentum integrals are not present in this calculation. This is due to the medium effects of Landau damping and Debye screening which cut off the long-range/low energy gluon interactions between quarks \cite{Alford:2007xm, Son:1998uk,Pisarski:1998nh}. As a result, $\Delta$ and therefore the CFL phase are predictions of QCD. Then, by extension the extremely high $\mu_B$ region and low-$T$ region of the QCD phase diagram is arguably the best understood theoretically. It is perhaps then a frustrating yet typical contrast that this regime is one of if not the most inaccessible to experiment and observations. 

\begin{figure}
    \centering
    \includegraphics[width=0.75\textwidth]{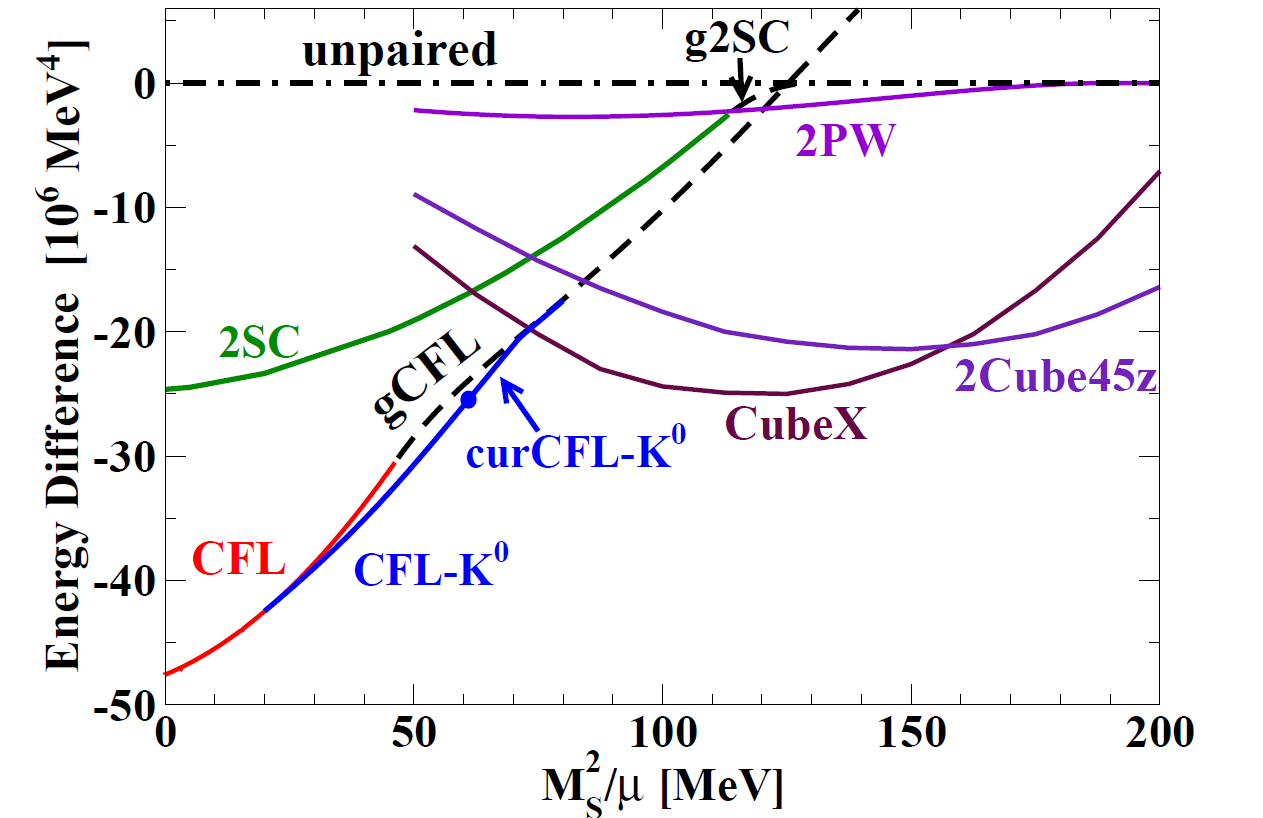}
    \caption{Free energy of different colour superconducting phases with $\Delta=25\,\rm{MeV}$ as a function of $M_s^2/\mu_q$. The homogeneous phases considered are CFL, 2SC, gapless CFL and 2SC (gCFL and g2SC respectively) and the kaon-condensed CFL (CFL-$K^0$). The other phases are CFL-$K^0$ with meson supercurrents (curCFL-$K^0$) which is anisotropic
    and the crystalline phases 2PW, CubeX and 2Cube45z which are inhomogeneous. Also labelled is the free energy of the unpaired quark phase. Taken from \cite{Alford:2007xm}.}
    \label{fig:CSCphases}
\end{figure}

The problem of course comes when we move out of this limit, where the weak-coupling approximation cannot be applied. Indeed, it is estimated that the weak coupling results are only valid for $\mu_q\gtrsim 10^8 \rm{MeV}$ which is alarmingly large. To see whether CFL is preferred at lower densities, extrapolations from high $\mu_B$ calculations and from effective models at low $\mu_B$ can be compared. They show some agreement and suggest $\Delta\sim 20-100\,\rm{MeV}$ \cite{Rajagopal:2000wf} for $\mu_q\approx 400-500\,\rm{MeV}$. While this should not be taken too seriously, the agreement is reassuring. These considerations imply the CFL phase may survive to much lower $\mu_B$ realistic for NSs. However, the estimated range of values for $\Delta$ at these intermediate densities is large enough to admit the possibility of other phases being preferred over CFL. CFL is preferred at high $\mu_B$ due to how large $\Delta$ is, since a large gap corresponds to a lower free energy, but $\Delta$ decreases as a function of $\mu_B$. Therefore, near NS densities, other superconducting phases may have larger gaps and would be favoured over CFL. This is demonstrated in Fig.\,\ref{fig:CSCphases}. In this figure, $\Delta=25\,\rm{MeV}$ and we see the free energy of some homogeneous phases become lower than that of CFL as we vary $\mu_B$ (note, recall the effective mass $M_s$ is also $\mu_B$ dependent). On the other hand at $\Delta\simeq100\,\rm{MeV}$ the CFL free energy is lower everywhere in this figure \cite{Alford:2007xm}. The figure also shows the large amount of possible colour-superconducting phases, demonstrating the potential richness of the CSC region. The emergence of these phases can be appreciated by the consideration that non-zero $M_f$ disfavours the CFL phase. Massless quarks become a worse approximation at lower $\mu_B$ and thus the symmetry of CFL becomes more approximate. In the CFL phase, all quarks participate equally in pairing, meaning the states at the Fermi surface for each flavour must be close together. As we decrease $\mu_B$ the difference in the Fermi surfaces becomes larger, which is most pronounced for the strange quark given its comparatively larger bare mass. Therefore, different pairing patterns where not all quarks participate begin to be relatively more favourable. 

Ignoring colour charge and same flavour pairing (generally disfavoured due to larger spin), we can form three diquark condensates. These are formed of $u$ and $d$ ($ud$) quark pairs, $d$ and $s$ ($ds$) quark pairs or $s$ and $u$ ($su$) quark pairs. Each condensate has a corresponding superconducting gap/condensation energy. Denoting these $\Delta_i$ ($\neq \Delta$) where $i=ud,ds,su$, we can analyse the ``melting pattern'' of the condensates i.e.\ at constant $\mu_B$, which phase transitions we undergo as we increase the temperature from $T=0$. Using an NJL model \cite{Fukushima:2004zq} and Cornwall-Jackiw-Tomboulis (CJT) formalism \cite{Ruester:2004eg}, it can be shown that $\Delta_{ud}(T)>\Delta_{ds}(T)>\Delta_{su}(T)$. Assuming the CFL is not favoured for all $\mu_B$, this would imply that at a constant $\mu_B$ (close to NS values say) as we increase $T$ we would undergo phase transitions in the order CFL$\rightarrow$dSC $\rightarrow$2SC$\rightarrow$NQM. Here, dSC is the phase where the $ud$ and $ds$ condensate are present, 2SC stands for two flavour pairing where there is only the $ud$ condensate\footnote{It can be shown that this phase has the same $T_c$ as the CFL phase in the weak-coupling regime. In Eq.\,\eqref{TcCFL}, dropping the numerical factor gives us the 2SC result which is the same as in BCS theory. The 2SC gap however is smaller than $\Delta$ by the same numerical factor, yielding the same $T_c$. }, and NQM stands for ``normal quark matter'' where there are no diquark condensates, which we might equate with the QGP (at least at high temperatures). Depending on the model the pattern may change. For instance, uSC in place of dSC, where in the uSC phase there are only $ud$ and $us$ condensates. The melting pattern from Ref.\,\cite{Fukushima:2004zq,Ruester:2004eg} agrees with a similar analysis in GL theory \cite{Iida:2003cc} (see Sec.\,(\ref{Background:subsec:GL})), which can give us some other insights into the phase structure at intermediate $\mu_B$. Many aspects of CSC can be analysed within this framework, such as vortices and the effect of non-zero quark masses (see Chapter \ref{chpt:Project1}). 

Aside from a transition from the CSC region to the QGP phase (or NQM phase at low temperatures, as one may hesitate to call matter ``QGP'' there), we can also consider the possibility of a transition between the confined region and the CSC region as depicted in Fig.\,\ref{fig:TmuPlane}. In this scenario, at low temperatures the deconfinment/chiral transition leads from the confined phase to the CSC region as we increase $\mu_B$. In our earlier discussion, these transitions were first order at low $T$. However, if the CFL phase is favoured all the way down to nuclear matter densities over other CSC phases, then this could lead to a crossover rather than a first-order transition, signalling a quark-hadron continuity \cite{Schafer:1998ef}. This is due to common symmetries between the CFL phase and the confined phase. Chiral symmetry is broken in both phases, which means chiral restoration would not occur between the confined phase and CFL. Furthermore, nuclear matter at low temperatures is superfluid where the exact $U(1)_B$ symmetry of QCD is broken like the CFL phase. Therefore, we could have another critical point near the critical temperature of nucleonic superfluidity below which the deconfinment and chiral transition become crossovers once more. Of course this entirely depends on CFL being preferred over other colour-superconducting phases at nuclear matter densities. Not all colour-superconducting phases share these broken symmetries. For instance, the 2SC phase is chirally restored. Another consideration is the restoration of $U(1)_A$ symmetry near the deconfinement/chiral transition at low temperatures. This is only an approximate symmetry (but then again so is chiral symmetry), broken by the chiral anomaly in full QCD. This possible crossover and critical point can be analysed from the perspective of the $U(1)_A$ anomaly using the GL approach once again \cite{Hatsuda:2006ps,Yamamoto:2007ah,Schmitt:2010pf}. Recently, an order parameter that could distinguish between the CFL and nuclear matter superfluid phases has been proposed \cite{Cherman:2020hbe}, continuing the debate on the existence of a quark-hadron continuity.

To finish the theoretical discussion of this plane, we point to more specialised and in-depth reviews, some of which have already been referenced. For general references on the theoretical discussion of the QCD phase diagram, see Refs.\,\cite{Fukushima:2010dr,Fukushima:2013rx} and for reviews on specific topics discussed see Refs.\,\cite{Rajagopal:1999cp,Stephanov:2004xs,Alford:2007xm,Guenther:2020jwe,Ghiglieri:2020dpq,Philipsen:2021qji}.

\subsection{Interlude: Experimental overview}
\label{Background:subsec:exp}

Having discussed the theoretical knowledge of the $\mu_B$-$T$ plane at $B=0$, we now summarise highlights of the experimental side of our understanding of this plane. We will focus on the results from HIC experiments and observation of NSs and some of the discussion will apply to the other planes discussed in this section as well. It should be noted that the conditions in HICs and NSs are not perfectly reflected in any of the planes we discuss. In HICs, there is finite $\mu_B$, $T$ and $B$ simultaneously. While NS (but not mergers) naturally lie in the $\mu_B$-$B$ plane at $T=0$, they also have a non-zero $\mu_I$. Furthermore, both NSs and HICs involve rotation and are thus relevant for non-zero $\nu$. Therefore, they concern higher dimensional planes than the ones discussed here in reality.

\begin{figure}
    \centering
    \includegraphics[width=0.5\textwidth]{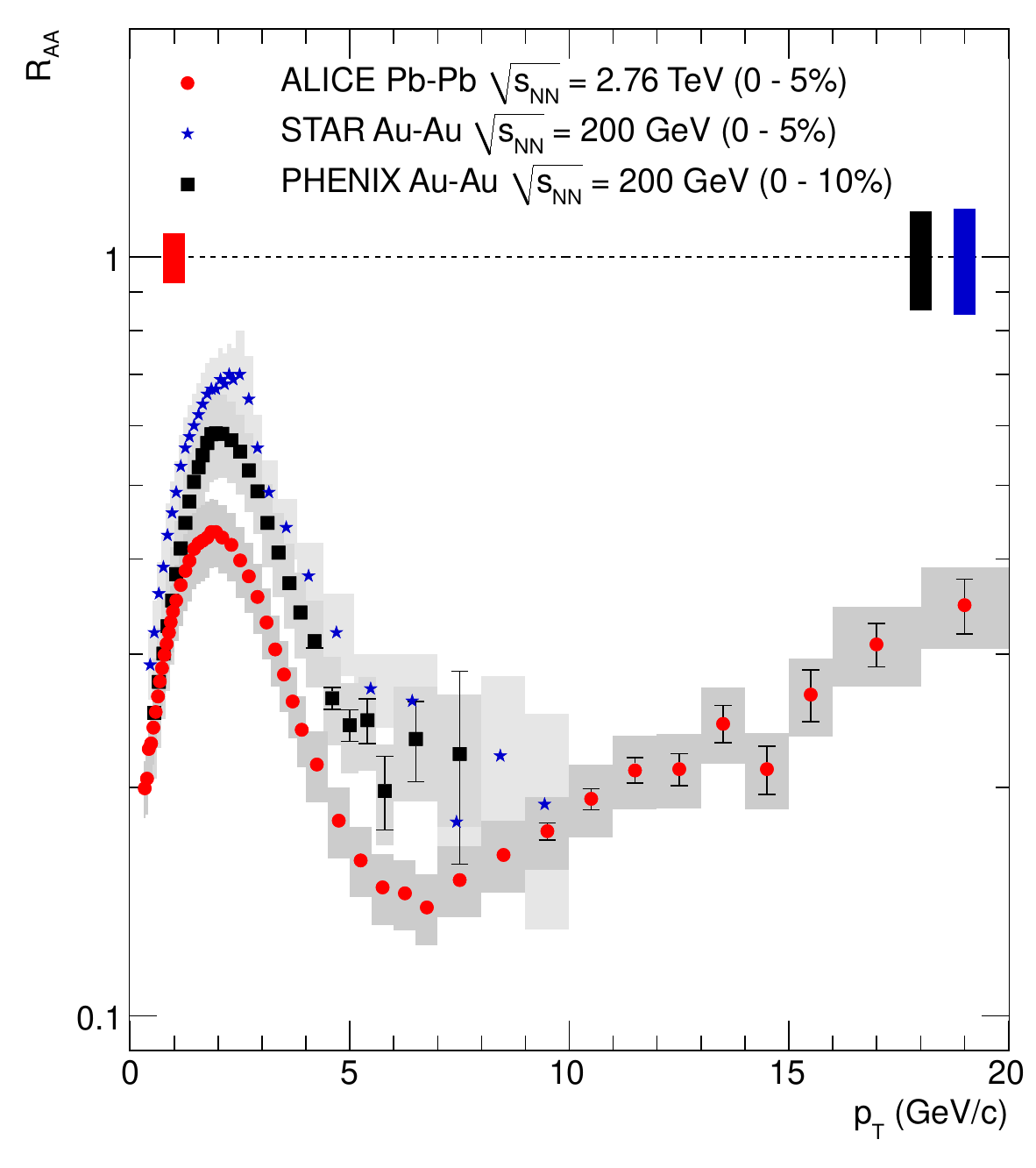}\includegraphics[width=0.5\textwidth]{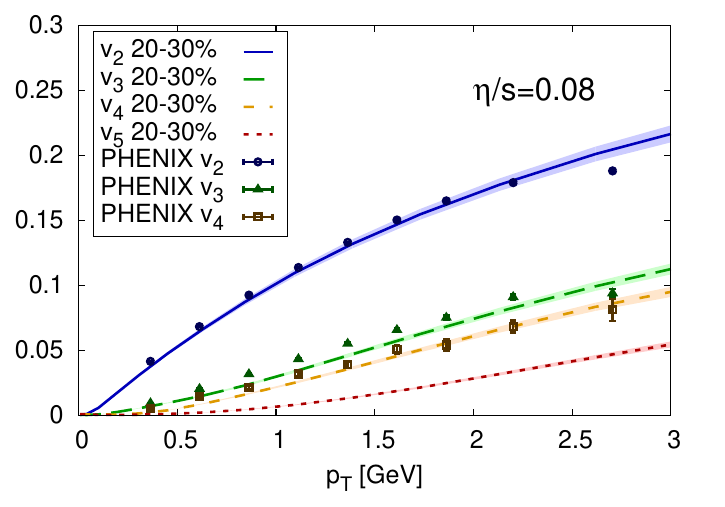}
    \caption{Signatures of the QGP in HIC experiments. \textit{Left panel:} Plot of $R_{\rm{AA}}$ from experiment as a  function of the transverse momentum $p_T$, the momentum component of partons perpendicular to the beam line (i.e.\ the colliding particles). Taken from \cite{STAR:2003fka,PHENIX:2003djd,ALICE:2010yje}. \textit{Right panel:} Hydrodynamic calculations of the coefficients $v_n$ of the Fourier decomposition in momentum space of the azimuthal distribution of particles compared to extracted values from experimental data also as a function of the transverse momentum $p_T$. The ratio of the shear viscosity $\eta$  and entropy density $s$ is quoted. Calculations are done at different ratios and then compared to data. We see that particularly low value $\eta/s=0.08$ gives good agreement. Taken from \cite{PHENIX:2011yyh,Schenke:2011bn}.
    }
    \label{fig:QGPsign}
\end{figure}

Experimental evidence for the deconfinement/chiral transition, the critical point and QGP comes from HIC experiments. The current heavy-ion colliders which can be used to probe this plane are the Large Hadron Collider (LHC) and Relativistic Heavy Ion Collider (RHIC). A key observation is that varying the centre of mass energy $\sqrt{s_{\rm{NN}}}$ in HICs has the opposite effect on both $T$ and $\mu_B$ - $T$ increases with $\sqrt{s_{\rm{NN}}}$, while $\mu_B$ decreases \cite{Andronic:2005yp}. Perhaps the place to begin is evidence for the existence of the QGP itself \cite{BRAHMS:2004adc,PHENIX:2004vcz,PHOBOS:2004zne,STAR:2005gfr}. The QGP has many signatures \cite{Niida:2021wut,Elfner:2022iae} whose detection have made its existence established (see \cite{Rafelski:2015cxa} for the historical context). To give an idea, we briefly discuss two of these. First is ``jet-quenching'' which refers to the energy loss of particles produced in HICs (partons) by traversing the QGP medium. The quark and gluon partons produced in the HIC form hadrons which, emerging in ``jets'', are detected. If there is a smaller number of high momentum hadrons detected than in other kinds of collisions (e.g.\ proton-proton), it is a sign that the hadrons of the jet were ``quenched'', possibly due to the presence of QGP. The extent of this quenching can be measured by the nuclear modification factor $R_{\rm{AA}}$. If $R_{\rm{AA}}<1$ then it indicates that jet-quenching has taken place in a HIC. This is clearly shown to occur by several different experimental collaborations in the left plot of Fig.\,\ref{fig:QGPsign}. The second signature is the collective non-viscous flow of partons from the QGP. Interestingly, hydrodynamics is a good approximation to the fluid mechanics of QGPs in HICs for a period of its evolution after the collision (see the recent review Ref.\,\cite{Schenke:2021mxx} and references therein). The viscosity of the strong elliptic flows detected are inferred from the $v_2$ coefficient of the Fourier decomposition in momentum space of the azimuthal distribution of particles. The right panel of Fig.\,\ref{fig:QGPsign} shows the strong agreement between predictions of low-viscosity hydrodynamics and experimental data. 

Experimental evidence of the QGP is evidence not only of a new phase of matter but also of a phase transition. Questions then remain about where this occurs and of its nature. Theoretically, we expect a crossover region that becomes first-order at low $\mu_B$ beyond a critical point. What can be said from the experimental side? It is thought that chemical freeze-out is an indication that the phase transition has happened. Chemical freeze-out refers to the moment when the formed hadrons in HICs stop interacting with each other. (Kinetic freeze-out refers to when the elastic scattering between them also stops.) As a result, the hadron abundances become fixed implying that the QGP has ``hadronised'' i.e. the moment the QGP turns into hadronic matter, where there are no longer any free quarks and gluons and thus no QGP. This is taken as a sign that the confinement transition has occurred. Therefore, extracting the $T$ and $\mu_B$ when chemical freeze-out is achieved could be taken as estimates of the location of the phase transition in the $\mu_B$-$T$ plane. Using statistical models of experimental data, chemical freeze-out is measured to occur at $T \simeq 156\,\rm{MeV}$ when $\mu_B\approx 0$ \cite{Stachel:2013zma,Andronic:2017pug}. Figure 5 of Ref.\,\cite{Andronic:2017pug} also provides predictions of the model at higher $\mu_B$ which includes comparison to lattice QCD. These results show good agreement with the lattice QCD results quoted there and in our theoretical discussion \cite{Borsanyi:2010bp,HotQCD:2018pds}.

Investigations into the nature of this transition are also being conducted. In particular, are efforts searching for the critical point. It is expected that conserved quantities like the net-electric charge, net-baryon number and net-strangeness would experience larger fluctuations in its vicinity. Specifically, higher-order cumulants of net-proton distribution are taken as a proxy for the net-baryon number, and show non-monotonic behaviour in experiments as a function of $\sqrt{s_{\rm{NN}}}$ \cite{HADES:2020wpc,STAR:2020tga}. Although the uncertainties are large, this points to fluctuations in the baryon number as we increase $T$ and decrease $\mu_B$ (and vice versa).

Finally, we point to a fairly recent summary of progress from HICs and theory in mapping the $\mu_B$-$T$ plane of the diagram given in Table 3 of Ref.\,\cite{Mohanty:2013yca}. While the existence of the QGP seems well established, determining the transition line, nature of the phase transition and the location of the critical point are still ongoing pursuits. Future HIC experiments aim to further our knowledge on these topics by exploring lower beam energies, corresponding to lower temperatures and higher densities. These include the FAIR \cite{Ablyazimov:2017rf},  NICA \cite{Kekelidze:2017ual}, HIAF \cite{Xiaohong:2018weu} and J-PARC \cite{Hachiya:2020bjg} experiments which are expected to continue probing these questions.

The other source of experimental data to probe the phase structure comes from astrophysical observations of NSs and their mergers. A NS is a very dense object compared to everyday standards, with $\mu_B\lesssim 1.5 \,\rm{GeV}$ \cite{Schmitt:2010pn} in the core. Although also extremely hot by everyday standards, 
on QCD scales $T\sim 10^6 \,K\sim 10^{-6}\,\rm{MeV}\simeq 0\,\rm{MeV}$ for the majority of their lifetime (newly formed NSs lie in the $10^{11}-10^{12}\,K\sim 0.1-1 \,\rm{MeV}$ range but quickly drop in temperature \cite{Schmitt:2010pn,Alford:2007xm}). Neutron star mergers can breach into the $T\sim 10-100 \,\rm{MeV}$ range according to simulations (e.g.\ see Ref.\,\cite{Prakash:2021wpz}) such that NSs and their mergers probe primarily the low temperature, intermediate density range of the $\mu_B$-$T$ plane. This region is challenging theoretically, as this is a regime where few first-principle calculations can reach e.g.\ lattice simulations and pQCD. It is expected to include high density nuclear matter, the deconfinement and chiral transition near $T=0$, and the onset of theorised deconfined phases beyond. Neutron stars could be the only naturally occurring system where CSC exists. We give selected insights into the phase structure from these astrophysical systems.

\begin{figure}
    \centering
    \includegraphics[width=0.5\textwidth]{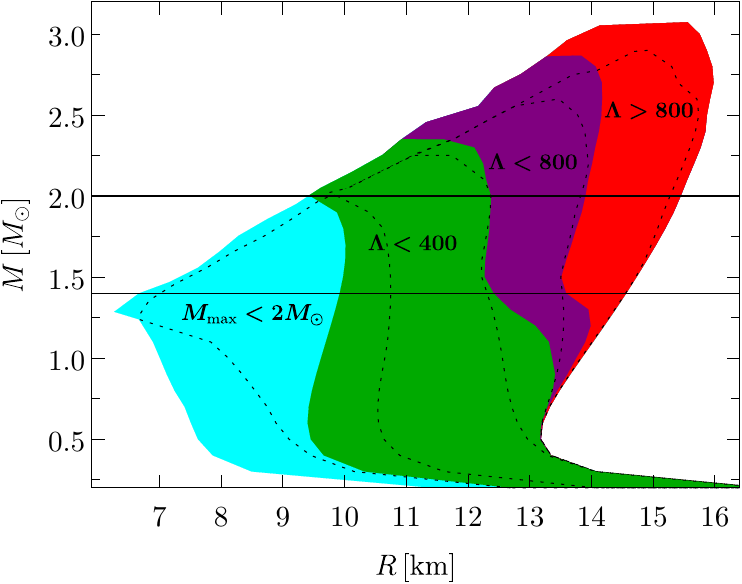}\includegraphics[width=0.5\textwidth]{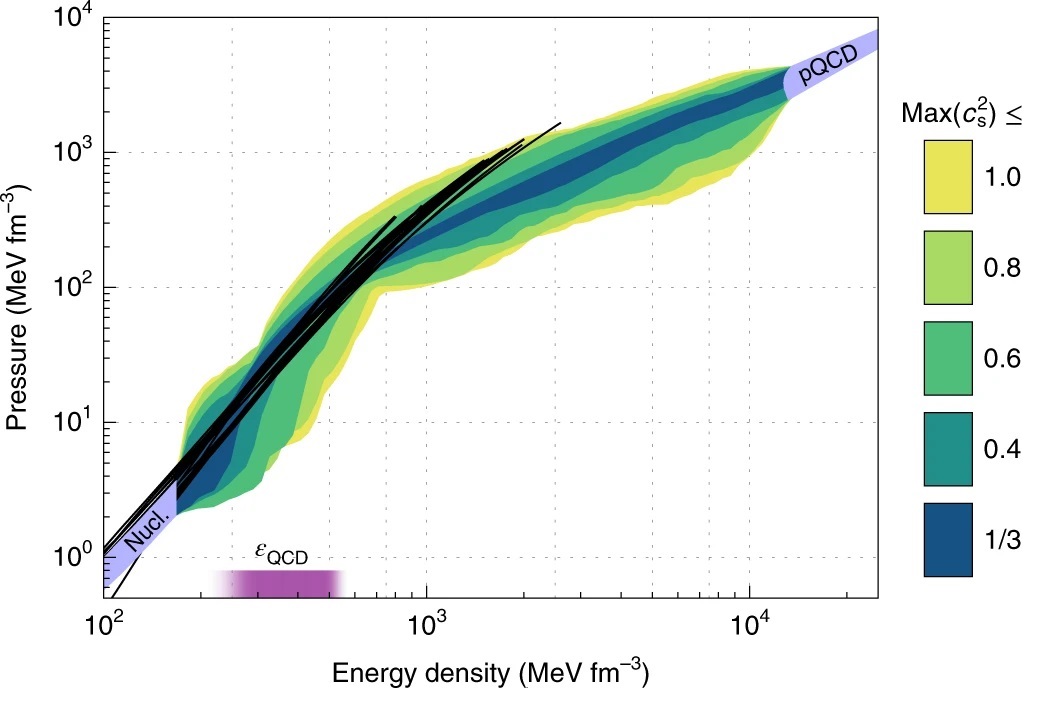}
    \caption{\textit{Left panel:} Plot of the constrained regions in the $R$-$M$ plane. $M$-$R$ curves lying in the coloured regions indicate they obey the condition labelling that region. Observations of $M> 2M_{\odot}$ NSs and $\Lambda_{1.4}<800$ implies that EoSs that produce $M$-$R$ curves in the cyan or red regions are not realistic for NSs. Dotted lines are regions obtained by a different interpolation method. Taken from \cite{Annala:2017llu}. \textit{Right panel:} Plot showing constraints on the EoSs, showing the allowed region between hadronic EoSs from chiral effective models \cite{,Hebeler:2013nza} i.e. nuclear theory and pQCD EoS \cite{Kurkela:2009gj} (light purple bands near bottom left and top right corner respectively). Dark lines are predictions of hadronic EoSs from Ref.\,\cite{Lattimer:2000nx,Gandolfi:2011xu,Fortin:2016hny}. The colour scheme corresponds to the maximum allowed speed of sound in that region. The rough region of the deconfinement transition from hot QGP is denoted by the purple strip labelled by $\epsilon_{\rm{QGP}}$.
    Taken from \cite{Annala:2019puf}. }
    \label{fig:EoSConstrain}
\end{figure}

The interior of a NS is usually separated into two layers, the crust and the core, which can each be divided further into ``inner'' and ``outer''. The outer crust is formed of neutron-rich nuclei and free electrons whereas superfluid neutrons are additionally present in the inner crust. In the outer core, the nuclei have completely disassembled to leave superfluid neutrons and superconducting protons with the possible addition of free muons as well as electrons. The composition of the inner core is less established. Here there is a myriad of possibilities, reflecting the situation between the confined and CSC regions of the $\mu_B$-$T$ plane. Hyperonic matter, deconfined quarks, CSC and QGP are all hypothesised along with the simple case where the composition is the same as the outer core. See Ref.\,\cite{Schmitt:2010pn} and the collection of reviews at the end of this section for more on the composition. It is the composition of the inner core which holds the most relevance to the phase diagram and is expected to effect several observables. This includes the mass $M$ and radius $R$ of NSs, and the relation between them depicted in $M$-$R$ curves. These are dependent on the equation of state (EoS) which is in turn dependent on the matter in question. Constraints from observational data on the $M$-$R$ curve and EoS can therefore give us insight into the phases we can expect at NS densities. The simplest constraint is the maximum observed mass. It can be confidently claimed that the maximum mass $M_{\rm{max}}>2M_{\odot}$ from electromagnetic radiation emitted from pulsars \cite{Demorest:2010bx,Antoniadis:2013pzd,NANOGrav:2019jur,Fonseca:2021wxt} where $M_{\odot}$ is the solar mass. The recent gravitational wave measurement GW170817 of a NS merger limits the tidal deformability of a $1.4\,M_{\odot}$ mass NS, $\Lambda_{1.4}$, to $\Lambda_{1.4}<800$ \cite{LIGOScientific:2017vwq} with high credence. In addition, the even more recent simultaneous $M$ and $R$ measurements of PSR J0030+0451 \cite{Riley:2019yda,Miller:2019cac} and PSR J0740+6620 \cite{Riley:2021pdl,Miller:2021qha} from the NICER mission provides additional constraints. With all these consideration, the constraints on the EoS in the intermediate density range $1.5\,n_0\leq n_B \leq 3\,n_0$ has been improved, and it is estimated that the radii of NSs now lie in the range $11.6\,\mathrm{km} \leq R\leq 13.1\,\mathrm{km}$ \cite{Miller:2021qha}. This leads to a far narrower band of viable EoSs and $M$-$R$ curves than previously considered. Another constraint is the speed of sound, $c_s$. Clearly this must obey $c_s\leq 1$ (in natural units) but one can also show from pQCD that it must approach $c_s^2=1/3$ for high density matter \cite{Kurkela:2009gj}\footnote{The $P\sim \mu_B^4$ result of Ref.\,\cite{Kurkela:2009gj} for high density quark matter implies this via $c_s^2= dP/d\varepsilon=(n_B/\mu_B) d \mu_B/d n_B=\mu_B^{-1}\partial P/\partial \mu_B\left(\partial^2 P/\partial \mu_B^2\right)^{-1}$
where $P$ and $\varepsilon$ are the pressure and (internal) energy density respectively. This can even be inferred via the simple consideration of non-interacting quarks.}. In  Ref.\,\cite{Annala:2019puf}, the speed of sound alongside other constraints including the tidal deformability and $M_{\rm{max}}$ bound was used to constrain the EoS in a model-independent way. Their conclusions support the existence of quark matter in NS cores. The results from this work and another by the same authors \cite{Annala:2017llu} are shown in Fig.\,\ref{fig:EoSConstrain}.

This analysis provides guidance for constructing the EoS in the gulf of densities $1\,n_0 \lesssim n_B \lesssim 40\,n_0$ (for reference, $n_B\sim 50\,n_0$ corresponds to $\mu_q\sim 1\,\rm{GeV}$ \cite{Kojo:2020krb}), where neither low energy effective models nor pQCD is valid. An insight from these constraints is that EoSs are required to be soft (high compressibility) at low densities, and stiff (low compressibility) at high densities. This change in property implies that there may be a phase transition from nuclear matter to quark matter at NSs densities. However, it is still up to debate whether quark matter cores exist in heavy NSs, or if all NSs are hadronic from crust to core. As already discussed from the theory side, this phase transition could be first order or a crossover. Currently, EoSs that incorporate either type of transition are consistent with current constraints, see Ref.\,\cite{Kojo:2020krb} for more on this topic. It should be noted in Fig.\,\ref{fig:EoSConstrain} that although perhaps it is implied that EoSs where $c_s^2=1/3$ is the maximum are favoured, this need not be the case. The breaching of this limit does not violate the need to approach it from below. To explore the possibility of the quark-hadron transition at higher temperatures, gravitational wave signals from NS mergers can once again provide insight. Numerical simulations in Ref.\,\cite{Most:2018eaw} point to possible signatures that such a transition takes place. In this reference, the presence of quarks and hyperons were included and contrasted with studies of EoSs where hyperons were the only exotic degree of freedom \cite{Sekiguchi:2011mc,Radice:2016rys}, noting a clear difference in the predicted gravitational wave signals. Gravitational waves from NS mergers could therefore possibly determine whether there is a transition to either hyperons or quarks or to both in the $T$-$\mu_B$ plane at NS merger temperatures and densities.

While the EoS can tell us much about the bulk properties of NS matter, it can be blind to some microphysical effects which could give us clues about the phases of matter at these densities. Thus, it is important also to look at the transport properties for which the low energy excitations are significant. Cooper pairing for instance has little effect on the bulk thermodynamics but can impact transport coefficients in the star. From the perspective of the $\mu_B$-$T$ plane it is perhaps best to discuss when these properties derived from hadronic and nuclear matter deviate from observations, pointing to different matter existing at NSs densities. Typically, this means strange and/or quark matter, including hyperons and colour-superconducting phases. We will briefly discuss the effects of these on select transport properties and observations.

An important property where Cooper pairing and thus the presence of CSC becomes significant is the neutrino emissivity and heat capacity. Less than a minute after its birth, escaping neutrinos from the NS interior become the primary way in which it radiates heat and it remains this way for the next few million years \cite{Alford:2007xm}. Thus, the neutrino emission of NSs and their heat capacity can be deduced from measurements of their temperature and age. Cooling tends to be dominated by whichever phase of matter in the NS has the highest emissivity and heat capacity. Ordinary nuclear matter emits neutrinos via the modified Urca process and the resulting emissivity (including the effects of the nucleonic superfluid) can be used to produce a family of cooling curves \cite{Yakovlev:2000jp,Page:2004fy,Yakovlev:2004yr}. Most NSs are consistent with these curves, but there are outliers which cool far more quickly (see \cite{Page:2004fy,Page:2006ud,Heinke:2010cr,Elshamouty:2013nfa} for NSs that are too cold for their age). It is hypothesised that the faster cooling must occur via the direct Urca process which leads to greater emissivity than modified Urca. Direct Urca can occur in hyperonic matter, unpaired quark matter, and many of the colour-superconducting phases except CFL. Therefore, these rapidly cooling stars may house exotic phases. Interestingly, the CFL phase has a comparatively low emissivity and specific heat capacity. This means that the cooling properties of other matter would dominate in a NS. Therefore, a NS with a CFL matter core could not be distinguished from a NS comprised of only ordinary nuclear matter by observing NS cooling. See Ref.\,\cite{Alford:2007xm} for more detailed discussion and references. 

A pulsar is a rotating NS  emitting electromagnetic radiation. These are known to ``glitch'' where their rotational speed suddenly increases (or they ``spin up''). Glitches have long been thought to be caused by unpinning of vortices in the neutron superfluid from the crust \cite{Anderson:1975zze}. While this is the standard explanation, there are still some unanswered questions (see \cite{Manchester_2011,Andersson_2012,Piekarewicz_2014} for instance). An alternate model which has the potential to address some of these issues is the possibility of crystalline CSC (see \cite{Anglani:2013gfu} for a recent review) in the core, a hypothesised colour-superconducting phase which has both superfluid
and rigid body properties \cite{Rajagopal:2006ig,Mannarelli:2008cf}. Crystalline CSC and other colour-superconducting phases could also be detectable via gravitational waves from ``mountains'' on the surface of NSs. The former could sustain such deformations by virtue of its large shear modulus \cite{Haskell:2007sh,Lin:2007rz}. The other possibility is flux tubes in a type-II colour-superconducting quark matter core sustaining these deformations \cite{Glampedakis:2012qp}. However, there is still much work to be done in these directions. 

\begin{figure}
    \centering
    \includegraphics{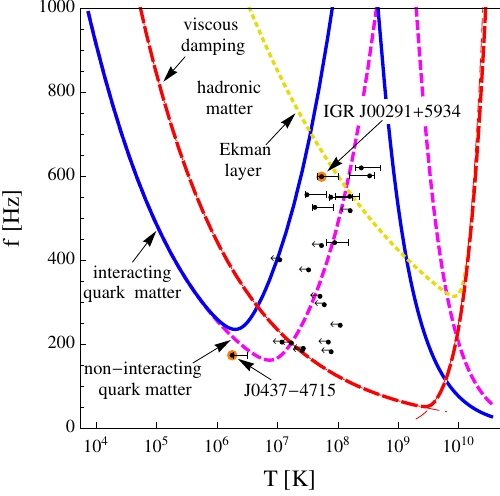}
    \caption{Instability windows depending on bulk viscosities from different types of matter. Vertical axis is the spin frequency of the star, horizontal axis is the surface temperature. Dots with error bars are the locations of stars on this plot from observational data. Curves are the lines of instability, with the area above the curves denoting the regions where NSs become unstable to $r$-modes. Neutron stars should spin down quickly due to the instability, and thus we do not expect many to be in this region at a given time. Taken from \cite{Alford:2013pma,Schmitt:2017efp}.}
    \label{fig:InstabilityWindow}
\end{figure}

Typically, NS matter is treated as a fluid. Fluid oscillations of NSs is another phenomenon that probes their transport properties. Specifically, $r$-modes, bulk flows which become unstable in rotating NSs. The Coriolis force acts as a restoring force to this flow which causes the NSs to lose angular momentum (and ``spin down'') via the emission of gravitational waves \cite{Andersson:1997xt}. This instability is limited by viscous damping and therefore the shear and bulk viscosity of NSs. Both shear and bulk viscosity rely on the microphysics and composition of the star. In particular, there is an ``instability window'' obtained from the shear and bulk viscosity of ordinary nuclear matter. Stars are not expected to remain for long in this region (and thus be unstable to $r$-modes) and so we do not expect to find many in the window at a given time. From temperature and frequency measurements, we see that many NS lie in this window (see \cite{Haskell:2012vg} and figures therein) in contradiction to the theoretical calculations based on ordinary nuclear matter. One possible explanation is that NSs are not only composed of nuclear matter and the bulk viscosity must be revised. Hyperons in the core may increase the bulk viscosity, restricting the window \cite{Reisenegger:2003cq,Haskell_2010}. Moreover, the bulk viscosity calculated with deconfined interacting quark matter in the core of Ref.\,\cite{Alford:2013pma} (see Fig.\,\ref{fig:InstabilityWindow}) is seen to provide a window in which no stars are found i.e. in agreement with expectations and observations, which is only achieved by hadronic matter by considering additional damping mechanisms, hinting at the possibility of exotic states at NS densities.

In the future, there are many new sources of observational data to look forward to. The already existing gravitational wave detectors like LIGO \cite{LIGOScientific:2014pky}, Virgo \cite{VIRGO:2014yos} and KAGRA \cite{KAGRA:2018plz} are expected to be joined by the Einstein Telescope \cite{Maggiore:2019uih} and Cosmic Explorer \cite{Reitze:2019iox} as well as the space-based detector of the LISA mission \cite{LISA:2017pwj} to name some in the not too distant future. These new experiments alongside sensitivity improvements to the already existing ones will provide more data and greater constraints on the properties of dense matter. Data from new \cite{WEISSKOPF20161179} and upcoming X-ray telescopes \cite{eXTP:2018anb,STROBE-XScienceWorkingGroup:2019cyd} is expected to provide additional insight into transport in NSs. For more on the future of nuclear astrophysics, see Ref.\,\cite{Schatz:2022vzq}. For more in-depth and specialised reviews, for HICs see Refs.\,\cite{Boyanovsky:2006bf,Mohanty:2013yca,Niida:2021wut,Schenke:2021mxx,Elfner:2022iae,} and for NSs Refs.\,\cite{Anglani:2013gfu,Schmitt:2017efp,Kojo:2020krb,Raduta:2021coc}.

\subsection{The $B$-$T$ plane at $\mu_B=0$}
\label{Background:subsec:BTPlane}

Perhaps the second most well-explored plane of the full phase diagram is the temperature and external magnetic field plane at $\mu_B=0$. Motivation for studying this plane is driven theoretically by the ability of lattice QCD to in principle explore the whole parameter space, and experimentally once again by HICs and expected conditions in the early Universe where extremely large magnetic fields are estimated to be present \cite{Kharzeev:2007jp,STAR:2015wza,Grasso:2000wj,Vachaspati:1991nm}. The magnetic fields need to be (very) large in order to affect the phase structure. Put another way, $eB\sim \Lambda_{\rm{QCD}}^2$. In Gaussian units, this translates to $eB\sim 0.1\,\rm{GeV^2}\sim 10^{18}$-$10^{19}\,\rm{G}$, at least $10^4$ times greater than the largest magnetic fields expected at the surface of NSs \cite{Duncan:1992hi}. It is estimated that these strengths are reached in peripheral HICs if only very briefly. In the presence of these magnetic fields, the chiral anomaly leads to the non-conservation of the axial chiral current. This in turn induces an electromagnetic current that is proportional and aligned with the magnetic field. 
This is known as the chiral magnetic effect (CME). In particular, signatures of the CME can indicate the occurrence of chiral restoration \cite{Fukushima:2011jc}. The CME only occurs with massless fermions and so reflects when the lightest quarks become effectively massless and therefore chiral symmetry restored. There is experimental effort into detecting these signatures but current results currently are only  qualitatively. See Refs.\,\cite{Kharzeev:2015znc,Li:2020dwr} for a summary of these results. Therefore, at $\mu_B\approx0$ this plane has some relevance for low baryon density HIC programs e.g.\ at the LHC. From the side of cosmology, the primordial magnetic field is expected to reach the aforementioned scales between the electroweak and deconfinement phase transitions \cite{Vachaspati:1991nm,Grasso:2000wj} of the early Universe, with the estimated range $10^8$-$10^{23}\,\rm{G}$. Results from other theoretical tools beside lattice QCD also exist, such as the NJL model and ChPT. Most investigations into this plane focus on the deconfinement and chiral transition. Since pQCD is limited in its usefulness in describing these transitions, it will not be discussed.

\begin{figure}
    \centering
    \includegraphics[width=0.5\textwidth]{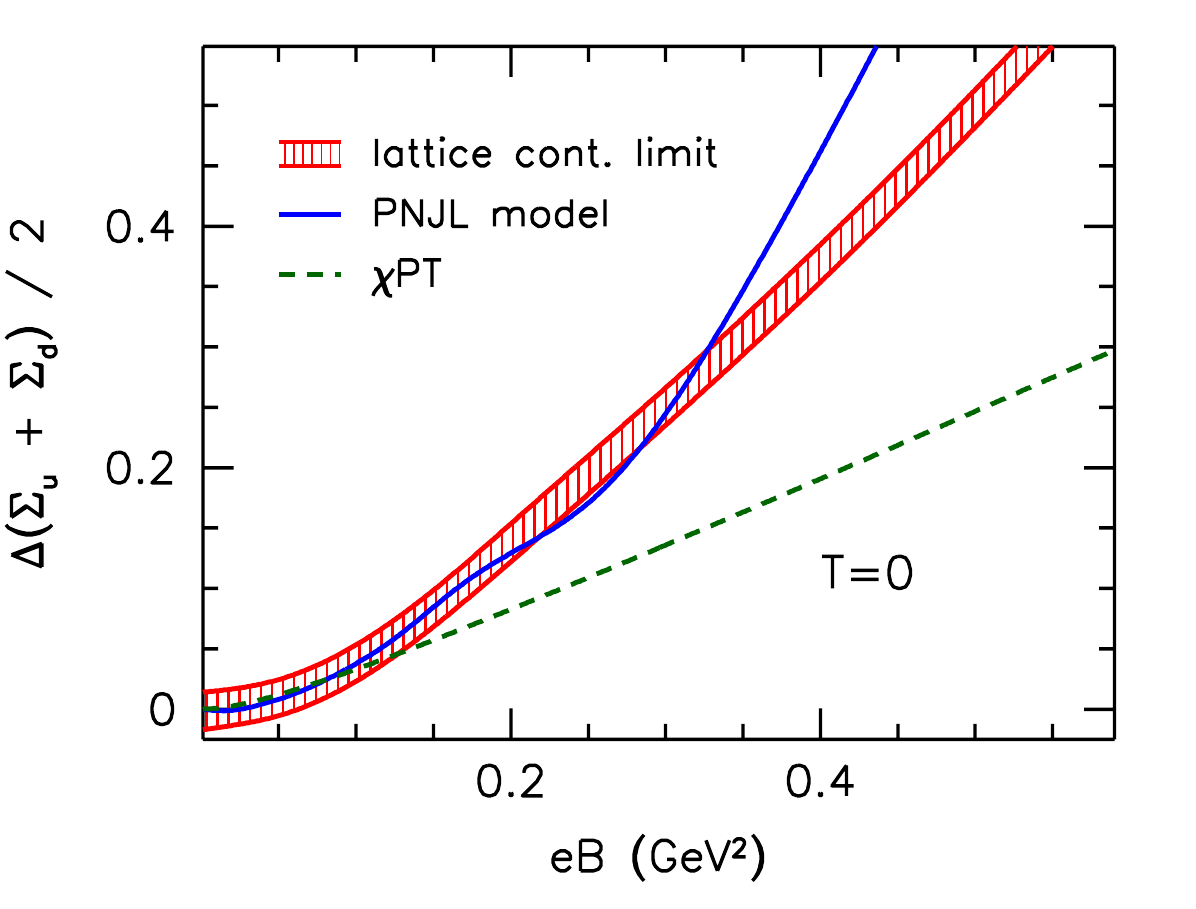}\includegraphics[width=0.5\textwidth]{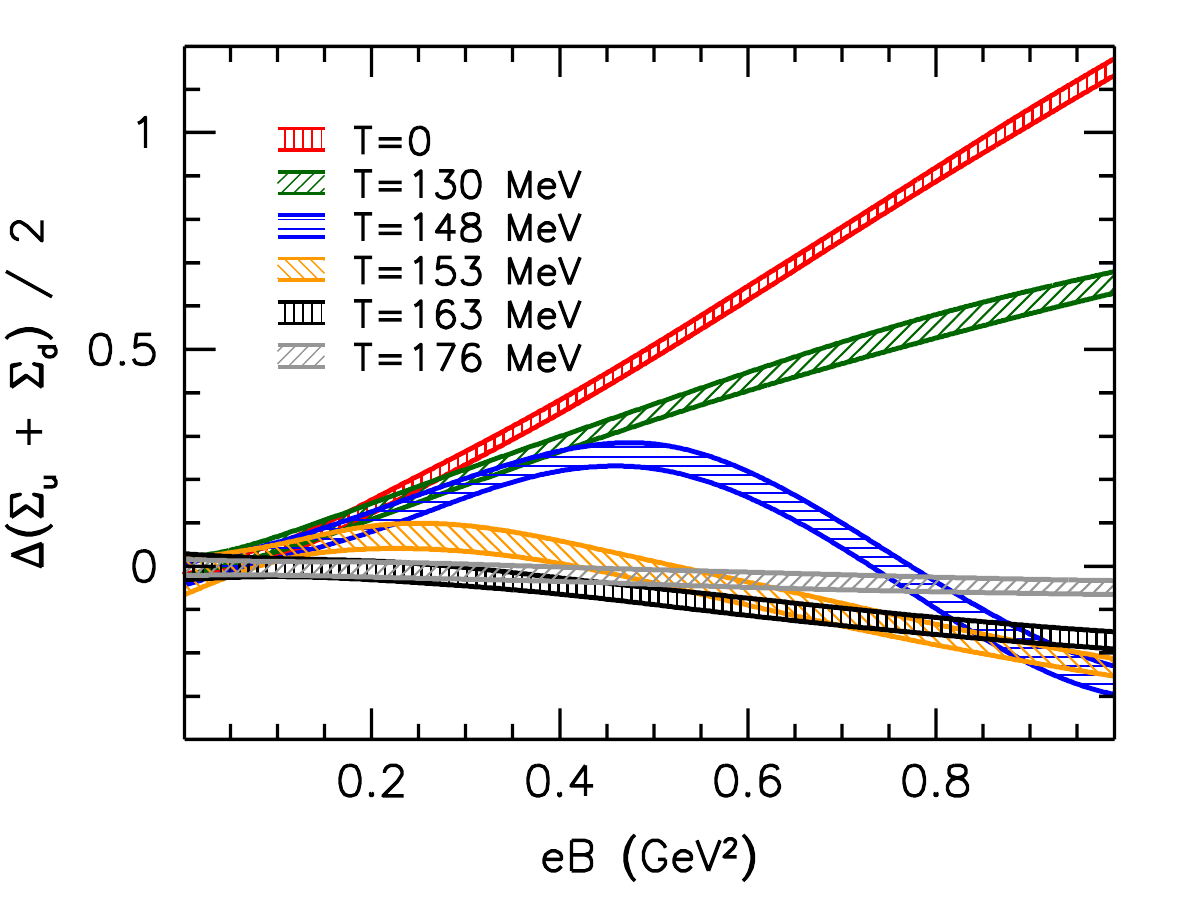}
    \caption{ Plots of the shifted chiral condensate as function of $B$ both taken from \cite{Bali:2012zg}. \textit{Left panel:} Prediction comparison of magnetic catalysis from lattice data in the continuum limit from \cite{Bali:2012zg}, ChPT from \cite{Cohen:2007bt,Andersen:2012dz,Andersen:2012zc} and (P)NJL model from \cite{Gatto:2010pt} at $T=0$ (the ordinary \cite{Mattos:2021alf} PNJL model reduces to the NJL model at $T=0$). \textit{Right panel:} Predictions from the lattice measured at different temperatures. Note that around $T_{pc}$ for the chiral crossover in Fig.\,\ref{fig:TmuPlane} a suppression of the condensate emerges, indicating inverse magnetic catalysis.}
    \label{fig:Catalysis}
\end{figure}

\begin{figure}
    \centering
    \includegraphics[width=0.5\textwidth]{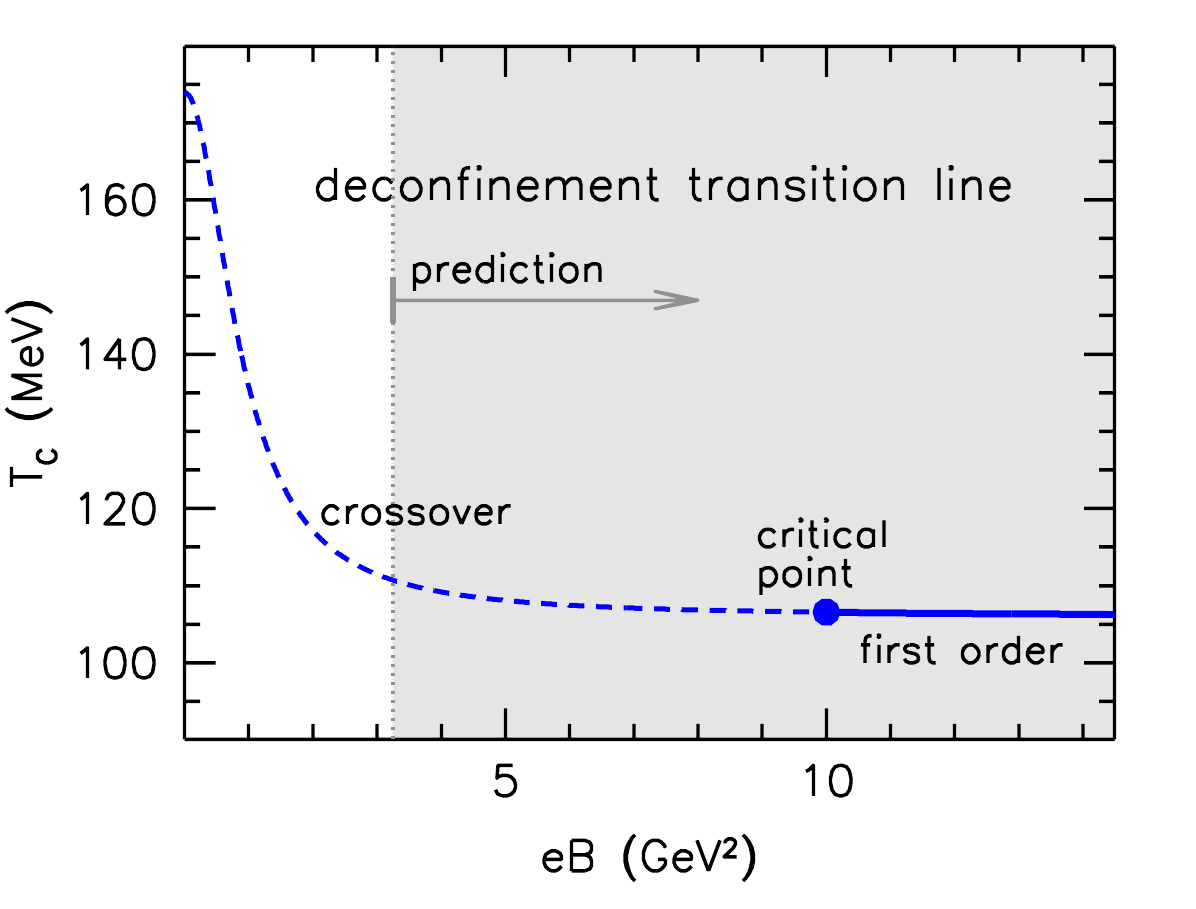}\includegraphics[width=0.5\textwidth]{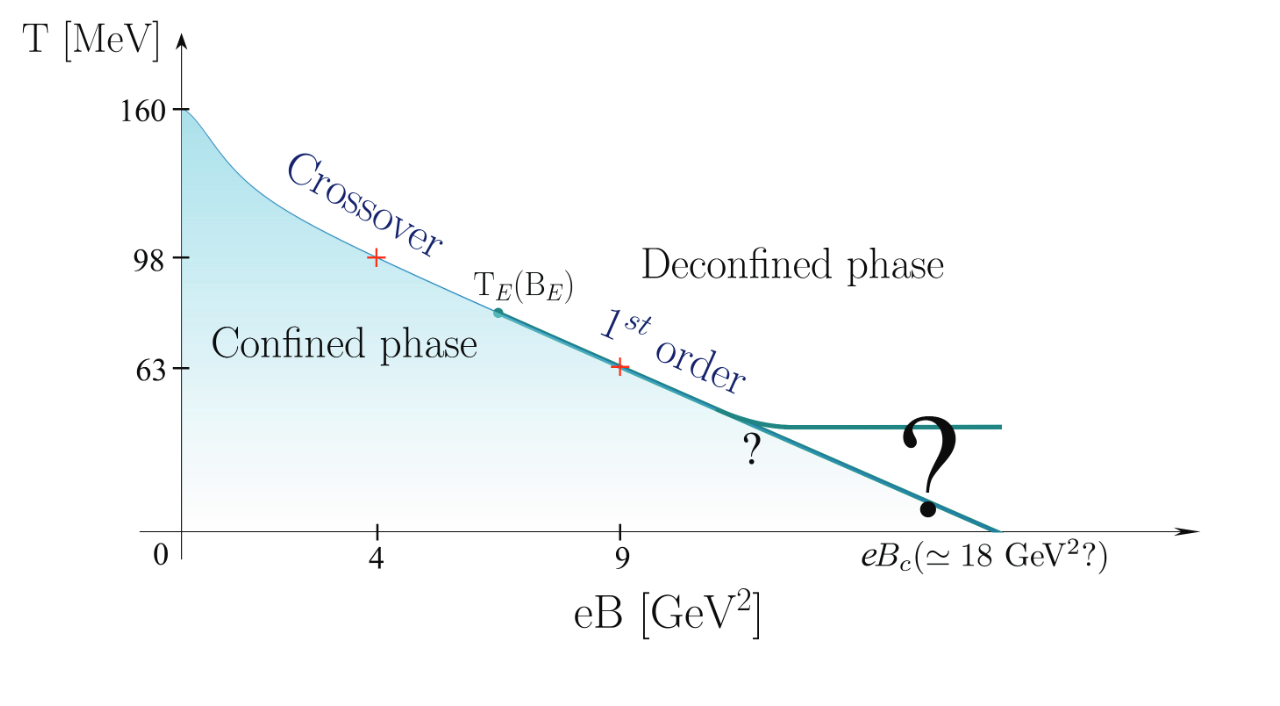}
    \caption{ Plots of the deconfinement transition and a rough picture of the $B$-$T$ plane. \textit{Left panel:} Shows a lattice results up until the grey region,
    beyond which is a prediction from (anisotropic) pure gauge theory. Taken from \cite{Endrodi:2015oba}. \textit{Right panel:} Interpolated plot of the plane from lattice simulations performed at two different values of $eB$ denoted by the red crosses. It is a crossover at one value, first order at the other, indicating a critical point $(B_E,T_E)$ in the range $4\,\text{GeV}^2<eB<9\,\text{GeV}^2$. The possibilities beyond this point are shown. Taken from \cite{DElia:2021yvk}.}
    \label{fig:BTplane}
\end{figure}

The QCD coupling strength is expected to decrease with magnetic field \cite{Miransky:2002rp}, thus we would expect a deconfinement/chiral transition between hadronic matter and QCD vacuum to a QGP phase. In contrast, there are other possible order parameters than the chiral condensate and Polyakov loop one can consider in the presence of a finite magnetic field. These are the magnetisation and magnetic susceptibility which could vary depending on the change in the response of the QGP and hadronic phases to an external magnetic field \cite{Yamamoto:2021oys}. For our discussion, we will focus on the usual order parameters, the chiral condensate and the Polyakov loop, but it is interesting to note that studies in this plane consider these additional order parameters as well (for instance, see \cite{Bali:2013esa,Bonati:2013lca}). At $T=0$, the chiral condensate is found to increase as a function of $B$ via magnetic catalysis (MC) in lattice simulations \cite{DElia:2010abb} and analytical calculations in ChPT \cite{Shushpanov:1997sf} and the NJL model \cite{Klevansky:1989vi}. Put in an alternative way, the magnetic field works to break chiral symmetry \cite{Shovkovy:2012zn}, increasing the effective mass of quarks \cite{Miransky:2002rp}. From ChPT and model calculations we would expect the chiral transition temperature i.e.\ $T_{pc}$ in this plane to increase with $B$ due to this effect \cite{Cohen:2007bt,Gatto:2010pt,Andersen:2012dz,Andersen:2012zc}. However, lattice simulations find the opposite - $T_{pc}$ decreasing as a function of $B$ \cite{Bali:2012zg} like it does in the $\mu_B$-$T$ plane. This is the so-called ``inverse magnetic catalysis'' (IMC). This discrepancy is shown in Fig.\,\ref{fig:Catalysis} where the chiral condensate is suppressed near the transition. The analysis of Ref.\,\cite{Bruckmann:2013oba} also links this behaviour to the Polyakov loop and it is expected that the chiral and deconfinement transition behave in the same manner as a function of $B$, replicating the behaviour in Fig.\,\ref{fig:BTplane}.

After the discovery of the IMC from the lattice and the failure of model calculations to reproduce this effect, a number of different approaches were introduced within these models in order to account for it \cite{Andersen:2014xxa}. A common approach is to introduce some $B$-dependence into parameters, such as the coupling and the Polyakov-loop potential \cite{Ferreira:2013tba,Farias:2014eca,Endrodi:2019whh}, with varying success at replicating what is observed by the lattice. See Ref.\,\cite{Andersen:2014xxa,Andersen:2021lnk} and references therein for details on these approaches. 

Lattice simulations show that the both chiral and deconfinement transitions are crossovers and are expected to become first order at some critical point (see Ref.\,\cite{Cohen:2013zja} for theoretical arguments for a critical point). In Ref.\,\cite{Endrodi:2015oba}, lattice results for multiple observables, including the chiral condensate and Polyakov loop, were extrapolated to large $B$ using an appropriate effective theory for this limit. They predicted that the critical point was around $eB\sim 10 \,\rm{GeV^2}$, $T\sim 100 \,\rm{MeV}$. Evidence from Ref.\,\cite{DElia:2021yvk} supports this for the chiral transition, placing the critical point in the range $4\,\text{GeV}^2<eB<9\,\text{GeV}^2$, $65\,\text{MeV}<T<95\,\text{MeV}$ as can be seen from the right panel of Fig.\,\ref{fig:BTplane}. It can also be seen that a simple extrapolation predicts for the transition line to end at some finite (but very large) $B$. However, this needs more investigation and it is in tension with the predictions of Ref.\,\cite{Endrodi:2015oba}, hence the question marks in the right panel of Fig.\,\ref{fig:BTplane}.

A possibly interesting phenomenon which belongs to this plane is vacuum superconductivity \cite{Chernodub:2010qx}, which bares mentioning due to the topic of this thesis. The QCD vacuum can be thought of as a soup of virtual quarks and anti-quarks arising from quantum fluctuations. Thus, charged, albeit virtual fermions are present. With gluons providing an attractive interaction between them (at low energies), the potential ingredients for Cooper pairing are present in the QCD vacuum. At strong enough magnetic fields, the virtual quarks and anti-quarks can become real and pair via the strong interaction to form mesons. If these mesons condense, then they could form a superconducting and/or superfluid phase. By considering $u$ and $d$ quarks and anti-quarks, it was hypothesised that the vacuum becomes unstable to rho meson condensation at large magnetic fields, leading to a superconducting phase emerging from the vacuum.
This phase is unlike conventional superconductivity as increasing the magnetic field tends to induce the condensate rather than depleted it. However, thermal fluctuations are still unfavourable towards its thermodynamic stability.
As such, it is expected there is a second-order phase transition out of this phase at temperatures near the QCD scale, $T\sim \Lambda_{\rm{QCD}}\sim 100\,\rm{MeV}$. Thus, one can draw a region in the lower left corner of the diagram where we may expect the vacuum superconductivity phase (see \cite{Chernodub:2011mc} for a schematic plot). This phase has been explored using GL theory \cite{Chernodub:2010qx}, an NJL model \cite{Chernodub:2011mc} with supporting evidence found in holography \cite{Callebaut:2011ab,Ammon:2011je}, and lattice QCD \cite{Braguta:2011hq,Braguta:2014ksa}. However, there is uncertainty surrounding this novel phase, see Refs.\,\cite{Liu:2016vuw,Ding:2020hxw,Carlomagno:2022arc} and discussions in Refs.\,\cite{Miransky:2015ava,Cao:2021rwx}. For more details on the broader topics of this section, see the reviews in Refs.\,\cite{Kharzeev:2013jha,Andersen:2014xxa,Guenther:2020jwe,Yamamoto:2021oys,} and references therein.

\subsection{The $\mu_B$-$B$ plane at $T=0$}
\label{Background:subsec:muBPlane}

Now we turn to the (very) conjectural $\mu_B$-$B$ plane of the QCD phase diagram at $T=0$. The discussion of the $\mu_B$-$T$ and $B$-$T$ planes act as a good foundation for our expectations as some concepts, phases and phenomena from them return. Furthermore, they can act as boundary conditions for a future schematic construction if we assume continuity between them all. This includes the scale of the axes to some extent, with both $eB$ and $\mu_B$ around the $\rm{GeV}^2$ ($10^{18}-10^{19}\,\rm{G}$) and $\rm{GeV}$ scale respectively. It should be noted that the term ``conjectural'' is used in a stronger sense here, as we have little input from first-principle calculations and there is less experimental motivation to probe the phase structure. With $T \sim 100 \,\rm{MeV}$ in HICs our only observational tools are compact stars, which are limited to intermediate values of $\mu_B$ \cite{Schmitt:2010pn}. For details on the experimental input from compact stars see the experimental discussion of Sec.\,(\ref{Background:subsec:exp}). Our primary source of knowledge will therefore come from models and effective theories. 

\begin{figure}
    \centering
    \includegraphics[width=0.5\textwidth]{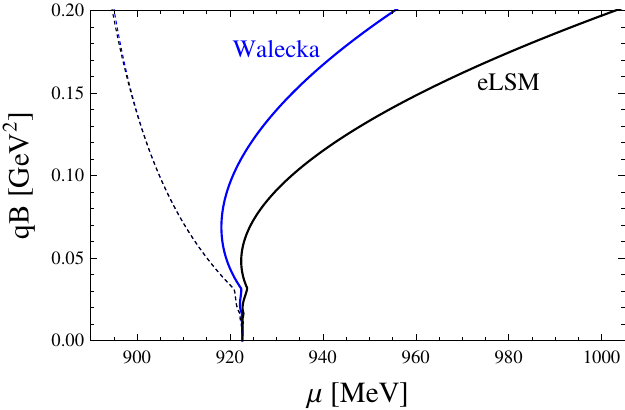}\includegraphics[width=0.5\textwidth]{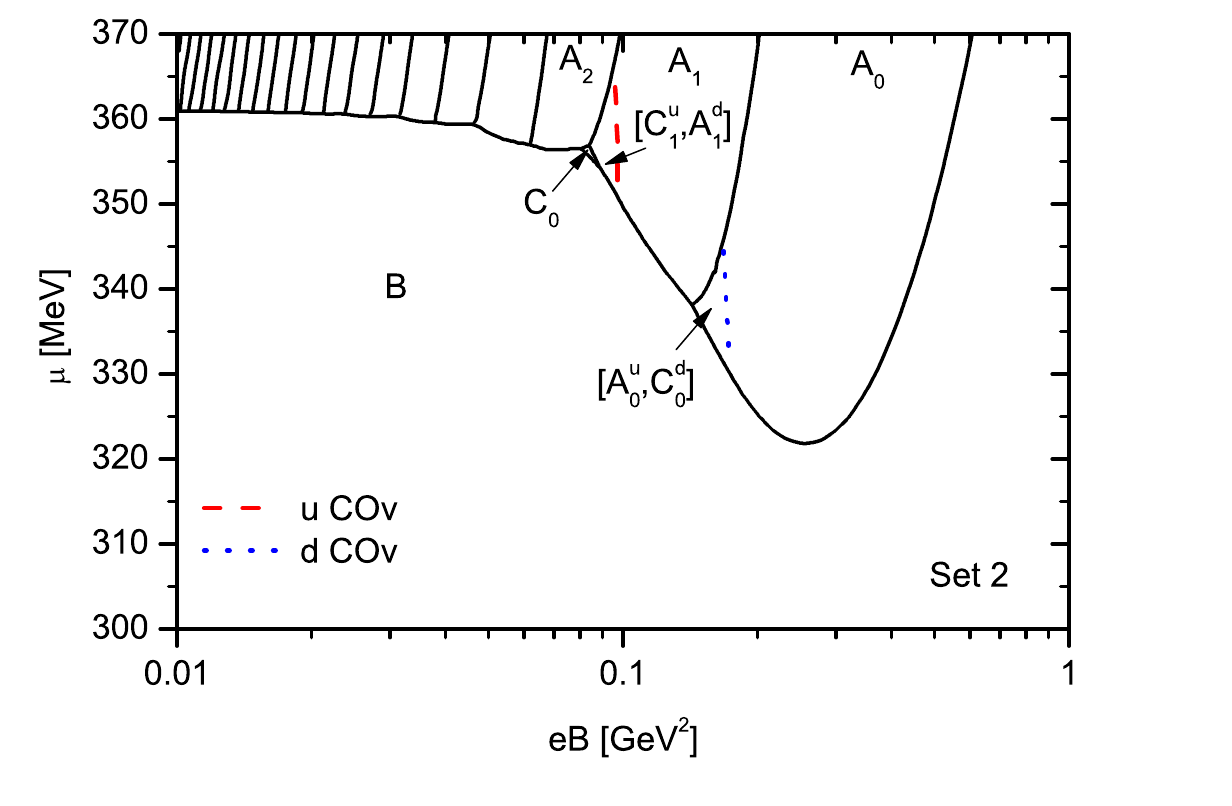}
    \caption{\textit{Left panel:} Plot of the baryon onset in the $\mu_B$-$B$ plane in both the Walecka model \cite{Walecka:1974qa} and an extended linear sigma model. The two dotted (barely distinguishable) lines are the onset when MC is not taken into account. Taken from Ref.\,\cite{Haber:2014ula}. \textit{Right panel:} Plot of the chiral transition in the $eB$-$\mu_q$ plane at $T=0$ from an $SU(3)$ NJL model. As mentioned in the text, B is a chirally broken phase and $\text{A}_i$ are chirally restored phases with $i$ denoting the associated Landau level. The B phase is shown to be favoured at large $B$, meaning chiral symmetry is broken at large $B$ and $\mu_q\sim 350\,\text{MeV}$. For full information on the labelling scheme, see Ref.\,\cite{Ebert:1999ht} and Ref.\,\cite{Grunfeld:2014qfa}. This plot is taken from the latter of these references.}
    \label{fig:BaryonOnsetmuB}
\end{figure}

As touched upon, at low energies our second-best tool to a first principle approach is ChPT, which can reliably investigate energy scales below the chiral restoration. It has been successfully applied to investigate inhomogeneous pion phases in this region \cite{Son:2007ny,Brauner:2016pko}. This will be related to the topic of Chapter \ref{chpt:Project2} and so we relegate the discussion of the region $\mu_B\lesssim m_N$ to that chapter. Going to higher $\mu_B$ requires different tools. Perhaps our first port of call is the baryon onset. In the $\mu_B$-$T$ plane at $B=0$, the liquid-gas transition ends at $T=0$, which implies it may stretch into other planes involving $\mu_B$. This was investigated in Ref.\,\cite{Haber:2014ula} where the authors used a Walecka model with self-interactions \cite{Walecka:1974qa,Boguta:1977xi} and an extended linear sigma model (see reference for specifics of the model). Importantly, it takes into account the effects of MC and the results can be found in the left panel of Fig.\,\ref{fig:BaryonOnsetmuB}. Similarly to the $\mu_B$-$T$ plane at $B=0$, the transition is first-order (in both models) and occurs at $\mu_B = 923 \,\rm{MeV}$ fulfilling the boundary condition with that plane. In contrast to this plane, its behaviour is very non-monotonic, with the critical magnetic field for the onset increasing as a function of $\mu_B$ for $eB\gtrsim 0.1\,\text{GeV}^2$. This is a consequence of the nucleon mass increasing as a function of $B$ at these higher values, exhibiting MC. However, it should be noted that the calculation of Ref.\,\cite{Haber:2014ula} does not take the anomalous magnetic moment of nucleons into account, which could work to counteract this effect \cite{Andreichikov:2013pga}. Up to fields of $eB\approx 0.04\,\rm{GeV^2}$ this is shown to be the case in the Walecka model \cite{Mukherjee:2018ebw}, but it remains to be seen whether this extends to larger $B$. 

On the topic of nuclear matter, there is also the nucleonic superfluidity expected at low temperatures. Studies of coupled neutron superfluids and proton superconductors in large magnetic fields and high densities using GL theory have been done in Ref.\,\cite{Haber:2017kth,Wood:2020czv,Wood:2022smd}. While there is interesting phase structure to be found, the magnetic field in these references is orders of magnitude lower than our expected scale. Neutron superfluidity is expected to be destroyed for $B\gtrsim 10^{17} \text{G}$ in magnetars and this threshold is expected to be lower for proton superconductivity \cite{Stein:2015bpa}. This means we may only expect these phases very near the $\mu_B$-axis.

As with the two previous planes, we expect there to be a chiral or deconfinement transition. Such a transition has been primarily analysed using NJL models, in the $U(1)$, $SU(2)$ \cite{Ebert:1999ht, Klimenko:1996vzw} and $SU(3)$ \cite{Grunfeld:2014qfa} cases, all with similar results. References \cite{Ebert:1999ht, Klimenko:1996vzw} introduce four types of phases labelled A, B and C. (There is occasionally also a D phase, but its definition changes depending on the focus of the model e.g.\ Refs.\,\cite{Frolov:2010wn,Coppola:2017edn}.) The A phase is a massless, chirally restored phase while the B (not to be confused with the magnetic field, $B$) and C phases are chirally broken. The difference between the B and C phase is the quark number density - in B it is zero, in C it is finite. For this discussion, we will simply refer to the chirally broken and chirally restored phases and the transition between them. It is generally predicted that the chiral transition happens for $300\lesssim\mu_q\lesssim 400\,\rm{MeV}$ and is first-order \cite{Ebert:1999ht}. This can also be seen in a chiral quark model \cite{Ferraris:2021vun}. A similar range is expected for the critical $\mu_q$ at $T=0$ for the deconfinement/chiral transition in the $\mu_B$-$T$ plane. Interestingly, the transition seems rather insensitive to the magnetic field at low $B$ but then becomes non-monotonic at higher values, resembling the behaviour of the baryon onset of Ref.\,\cite{Haber:2014ula}, exhibiting MC \cite{Grunfeld:2014qfa} as can be seen in the right panel of Fig.\,\ref{fig:BaryonOnsetmuB}. This can be seen by comparing the left and right panels of Fig.\,\ref{fig:BaryonOnsetmuB} though one should note that this similarity is only qualitative. It should also be kept in mind that the NJL model has a cutoff scale $\Lambda$ above which it is not valid. Commonly, $\Lambda\sim 600 \,\rm{MeV}$ which translates conservatively to $eB\sim 0.3\,\rm{GeV^2}$, which means the behaviour above this value should not be trusted too strongly. 

Comparison between the chiral and deconfinement transition can be discussed using the PNJL model. In the $T=0$ case, the PNJL model usually reduces to an ordinary NJL model, but recent efforts have found a method where the Polyakov loop can be analysed in the $T\rightarrow 0$ limit by using Polyakov loop dependent couplings \cite{Mattos:2019cmp,Mattos:2021alf}. While this method may not be well established yet, it does provide some interesting results. Using this new approach (dubbed the ``PNJL0'' model), the deconfinement transition can be analysed in the same model as the chiral transition \cite{Wang:2022xxp}. These transitions are shown to split at $B>0$. The chiral transition line slightly decreases as we increase the magnetic field but the deconfinement transition increases with $B$, occurring at larger and larger $\mu_q$. This raises the possibility of some intermediate phase emerging, like the quarkyonic matter briefly discussed at the beginning of this chapter. Such matter is possible in the $N_c\rightarrow \infty$ limit of QCD. It is within this limit that holographic methods become available. These methods are based on the gravity/gauge duality or AdS/CFT correspondence \cite{Maldacena:1997re} where certain QFTs can be analysed within extra-dimensional models of gravity.  Their appliance in the other planes were not discussed due to the tenuous assumption required that the behaviour of large $N_c$ QCD is the same as real world QCD. However, due to the lack of first-principle and observational investigation methods available in this plane, we resolve to briefly mention some results here. In particular, their comparison to the chiral transition in $U(1)$ NJL and the Walecka model of Refs.\,\cite{Preis:2010cq, Preis:2012fh}. There is qualitative agreement in both cases and with the general picture described earlier. Furthermore, one can investigate the inclusion of baryons \cite{Preis:2011sp, Preis:2012fh} to compare with Fig.\,\ref{fig:BaryonOnsetmuB} and Ref.\,\cite{Haber:2014ula}. The magnetic field seems to increase the critical chemical potential in both cases at large $B$.

As an aside, the holographic models do highlight the topic of IMC. The term ``inverse magnetic catalysis'' was in fact first discussed in this context \cite{Preis:2011sp}. Here it describes the same effect on the chiral condensate but with a different explanation to why it occurs to the one given for the effect in the $B$-$T$ plane given in Ref.\,\cite{Bruckmann:2013oba}. Crucially, this effect is absent in naive applications of the effective theories and models in that plane as already discussed. Though the suggested origin of the mechanism is different, it is still an effect that emerges due to non-zero $B$ and one may wonder if steps should be taken to include this effect in model calculations at finite $\mu_B$ and $T=0$. Many of the model results discussed in this section do not use approaches that explicitly include IMC. At best, the effect is implicit and the authors of Refs.\,\cite{Grunfeld:2014qfa,Ferraris:2021vun} interpret their results as manifesting IMC. Interestingly, the PNJL0 analysis of Ref.\,\cite{Wang:2022xxp} includes a $B$ dependent coupling, which is one of the aforementioned methods used to incorporate the effect \cite{Farias:2014eca}. Although the $B$-dependence of the coupling is not the same as the ones used to incorporate IMC in the $B$-$T$ plane (nor is it stated in Ref.\,\cite{Wang:2022xxp} to be included for this reason), the $B$ dependence does have a significant effect on the chiral transition line. There are also the recent examples Ref.\,\cite{Cao:2023bmk} and Ref.\,\cite{Mao:2022dqn} which incorporate the IMC via running of the coupling and anomalous magnetic moment of quarks in the (P)NJL model respectively. Both conclude that the chiral transition is disfavoured at low $\mu_B$, which is an indication that there is IMC at low $B$.

\begin{figure}
    \centering
    \includegraphics[width=0.5\textwidth]{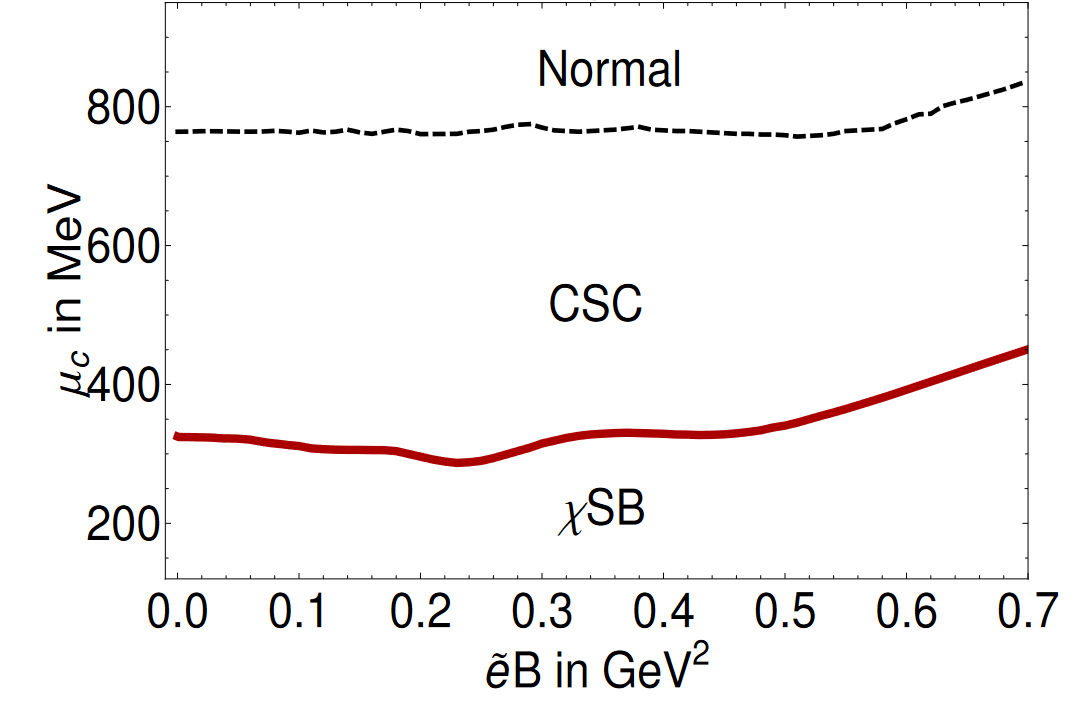}\includegraphics[width=0.5\textwidth]{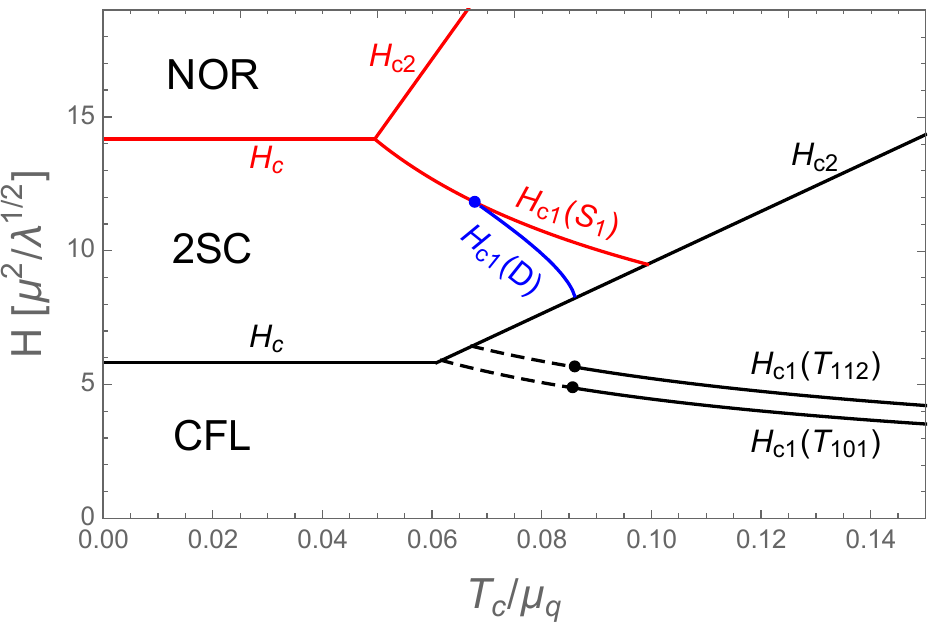}
    \caption{\textit{Left panel:} Plot of the critical chemical potential of the transition between the chirally broken phase ($\chi$SB) and a colour-superconducting phase (CSC) and the transition from the CSC region to the normal (or NQM) phase as a function $eB$. Taken from Ref.\,\cite{Fayazbakhsh:2010gc} where an $SU(2)$ NJL model was used. \textit{Right panel:} Phase diagram of the critical magnetic fields of three-flavour, type-I and type-II colour-superconducting quark matter from a GL approach in an external magnetic field, $H$ and $T_c/\mu_q$ plane. Labelled are the CFL, 2SC and normal phase (NOR) or NQM. The external magnetic field $H$ is measured in units of the parameters in Sec.\,(\ref{Project1:subsec:para}) but for reference, $H \sim 14 \mu^2/\sqrt{\lambda}\sim 0.2\,\rm{GeV^2}$. The critical fields $H_{c}$, $H_{c1}$ and $H_{c2}$ resemble those of conventional superconductors, see Sec.\,(\ref{Background:sec:SC}) and Fig.\,\ref{fig:type1and2}.  Different types of magnetic defects affect the type-II transition associated with $H_{c1}$. See Ref.\,\cite{Haber:2017oqb} from which this figure was taken for more details.}
    \label{fig:ChSBtoCSCmuB}
\end{figure}

Just beyond the chiral/deconfinement transition, the aforementioned NJL models show a cascade of massless phases labelled A$_i$ where $i$ denotes the preferred Landau level in the right panel of Fig.\,\ref{fig:BaryonOnsetmuB}. These do not account for certain phenomena such as the inhomogeneous chiral density wave (CDW) and CSC which can be incorporated into the NJL model \cite{Frolov:2010wn, Fayazbakhsh:2010gc,Fayazbakhsh:2010bh,Coppola:2017edn}. It is not currently known which is preferred, see discussion in Ref.\,\cite{Cao:2021rwx}. In the $SU(2)$ case, at most only two-flavour pairing is possible such that the 2SC phase is mainly considered. Reference \cite{Coppola:2017edn} investigated this phase under NS conditions and identified some of the massless A$_i$ phases in the $\mu_B$-$B$ plane to be in the 2SC phase (see Fig. 4 of the same reference). Colour superconducting\footnote{Recall that the colour-superconducting phases discussed are electromagnetic superconductors only in a very weak sense. To avoid confusion, the Meissner effect they exhibit mainly acts on the chromo-magnetic fields and only slightly screens the $B$ field. Thus, type-I CSC phases can appear in the diagram since $B\neq 0$ in those phases. Type-II electromagnetic superconductivity is also possible as long as we interpret $B$ as a spatial average, since the $B$ field is inhomogeneous in the vortex lattice phase.} phases are once again expected to dominate at large $\mu_B$. As in the $\mu_B$-$T$ plane at $B=0$, many different phases are possible, including the transitions between type-I and type-II superconducting phases (see the right panel of Fig.\,\ref{fig:ChSBtoCSCmuB}). As with the low $\mu_B$ regime, this possibility will be explored in more depth in a later section and so we relegate much of the discussion (especially the possibility of inhomogeneous phases, which is rather rich here \cite{Noronha:2007wg,Son:2007ny}), to Chapter\,\ref{chpt:Project1}. The right panel of Fig.\,\ref{fig:ChSBtoCSCmuB} is a phase diagram in the $T_c/\mu_q$ plane, (similar to the $\kappa$-$H$ phase diagram presented later in Fig.\,\ref{fig:type1and2}) where $H$ is an \textit{external} magnetic field measured in units of model parameters $\mu^2/\sqrt{\lambda}\propto T_c \mu_q$. Thus, $H\sim (T_c/\mu_q)\mu_q^2$ and knowing that $T_c$ tends to increase with $\mu_q$, the CSC region will get larger as we increase $\mu_B$ according to these results. Another difference is that the results are in terms of $H$ rather than $B$. We expect these to only differ by a constant offset however, which implies the general structure is not greatly effected. Furthermore, this phase diagram suggests the CSC at high $B$ is disfavoured to some normal conducting, deconfined quark matter like the NQM phase from the earlier discussion of CSC. This is supported by literature on the stability of CSC against transitions to ferromagnetic dense quark matter and spin density wave phases \cite{Tatsumi:2003bk,Iwazaki:2005nr,Tatsumi:2005ys,Iwazaki:2005ut,Wen:2013yra}. Due to the different type-I and type-II structure, the transition out of the colour-superconducting phases could go from second order to first order as we increase $\mu_B$. The CSC phases have further phase structure aside from the type-I/type-II transitions. For instance, one can posit further phase transitions between ``magnetic phases'' of CFL. This introduces the possibility of other crossover and first-order transitions in this region. See Ref.\,\cite{Ferrer:2012wa} for more details.

Making a schematic plot of the $\mu_B$-$B$ plane cannot be done with much confidence due to the lack of experimental input and first principle calculations. Thus, we will speculate on what it may look like. We do this by posing some questions with following discussion based on the literature reviewed above.

Where do we expect MC and IMC and how do they affect the baryon onset and chiral transition? From the literature, there is support for both the MC and IMC emerging in the phase diagram, disfavouring and favouring chiral restoration respectively. They may also influence the baryon onset. From the model calculations, it seems both the transition to nuclear matter and chirally restored phases are slightly disfavoured at low magnetic fields (indicating IMC) and then favoured at higher magnetic fields (indicating MC), though the scales at which this occurs seem different. Furthermore, it is not apparent if the IMC must be incorporated explicitly in model calculations like in the $B$-$T$ plane at $\mu_B=0$. What of the deconfinment transition? New methods developed recently allow the use of the PNJL model at $T=0$. The single work discussed seems to suggest the deconfinment transition behaves differently to the chiral transition. It would be interesting to see what further studies using the PNJL0 model can tell us about this transition.

Is there a quark-hadron continuity like in the $\mu_B$-$T$ plane at $B=0$? Given that it borders the $\mu_B$-$T$ plane at $B=0$, the possibility of a quark-hadron continuity (discussed in Sec.\,(\ref{Background:subsec:muTPlane})) in the $\mu_B$-$B$ plane at $T=0$ can be entertained. At $\mu_B$ where we expect nuclear matter (in both planes), the critical magnetic field ($\sim 10^{17}\,\text{G}$) of nucleonic superfluidity is about an order of magnitude or two smaller than the typical scale of $eB\sim\Lambda_{\text{QCD}}^2$ (corresponds to $10^{18}$-$10^{19}\,\text{G}$) where we would expect a deconfinment or chiral transition. Coincidentally, the critical temperature (around $1$-$10\,\text{MeV}$) is smaller than $T\sim\Lambda_{\text{QCD}}$ ($\sim 100\,\text{MeV}$) by roughly the same order of magnitude. A quark-hadron continuity in the $\mu_B$-$B$ plane at $T=0$ might therefore occupy a similarly sized region at the scales presented. Of course, this depends once more on the phase that follows the chiral transition, and whether it is chirally broken and/or is a superfluid. This leads to another question - what phases do we expect at NS densities? At large densities we might expect the CSC phase and the CFL phase at asymptotically high $\mu_B$. Do we expect this region to exist all the way down to the $\mu_B$ realistic for NS? It is possible that 2SC would be favoured here over CFL. The CDW is also a candidate at these densities. A further question is then which is preferred, CDW or CSC?

What phases are there at $eB>0.3\,\rm{GeV}^2$? While the calculations from the literature at least give us a vague idea of what the phase structure might look like at small magnetic fields, we seem to know even less about magnetic fields beyond $\sim 0.3\,\rm{GeV}^2$. The NJL models show that the chirally broken phase B is preferred, though at these scales one might question if we are reaching scales beyond the cutoff $\Lambda$ used in these models. With $\Lambda=600\,\rm{MeV}$, $eB=0.36\,\rm{GeV}^2$, making results at large fields qualitative at best. Similarly, the scale of the critical fields in the right panel of Fig.\,\ref{fig:ChSBtoCSCmuB} is around $0.1\,\rm{GeV}^2$ and predicts a transition to NQM. However, this phase could be competing with aforementioned chirally broken phase and other phases of deconfined quark matter. The chirally broken phase existing to large $eB$ is at least consistent with the situation expected from lattice calculations in the $B$-$T$ plane at $\mu_B=0$ (see Fig.\,\ref{fig:BTplane}) if one argues there should be continuity between planes. However, this only tells us that at low $\mu_B$ the chirally broken phase B may exist up to large $eB$ and not necessarily up to $\mu_B$ expected at NSs as the NJL models predict.

As far as we know there are no specific reviews of the literature to draw from in this plane, which may make this discussion of the $\mu_B$-$B$ plane at $T=0$ the first of its kind. While our summary has tried to be broad, we make no assertions that it is comprehensive nor the references cited exhaustive. Nevertheless, we hope it will be useful to others investigating this parameter space and as an early step in constructing a wider and more detailed $\mu$-$B$ plane at $T=0$.

\section{Superconductivity}
\label{Background:sec:SC}

Superconductivity is one of the great physics discoveries of the 20\textsuperscript{th} century. It is a phenomenon which exhibits collective quantum behaviour of particles at macroscopic scales. Alongside the similar superfluidity, it serves to remind us that we live in a quantum mechanical world. It was discovered by Heike Kamerlingh Onnes in 1911 when they observed the resistance of mercury drop to zero at a very low but finite temperature \cite{Onnes:1911DiscSC}. A further experimental milestone was the discovery of the Meissner effect in 1933 \cite{Meissner1933EinNE}, where superconductors were found to expel magnetic fields from their interior. The ``supercurrent'' - the dissipationless flow of charge carriers due to the vanishing resistivity of superconductors - and the Meissner effect became iconic properties of superconductors. The first milestone in the theoretical description of superconductors came in the form of the London equations in 1935 \cite{London1935TheEE}. From two simple equations relating the supercurrent $\bm{j}$ to the electric field and magnetic field $\bm{E}$ and $\bm{B}$,
\begin{subequations}
    \begin{align}
            \frac{\partial \bm{j}}{\partial t} = \frac{1}{\Lambda^2}\bm{E} \,,
            \label{LondonE}
            \\[2ex]
            \nabla \times \bm{j} = -\frac{1}{\Lambda^2}\bm{B}\,,
            \label{LondonB}
    \end{align}
    \label{London}%
\end{subequations}
they successfully explained the electrodynamics of superconductors alongside the Maxwell equations. With these equations, the Meissner effect can be expressed mathematically as 
\begin{equation}
    \nabla^2\bm{B}=\frac{1}{\Lambda^2} \bm{B}\,.
    \label{Meissner}
\end{equation}
Here, $\Lambda$ is the penetration depth, which determines the decay scale for an external magnetic field penetrating into the superconductor. 

Following this came Ginzburg-Landau (GL) theory \cite{Landau:1950lwq} and Bardeen-Cooper-Schrieffer theory (BCS) \cite{Bardeen:1957mv, Bardeen:1957kj} in the 1950s, the standard theoretical tools of superconductivity. The latter introduced the concept of Cooper pairing and provided a microscopic picture of superconductivity. The former is a phenomenological approach which is sometimes preferable to use over the microscopic description e.g.\ when dealing with inhomogeneous condensates for instance. Superconductors are known to fall into two categories: type-I and type-II. These can be differentiated by the behaviour of their internal magnetic field $B$ or, equivalently, by their response to an external magnetic field $H$. Assuming we are below the critical temperature i.e.\ in the superconducting phase, type-I superconductors exhibit the Meissner effect up to an external critical magnetic field $H_c$, above which they undergo a first-order phase transition to the normal conducting phase (or the ``normal'' phase). In type-II superconductors, they too begin in a phase which exhibits the Meissner effect (the ``Meissner'' phase) up to a certain critical magnetic field. At this point, instead of the superconducting condensate vanishing there is a second-order transition to a new phase where magnetic flux partially penetrates the system and the condensate becomes inhomogeneous. These magnetic defects are usually vortices sometimes also called flux tubes. As we increase the $H$ we create more vortices. Since their interaction is repulsive, they order themselves into a lattice in the plane perpendicular to the applied field. We therefore call this the ``vortex lattice'' phase. Eventually, it is more energetically favourable to allow the magnetic field to fully penetrate like in type-I, leading to another second-order transition to the normal phase. Type-II superconductors are therefore characterised by two second-order phase transitions occurring at the two different, critical, external magnetic fields, $H_{c1}$ and $H_{c2}$ often called the ``lower'' and ``upper'' critical magnetic fields respectively. We can also define the superconducting to normal transitions in terms of $B$. In type-I, the critical magnetic field $B_c$ is the value $B$ jumps to from zero after the first-order phase transition and the minimal possible $B$ in the normal phase. Similarly, one can define $B_{c2}$ as the $B$ below which the system becomes superconducting.

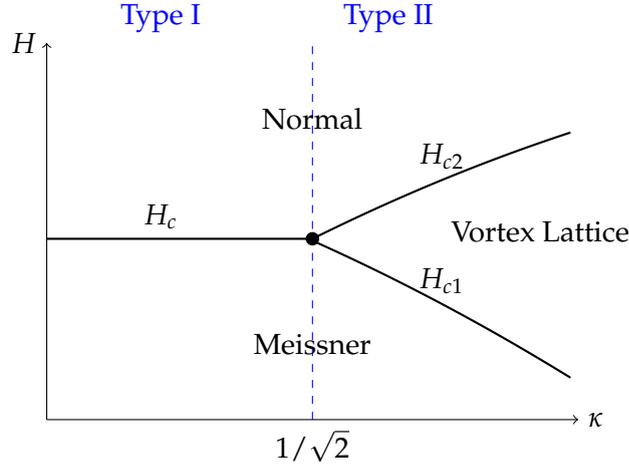
\begin{figure}
    \centering
    \begin{tikzpicture}
        \draw[->] (0.5,0) -- (7.5,0);
        \draw[->] (0.5,0) -- (0.5,5);
    
        \draw[black, thick](0.5,2.4)--(4,2.4);
        \draw[black, thick, domain=4:7.4, samples=100]
        plot(\x,{-(\x+5)^2/40+4.4});
        \draw[black, thick, domain=4:7.4, samples=100]
        plot(\x,{-(\x-14)^2/40+4.9});
        \draw[blue, dashed](4.0,0.0)--(4.0,5.0);
    
        \node[right] at (7.5,0){$\kappa$};
        \node[left] at (0.5,5.0){$H$};
        \node[below] at (4.0,0.0){$1/\sqrt{2}$};
        \node[above] at (2,2.4){\textcolor{black}{$H_{c}$}};
        \node[above] at (5.7,3.2){\textcolor{black}{$H_{c2}$}};
        \node[above] at (5.7,1.54){\textcolor{black}{$H_{c1}$}};
        \node[above] at (2,5){\textcolor{blue}{Type I}};
        \node[above] at (5,5){\textcolor{blue}{Type II}};
        \filldraw[thick, black](4,2.4) circle (0.075);
        \node at (4.0,4){Normal};
        \node at (4.0,1){Meissner};
        \node at (7.0,2.5){\textcolor{black}{Vortex Lattice}};
    
    \end{tikzpicture}    
    \caption{Schematic phase diagram in the $\kappa$-$H$ plane, demonstrating the different phase structure of type-I and type-II superconductors. We have also labelled the critical point where the transition lines meet indicating the change from first-order, to second-order phase transition(s).}
    \label{fig:type1and2}
\end{figure}

\begin{figure}
    \centering
        \begin{tikzpicture}
    
        \draw[->] (0,0) -- (7,0);
        \draw[->] (0,0) -- (0,5);
    
        \draw[blue, thick, domain=0:7, samples=100]
        plot(\x,{3.8*tanh(\x/1.5)});
        \draw[red, thick, domain=0:7, samples=100]
        plot(\x,{4.5/cosh(\x/1.5)});
    
        \draw[blue,<->] (0,2.59) -- (1.25,2.59);
        \draw[red, <->] (0,1.64) -- (2.5,1.64);
    
        \node[right] at (7,0){$r$};
        \node[above] at (6,3.8){\textcolor{blue}{$\varphi(r)$}};
        \node[above] at (1,4.0){\textcolor{red}{$B(r)$}};
        \node[above] at (0.625,2.59){\textcolor{blue}{$\xi$}};
        \node[above] at (1.25,1.64) {\textcolor{red}{$\Lambda$}};
    
    \end{tikzpicture}
    \caption{Schematic plot of a flux tube profile where the inhomogenous magnetic field (red) and condensate (blue) are plotted as a function of the radial coordinate $r$. We see that the maximum of the magnetic field is at $r=0$. It then appreciably decays over length scale $\Lambda$, the penetration depth, and vanishes at large $r$. The condensate vanishes in the core and then increases appreciably over the coherence length $\xi$ until reaching a constant value $\varphi\simeq\varphi_{\rm{hom}}$ far away from the flux tube.}
    \label{fig:Fluxtube}
\end{figure}
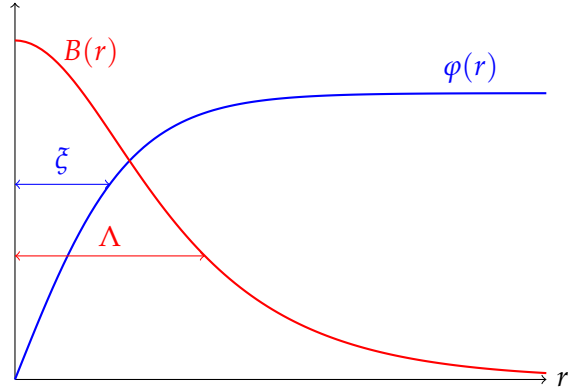

The introduction of vortices makes both the condensate and $B$ inhomogeneous, and therefore is often studied within GL theory. Indeed, the vortex lattice phase of type-II was first studied within GL by Abrikosov \cite{Abrikosov:1956sx}. Furthermore, the parameter that determines whether the superconductor is type-I or type-II is the Ginzburg-Landau parameter, $\kappa$, defined as
\begin{equation}
    \kappa = \frac{\Lambda}{\xi}\,,
    \label{GLparam}
\end{equation}
where $\xi$ is the coherence length, the natural length scale of an inhomogeneous condensate. When $\kappa<1/\sqrt{2}$ the superconductor is type-I, when $\kappa>1/\sqrt{2}$ the superconductor is type-II. A schematic phase diagram of type-I and type-II superconductors is found in Fig.\,\ref{fig:type1and2}. Furthermore, the inhomogeneity of the condensate and magnetic field is demonstrated by the flux tube profile in Fig.\,\ref{fig:Fluxtube}. A common theme of the work presented in this thesis is superconducting phase transitions and magnetic defects, in particular flux tubes. As a result, we will in essence employ the GL framework throughout to analyse the phase structure and the properties of these defects. Thus, this section outlines GL theory of type-I and type-II superconductors, acting as a warm-up for Chapters \ref{chpt:Project1} and \ref{chpt:Project2}. In Sec.\,(\ref{Background:subsec:GL}), we derive the equations of motion and free energy beginning from a Lagrangian in classical field theory formalism. The following sections, Sec.\,(\ref{Background:subsec:Hc}), (\ref{Background:subsec:Hc1}) and (\ref{Background:subsec:Hc2}) will discuss the critical magnetic fields $H_c$, $H_{c1}$ and $H_{c2}$ respectively. The last section will also determine the preferred configuration of the vortex lattice and run through some details of the calculations first done in Refs.\,\cite{Abrikosov:1956sx,PhysRev.133.A1226} which we will use in Chapter \ref{chpt:Project2}.

\subsection{Ginzburg-Landau theory}
\label{Background:subsec:GL}

The fundamental assumption of Landau theory \cite{Landau:1937obd} is that one can express the free energy of the system as a power series of an order parameter $\varphi$ and its gradient $\nabla\varphi$ near a second-order phase transition. During a second-order phase transition, $\varphi$ is expected to continuously go to zero and is small in its vicinity, allowing one to perform an expansion in $\varphi$. One writes down the most general free energy, disregarding terms that do not obey the symmetries of the system. GL theory is the Landau theory for superconductors. While Landau theory is phenomenological and in general not derived from first-principles, the GL theory can in fact be derived from the microscopic BCS theory, which perhaps explains its success. It is an expansion of the free energy near the critical temperature $T_c$ and the GL free energy is temperature dependent. One can therefore analyse the second order temperature transition within GL. However, we will be interested in the transitions due to magnetic fields, assuming we are at zero temperature and will extrapolate results from this limit. Our starting point will be writing down the expansion in terms of a Lagrangian rather than the free energy. To this end, the order parameter(s) will be the expectation value of a complex scalar field(s). If we interpret the density of Cooper pairs as being proportional to $|\varphi|^2$, then it is clear that when this vanishes, so does the superconducting state.

We consider the following Lagrangian for a complex scalar field $\varphi$
with mass $m$, electric charge $e$, and coupling constant\footnote{The symbol $\lambda$ is usually reserved for the penetration depth in conventional superconductivity but is also commonly used as the coupling constant in field theory. In an attempt to avoid confusion, we have chosen the upper case $\Lambda$ for the penetration depth and lower case $\lambda$ for the coupling constant.} $\lambda$,
\begin{equation}
    \mathcal{L}=D_{\mu}\varphi \left(D^{\mu}\varphi \right)^* - m^2|\varphi|^2 -\lambda|\varphi|^4 +\mathcal{L}_{\rm{em}}\,,
    \label{ConventionalLagrangian}
\end{equation}
where the covariant derivative is $D^{\mu} \equiv \partial^{\mu}+ ieA^{\mu}$ with the electromagnetic gauge field $A^\mu$. The electromagnetic part is 
\begin{equation}\label{Lem}
    {\cal L}_{\rm em} = -\frac{1}{4}F_{\mu\nu}F^{\mu\nu} \, ,
\end{equation} 
where $F^{\mu\nu}=\partial^\mu A^\nu - \partial^\nu A^\mu$ is the electromagnetic field strength tensor. The equations of motion for $\varphi^*$ and $A^{\mu}$ are
\begin{subequations}
    \begin{align}
        \left(D_{\mu}D^{\mu} +m^2 +2\lambda|\varphi|^2\right)\varphi&=0\,,
        \label{varphiEoM}
        \\[2ex]
        \partial_{\mu}F^{\mu\nu} &=j^{\nu}\,,
    \end{align}
\end{subequations}
where 
\begin{equation}
    j^{\nu}= ie\left(\varphi^*\partial^{\nu}\varphi -\varphi\partial^{\nu}\varphi^*\right) -2e^2A^{\nu}|\varphi|^2\,,
\end{equation}
is the electromagnetic four-current. Condensation of the complex field is induced by a chemical potential $\mu$, which we introduce via the temporal component of the gauge field, $A^{\nu}=(\mu/e,\bm{A})$. Here, $\mu$ is the chemical potential coupled to the global $U(1)$ symmetry with the associated unit charge $e$. Although here we have kept the starting point general, it will be useful to make the comparison back to the conventional superconductor, where the Cooper pairs are formed of two electrons and we have a lattice of positive ions.

In the static limit the equations of motion become 
\begin{subequations}
    \begin{align}
        0&=\left( K  -2\lambda|\varphi|^2\right)\varphi\,, \label{varphiEoM2}
        \\[2ex]
        \nabla\cdot\bm{E}  &= \rho= -2e^2\mu|\varphi|^2\,,
        \label{EoMDivE}
        \\[2ex]
        \nabla\times\bm{B} &= \bm{j} =  -ie\left(\varphi^*\nabla\varphi -\varphi\nabla\varphi^*\right) -2e^2\bm{A}|\varphi|^2  \, , 
        \label{EoMCurlB}
    \end{align}
    \label{EoM2}%
\end{subequations}
where
\begin{equation}
    K \equiv \nabla^2-2ie\bm{A}\cdot\nabla-ie\nabla\cdot\bm{A}-e^2A^2+\mu^2-m^2 \, . 
    \label{Koperator}
\end{equation}
We will ignore the contributions of fluctuations to the thermodynamics, so the grand canonical potential density is simply given by $\Omega=-{\cal L}$. This gives
\begin{equation}
    \Omega (\bm{x}) =  |\left(\nabla -ie\bm{A}\right)\varphi|^2 -\left(\mu^2- m^2\right)|\varphi|^2 +\lambda|\varphi|^4 + \frac{B^2}{2}\,,
    \label{ThermoPot}
\end{equation}
which in general can be a function of space and depend on the set of coordinates $\bm{x}$. Equation \eqref{EoMCurlB} is a generalisation of the London equation \eqref{London}. With solutions $\varphi(\bm{x})=|\varphi|e^{i\phi(\bm{x})}$ such that $|\varphi|^2$ is homogeneous, taking the curl of $\bm{j}$ reproduces the London equation \eqref{LondonB}. By comparing these equations we find
\begin{equation}
    \Lambda=\frac{1}{\sqrt{2}e\varphi_{\rm{hom}}}\,,
    \label{Lambda}
\end{equation}
where $\varphi_{\rm{hom}}=|\varphi|$ is the magnitude of the homogeneous condensate. An important thing to note is that due to our choice of $A^{\nu}$
\begin{equation}
    \bm{E}=-\nabla A^0 -\partial_t\bm{A}=\bm{0}\,,
    \label{vanishE}
\end{equation}
since we have taken $A^0=\mu/e$ as constant and $\bm{A}$ to be time independent. This is the reason there is no electric contribution to Eq.\,\eqref{ThermoPot}. However, Eq.\,\eqref{vanishE} conflicts with Eq.\,\eqref{EoMDivE}. To fix this, we assume the presence of a background charge cancelling the charge of the complex scalar field, such that there is no net electric field. Thus, there would be another contribution to  Eq.\,\eqref{EoMDivE} which makes the right-hand side $0$. In the conventional scenario, the lattice of positive ions neutralises the combined charge of the Cooper pairs, leading to a zero net charge. Essentially, it is a statement of charge neutrality. The free energy density can be written as 
\begin{equation}
    \mathcal{F}=\frac{1}{V}\int d^3\bm{x}\, \Omega(\bm{x}) = \frac{1}{V}\int d^3\bm{x}\, \left( \frac{B^2}{2} -\lambda|\varphi|^4\right)\,, 
    \label{F}
\end{equation}
where $V$ is the volume of the system, and we consider the limit $V\rightarrow\infty$. In the last step we have derived a general expression by using the equation of motion \eqref{varphiEoM2} and dropping surface terms. It will be convenient to calculate the Gibbs free energy density 
\begin{equation}
    \mathcal{G}= \mathcal{F} -\frac{1}{V}\int d^3\bm{x}\,  \bm{H}\cdot\bm{B}\, ,
    \label{Gibbs}
\end{equation}
in some situations, where $\bm{H}$ is the vector external magnetic field with magnitude $H$. This will allow us to study the phase structure as we vary the external magnetic field $H$ rather than $B$.

\subsection{The Meissner phase and critical field $H_c$}
\label{Background:subsec:Hc}

To start with, we look at the two simplest scenarios; the Meissner and normal phase. In the former we have a spatially homogeneous condensate and no magnetic or gauge field. The homogeneous thermodynamic potential $\Omega_{\rm{hom}}$, is then
\begin{equation}
    \Omega_{\rm{hom}} = -\left(\mu^2- m^2\right)|\varphi|^2 +\lambda|\varphi|^4 \,.
    \label{OmegaHom}
\end{equation}
Solving the equation of motion \eqref{varphiEoM2}, either $\varphi=0$ or
\begin{equation}
    |\varphi|^2 = \frac{\mu^2-m^2}{2\lambda}\,.
\end{equation}
Inserting both solutions into the free energy \eqref{F}, we see that the second gives a lower result provided $\mu>m$. With $\varphi=0$ there is no superconductivity, thus we conclude that we have a superconducting phase only when $\mu>m$ and
\begin{equation} \varphi = \sqrt{\frac{\mu^2-m^2}{2\lambda}}e^{i\delta}\equiv \varphi_{\rm{hom}}e^{i\delta}\,, \label{varphi0}
\end{equation}
with phase $\delta$ and associated free energy
\begin{equation}
    \mathcal{F}_{\rm{hom}}\equiv-\frac{\left(\mu^2-m^2\right)^2}{4\lambda}\,.
\end{equation}
Here we have defined $\varphi_{\rm{hom}}$ (first appearing Eq.\,\eqref{Lambda}) and $\mathcal{F}_{\rm{hom}}$ as the magnitude of the homogeneous solution and free energy of the homogeneous phase respectively. 

Sketching the potential \eqref{OmegaHom} as a function of the real and imaginary components of $\varphi$, the minimum is at $\varphi=0$ for $\mu<m$. For $\mu>m$, the minimum is a circle described by solution \eqref{varphi0} for $\mu>m$ . When $\mu>m$, this potential is commonly referred to as the ``Mexican Hat'' potential due to its shape and is often used pedagogically to explain spontaneous symmetry breaking, which has already been discussed in previous sections to some extent. Here we will give a brief discussion. Although the Lagrangian density \eqref{ConventionalLagrangian} is invariant under the $U(1)$ transformation $\varphi\rightarrow \varphi e^{i\phi}$, the ground state \eqref{varphi0} is not. Graphically, this transformation is the same as moving an angle $\phi$ along the circle at the minimum of the potential. The potential has the same value but we have moved to a different ground state. While all ground states have equal free energy, the system can only be in one of them. Thus, the solution is said to ``spontaneously'' choose a point along this minimum at the point when $\mu>m$, ``spontaneously'' breaking the symmetry. Superconductivity is therefore an example of spontaneous symmetry breaking. Other examples include the aforementioned chiral symmetry breaking with massless quarks. Just like the chiral condensate breaks the chiral symmetry, $\varphi$ breaks the $U(1)$ symmetry of our system. The field $\varphi$ is said to condense for $\mu>m$. As this field is bosonic, it is an example of Bose-Einstein Condensation (BEC).  

The normal phase is the opposite scenario, where there is no superconducting condensate and the magnetic field fully penetrates. Thus, the free energy density of the normal phase is just
\begin{equation}
    \mathcal{F}_{\rm{nor}}\equiv\frac{B^2}{2}\,.
\end{equation}
To determine the critical magnetic field $H_c$, we must compare the difference in Gibbs free energy. By minimising the free energy with respect to $B$ we find 
\begin{equation}
    \mathcal{G}_{\rm{hom}}\equiv\mathcal{F}_{\rm{hom}}\,, \qquad \mathcal{G}_{\rm{nor}}\equiv-\frac{H^2}{2}\,.
    \label{Ghom+nor}
\end{equation}
The normal phase is preferred over the Meissner phase when $\mathcal{G}_{\rm{hom}} \geq\mathcal{G}_{\rm{nor}}$, from which we find the critical field
\begin{equation}
    H _{c}\equiv\frac{\mu^2-m^2}{\sqrt{2\lambda}}=\sqrt{2\lambda}\varphi_{{\rm{hom}}}^2\,,
    \label{Hc}
\end{equation}
above which we are in the normal phase. Note that the normal phase is automatically preferred when $\mu<m$. The competition between the two only matters when we have the non-trivial solution for $\varphi$ and the field condenses. When minimising the Gibbs free energy, we find $B=H$ in the normal phase and therefore $B_c=H_c$. The equivalence of the critical fields in $B$ and $H$ is due to the fact that the homogeneous phase and normal phase both coexist at the first-order transition.

\subsection{Asymptotics of single vortex solutions and lower critical field $H_{c1}$}
\label{Background:subsec:Hc1}

We now move on to situations where we allow the condensate and the gauge field (and thus the magnetic field) to be inhomogeneous. This is relevant for the type-II regime where $\kappa>1/\sqrt{2}$. This section will discuss $H_{c1}$ and solving the equations of motion for a single flux tube. In type-II superconductors above $H_{c1}$, instead of entering the  normal phase, we expect the magnetic field to partially penetrate the superconductor via vortices. At the critical field, we expect a single vortex to enter the system through which the magnetic field partially penetrates. In terms of free energy, this happens at the point when the Gibbs free energy density of the system with a single vortex becomes equal to $\mathcal{G}_{{\rm{hom}}}$, the Gibbs free energy density in the Meissner phase. The Gibbs free energy density of the system with a single vortex can be expressed in terms of free energy densities $\mathcal{F}$ as
\begin{equation}
    \mathcal{G}=\mathcal{F}_{{\rm{hom}}} + \mathcal{F}_{\rm{vort}} -\frac{1}{V}\int d^3\bm{x}\, \bm{H}\cdot\bm{B}\,,
    \label{Gvort}
\end{equation}
where $\mathcal{F}_{\rm{vort}}$ is the free energy of the vortex. Here we have separated the contribution from the vortex $\mathcal{F}_{\rm{vort}}$ and from the remainder of the superconductor. Away from the vortex, the superconductor is essentially in the Meissner phase and so we can use $\mathcal{F}_{{\rm{hom}}}$ from the previous section. Since the magnetic field is non-zero in the presence of a vortex, the Legendre transformation term from Eq.\,\eqref{Gibbs} remains. Setting Eq.\,\eqref{Gvort} equal to $\mathcal{G}_{{\rm{hom}}}=\mathcal{F}_{{\rm{hom}}}$, we can re-arrange for $H_{c1}$. Assuming cylindrical symmetry with $\bm{B}$ and $\bm{H}$ aligned with the $z$-axis,
\begin{equation}
    H_{c1}= \frac{V\mathcal{\mathcal{F}_{\rm{vort}}}}{L_{z}\Phi}\,,
    \label{Hc1cond}
\end{equation}
with
\begin{equation}
    \int_S d\bm{S}\cdot \bm{B} = \oint_{s} d\bm{s} \cdot \bm{A} = \Phi\,,
    \label{TotalFlux}
\end{equation}
where we have adopted cylindrical coordinates $(r,\theta,z)$ in anticipation of the vortex geometry and used Stokes' theorem. In the above, $\Phi$ is the total magnetic flux, $L_{z}$ is the length of the system in the $z$-direction, and $S$ is the surface of the plane perpendicular to the vortex. The closed curve $s$ is around the vortex. Both $s$ and $S$ should then in principle be taken at $r=\infty$ to capture all the magnetic flux in the system since $V\rightarrow\infty$.

We must determine $\mathcal{F}_{{\rm{vort}}}$ to find $H_{c1}$. Usually, one would solve the equations of motion to find $\varphi(\bm{x})$ and $\bm{A}(\bm{x})$ and evaluate this free energy. However, the equations of motion are not analytically solvable in general, and this remains the case for a single vortex. One can determine $H_{c1}$ in the limit where we are deep in the superconducting phase and $\kappa\gg 1$ and $\Lambda\gg\xi$. The superconductor can then be treated as mostly homogeneous except for the vortex with a very small core $0<r<\xi$. Running through the calculation will not be particularly useful for the purposes of the following chapters. For the sake of completeness, we quote the result 
\begin{equation}
     H_{c1}=  \frac{\Phi_0}{4\pi \Lambda^2} \ln{\kappa} =\frac{H_c}{\sqrt{2}\kappa}\ln{\kappa}\,, \quad \kappa\gg 1\,,
    \label{Hc1}
\end{equation}
and refer those interested to Ref.\,\cite{tinkham2004introduction} for more details. Here, $\Phi_0$ is the unit magnetic flux through a vortex which we will determine later in this section. It is expressed in terms of $H_c$ and $\kappa$ using Eqs.\,\eqref{Lambda}, \eqref{Hc} and \eqref{GLparam2} and shows $H_{c1}$ is a decreasing function of $\kappa$ consistent with Fig.\,\ref{fig:type1and2}.

For the remainder of this section, we will look at asymptotic solutions of the equations of motion. In doing this we will derive some useful relations and gain some insight into the behaviour of the full solutions which is useful for the implementation of a numerical computation. We use the ansatz
\begin{equation}
    \varphi(r,\theta) = \varphi_{\rm{hom}} f(r) e^{i n\theta}\,, \qquad \bm{A}= A_{\theta}(r)\hat{\bm{e}}_{\theta}\,,
    \label{ansatzHc1}
\end{equation}
where $n$ is an integer called the ``winding number''. By construction, this choice of gauge field yields $\bm{B}=B(r)\bm{\hat{e}}_z$ and $\nabla\cdot \bm{A}=0$. We impose the boundary conditions $f(0)=0$ and $f(\infty)=1$ such that in the core of the flux tube the superconducting condensate is zero and far away from the superconductor $|\varphi|^2=\varphi_{\rm{hom}}^2$. Furthermore, $B(\infty)=0$  such that the magnetic field should vanish far away from the vortex where it is expected to be almost completely expelled. These boundary conditions reflect the schematic picture in Fig.\,\ref{fig:Fluxtube}. First looking at the gauge field, using our ansatz the equation of motion \eqref{EoMCurlB} becomes 
\begin{equation}
    \left(\partial_{\bar{r}}^2 +\frac{1}{\bar{r}}\partial_{\bar{r}} -\frac{1}{\bar{r}^2} -f^2\right)A_{\theta} = -\frac{n}{e\Lambda \bar{r}}f^2\,,
    \label{EoMA}
\end{equation}
where we have rescaled $r=\Lambda \bar{r}$ using the penetration depth from Eq.\,\eqref{Lambda}. In the limit $r\rightarrow\infty$, Eq.\,\eqref{TotalFlux} implies that
\begin{equation}
    A_{\theta}(r)\sim\frac{\Phi}{2\pi r}\,.
    \label{InftyA}
\end{equation}
This asymptotic form can be substituted into Eq.\,\eqref{EoMA} to check that it obeys the equation of motion within this limit. We find that it does provided $\Phi=2\pi n /e$\,. This is the total flux of the vortex with winding number $n$. Setting $n=1$, we define the unit flux of a vortex in Eq.\,\eqref{Hc1}
\begin{equation}
    \Phi_0 \equiv \frac{2\pi}{e}\,,
    \label{FluxQuantum}
\end{equation}
 also known as the flux quantum. In the opposite limit $r\rightarrow 0$, it can be shown that 
\begin{equation}
    A_{\theta}(r) \sim \frac{B(0)}{2} r
    \label{CoreA}
\end{equation}
by using a power series solution in Eq.\,\eqref{EoMA}. We have determined the integration constant by using $\bm{B}=\nabla\times\bm{A}$. 

We can also analytically determine the asymptotic behaviour of $f(r)$ from the differential equation
\begin{equation}
    \left\{ \partial_{\tilde{r}}^2 +\frac{1}{\tilde{r}}\partial_{\tilde{r}} -\left[\frac{n}{\tilde{r}} - \xi eA_{\theta}(\tilde{r})\right]^2 + 1-f^2\right\}f=0\,,
\end{equation}
obtained from Eq.\,\eqref{varphiEoM2}. Note, here we have again re-scaled such that $r= \xi\tilde{r}$ where
\begin{equation}
    \xi\equiv\frac{1}{\sqrt{2\lambda}\varphi_{\rm{hom}}}\,,
    \label{xi}
\end{equation}
defining the coherence length from Eq.\,\eqref{GLparam}. With this, we can show using Eqs.\,\eqref{Lambda} and \eqref{GLparam} that
\begin{equation}
    \kappa = \frac{\sqrt{\lambda}}{e}\,.
    \label{GLparam2}
\end{equation}
According to our boundary condition, at large $r$ we expect $f(r)$ to be near unity, so we expand $f(r)\approx 1 +g(r)$ where $g(r)$ is a small perturbation. Using the asymptotic form of $A_{\theta}(r)$ from Eq.\,\eqref{InftyA} and neglecting terms of $\mathcal{O}(g^2)$ the differential equation becomes
\begin{equation}
    r'^{\,2}\partial_{r'}^2g +r'\partial_{r'}g -  r'^{\,2} g=0\,,
    \label{BesselEqn}
\end{equation}
where we have re-scaled (again) with $r'=\sqrt{2}\tilde{r}$ and placed it in the form of the modified Bessel's equation. The solution we can then immediately write:
\begin{equation}
    g(r') = C K_0(r')\,,
\end{equation}
where $K_n$ are the modified Bessel functions of the second kind and $C$ is a constant. The full solution also includes the modified Bessel functions of the first kind but we discard these as they diverge in the limit $r\rightarrow\infty$. In this same limit, $K_0(r')\sim \sqrt{\pi/2r'}e^{-r'}$, thus
\begin{equation}
    f(r)\approx 1 +C \sqrt{\frac{\pi\xi}{2\sqrt{2} r}}e^{-\sqrt{2} r/\xi} \,,
    \label{assymbf}
\end{equation}
for large $r$. While $C$ needs to be determined numerically, clearly $f(r)> 1$ if $C$ is positive, yet the maximum should be $f(\infty)=1$. So we can at least deduce $C<0$.

For small $r$, near the core of the vortex, we try a power series solution for $f(r)$ of the form
\begin{equation}
    f(\tilde{r})=\sum_{m=0}^{\infty} c_{m}\tilde{r}^{m}\,,
\end{equation}
and substitute in the asymptotic expression \eqref{CoreA}. After simplifying, we get the following relation,
\begin{equation}
    \sum_{m=0}^{\infty} \left\{\frac{m^2-n^2}{\tilde{r}^2}+ \xi^2eB(0)n + 1 -\left[\frac{\xi^2 e B(0) }{2}\right]^2\tilde{r}^2\right\}c_m \tilde{r}^m + f^3  = 0\,.
    \label{reoccurance1}
\end{equation}
Note that we have not substituted the expansion in for $f^3$. In principle, there are terms from $f^3$ that contribute at lowest order, but we will see that these become irrelevant. The term $\left(m^2-n^2\right)/r^2$ within the sum in Eq.\,\eqref{reoccurance1} is divergent at $r=0$ unless $m=n$ for $m=0,1$. Unavoidably, $c_0=0$ if we seek solutions where $n\neq0$ (the situation where $n=0$ would mean no vortex or magnetic flux). The same can be said for $c_1$ unless $n=1$. This can be generalised to larger $n$. While there is no divergence for $m>1$, the $\left(m^2-n^2\right)/r^2$ term within the sum in Eq.\,\eqref{reoccurance1} will always be the lowest order term in $r$. As a result, all coefficients $c_m=0$ for $m<n$ and the first non-zero coefficient will be $c_n$, which we cannot determine analytically. Therefore, 
\begin{equation}
    f(\tilde{r})= c_n \tilde{r}^n + \dots\,,
    \label{fn}
\end{equation}
in general. 
In conventional superconductors, $n=1$ is preferred. This is plausible from our asymptotic solutions. The free energy in Eq.\,\eqref{F}, has two contributions. There is a positive $B^2$ contribution and a negative $|\varphi|^4\propto f^4$ contribution. At large $r$, we see from our asymptotic solution \eqref{InftyA} that $B(r)\propto n$ and $f(r)$ does not depend on $n$. For the lowest free energy, we therefore want the lowest $n$ as $B$ contributes positively to the free energy. At small $r$, $f^4\sim r^{4n}$. Since this is a negative contribution this would lower the free energy most when $n$ is small for small $r$. Our asymptotic solution for the gauge field \eqref{CoreA} tells us the magnetic field is approximately constant with no (explicit) $n$ dependence at small $r$. Therefore, from our asymptotic solutions at small and large $r$, smaller winding number gives a lower free energy. Of course, this is not conclusive as the free energy is an integral over all the whole system and we don't know the $n$-dependence of the condensate and magnetic field at intermediate $r$. One could also argue that the undetermined constants could change with $n$. Nonetheless, it shows that lower winding numbers yielding lower free energies is plausible without doing a full numerical computation. The lowest value for $n$ where we have non-trivial solutions is $n=1$. However, we will see later in Chapter \ref{chpt:Project1} that this can change when we consider more complicated and exotic systems. Focusing on the conventional case, we set $n=1$ and proceed from there. With the divergences dealt with, we can proceed to determine the next term in the expansion from
\begin{equation}    
\left\{\left[(m+2)^2-1\right]c_{m+2}+\left(\xi^2 eB(0) +1\right)c_m \right\} \tilde{r}^m +\mathcal{O}(\tilde{r}^3) = 0\,,
\label{reoccurance2}
\end{equation}
where we have shifted the index $m$ in the previously divergent term within the sum in Eq.\,\eqref{reoccurance1}. It can be appreciated from this recurrence relation that all even $c_m$ are going to vanish due to $c_0=0$ and all odd $c_n$ can be expressed in terms of $c_1$. Looking at terms of order $r$, we use Eq.\,\eqref{reoccurance2} to determine $c_3$ and find
\begin{equation}
    f(r)\approx c_1 r\left\{1 -\left[1 +\xi^2eB(0)\right]\frac{r^2}{8\xi^2}\right\}\,,
    \label{assymbf0}
\end{equation}
where we have absorbed a factor of $\xi$ into $c_1$. To find higher order terms, the terms of order $r^3$ become relevant i.e.\ the last two terms of Eq.\,\eqref{reoccurance1} must also be considered (with $n\geq 1$, $f^3\sim r^{3n}$). We see that both asymptotic expressions \eqref{assymbf} and \eqref{assymbf0} have constants that we must determine numerically.

\subsection{The upper critical field $H_{c2}$ and the Abrikosov Vortex Lattice}
\label{Background:subsec:Hc2}

Finally, we derive expressions for $B_{c2}$ and $H_{c2}$ and show that the vortex lattice phase is preferred over the normal phase below $B_{c2}$. Then, we will proceed to determine the configuration of the vortices which minimises the free energy. These tasks follow the work of Ref.\,\cite{Abrikosov:1956sx,PhysRev.133.A1226}. For this section we will work in Cartesian coordinates. Although clearly preferable when dealing with a single vortex, cylindrical coordinates are less likely to be advantageous when analysing many vortices arranged in a lattice. 

Starting from the normal phase, at the critical field $B_{c2}$ we expect a continuous transition from the non-condensed phase $\varphi=0$ to a superconducting phase. To determine this transition, we temporarily restore the time dependence and linearise around $\varphi=0$. With the ansatz $\varphi(t,\bm{x})=e^{i\omega t}f(\bm{x})$ this allows us to compute the dispersion relation of the fluctuations in the non-superconducting state in the presence of the magnetic field. The equation of motion \eqref{varphiEoM} becomes 
\begin{equation}
    \left(\omega+ \mu\right)^2f(\bm{x}) = -\left(\nabla^2 -2ie\bm{A}\cdot\nabla -ie\nabla\cdot\bm{A}  -e^2A^2 -m^2\right)f(\bm{x})\,.
\end{equation}
Aligning the $z$-axis with the magnetic field, $\bm{B}=B\hat{\bm{e}}_z$, we may choose the gauge 
$\bm{A}=Bx\hat{\bm{e}}_y$ (obeying $\nabla\cdot\bm{A}=0$) and make the ansatz $f(\bm{x})=e^{ik_{y}y}e^{ik_z z}\psi(x)$ to obtain 
\begin{equation}
    \left[\left(\omega+\mu\right)^2-k_z^2-m^2\right]\psi(x) = \left[-\partial_{x}^2 +e^2B^2\left(x-\frac{k_y}{eB}\right)^2\right]\psi(x)\,.
    \label{omegapsi}
\end{equation}
This equation has the form of the Schr\"{o}dinger equation for the one-dimensional harmonic oscillator and its solution gives the usual Landau levels labelled by the integer $\ell$, 
\begin{equation}
    \omega = \sqrt{\left(2\ell+1\right)eB +m^2 +k_{z}^2}-\mu\,.
\end{equation}
This energy  is positive for all $\ell$ and $k_z$ if $B$ is sufficiently large. A negative energy, and thus the indication of an instability, first occurs for $\ell=k_z=0$ at the critical field
\begin{equation}
   B_{c2}  \equiv \frac{\mu^2-m^2}{e} = \frac{1}{e\xi^2}\, .
   \label{bc2}
\end{equation}
This implies the system is unstable to the condensation of the charged condensate. Furthermore, we can show that
\begin{equation}
    B_{c2}=\sqrt{2}\kappa B_c\,,
    \label{Bc2Bc}
\end{equation}
using Eqs.\,\eqref{Hc} and \eqref{GLparam2}. Similar to $B_c$, $B_{c2}$ can be interpreted as the field below which the system goes superconducting. Therefore, approaching from the normal phase, $B=H$ and $B_{c2}=H_{c2}$. As a result, the above relation holds with $B_{c2}$ and $B_c$ switched for $H_{c2}$ and $H_c$ respectively. The external upper critical magnetic field is therefore an increasing function of $\kappa$ and is consistent with the phase diagram sketch in Fig.\,\ref{fig:type1and2}. Once again we require $\mu>m$, for $B_{c2}$ to be positive. Otherwise $B_{c2}<0$, signalling that the normal phase is preferred everywhere.

Due to the instability, we thus expect a superconducting phase with a charged condensate to take over when $B<B_{c2}$. To construct this phase just below $B_{c2}$ we employ an expansion in $\epsilon\sim\sqrt{B_{c2}-B}$,
\begin{equation}
    \varphi  = \varphi_0 +\delta\varphi +\ldots \,, \qquad \bm{A} = \bm{A}_0 +\delta\bm{A} + \ldots \,,
\end{equation}
where $\varphi_0$ and $\delta\varphi$ are of order $\epsilon$ and $\epsilon^3$, while $\bm{A}_0$ and $\delta\bm{A}$ are of order 1 and $\epsilon^2$. The logic is that near the second-order phase transition to the normal phase, $\varphi$ is very small and $B\approx B_{c2}$. Hence, for $\epsilon\to 0$ we approach the critical field and need $\nabla \times\bm{A}_0 = \bm{B}_{c2}$, which we can satisfy with $\bm{A}_0 = x B_{c2}\hat{\bm{e}}_y$. We may thus write the expansion of the magnetic field as
\begin{equation}
    \bm{B} = \bm{B}_{c2} + \delta\bm{B} + \ldots \, , 
\end{equation}
with $\nabla\times \delta\bm{A} = \delta\bm{B}$. The $\varphi^*$ equation of motion \eqref{varphiEoM} yields the order $\epsilon$ and $\epsilon^3$ equations
\begin{subequations}
    \begin{align}
        K_0 \varphi_0&=0 \, , \label{appBphi0EoMEx} \\[2ex]
        K_0 \delta\varphi &= \left(2ie\delta\bm{A}\cdot\nabla+2e^2\bm{A}_0\cdot\delta\bm{A}+ie\nabla\cdot\delta\bm{A}+2\lambda|\varphi_0|^2\right)\varphi_0 \, ,  \label{appBdeltaphiEoMEx}
    \end{align}
\end{subequations}
where $K_0$ is the operator $K$ \eqref{Koperator} with $\bm{A}$ replaced by its lowest order contribution $\bm{A}_0$. The lowest-order contribution to the equation of motion for $\bm{A}$ \eqref{varphiEoM2} simply gives $\nabla\times \bm{B}_{c2}=0$, which is trivially solved since the magnetic field is constant at (and above) the critical value. The order $\epsilon^2$ contribution gives
\begin{equation}
\nabla\times\delta\bm{B} = -ie\left(\varphi_0^*\nabla\varphi_0 -\varphi_0\nabla\varphi_0^*\right) -2e^2\bm{A}_0|\varphi_0|^2 \, 
\label{appBdeltaAEoMEx}.
\end{equation}
For later, it is useful to combine Eq.\,\eqref{appBdeltaphiEoMEx} with Eq.\,\eqref{appBdeltaAEoMEx} as follows. We multiply Eq.\,\eqref{appBdeltaphiEoMEx} from the left with $\varphi^*_0$ and multiply Eq.\,\eqref{appBdeltaAEoMEx} with $\delta\bm{A}$. In both resulting equations we have created a term $2e^2\bm{A}_0\cdot\delta \bm{A}|\varphi_0|^2$, and so we can insert one equation into the other to obtain
\begin{equation} \label{phiDphi}
    \varphi^*_0K_0\delta\varphi = ie\nabla\cdot(\delta\bm{A}\,|\varphi_0|^2)-\delta\bm{A}\cdot(\nabla\times\delta\bm{B})+2\lambda|\varphi_0|^4 \, .
\end{equation}
With partial integration,  dropping the surface term, and using the equation of motion $K_0^*\varphi_0^*=0$, the integral over the left-hand side vanishes, 
\begin{equation}
    \int d^3\bm{x} \, \varphi_0^*K_0\delta\varphi = 0\,.
\end{equation}
Consequently, the integral over the right-hand side of Eq.\,\eqref{phiDphi} must vanish as well. Dropping the boundary term, this yields the useful relation 
\begin{equation} \label{appBorthogonal}
    0=\int d^3\bm{x}\,\left[2\lambda|\varphi_0|^4-\delta\bm{A}\cdot(\nabla\times\delta\bm{B})\right]  \, . 
\end{equation}

To solve the equations of motion explicitly we first note that Eq.\,\eqref{appBphi0EoMEx} can be brought in the form of Eq.\,\eqref{omegapsi} with $\omega=0$ and $B=B_{c2}$ and, assuming no variation in the $z$-direction, $k_z = 0$. The solution for the lowest Landau level, $\ell=0$,  is a Gaussian, and in order to construct periodic solutions we set $k_y=n q$, where $n$ is an integer, and consider the superposition of Gaussians
\begin{equation} \label{phi0xy}
    \varphi_0(x,y) =\sum_{n=-\infty}^{\infty}C_{n}e^{in qy}\psi_n(x)\,, \qquad \psi_n(x) = e^{-\frac{(x-x_n)^2}{2\xi^2}} \, , 
\end{equation}
with complex coefficients $C_n$ and $x_n \equiv n q\xi^2$. We have also used Eq.\,\eqref{bc2} to express our solutions in terms of $\xi$. Next, we need to compute $\delta\bm{A}$ and thus $\delta\bm{B}$ from the equation of motion \eqref{appBdeltaAEoMEx}. We will see that it is consistent to restrict the correction to the gauge field to the $y$-direction, $\delta \bm{A}=\delta A_y(x,y)\hat{\bm{e}}_y$, and we can write $\delta\bm{B} = \delta B(x,y)  \,\hat{\bm{e}}_z$. One first derives the following useful identities with the help of the explicit solution \eqref{phi0xy},
\begin{subequations} \label{iphi0}
    \begin{align}
        i(\varphi_0^*\partial_x\varphi_0-\varphi_0\partial_x\varphi_0^*)&= \partial_y|\varphi_0|^2 \, , \\[2ex]
        i(\varphi_0^*\partial_y\varphi_0-\varphi_0\partial_y\varphi_0^*)+2e xB_{c2}|\varphi_0|^2 &= -\partial_x|\varphi_0|^2 \, .
    \end{align}
\end{subequations}
Consequently, the non-trivial components of Eq.\,\eqref{appBdeltaAEoMEx} take the simple form
\begin{subequations}
    \begin{align}
        \partial_y\partial_x \delta A_y&= -e\partial_y|\varphi_0|^2 \, , \\[2ex]
        \partial_x^2\delta A_y &= -e\partial_x|\varphi_0|^2 \, .
    \end{align}
\end{subequations}
The first equation gives $\partial_x\delta A_y = -e|\varphi_0|^2+{\rm const}$, and the second equation implies that the integration constant is indeed a constant that does not depend on $x$. We express the integration constant in terms of the spatial average of the magnetic field $\langle B\rangle$,  which we choose as our independent thermodynamic variable. We cannot use $B$ since it is now space dependent. To this end, we define the spatial average of a space dependent function $f(\bm{x})$ as
\begin{equation}
    \langle f(\bm{x})\rangle \equiv \frac{1}{V}\int_{V} d^3\bm{x}\, f(\bm{x})\,.
    \label{avdef}
\end{equation}
Requiring $\langle B\rangle = B_{c2}+\langle \delta B \rangle$, we obtain
\begin{equation}
\delta B = \partial_x\delta A_y = \langle B\rangle - B_{c2}+e\left(\langle|\varphi_0|^2\rangle-|\varphi_0|^2 \right) \, .
\end{equation}
We can use this expression to compute 
\begin{equation}
\delta\bm{A}\cdot(\nabla\times\delta\bm{B}) = -e\left(\langle B\rangle - B_{c2} +e \langle|\varphi_0|^2\rangle\right) |\varphi_0|^2 +e^2 |\varphi_0|^4 +\mbox{total derivatives} \, .
\end{equation}
Inserting this result into Eq.\,\eqref{appBorthogonal} and dropping the boundary terms gives
\begin{equation}
    e\langle |\varphi_0|^2\rangle = \frac{B_{c2}-\langle B\rangle}{\left(2\kappa^2-1\right)\beta +1}\,,
    \label{varphi0BKbeta}
\end{equation}
where
\begin{equation}
    \beta \equiv \frac{\langle|\varphi_0|^4\rangle}{\langle|\varphi_0|^2\rangle^2}\,.
    \label{beta}
\end{equation}
We can see from Eq.\,\eqref{varphi0BKbeta} that the spatial average of $\varphi_0$ goes to zero as $\langle B\rangle \rightarrow B_{c2}$ which is to be expected. With these preparations we can now go back to the free energy \eqref{F}. To express the result in terms of our thermodynamic variable $\langle B\rangle$, we need to rewrite the magnetic energy with the help of
\begin{equation}
\langle B^2\rangle =  \langle B\rangle^2 +e^2\left(\langle |\varphi_0|^4\rangle-\langle |\varphi_0|^2\rangle^2\right) \, ,
\end{equation}
such that, using Eq.\,\eqref{varphi0BKbeta}, we obtain the free energy density  
\begin{equation}
   {\cal F}= \frac{\langle B\rangle^2}{2} -\frac{1}{2}\frac{\left(B_{c2} -\langle B\rangle\right)^2}{\left(2\kappa^2-1\right)\beta+1}\,.
    \label{FLattice}
\end{equation}
This form is very useful since all the details of the lattice structure are captured by the parameter $\beta$, which was first introduced by Abrikosov \cite{Abrikosov:1956sx}. We can identify the first term with the free energy of the normal phase $\mathcal{F}_{\rm{nor}}$. Thus, provided the second term remains negative, the vortex lattice phase is preferred over the normal phase. The structure of the denominator in the second term is such that we can recover the free energy at the type-I transition when $\varphi_0\rightarrow\varphi_{\rm{hom}}$. With $\beta=1$ and setting $\langle B\rangle =B_c$, we can use Eq.\,\eqref{Bc2Bc} and let $\kappa\rightarrow 1/\sqrt{2}$ to find $\mathcal{F}=B_c^2/2$. This is the free energy of the normal and homogeneous phase at the type-I transition, indicating that the vortex lattice phase, normal phase and Meissner phase are equally preferred at $\kappa=1/\sqrt{2}$ i.e.\ the critical point in Fig.\,\ref{fig:type1and2}. 

It is instructive to apply a Legendre transformation and instead of $\langle B\rangle$ use the external magnetic field, $H$, as a thermodynamic variable, using the relation
\begin{equation}
H = \frac{\partial{\cal F}}{\partial\langle B\rangle} \, . 
\end{equation}
With $H_{c2}=B_{c2}$ this yields the Gibbs free energy density
\begin{equation}
    {\cal G} =  -\frac{H^2}{2} -\frac{1}{2}\frac{(H_{c2}-H)^2}{(2\kappa^2-1)\beta}\,.
    \label{GLattice}
\end{equation}
Similar to Eq.\,\eqref{FLattice}, we identify the first term with $\mathcal{G}_{\rm{nor}}$, the corresponding Gibbs free energy of the normal phase. In this form we see that the free energy is lowered by a flux tube lattice if $\kappa>1/\sqrt{2}$ (since $\beta>0$), which is exactly the condition for type-II superconductivity. As with Eq.\,\eqref{FLattice} we can recover the Gibbs free energy at the type-I transition with $\beta=1$, $H=H_c$ and $\kappa\rightarrow 1/\sqrt{2}$, demonstrating the equality in Gibbs free energy between the vortex lattice phase, Meissner phase and normal phase at the critical point as well. Since $H$ is now the thermodynamic variable, the link to the critical point in Fig.\,\ref{fig:type1and2} is perhaps clearer.

As just discussed, for $\kappa>1/\sqrt{2}$ the free energy \eqref{GLattice} is lower than that of the normal phase for $\beta>0$. We can go further and determine how much lower by minimising $\beta$. It turns out this process is equivalent to determining the preferred configuration of the vortex lattice. This will involve computing $\beta$. To this end, we introduce dimensionless variables with the help of the coherence length,
\begin{equation}
x\to \xi x  \, , \qquad y\to \xi y  \,, \qquad q\to \frac{q}{\xi} \,. 
\label{dimless}
\end{equation}
Since the $z$-dependence of our system is trivial, we can write the spatial average \eqref{avdef} as 
\begin{equation}
    \langle f(x,y) \rangle =  \frac{1}{L_xL_y}\int_{0}^{L_{x}} \int_{0}^{L_y}dydx\,f(x,y) \, ,
    \label{avdef2}
\end{equation}
where $L_{x}$ and $L_y$ are arbitrary lengths in $x$, $y$ respectively. We need to determine the spatial averages of $|\varphi_0|^2$ and $|\varphi_0|^4$ to proceed. As it stands,
\begin{subequations}
    \begin{align}
        \langle |\varphi_0|^2 \rangle =& \sum_{m,n}  \frac{C_m^* C_n}{L_x L_y} \int_{0}^{L_y} \int_{0} ^{L_x} dy dx\, e^{i(n-m)qy} \psi_m(x)\psi_n(x)  \,,
        \\[2ex]
        \begin{split}
            \langle |\varphi_0|^4 \rangle =& \sum_{l,p,m,n}  C_l^* C_p \frac{C_m^* C_n}{L_x L_y} \int_{0}^{L_y}  \int_{0} ^{L_x} dydx\, e^{i\left(n+p-(m+l)\right)qy} \psi_l(x) \psi_p(x)\psi_m(x)\psi_n(x) \,,
        \end{split}
    \end{align}
\end{subequations}
where the sums for each index go from $-\infty$ to $\infty$ (which from now on will be taken as implicit unless specified otherwise). For now, we allow our lattice to have an unspecified periodicity by setting $L_y=2\pi N_y/q$ and $L_x= N_x q$ where $N_y$ and $N_x$, which we assume to be finite integers and non-zero, are the periodicities in the $y$ and $x$ directions respectively. Both $y$-integrals can be expressed in terms of the Kronecker delta and, interestingly, the $N_y$ factors cancel from the expression, removing the dependence on the periodicity in $y$. The above then becomes
\begin{subequations}
    \begin{align}
        \langle |\varphi_0|^2 \rangle =& \frac{1}{N q} \sum_n  |C_n|^2  \int_{0} ^{N q} dx \, e^{-\left( x-nq \right)^2 } \,,\label{AvgVarphi}
        \\[2ex]
        \langle |\varphi_0|^4 \rangle =& \frac{1}{Nq} \sum_{n_1,n_2,n_3} C^*_{n_1}  C_{n_1+n_2} C_{n_1+n_3} C^*_{n_1+n_2+n_3} e^{-\frac{\left(n_2^2+ n_3^2\right)q^2}{2}}
       \int_{0}^{Nq} dx\, e^{- 2\left( x-\frac{\left(2n_1+n_2+n_3\right) q}{2} \right)^2} \,, \label{AvgVarphi4}
    \end{align}
    \label{GenAvgs}%
\end{subequations}
where we have set $N_{x}\equiv N$ for brevity, introduced the new indices
\begin{equation}
    n_1=m \,, \qquad n_2=l-p \,, \qquad n_3=p-m \,,
    \label{n123def}
\end{equation}
for $\langle|\varphi_0|^4\rangle$, and manipulated the exponential functions to be in the form of a single Gaussian by completing the square. 

\begin{figure}
    \centering
    \includegraphics[width=\textwidth]{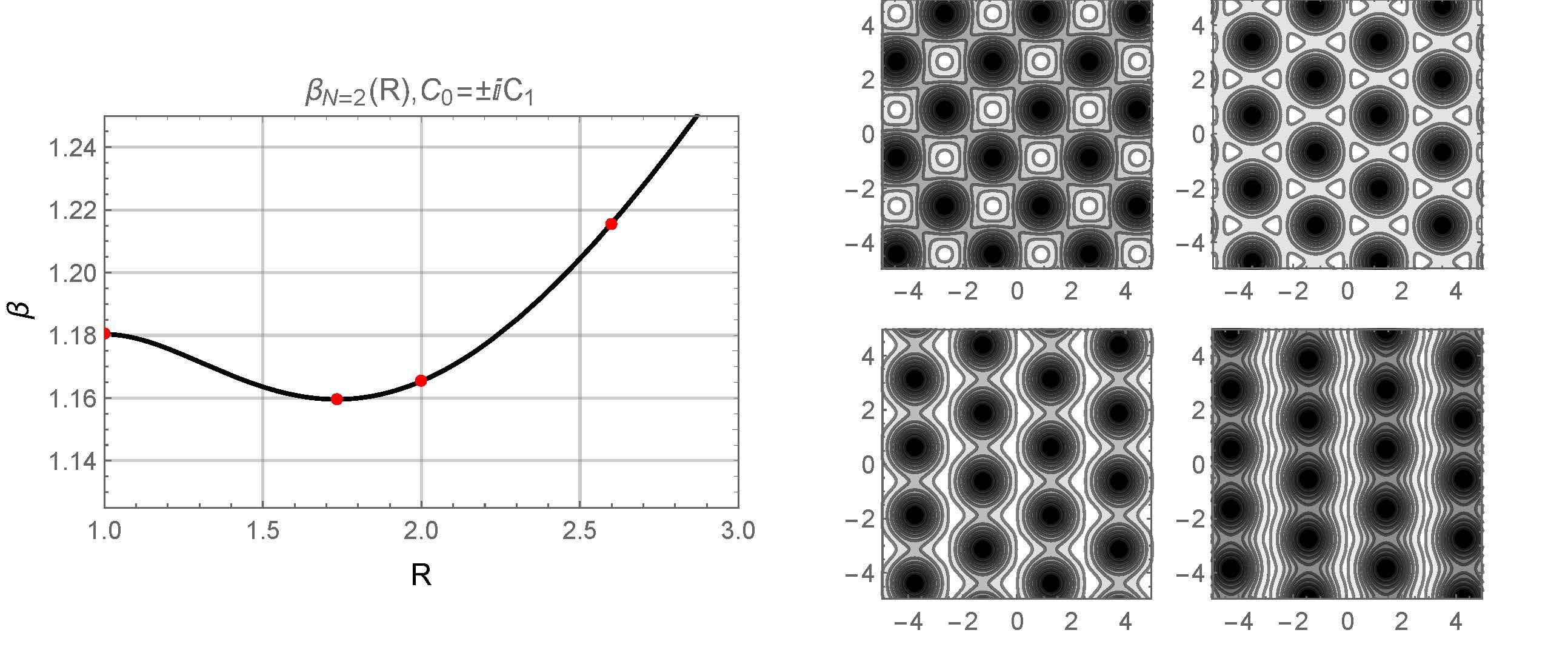}
    \caption{\textit{Left panel}: Plot of $\beta$ for the $N=2$ lattice as a function of $R$. There are two minima at $R=\sqrt{3}$ and $1/\sqrt{3}$ (not depicted). The red dots correspond to the values of $R$ for which contour plots of $|\varphi_0(x,y)|^2$ in the $x$-$y$ plane are displayed in the right panel. \textit{Right panel}: Contour plots of the condensate for values of $R$ denoted by red dots in the left panel. Starting from the leftmost dot in the $\beta(R)$ plot, we go from top to bottom, left to right. The rectangular lattice at $R=1$ is in the top left, the hexagonal lattice for $R=\sqrt{3}$ is in the top right. Darker regions correspond to vortices, i.e.\ where the condensate goes to zero. The axes are in the dimensionless $x$, $y$ and so measured in units of the coherence length $\xi$. 
    }
    \label{fig:betaRak}
\end{figure}

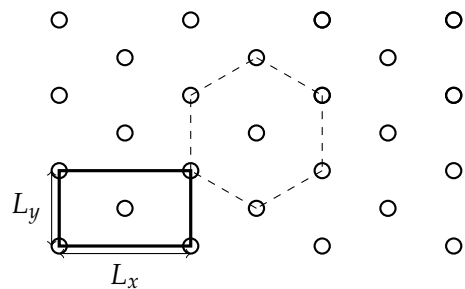
\begin{figure}
    \centering
    \begin{tikzpicture}
    \draw[thick](0,0) circle (0.1);
    \draw[thick](1.732,1) circle (0.1);
    \draw[thick](0.866,0.5) circle (0.1);
    \draw[thick](1.732,0) circle (0.1);
    \draw[thick](0,1) circle (0.1);
    
    \draw[thick](0.866,1.5) circle (0.1);
    \draw[thick](1.732,2) circle (0.1);
    \draw[thick](0,2) circle (0.1);
    
    \draw[thick](0.866,2.5) circle (0.1);
    \draw[thick](1.732,3) circle (0.1);
    \draw[thick](0,3) circle (0.1);
    
    \draw[thick](2.598,0.5) circle (0.1);
    \draw[thick](3.464,0) circle (0.1);
    \draw[thick](3.464,1) circle (0.1);
    
    \draw[thick](4.33,0.5) circle (0.1);
    \draw[thick](5.196,0) circle (0.1);
    \draw[thick](5.196,1) circle (0.1);
    
    \draw[thick](2.598,1.5) circle (0.1);
    \draw[thick](3.464,2) circle (0.1);
    \draw[thick](3.464,2) circle (0.1);
    
    \draw[thick](2.598,2.5) circle (0.1);
    \draw[thick](3.464,3) circle (0.1);
    \draw[thick](3.464,3) circle (0.1);
    
    \draw[thick](4.33,1.5) circle (0.1);
    \draw[thick](5.196,2) circle (0.1);
    \draw[thick](5.196,2) circle (0.1);
    
    \draw[thick](4.33,2.5) circle (0.1);
    \draw[thick](5.196,3) circle (0.1);
    \draw[thick](5.196,3) circle (0.1);
    
    \draw[<->] (0,-0.1) -- (1.732,-0.1);
    \draw[<->] (-0.1,0) -- (-0.1,1);
    \node[below] at (0.866,-0.1){$L_{x}$};
    \node[left] at (-0.1,0.5){$L_{y}$};
    
    \draw[very thick] (0,0) rectangle (1.732,1);
    
    \draw[dashed] (2.598,0.5) --(1.732,1);
    \draw[dashed] (1.732,1) --(1.732,2);
    \draw[dashed] (1.732,2) --(2.598,2.5);
    \draw[dashed] (2.598,2.5)--(3.464,2);
    \draw[dashed] (3.464,2)--(3.464,1);
    \draw[dashed] (3.464,1)--(2.598,0.5);
    
\end{tikzpicture}
    \caption{Hexagonal vortex lattice in the $x$-$y$ plane. A vortex is designated by a circle. It has translational symmetry $(x,y)\rightarrow(x+(2M_x+Z)L_{x}/2, y + (2M_y+Z)L_y/2)$ where $M_{x}$, $M_y$, $Z$ are integers and $C_0=\pm iC_1$ in $\varphi_0(x,y)$. In other words, the system is invariant when we take even or odd steps simultaneously in both directions (i.e.\ when $N_1=2M_x+Z$ and $N_2=2M_y+Z$ are both odd or both even). The cell which can be repeated to build the lattice and over which we take the spatial average is shown by the solid lines. The hexagonal pattern is indicated by the dashed lines. Note that here $R=\sqrt{3}$ (see text and Fig.\,\ref{fig:betaRak}), but the symmetry and condition \eqref{TriangCond} are common to all lattices encapsulated by the $\beta$ plotted in Fig.\,\ref{fig:betaRak}.}
    \label{fig:HexLattice}
\end{figure}

It is reasonable to assume that a periodic lattice is energetically favoured over a randomly assorted one. We introduce $C_{n}=C_{n+N}$ which partially determines the arrangement of vortices in the superconductor. We expand the sum in \eqref{AvgVarphi} for general periodicity $N$ and collect coefficients using $C_{n+N}=C_n$. One finds that the integrals can be combined to give $N$ Gaussian integrals over $x\in[-\infty, \infty]$ for all $C_n$ between $C_0$ and $C_{N-1}$. Similar results are obtained for \eqref{AvgVarphi4} once we notice that $C^*_{n_1+N} C_{n_1+N+n_2} C_{n_1+N+n_3} C^*_{n_1+N+n_2+n_3} = C_{n_1+n_2} C_{n_1+n_3} C^*_{n_1+n_2+n_3}$ (since the periodicity is carried over from the original indices). Evaluating the integrals, Eqs.\,\eqref{GenAvgs} become
\begin{subequations}
    \begin{align}
        \langle|\varphi_0|^2\rangle &=\frac{\sqrt{\pi}}{N q}\sum_{n=0}^{N-1}|C_{n}|^2
        \label{NAvgVarPhi}
        \,,
        \\[2ex]
        \langle|\varphi_0|^4\rangle &=\frac{1}{N q}\sqrt{\frac{\pi}{2}}\sum_{n_3} \sum_{n_2} e^{-\frac{\left(n_2^2 +n_3^2\right)q^2}{2}}\sum_{n_1=0}^{N-1}C^*_{n_1} C_{n_1+n_2} C_{n_1+n_3} C^*_{n_1+n_2+n_3}\,.
        \label{NAvgVarPhi4}
    \end{align}
    \label{NCaseResults}%
\end{subequations}
More explicit derivations of these results can be found in Appendix \ref{app:LatticeN}. An analysis for general $N$ is likely a large and complex task. Instead, we follow the example of Abrikosov in Ref.\,\cite{Abrikosov:1956sx} and resort to looking at the simplest configurations, assuming these would be favoured. Famously, he incorrectly concluded that a rectangular arrangement of vortices, where $C_n=C_{n+1}$, was the favoured configuration. We can recover this result by setting $N=1$ in Eqs.\,\eqref{NCaseResults}. However, Kleiner, Roth and Autler in Ref.\,\cite{PhysRev.133.A1226} demonstrated that a rectangular lattice is unstable to a hexagonal one, which has a lower free energy\footnote{Experimental observation of the hexagonal lattice then confirmed that a hexagonal lattice is the preferred configuration in general (e.g. see Ref.\,\cite{ESSMANN1967526})}. They achieved this by deriving a way to compare many different configurations. The derivation begins by looking at a general triangular lattice (of which a rectangular lattice is a special case) where $C_{n}=C_{n+2}$. Following suit, we set $N=2$ such that there are only two unique coefficients - $C_0$ and $C_1$. From Eqs. \eqref{NCaseResults} we obtain
\begin{subequations}
    \begin{align}
        \langle |\varphi_0|^2 \rangle =& \frac{\sqrt{\pi}}{2q} \left( |C_0|^2 +  |C_1|^2 \right) \,,
        \label{N=2AvgVarPhi}
        \\[2ex]
        \langle |\varphi_0|^4 \rangle =& \frac{1}{2q} \sqrt{\frac{\pi}{2}} \left\{ \left(|C_0|^4 + |C_1|^4\right) f_0^2 + 4 |C_0|^2 |C_1|^2 f_0 f_1 +  \left[ \left(C^*_{0} C_1\right)^2 + \left(C^*_{1} C_0\right)^2 \right] f_1^2 \right\}\,,     \label{N=2AvgVarphi4}
    \end{align}
    \label{N=2VarphiAvgs}%
\end{subequations}
where
\begin{equation}
     f_{s}=\sum_{r} e^{-\frac{(2r+s)^2q^2}{2}} \,,
\end{equation}
which with $s=0,1$ are respectively the even and odd parts of both the $n_2$ and $n_3$ sums in Eq.\,\eqref{NAvgVarPhi4}. These functions can be expressed as  
\begin{equation}
    f_0 = \vartheta_3(0,e^{-2q^2}) \, , \qquad f_1 = e^{-q^2/2}\vartheta_3(iq^2,e^{-2q^2}) \, , 
\end{equation}
with the Jacobi theta function 
\begin{equation}
    \vartheta_3(z,x) = \sum_{n=-\infty}^\infty x^{n^2}e^{2niz}  \, .
    \label{Jth}
\end{equation} 
The $N=2$ lattice encapsulates a continuum of geometries whose common feature is $C_{n}=C_{n+2}$ in the solution, such that $\varphi_0(x,y)$ can be split into odd an even parts. In the geometric picture, this corresponds to some skewed lattice, of which the rectangular and triangular lattice are special cases. Further assuming that more symmetrical configurations will be preferred over less symmetrical configurations, the general, discrete translational symmetry $|\varphi_0(x,y)|^2=|\varphi_0(x+N_1 L_x/2,y+N_2 L_y/2)|^2$ where $N_1$ and $N_2$ are integers can be investigated. By exploring different combinations of even and odd $N_1$ and $N_2$, we find the translational symmetry which gives the minimal $\beta$ for $N=2$ is $N_1=2M_x+Z$ and $N_2=2M_y+Z$ for any integer $M_x$, $M_y$, $Z$ with
\begin{equation}
    C_0=\pm i C_1\,.
    \label{TriangCond}
\end{equation}
This analysis is done in Fourier space in Appendix \ref{app:LatticeSymmetry}. Inserting the condition Eq.\,\eqref{TriangCond} into $\beta$, we obtain 
\begin{equation}
    \begin{split}
        \beta &=\frac{q}{\sqrt{2\pi}}(f_0^2+2f_0f_1-f_1^2)
        \\
        &= \sqrt{\frac{R}{2}} \Big\{ \left[\vartheta_3(0,e^{-2\pi R})\right]^2+2e^{-\frac{\pi R}{2}} \vartheta_3(0,e^{-2\pi R})\vartheta_3(i\pi R,e^{-2\pi R})
        \\
        &-e^{-\pi R}\left[\vartheta_3(i\pi R,e^{-2\pi R})\right]^2\Big\}\, , 
    \label{betaa}
    \end{split}
\end{equation}
where we have introduced the variable 
\begin{equation} 
    R \equiv \tan\theta= \frac{L_x}{L_y} = \frac{q^2}{\pi} \, .
    \label{Rq}
\end{equation}
(Recall that $q$ is dimensionless here, in terms of dimensionful quantities $R=q^2\xi^2/\pi$.) This variable parameterises a continuum of triangular lattices, where $R=1$ and thus $\theta=\frac{\pi}{4}$ corresponds to a quadratic lattice, while $R=\sqrt{3}$ and thus $\theta = \frac{\pi}{3}$ gives a hexagonal lattice (or $R=1/\sqrt{3}$, which gives the same lattice with $x$- and $y$-directions swapped).
By plotting $\beta$ as a function of $R$, we see that it is minimised by the hexagonal structure, for which $\beta\simeq 1.1596$, while $\beta\simeq 1.1803$ for the quadratic lattice, which is shown in Fig.\,\ref{fig:betaRak}. The translational symmetry of the hexagonal lattice is demonstrated in Fig\,\ref{fig:HexLattice}.

\chapter{Magnetic defects in the 2SC phase with strange quark mass corrections}
\label{chpt:Project1}
In Chapter \ref{chpt:Background}, Sec.\,\ref{Background:sec:QCD} we introduced the concept of CSC. Its possible presence in the $\mu_B$-$B$ plane at $T=0$ and NS cores was also discussed. Of relevance to both these discussions is the work done in Ref.\,\cite{Haber:2017oqb}, where the authors analysed dense, three-flavour quark matter at zero temperature in a large external magnetic field. This was done by considering three different condensates and gauge fields in a GL approach \cite{Blaschke:1999fy,Iida:2000ha,Iida:2001pg,Iida:2002ev,Sedrakian:2002mk,Giannakis:2003am,Hatsuda:2006ps}. The authors focused on the CFL and 2SC phases, both of which act as very weak electromagnetic superconductors, expelling a small part of the ordinary\footnote{``Ordinary'' used here means ``the magnetic part of the electromagnetic field'' for brevity.} magnetic field. Determining the critical fields for both type-I and type-II regimes, they constructed the phase diagram displayed in the right panel of Fig.\,\ref{fig:ChSBtoCSCmuB}. To calculate $H_{c1}$ for each phase required analysing the magnetic defects that emerge above this transition. Interestingly, the authors determined that the flux tubes in the 2SC phase become domain walls. To see why this might be the case, we need to first look at the pairing in the 2SC phase.

In the 2SC phase, pairing only occurs between two of the three quarks flavours and two of the three colours, forming a single condensate. When quarks are massive, pairing with the strange quark is disfavoured due to the comparatively large $M_s$. Therefore, 2SC forms a $ud$ condensate and all $s$ quarks do not pair. When quarks are massless, the difference in Fermi surfaces is small and therefore forming pairs of any kind is equally favourable. Ignoring colour, this means there are three different possible options for a 2SC phase, one for each Cooper pair condensate. Hence, the 2SC phase can be split according to which quarks pair; 2SC$_{ud}$, 2SC$_{us}$ and 2SC$_{ds}$ (subscript denotes the paired quarks). For comparison, the CFL phase involves a coexistence of the three aforementioned Cooper pair condensates, each with a specific colour charge. With this in mind, CSC warrants the examination of multi-component superconductors.

In Ref.\,\cite{Haber:2017oqb}, the quarks are treated as massless. This results in the 2SC$_{ud}$ and 2SC$_{us}$ phases having the same free energy (the 2SC$_{ds}$ has a larger free energy). Equal free energy means equal pressure and thus the two condensates form a domain wall. The aim of the work presented in this chapter is to improve upon the results of Ref.\,\cite{Haber:2017oqb} by incorporating the mass of the heaviest quark in the three-flavour regime i.e.\ $m_s$. Near compact star densities $\mu_q\sim 400 \,\mathrm{MeV}$, while $m_u\approx2\,\mathrm{MeV}$, $m_d\approx 5 \,\mathrm{MeV}$ and  $m_s \approx 100 \,\mathrm{MeV}$. The mass of the $u$ and $d$ quarks are small compared to $\mu_q$, but $m_s\sim\mu_q$, making the massless quarks approximation less justifiable in this astrophysical context. Hence, with the motivation of a more realistic calculation, we keep $m_u=m_d=0$ and include the effects of $m_s\neq 0$ by adding correction terms in the GL free energy, following the examples of \cite{Iida:2003cc,Iida:2004cj,Schmitt:2010pf}. A primary goal is determining the fate of the domain walls within this semi-massive regime and, if they do not survive, if some other defect takes its place. Other objectives include constructing phase diagrams for both homogeneous and inhomogeneous systems and comparing them to the massless results presented in Ref.\,\cite{Haber:2017oqb}. We do this by obtaining the Gibbs free energy (derived from the GL potential) for each phase, through which we find the critical fields. For the critical field $H_{c1}$, this involves determining the flux tube and magnetic field profiles numerically.

In Sec.\,(\ref{Project1:sec:setup}) our GL formalism including $m_s$ is introduced. The following Sec.\,(\ref{Project1:sec:hom}) discusses the homogeneous phases in preparation for the next section (Sec.(\,\ref{Project1:sec:Hc})), which concerns the critical magnetic fields. Here, all critical fields $H_c$ of the homogeneous phases are presented, and the upper critical fields ($H_{c2}$) of the inhomogeneous phases are determined analytically. The numerical calculation for the lower critical field ($H_{c1}$) is then laid out. Following this is the final section, Sec.\,(\ref{Project1:sec:results}), presenting and analysing our numerical results. A summary and wider discussion of the results are reserved for Chapter \ref{chpt:Outlook}.


\section{Setup}
\label{Project1:sec:setup}
The following lays out the framework in which we derive our results. We work in the previously introduced GL regime, which has been used to study CSC in the past e.g.\ se Refs.\,\cite{Iida:2000ha,Iida:2001pg,Iida:2002ev,Giannakis:2003am,Hatsuda:2006ps}. Our starting expression is the same as the one in Ref.\,\cite{Haber:2017oqb} but with strange quark mass correction terms adopted from the potential in Refs.\,\cite{Iida:2003cc,Iida:2004cj}. While the same as the framework introduced in Sec.\,(\ref{Background:subsec:GL}), the inclusion of additional gauge fields and possibility of multiple non-zero condensates adds more complexity. Our focus will be primarily on the 2SC phase so we rotate the colour and electromagnetic gauge fields in a way which is convenient for our purposes. Also discussed is the computation of the Gibbs free energy using the minimised GL free energy density and our choices of parameters, which elaborates on the inclusion of non-zero $m_s$.

\subsection{Ginzburg-Landau potential}
\label{Project1:subsec:GL}

In three-flavour quark matter with sufficiently small mismatch in the Fermi momenta of the different fermion species, Cooper pairing predominantly occurs in the spin-zero channel and in the antisymmetric anti-triplet channels in colour and flavour space, $[\bar{3}]_c$ and $[\bar{3}]_f$. As pairing is assumed to occur between fermions of the same chirality, the flavour channel stands for either left-handed or right-handed fermions. Therefore, the order parameter for Cooper pair condensation can be written as 
\begin{equation} \label{Psi}
\Psi = \Phi_{ij} J_i\otimes I_j \in [\bar{3}]_c \otimes [\bar{3}]_f \, , 
\end{equation}
where the anti-symmetric $3\times 3$ matrices  $(J_i)_{jk} = -i\epsilon_{ijk}$ and $(I_j)_{k\ell} = -i\epsilon_{jk\ell}$ form bases of the three-dimensional spaces $[\bar{3}]_c$ and $[\bar{3}]_f$, respectively. As a consequence, we can characterise a colour-superconducting phase by the $3\times 3$ matrix $\Phi$, which has one (anti-)colour and one (anti-)flavour index. We shall put the colours in the order $(r, g, b)$ 
and the flavours in the order $(u, d, s)$. The colour charges are only labels and thus their order is not crucial, but the order of the flavours matters since electric charge and quark masses break the flavour symmetry\footnote{For the CFL phase the order $(d,s,u)$ is often used, which is convenient since then the generator of the electromagnetic gauge group is proportional to the eighth generator of the colour gauge group. Since our main focus is on the 2SC phase, this re-ordering is not necessary and we work with the more common order $(u,d,s)$.}. In this convention, for instance, $\Phi_{11}$ carries anti-indices $\bar{r}$ and $\bar{u}$ and thus describes pairing of $gd$ with $bs$ quarks and of $gs$ with $bd$ quarks. 

We consider the following GL potential up to fourth order in $\Phi$,
\begin{equation} \label{UPhi}
    \begin{split}
        \Omega =& -12\Big\{\mathrm{Tr}[(D_0\Phi)^{\dagger}(D_0\Phi)]+u^2\mathrm{Tr}[(D_i\Phi)^\dagger(D^i\Phi)]\Big\}-4\mu^2 \mathrm{Tr}[\Phi^\dagger\Phi]
        \\[2ex]
        &+4\sigma\mathrm{Tr}\left[\Phi^\dagger \left(\frac{M^2}{2\mu_q}+\mu_eQ\right)\Phi\right] 
        +16(\lambda+h)\mathrm{Tr}[(\Phi^\dagger\Phi)^2]-16h (\mathrm{Tr}[\Phi^\dagger\Phi])^2 
        \\[2ex]
        &+\frac{1}{4}F_{\mu\nu}^aF_a^{\mu\nu}+\frac{1}{4}F_{\mu\nu}F^{\mu\nu}
        \,. 
    \end{split}
\end{equation}
Apart from the mass correction, proportional to the parameter $\sigma$, this is exactly the same potential, and the same notation, as in Ref.\ \cite{Haber:2017oqb}, where the starting point was the potential for $\Psi$, based on previous works \cite{Iida:2000ha,Iida:2001pg,Iida:2002ev,Giannakis:2003am}. Due to the broken Lorentz invariance in the medium, the temporal and spatial parts of the kinetic term have different prefactors, $u=1/\sqrt{3}$. The covariant derivative is given by 
\begin{equation}
    D_\mu \Phi \equiv \partial_\mu\Phi-ig A_\mu^a T_a^T\Phi+ieA_\mu\Phi Q \,.
\end{equation}
where $g$ and $e$ are the strong and electromagnetic coupling constants, respectively. The colour gauge fields are denoted by $A_\mu^a$, $a=1,\ldots, 8$, and $A_\mu$ is the electromagnetic gauge field. Furthermore, $T_a=\lambda_a/2$, with the Gell-Mann matrices $\lambda_a$, are the generators of the colour gauge group $SU(3)$, and the electric charge matrix for the Cooper pairs $Q = {\rm diag}(q_d+q_s,q_u+q_s,q_u+q_d) = {\rm diag}(-2/3,1/3,1/3)$ is the generator of the electromagnetic gauge group $U(1)$. Here, $q_u$, $q_d$, $q_s$ denote the individual quark charges in units of $e$. The field strength tensors are  $F_{\mu\nu}^a = \partial_\mu A_\nu^a-\partial_\nu A_\mu^a+gf^{abc}A_\mu^bA_\nu^c$ for the colour sector, where $f^{abc}$ are the $SU(3)$ structure constants, and $F_{\mu\nu} = \partial_\mu A_\nu-\partial_\nu A_\mu$ for the electromagnetic sector. The constants in front of the quadratic and quartic terms in $\Phi$ are written conveniently as combinations of $\mu$, $\lambda$ and $h$, whose physical meaning will become obvious after performing the traces (see Eq.\,\eqref{U1}).

We have incorporated a mass correction to lowest order, with the (fermionic) quark chemical potential $\mu_q$ and the mass matrix for the Cooper pairs $M={\rm diag}(m_d+m_s,m_u+m_s,m_u+m_d)\simeq {\rm diag}(m_s,m_s,0)$, i.e., we shall neglect the masses of the light quarks, $m_u\simeq m_d\simeq 0$, and keep the strange quark mass $m_s$ as a free parameter. We have also included the contribution of the electric charge chemical potential $\mu_e$, which is of the same order for small quark masses, $\mu_e\propto m_s^2/\mu_q$. The correction term is identical to the one used in Refs.\ \cite{Iida:2003cc,Iida:2004cj}, which is easily seen by an appropriate rescaling of $\Phi$. 

In the following we restrict ourselves to diagonal order parameters, $\Phi=\frac{1}{2}{\rm diag}(\phi_1,\phi_2,\phi_3)$
with $\phi_i\in \mathbb{C}$, where $\phi_1$ corresponds to $ds$ pairing, $\phi_2$ to $us$ pairing, and $\phi_3$ to $ud$ pairing. 
If flavour symmetry was intact, off-diagonal order parameters could always be brought into a diagonal form by an appropriate rotation in colour and flavour space. This is no longer true when flavour symmetry is explicitly broken, and thus our restriction to diagonal order parameters is a simplification of the most general situation \cite{Rajagopal:2005dg}. Even in this simplified case, $2^3=8$ qualitatively different homogeneous phases have to be considered
in principle, accounting for each of the three condensates to be either zero or non-zero. The restriction to diagonal matrices $\Phi$ allows us to consistently set all gauge fields with off-diagonal components to zero, $A_\mu^1=A_\mu^2=A_\mu^4=A_\mu^5=A_\mu^6=A_\mu^7=0$, such that the only relevant gauge fields are the two colour gauge fields $A_\mu^3$, $A_\mu^8$, and the electromagnetic gauge field $A_\mu$. Moreover, we are only interested in static solutions and drop all electric fields, i.e., we only keep the spatial components of the gauge fields, giving rise to the magnetic fields $\bm{B}_3 = \nabla\times \bm{ A}_3$, $\bm{ B}_8 = \nabla\times \bm{ A}_8$, and $\bm{ B} = \nabla\times \bm{ A}$. Within this ansatz, performing the traces in Eq.\,\eqref{UPhi} yields
\begin{equation} \label{U1}
    \begin{split}
        \Omega =& \left|\left(\nabla+\frac{ig\bm{ A}_3}{2}+\frac{ig\bm{ A}_8}{2\sqrt{3}}+\frac{2ie\bm{ A}}{3}\right)\phi_1\right|^2 +\left|\left(\nabla-\frac{ig\bm{ A}_3}{2}+\frac{ig\bm{ A}_8}{2\sqrt{3}}-\frac{ie\bm{ A}}{3}\right)\phi_2\right|^2 
        \\[2ex]
        &+ \left|\left(\nabla-\frac{ig\bm{ A}_8}{\sqrt{3}}-\frac{ie\bm{ A}}{3}\right)\phi_3\right|^2 
        -(\mu^2-m_1^2)|\phi_1|^2-(\mu^2-m_2^2)|\phi_2|^2  
        \\[2ex]
        &-(\mu^2-m_3^2)|\phi_3|^2 +\lambda(|\phi_1|^4+|\phi_2|^4+|\phi_3|^4) 
        \\[2ex]
        &-2h(|\phi_1|^2|\phi_2|^2+|\phi_1|^2|\phi_3|^2+|\phi_2|^2|\phi_3|^2)+\frac{\bm{ B}_3^2}{2}+\frac{\bm{ B}_8^2}{2}+\frac{\bm{ B}^2}{2} \, .
    \end{split}
\end{equation}
This potential can be viewed as a generalised version of our textbook superconductor which has a single condensate coupled to a single gauge field from Sec.\, (\ref{Background:subsec:GL}). Here we have three condensates with identical (bosonic) chemical potential $\mu$, with self-coupling $\lambda$, cross-coupling $h$, and three different effective masses (squared)
\begin{eqnarray} \label{m123}
m_1^2 &=&  \sigma\left(\frac{m_s^2}{2\mu_q}-\frac{2\mu_e}{3}\right)     \, , \qquad 
m_2^2 =  \sigma\left(\frac{m_s^2}{2\mu_q}+\frac{\mu_e}{3}\right)  \, , \qquad 
m_3^2 = \sigma\frac{\mu_e}{3} \, .
\end{eqnarray}
All three gauge fields couple to the condensates. We can simplify the potential by a suitable rotation of the gauge fields.

\subsection{Rotated electromagnetism and Gibbs free energy density}
\label{Project1:subsec:rotate}

We apply the double rotation from Ref.\ \cite{Haber:2017oqb}, which, denoting the rotated gauge fields by $\tilde{A}_\mu^3$, $\tilde{A}_\mu^8$, and $\tilde{A}_\mu$, reads, 
\begin{equation} \label{rotatetwice}
    \left(\begin{array}{c} \tilde{A}_\mu^3 \\ \tilde{A}_\mu^8 \\ \tilde{A}_\mu \end{array}\right) =  
    \left(\begin{array}{ccc} \cos\vartheta_2 & 0 &\sin\vartheta_2 \\ 0&1&0 \\ -\sin\vartheta_2&0&\cos\vartheta_2  \end{array}\right)\left(\begin{array}{ccc} 1 & 0 &0 \\ 0& \cos\vartheta_1 & \sin\vartheta_1 \\ 0 & -\sin\vartheta_1 & \cos\vartheta_1   \end{array}\right)\left(\begin{array}{c} A_\mu^3 \\ A_\mu^8 \\ A_\mu \end{array}\right)  \, ,
\end{equation}
where the mixing angles $\vartheta_1$ and $\vartheta_2$ are given by
\begin{subequations} \label{thetamix}
    \begin{eqnarray}
        \sin\vartheta_1 &=& \frac{e}{\sqrt{3g^2+e^2}} \, , \qquad \cos\vartheta_1 = \frac{\sqrt{3}g}{\sqrt{3g^2+e^2}} \, , 
        \\[2ex] 
        \sin\vartheta_2 &=& \frac{\sqrt{3}e}{\sqrt{3g^2+4e^2}} \, , \qquad \cos\vartheta_2 = \frac{\sqrt{3g^2+e^2}}{\sqrt{3g^2+4e^2}} \, .
    \end{eqnarray}
\end{subequations}
This rotation is the most convenient choice for our purpose of calculating flux tube profiles in the 2SC phase: The first rotation, with the usual ``2SC mixing angle'' $\vartheta_1$ ensures that in the homogeneous 2SC phase, where only the condensate $\phi_3$ is non-zero, only the $\tilde{\bm{ B}}_8$ field is expelled. The other two rotated fields penetrate the superconductor unperturbed (assuming zero magnetisation from the unpaired quarks). If we were only interested in the homogeneous 2SC phase, this rotation would be sufficient. However, we will allow for $\phi_1$ and $\phi_2$ to be induced in the core of the flux tube. Therefore, we apply a second rotation with mixing angle $\vartheta_2$. This rotation leaves $\tilde{\bm{ B}_8}$ invariant and creates a field, namely $\tilde{\bm{ B}}$, which is unaffected by the superconductor even if all three condensates are non-zero. Thus, $\tilde{\bm{ B}}$ simply decouples from the condensates and can be ignored in the  calculation of the flux tube profiles. For $g\gg e$ both mixing angles are small and thus in this case $\tilde{A}_\mu^3$ and $\tilde{A}_\mu^8$ are ``almost'' gluons with a small admixture of the photon, while $\tilde{A}_\mu$ is ``almost'' the photon with a small gluonic admixture. For consistency, we shall work with the rotated fields \eqref{rotatetwice} throughout this chapter, including the discussion of the homogeneous phases in Sec.\,(\ref{Project1:sec:hom}).  

Applying the gauge field rotation and writing the complex fields in terms of their moduli and phases,
\begin{equation}\label{PolarPhi}
\phi_i(\bm{ r}) = \frac{\rho_i(\bm{ r})}{\sqrt{2}}e^{i\psi_i(\bm{ r})} \, , \qquad i =1,2,3,
\end{equation}
the GL potential \eqref{U1} becomes
\begin{equation} \label{UU0}
\Omega = \Omega_0 +\frac{\tilde{\bm{ B}}_3^2}{2}+\frac{\tilde{\bm{ B}}_8^2}{2} +\frac{\tilde{\bm{ B}}^2}{2} \, , 
\end{equation}
with 
\begin{equation} \label{U0}
\begin{split}
    \Omega_0 =&\left(\nabla\psi_1+\tilde{q}_3\tilde{\bm{ A}}_3+\tilde{q}_{81}\tilde{\bm{ A}}_8\right)^2\frac{\rho_1^2}{2} +\left(\nabla\psi_2-\tilde{q}_3\tilde{\bm{ A}}_3+\tilde{q}_{82}\tilde{\bm{ A}}_8\right)^2\frac{\rho_2^2}{2}  
    \\[2ex]
    &+\left(\nabla\psi_3-\tilde{q}_{83}\tilde{\bm{ A}}_8\right)^2\frac{\rho_3^2}{2}+\frac{(\nabla\rho_1)^2}{2}+\frac{(\nabla\rho_2)^2}{2}+\frac{(\nabla\rho_3)^2}{2} -\frac{\mu^2-m_1^2}{2}\rho_1^2
    \\[2ex]
    &-\frac{\mu^2-m_2^2}{2}\rho_2^2-\frac{\mu^2-m_3^2}{2}\rho_3^2 +\frac{\lambda}{4}(\rho_1^4+\rho_2^4+\rho_3^4)-\frac{h}{2}(\rho_1^2\rho_2^2+\rho_1^2\rho_3^2+\rho_2^2\rho_3^2)  \, , 
\end{split}
\end{equation}
where we have introduced the rotated charges
\begin{subequations}
\begin{eqnarray}
\tilde{q}_{81}&\equiv& \frac{g}{2\sqrt{3}}\cos\vartheta_1+\frac{2e}{3}\sin\vartheta_1 = \frac{3g^2+4e^2}{6\sqrt{3g^2+e^2}} \, , 
\\[2ex]
\tilde{q}_{82}&\equiv& \frac{g}{2\sqrt{3}}\cos\vartheta_1-\frac{e}{3}\sin\vartheta_1 = \frac{3g^2-2e^2}{6\sqrt{3g^2+e^2}} \, , 
\\[2ex]
\tilde{q}_{83}&\equiv& \frac{g}{\sqrt{3}}\cos\vartheta_1+\frac{e}{3}\sin\vartheta_1 = \frac{\sqrt{3g^2+e^2}}{3} \, , 
\\[2ex]
\tilde{q}_3 &\equiv& \frac{g}{2}\cos\vartheta_2+\frac{e}{2}\cos\vartheta_1\sin\vartheta_2= \frac{g}{2}\frac{\sqrt{3g^2+4e^2}}{\sqrt{3g^2+e^2}} \, .
\end{eqnarray}
\end{subequations}
In Eq.\,\eqref{UU0} we have separated the quadratic contributions of the magnetic fields, which is notationally convenient for the following. 

We shall be interested in the phase structure at fixed external magnetic field $\bm{ H}$, which we assume to be homogeneous and along the $z$-direction, $\bm{ H}(\bm{ r})=H\hat{\bm{ e}}_z$. Therefore, we need to consider the Gibbs free energy density for which we use Eq.\,\eqref{Gibbs} and compute $\mathcal{G}$ . We can obviously assume that all induced magnetic fields have only $z$-components as well. Denoting the $z$-components of the rotated fields by $\tilde{B}_3$, $\tilde{B}_8$, and $\tilde{B}$, we have  $\bm{ H}\cdot\bm{ B} = H[\sin\vartheta_1\, \tilde{B}_8+\cos\vartheta_1(\sin\vartheta_2\, \tilde{B}_3+\cos\vartheta_2\,\tilde{B})]$. Since $\tilde{\bm{ B}}$ does not couple to any of the condensates, it remains homogeneous even in the presence of flux tubes. Consequently, the equation of motion for $\tilde{\bm{ A}}$ is trivially fulfilled, and we determine $\tilde{B}$ by minimising the Gibbs free energy, which, using Eq.\,\eqref{UU0}, yields
\begin{equation} \label{tildeB} 
\tilde{B} =H\cos\vartheta_1\cos\vartheta_2 \, .
\end{equation} 
Reinserting this result into $\mathcal{G}$, we obtain
\begin{equation}\label{gibbs}
    \begin{split}
        \mathcal{G}=& -\frac{H^2\cos^2\vartheta_1\cos^2\vartheta_2}{2}
        \\[2ex]
        &+\frac{1}{V}\int d^3\bm{x}\,\left[\Omega_0+\frac{\tilde{B}_3^2}{2}+\frac{\tilde{B}_8^2}{2}-H(\sin\vartheta_1\, \tilde{B}_8+\cos\vartheta_1\sin\vartheta_2\, \tilde{B}_3)\right] \, .
    \end{split}
\end{equation}

\subsection{Choice of parameters}
\label{Project1:subsec:para}

The potential \eqref{U0} depends on the parameters $\mu$, $\lambda$, $h$, $\sigma$. The discussion of the homogeneous phases in Sec.\,(\ref{Project1:sec:hom}) turns out to be sufficiently simple to keep these parameters unspecified and to investigate the general phase structure. Our main results, however, require the numerical calculation of the flux tube profiles, and a completely general study would be extremely laborious. Therefore, for the results in Sec.\,(\ref{Project1:sec:results}), we employ the weak-coupling values of these parameters \cite{Iida:2002ev,Giannakis:2003am,Iida:2003cc,Iida:2004cj,Iida:2004if,Eto:2013hoa},
\begin{equation} \label{weak}
    \begin{split}
    &\mu^2 = \frac{48\pi^2}{7\zeta(3)}T_c^2\left(1-\frac{T}{T_c}\right) \, , \qquad \lambda = \frac{72\pi^4}{7\zeta(3)}\frac{T_c^2}{\mu_q^2}\,,
    \\[2ex]
    &h= -\frac{36\pi^4}{7\zeta(3)}\frac{T_c^2}{\mu_q^2} \, , \qquad \sigma = -\frac{24\pi^2}{7\zeta(3)}\frac{T_c^2}{\mu_q}\ln\frac{T_c}{\mu_q} \, , 
\end{split}
\end{equation}

where $T_c$ is the critical temperature. The ratio $T_c/\mu_q$ can be understood as a measure of the pairing strength since $T_c$ is closely related to the pairing gap. For instance, at weak coupling, which is applicable at asymptotically large densities, the zero-temperature pairing gap is exponentially suppressed compared to $\mu_q$. It is related by a numerical factor of order one to $T_c$ \cite{Schmitt:2002sc}, and thus $T_c/\mu_q$ is also exponentially small. We shall extrapolate our results to strong coupling, having in mind applications to compact stars, where the densities are large, but not asymptotically large. In this case, model calculations as well as extrapolations of perturbative results suggest that $T_c/\mu_q \sim 0.1$. Besides the 
implicit dependence on $T_c/\mu_q$ our potential also depends on the ratio $m_s/\mu_q$. Since $m_s$ is medium dependent, its value at non-asymptotic densities is poorly known. It is expected to be somewhere between the current mass and the constituent mass within a baryon, $m_s \sim (100 - 500)\, {\rm MeV}$; for a concrete calculation within the NJL model see for instance Ref.\ \cite{Ruester:2005jc}. With the quark chemical potential in the core of a compact star of about $\mu_q \sim (400 - 500)\, {\rm MeV}$ we thus expect $m_s/\mu_q \sim (0.2 - 1)$. Finally, our potential depends on the electric charge chemical potential $\mu_e$. In a fermionic approach, this chemical potential would be determined from the conditions of beta-equilibrium and charge neutrality. Since our GL expansion is formally based on small values of the order parameter, we follow Refs.\ \cite{Iida:2003cc,Iida:2004cj} and use the value of $\mu_e$ in the completely unpaired phase. At weak coupling and to lowest order in the strange quark mass this value is (see for instance Refs.\ \cite{Alford:2002kj,Schmitt:2010pn})
\begin{equation} \label{mue}
\mu_e = \frac{m_s^2}{4\mu_q} \, .
\end{equation}
With this result, it is convenient to trade the dimensionful parameter $\sigma$ for the dimensionless ``mass parameter''
\begin{equation}\label{alpha}
\alpha \equiv \frac{\sigma m_s^2}{\mu^2\mu_q} = \frac{m_s^2}{2\mu_q^2}\left(1-\frac{T}{T_c}\right)^{-1}\ln\frac{\mu_q}{T_c} \, ,
\end{equation}
such that the complete dependence of our potential on the strange quark mass is absorbed in $\alpha$,
\begin{equation} \label{malpha}
m_1^2 = \frac{\mu^2}{3}\alpha \, , \qquad m_2^2 = \frac{7\mu^2}{12}\alpha \, , \qquad m_3^2 = \frac{\mu^2}{12}\alpha \, .
\end{equation}
We shall see that if we are only interested in homogeneous phases, the phase structure is most conveniently calculated in the space spanned by $\alpha$, $g$, the normalised dimensionless magnetic field $H/(\mu^2/\lambda^{1/2})$, and the ratio
\begin{equation} \label{eta}
\eta \equiv \frac{h}{\lambda} \, .
\end{equation}
 At weak coupling $\eta=-1/2$, as one can see from Eq.\,\eqref{weak}. Later, in our explicit calculation of the flux tube profiles and the resulting critical magnetic fields, we consider a fixed $g$ and the parameter space spanned by $H/(\mu^2/\lambda^{1/2})$, $T_c/\mu_q$, and $m_s/\mu_q$. To choose a value of $g$, realistic for 
compact star conditions, we observe that according to the two-loop QCD beta function (which should not be taken too seriously at such low densities), $\mu_q\simeq 400\, {\rm MeV}$ corresponds to $\alpha_s\simeq 1$ and thus $g=\sqrt{4\pi\alpha_s} \simeq 3.5$. Of course, choosing such a large value for $g$ in our main results is a bold   extrapolation, given that we work with the weak-coupling parameters from Eq.\,\eqref{weak}. Furthermore, we shall set $T=0$ in Eq.\,\eqref{alpha}. Strictly speaking this is inconsistent because the GL potential is an expansion in the condensates, and we use a value for $\mu_e$ \eqref{mue} that is only valid very close to a second-order transition to the unpaired phase. Choosing a different, non-zero temperature, would not change our result qualitatively because it only enters the relation between $m_s/\mu_q$ and $\alpha$. The definition of $\alpha$ \eqref{alpha} shows that the mass effect is smallest for zero temperature (i.e., in this case $\alpha$ is smallest for a given $m_s/\mu_q$). Therefore, by our choice $T=0$ in Eq.\,\eqref{alpha}  we will obtain an upper limit in $m_s/\mu_q$ for the presence of multi-winding 2SC flux tubes. Any $T>0$ in Eq.\,\eqref{alpha} will give a smaller $m_s/\mu_q$ up to which these exotic configurations exist. In any case, the temperature dependence in the present approach is somewhat simplistic to begin with because, firstly, in a multi-component superconductor there can be different critical temperatures for the different condensates,
resulting in temperature factors different from the standard GL formalism \cite{Haber:2017kth}. And, secondly, away from the asymptotic  weak-coupling regime the phase transition becomes first-order due to  gauge field fluctuations \cite{Giannakis:2004xt}, and thus at strong coupling the behaviour just below the phase transition would have to be modified in a more sophisticated approach. 

\section{Homogeneous phases}
\label{Project1:sec:hom}
This section summarises the Gibbs free energy density calculations of the spatially homogeneous phases possible within our ansatz, which can be considered as the type-I phases. These calculations are in preparation for constructing not only the homogeneous phase diagram but the more general 2SC phase diagram. The subsections are separated according to the number of non-zero order parameters in the calculation i.e. how many different Cooper pairs are present at once, which determines the type of phase under consideration. We begin with the case where there are no Cooper Pairs and progress to the situation where all three are present which is discussed in the final subsection. We assume the order parameters in Eq.\,\eqref{PolarPhi} to be uniform in space, therefore $\nabla \rho_i = \nabla \psi_i = 0$. For the gauge fields, we consider the ansatz $\tilde{\bm{A}}_3 = x\tilde{B}_3 \hat{\bm{e}}_y$, $\tilde{\bm{A}}_8 = x\tilde{B}_8 \hat{\bm{e}}_y$ in keeping with the magnetic fields $\tilde{\bm{B}}_3$ and $\tilde{\bm{B}}_8$ being parallel to the external field $\bm{H}$. With these accounted for in our potential \eqref{U0}, we use the Euler-Lagrange equations and obtain five equations of motion. Listed respectively, for the gauge fields $\tilde{\bm{A}}_3$ and $\tilde{\bm{A}}_8$,
\begin{subequations}\label{eoma}
\begin{eqnarray} 
    \rho_1^2\left( \tilde{q}_{3}\tilde{B}_3 + \tilde{q}_{81}\tilde{B}_8 \right) - \rho_2^2\left( -\tilde{q}_{3}\tilde{B}_3 + \tilde{q}_{82}\tilde{B}_8 \right) = 0 \, , \label{emoa3}
    \\[2ex] 
    \rho_1^2\tilde{q}_{81}\left( \tilde{q}_{3}\tilde{B}_3 + \tilde{q}_{81}\tilde{B}_8 \right) + \rho_2^2\tilde{q}_{82}\left( -\tilde{q}_{3}\tilde{B}_3 + \tilde{q}_{82}\tilde{B}_8 \right) + \rho_3^2\tilde{q}_{83}^2 \tilde{B}_8 = 0 \label{eoma8} \, ,
\end{eqnarray}
\end{subequations}
and for the condensates $\rho_1$, $\rho_2$ and $\rho_3$,
\begin{subequations} \label{eomrho}
\begin{align}
    \rho_1 \left[ \lambda \rho_1^2 - h(\rho_2^2 + \rho_3^2 ) - \mu^2 + m_1^2 + x^2\left( \tilde{q}_{3}\tilde{B}_3 + \tilde{q}_{81}\tilde{B}_8  \right)^2\right] = 0 \,, \label{eomrho1} \\[2ex]
    \rho_2 \left[ \lambda \rho_2^2 - h(\rho_1^2 + \rho_3^2 ) - \mu^2 + m_2^2 + x^2\left( -\tilde{q}_{3}\tilde{B}_3 + \tilde{q}_{82}\tilde{B}_8  \right)^2\right] = 0 \,, \label{eomrho2} \\[2ex]
    \rho_3 \left[ \lambda \rho_3^2 - h(\rho_1^2 + \rho_2^2 ) - \mu^2 + m_3^2 + x^2\tilde{q}_{83}^2\tilde{B}_8^2\right] = 0 \label{eomrho3} \,. 
\end{align}
\end{subequations}
The value of the condensates and magnetic fields for every phase are deduced from these equations. With these we calculate the corresponding $\Omega_0$ and Gibbs free energy density $\mathcal{G}$ (from Eq.\,\eqref{gibbs}). Note that our condensates should be $x$-independent if they're spatially homogeneous, and therefore all terms proportional to $x$ must vanish separately in Eqs.\ \eqref{eomrho} (except when a vanishing condensate that is imposed as a condition means they're trivially fulfilled). Free energies are expressed to first order in $\alpha$, which is equivalent to second order in $m_s/\mu_q$ according to Eq.\,\eqref{alpha}, consistent with our starting GL potential \eqref{UPhi}. The condensates are first expressed in the terms of the parameters in our potential \eqref{U0} and then re-written, as with the free energies, in a form such that the strange mass terms appear as corrections. Then by setting $\alpha=0$, the massless results can be easily recovered for comparison with the findings in Ref.\ \cite{Haber:2017oqb}, where a different rotation is used for the gauge fields. Thus, it is useful way of checking all of our expressions are consistent (after we re-insert the expressions for the rotations \eqref{thetamix}), as they should be independent of the rotations used.

\subsection{NOR}
\label{Project1:subsec:NOR}
Briefly, before looking at the superconducting phases, we will look at the non-superconducting phase where there are no Cooper pair condensates i.e.\ $\rho_i=0$ for $i=1,2,3$,  referred to as the NOR (Normal) phase. This is analogous to the normal phase discussed in Sec.\,(\ref{Background:sec:SC}). All the equations of motion are trivially satisfied and the expressions $\tilde{B}_3=H\sin\vartheta_1\cos\vartheta_2$ and $\tilde{B}_8=H\sin\vartheta_1$ are obtained by minimising the Gibbs free energy density \eqref{gibbs} with respect to the magnetic fields. Feeding these back into the Gibbs free energy density yields

\begin{equation}
    \mathcal{\mathcal{G}}_{\mathrm{NOR}} = -\frac{H^2}{2} \, .
\end{equation}
This result is as expected from the corresponding result in Sec.\,(\ref{Background:subsec:Hc}). As it is not a superconducting phase, there are no induced colour-magnetic fields and the external magnetic field fully penetrates thus, in the non-rotated basis, $\bm{B}_3=\bm{B}_8=\bm{0}$ and $\bm{B}=\bm{H}$. One can check this is the case by undoing the rotations \eqref{rotatetwice} on the minimised magnetic fields in the rotated basis. The Gibbs free energy density does not receive a mass correction from the strange quark as the masses of unpaired fermions are not explicitly accounted for in the GL formalism. In this sense, the NOR phase is the ``vacuum'' of our theory. Unlike this phase, all the following phases are superconducting and receive a mass correction.

\subsection{2SC}
\label{Project1:subsec:2SC}
The simple case of a single non-zero condensate is the 2SC phase where only Cooper pairs of one kind are present. This means that there are three separate candidate 2SC phases with differing condensate, free energies and induced magnetic fields. Each individual case is equivalent to the single component analysis of Sec.\,(\ref{Background:subsec:Hc}) and generally the condensates and free energy take on the form
\begin{align}
    \rho_i^2&=\frac{\left( \mu^2- m_i^2 \right)}{\lambda}, 
    \\[2ex]
    \Omega_{0,\rho_i}&=-\frac{\left( \mu^2- m_i^2 \right)^2}{4\lambda} \, ,
    \label{Gen2SCres}
\end{align}
where the additional subscript on $\Omega_0$ denotes which $\rho_i$ is non-zero. Since each case has different values for the magnetic fields they (and by extension the Gibbs free energy density) cannot be expressed in such a concise and simple way. In addition, we wish to know the behaviour with the respect to $\alpha$ to first order, therefore we will use Eqs.\ \eqref{malpha} and provide these details for each phase. Noting that the subscripts $i=1,2,3$ corresponds to the Cooper pairs $(ds,us,ud)$ respectively, we label the 2SC phases according to the Cooper pairs present.

Beginning with 2SC$_{ds}$, where $\rho_1 \neq 0$ and $\rho_2=\rho_3=0$, Eqs.\ \eqref{eomrho2} and \eqref{eomrho3} are trivially fulfilled and Eqs.\ \eqref{eoma} gives $\tilde{q}_3\tilde{B}_3=-\tilde{q}_{81}\tilde{B}_8$. Then
\begin{equation}
    \rho_1^2  = \rho_{ds}^2= \frac{\mu^2}{\lambda}\left(1 - \frac{\alpha}{3}\right) \, ,
\end{equation}
and 
\begin{equation}
    \Omega_{0,\rho_1} = -\frac{\mu^4}{4\lambda}\left(1-\frac{\alpha}{3}\right)^2 \simeq -\frac{\mu^4}{4\lambda}\left(1-\frac{2\alpha}{3}\right) \, .
\end{equation}
Inserting this into Eq.\,\eqref{gibbs}, we once again minimise the Gibbs free energy density. As a result $\tilde{B}_3=\tilde{B}_8=0$ and the free energy density is 
\begin{equation}
    \mathcal{G}_{ds} \simeq -\frac{H^2\cos^2\vartheta_1\cos^2\vartheta_2}{2} - \frac{\mu^2}{4\lambda}\left(1 - \frac{2\alpha}{3}\right)= -\frac{3g^2H^2}{2(3g^2+4e^2)} - \frac{\mu^2}{4\lambda}\left(1 - \frac{2\alpha}{3}\right) \, ,
\end{equation}
working to first order in $\alpha$ as previously mentioned. We have also expressed the final result in terms of the coupling constants using Eqs.\ \eqref{thetamix} and labelled the Gibbs free energy density according to the present Cooper pair condensate.

In the case where only $\rho_2 \neq 0$, we are in the 2SC$_{us}$ phase. Following the same procedure, the relation between the magnetic fields is $\tilde{q}_3\tilde{B}_3=\tilde{q}_{82} \tilde{B}_8$ and
\begin{equation}
    \rho_2^2 = \rho_{us}^2 = \frac{\mu^2}{\lambda}\left(1 - \frac{7\alpha}{12}\right) \, ,
\end{equation}
with
\begin{equation}
    \Omega_{0,\rho_2} = -\frac{\mu^4}{4\lambda}\left(1-\frac{7\alpha}{12}\right)^2 \simeq -\frac{\mu^4}{4\lambda}\left(1-\frac{7\alpha}{6}\right) \, .
\end{equation}
The minimisation leads to
\begin{equation}
    \tilde{B}_3=\frac{3eg(3g^2-2e^2)H}{2(3g^2+4e^2)^{1/2}(3g^2+e^2)^{3/2}}\,, \quad \tilde{B}_8=\frac{9eg^2 H}{2(3g^2+e^2)^{3/2}}\,,
\end{equation}
and
\begin{equation}
  \mathcal{G}_{us} \simeq -\frac{3g^2 H^2}{2(3g^2+e^2)} -\frac{\mu^4}{4\lambda}\left(1-\frac{7\alpha}{6}\right) \, .
\end{equation}

For only $\rho_3 \neq 0 $ one magnetic field vanishes, $\tilde{B}_8=0$ which can be seen from \eqref{eoma8}. The resulting expressions are
\begin{align} 
    \rho_3^2 &= \rho_{ud}^2= \frac{\mu^2}{\lambda}\left(1 - \frac{\alpha}{12}\right)\label{rho32SC} \, ,
    \\[2ex]
    \Omega_{0,\rho_3} &= -\frac{\mu^4}{4\lambda}\left(1-\frac{\alpha}{12}\right)^2 \simeq -\frac{\mu^4}{4\lambda}\left(1-\frac{\alpha}{6}\right) \,,
    \\[2ex]
    \tilde{B}_3&=\frac{3eg H}{\sqrt{3g^2+e^e}\sqrt{3g^2+4e^2}}\label{B32SC}\,,
    \\[2ex]
    \mathcal{G}_{ud} &\simeq -\frac{3g^2 H^2}{2(3g^2 + e^2)} -\frac{\mu^4}{4\lambda}\left(1-\frac{\alpha}{6}\right)\,.
\end{align}
One can begin to appreciate already that not all these phases will appear in the phase diagram. The 2SC$_{ud}$ phase has a lower free energy for all parameters (taking $\alpha$ to always be positive) and therefore we usually refer to it as simply the 2SC phase in the context of this project and set $\rho_{\rm{2SC}}\equiv\rho_{\rm{ud}} $, $\mathcal{G}_{\rm{2SC}}\equiv\mathcal{G}_{\rm{ud}}$. Furthermore, when we set $\alpha=0$ all the condensates become equal and so do the free energies of the 2SC$_{ud}$ and 2SC$_{us}$ phases ($\mathcal{G}_{\mathrm{2SC}_{ds}}$ still differs in the magnetic contribution). This is in agreement with Ref.\ \cite{Haber:2017oqb} and is expected due to $d$ and $s$ quarks being effectively indistinguishable in the massless case. This is a key factor for the formation of the domain walls in the (massless) inhomogeneous phase.

\subsection{fSC}
\label{Project1:subsec:fSC}
When there are two non-zero condensates, similar to 2SC, we have three possible candidate phases but one is favoured over the others. This phase we call dSC as in Ref.\ \cite{Iida:2004cj}, named so because all Cooper pairs present have constituent $d$ quarks (i.e. with $\rho_2=0$ only). Generalising this convention to fSC (with ``f'' standing for flavour), we also present the results for uSC ($\rho_1=0$) and sSC ($\rho_3=0$) for completeness. In all cases (where only one $\rho_i=0$), $\tilde{B}_3=\tilde{B}_8=0$ due to equations of motion \eqref{eoma}, one equation from \eqref{eomrho} is fulfilled, and the remaining two equations are solved simultaneously for the non-zero condensates. The results are summarised below and have been expressed in terms of $\alpha$ and $\eta$ using relations \eqref{malpha} and \eqref{eta} respectively. With only $\rho_1=0$,
\begin{subequations}
    \begin{align}
     \rho_2^2 =\frac{\lambda(\mu^2-m_2^2) + h(\mu^2-m_3^2)}{\lambda^2-h^2}=
    \frac{ \mu^2}{\lambda(1-\eta)}\left[ 1 - \frac{\alpha}{12}\left(\frac{7+\eta}{1+\eta}\right)\right] \,, 
    \\[2ex] 
    \rho_3^2 =\frac{\lambda(\mu^2-m_3^2) + h(\mu^2-m_2^2)}{\lambda^2-h^2} = 
    \frac{ \mu^2}{\lambda(1-\eta)}\left[ 1 - \frac{\alpha}{12}\left(\frac{1+7\eta}{1+\eta}\right)\right]\,,
    \end{align}
\end{subequations}
\begin{equation}
    \mathcal{G}_{\mathrm{uSC}} \simeq -\frac{3g^2H^2}{2(3g^2+4e^2)} - \frac{\mu^4}{2\lambda(1-\eta)}\left( 1- \frac{2\alpha}{3}\right) \,.
\end{equation}
\noindent
For only $\rho_2=0$,
\begin{subequations}
    \begin{align}
    \rho_1^2 = \frac{\lambda(\mu^2-m_1^2) + h(\mu^2-m_3^2)}{\lambda^2-h^2}
    =\frac{ \mu^2}{\lambda(1-\eta)}\left[ 1 - \frac{\alpha}{12}\left(\frac{4+\eta}{1+\eta}\right)\right] \,, 
    \\[2ex]  
    \rho_3^2 = \frac{\lambda(\mu^2-m_3^2) + h(\mu^2-m_1^2)}{\lambda^2-h^2} = \frac{ \mu^2}{\lambda(1-\eta)}\left[ 1 - \frac{\alpha}{12}\left(\frac{1+4\eta}{1+\eta}\right)\right] \,, 
\end{align}
\end{subequations}
\begin{equation}
    \mathcal{G}_{\mathrm{dSC}} \simeq  -\frac{3g^2H^2}{2(3g^2+4e^2)} - \frac{\mu^4}{2\lambda(1-\eta)}\left( 1- \frac{5\alpha}{12}\right) \, .
\end{equation}
\noindent
Finally, with $\rho_3 = 0$,
\begin{subequations}
    \begin{align}
    \rho_1^2 = \frac{\lambda(\mu^2-m_1^2) + h(\mu^2-m_2^2)}{\lambda^2-h^2} =
    \frac{ \mu^2}{\lambda(1-\eta)}\left[ 1 - \frac{\alpha}{12}\left(\frac{4+7\eta}{1+\eta}\right)\right]\,, 
    \\[2ex]
    \rho_2^2 = \frac{\lambda(\mu^2-m_2^2) + h(\mu^2-m_1^2)}{\lambda^2-h^2} =
    \frac{ \mu^2}{\lambda(1-\eta)}\left[ 1 - \frac{\alpha}{12}\left(\frac{7+4\eta}{1+\eta}\right)\right]\,,
\end{align}
\end{subequations}
\begin{equation}
    \mathcal{G}_{\mathrm{sSC}} \simeq -\frac{3g^2H^2}{2(3g^2+4e^2)} - \frac{\mu^4}{2\lambda(1-\eta)}\left( 1- \frac{11\alpha}{12}\right)\,.
\end{equation}
With $\alpha=0$ the results all become identical. None of these phases are favoured when the quarks are all massless. In contrast, the mass correction allows the dSC phase to appear in our homogeneous phase diagram as previously mentioned here and observed in Ref.\ \cite{Iida:2003cc,Iida:2004cj} in the absence of an external magnetic field.

\subsection{CFL}
\label{Project:subsec:homCFL}

In the CFL phase all condensates are non-zero. We find $\tilde{B}_3=\tilde{B}_8=0$ from imposing $x$-independence, leaving Eqs.\ \eqref{eomrho} as coupled equations which are solved simultaneously for $\rho_1$, $\rho_2$, $\rho_3$. The results for the $\rho_i$ are
\begin{subequations}
    \begin{eqnarray} 
    \rho_1^2 = \frac{\mu^2}{\lambda}\frac{1}{(1-2\eta)}\left[1-\frac{4\alpha}{12}\right]\,, 
    \\[2ex] 
    \rho_2^2 =\frac{\mu^2}{\lambda}\frac{1}{(1-2\eta)}\left[1-\frac{\alpha}{12}\left(\frac{7-2\eta}{1+\eta}\right)\right]\,, 
    \\[2ex]
    \rho_3^2 =\frac{\mu^2}{\lambda}\frac{1}{(1-2\eta)}\left[1-\frac{\alpha}{12}\left(\frac{1+10\eta}{1+\eta}\right)\right]\,.
\end{eqnarray}
\end{subequations}
Equation \eqref{U0} yields
\begin{equation}
    \Omega_{0,\mathrm{CFL}} \simeq- \frac{3\mu^4}{4\lambda(1-2\eta)}\left(1-\frac{2\alpha}{3}\right)\,,
\end{equation}
and from Eq.\,\eqref{gibbs} the Gibbs free energy density is then
\begin{equation}
    \mathcal{G}_{\mathrm{CFL}} \simeq -\frac{3g^2H^2}{2(3g^2+4e^2)} -\frac{3\mu^4}{4\lambda(1-2\eta)}\left(1-\frac{2\alpha}{3}\right)\,.
\end{equation}
Once again the final results expressed in terms of $\eta$ and $\alpha$. To check, setting $\alpha=0$ recovers the massless case where the condensates all become equal as expected\footnote{Since the three condensates are different due to the strange quark mass, this phase was termed modified CFL (mCFL) in Ref.\,\cite{Iida:2004cj}. Here, we simply keep the term CFL.}.

\section{Critical magnetic fields}
\label{Project1:sec:Hc}
Here is presented the analytical work on the critical magnetic fields in both the homogeneous and inhomogeneous regimes. Following from the previous section, we use the free energy results to determine the type-I critical fields $H_c$, and use them to construct the homogeneous phase diagrams. In the inhomogeneous case, we only consider the type-II transitions between 2SC and neighbouring phases since our main interest are the 2SC flux tubes i.e. the magnetic defects in 2SC. The upper critical magnetic fields $H_{c2}$ are derived analytically, while the numerical calculation of the lower critical field $H_{c1}$ is set up in preparation for the next section. 

\subsection{Critical field $H_{c}$}
\label{Project1:subsec:Hc}
In principle, we have eight possible phases to consider when constructing the homogeneous phase diagram. As touched upon, not all phases we have considered will be relevant. To re-iterate, the 2SC$_{ud}$ and dSC phases always have a lower free energy than the other respective 2SC and fSC phases, leaving us with four phases: NOR, 2SC$_{ud}$, dSC and CFL. From now on we will refer to 2SC$_{ud}$ as the 2SC phase in this section. Comparing the free energies of the remaining phases yields the criteria for when each phase is preferred over another. Since we are interested in quark matter in external magnetic fields, we equate the free energies pairwise and solve for the critical magnetic field $H=H_c$. We summarise and express these in terms of the variables $\eta$, $g$ and $\alpha$;
\begin{equation} 
    \frac{H_c^2}{\mu^4/\lambda}=
    \begin{cases}
        \frac{3g^2 + e^2}{2e^2}(1-\frac{\alpha}{6})  & \mathrm{NOR/2SC} \\
        \\
        \frac{3(3g^2 + 4e^2)}{8e^2}\frac{1}{1-2\eta}(1-\frac{2\alpha}{3}) & \mathrm{NOR/CFL} \\
        \\
         \frac{(3g^2 + 4e^2)(3g^2 + e^2)}{9g^2e^2}\frac{1+\eta}{1-2\eta}\left(1 -\frac{\alpha}{12} \left(\frac{11+2\eta}{1+\eta}\right)\right) & \mathrm{2SC/CFL}\\
        \\
        \frac{(3g^2+4e^2)(3g^2+e^2)}{18g^2e^2}\frac{1+\eta}{1-\eta}\left(1-\frac{\alpha}{6}\left(\frac{4+\eta}{1+\eta}\right)\right)&\mathrm{2SC/dSC}
    \end{cases}
    \,.
    \label{Hcs}
\end{equation}
The convention adopted here is ``Phase at $H>H_c$'' /``Phase at $H<H_c$''. Note that there is no critical field for CFL/dSC and NOR/dSC given. The former transition does not depend on the magnetic field. By comparing the free energies, instead one obtains the condition in terms of $\eta$ and $\alpha$
\begin{equation}
    \eta = \frac{7\alpha-6}{2(3+\alpha)} \quad \quad \mathrm{CFL/dSC}\,,
    \label{HomoCritEta}
\end{equation}
where the CFL phase is preferred above this value and the dSC phase below it. For the latter, the critical field is
\begin{equation}
    \frac{H_c^2}{\mu^4/\lambda}=\frac{3g^2+4e^2}{4e^2}\frac{1}{1-\eta}\left(1-\frac{5\alpha}{12}\right) \quad \quad\mathrm{NOR/dSC}\,,
\end{equation}
but at either side of this transition other phases are favoured over both NOR and homogeneous dSC phases. Hence, it will not appear in any of our phase diagrams and was excluded from the critical field(s) in Eq.\,\eqref{Hcs}. 

\begin{figure} [t]
\begin{center}
\hbox{\includegraphics[width=0.5\textwidth]{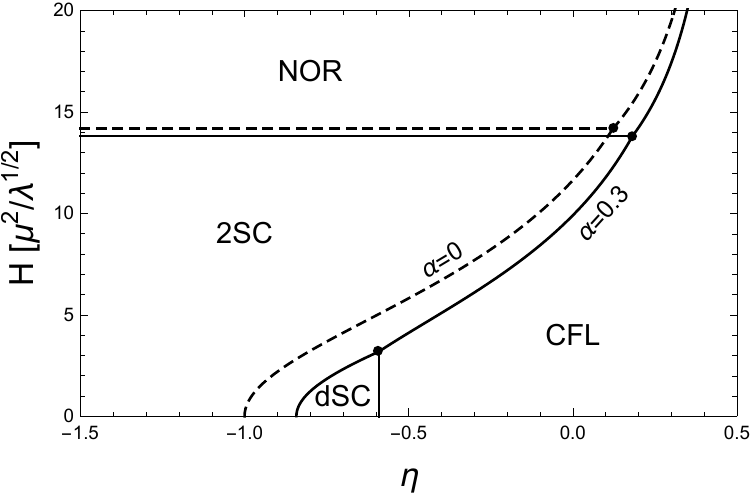}
\includegraphics[width=0.5\textwidth]{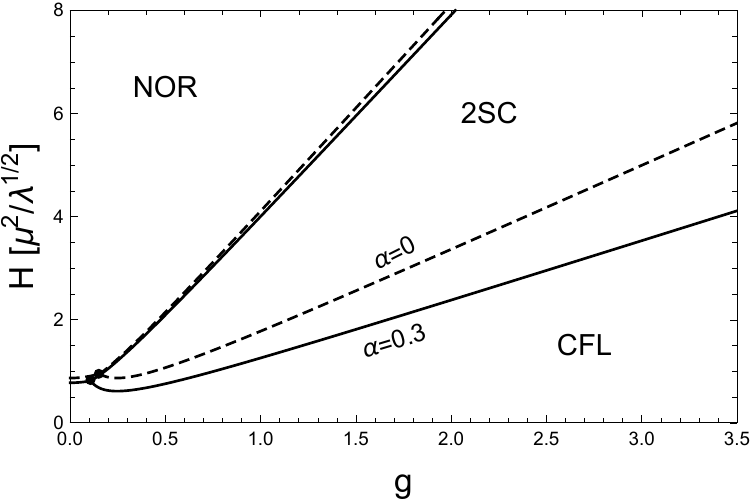}}
\caption{Homogeneous phases at non-zero external magnetic field $H$ in the plane spanned by $H$ and the parameter ratio  $\eta=h/\lambda$ for fixed strong coupling constant $g=3.5$ (left) and in the $g$-$H$ plane for fixed $\eta=-1/2$ (right). In both panels, the solid curves are for mass parameter $\alpha=0.3$, while the dashed curves correspond to the massless limit $\alpha=0$. 
}\label{fig:Heta}
\end{center}
\end{figure}

With $H_c$ in units of $\mu^2/\sqrt{\lambda}$ it reduces the parameter space to $H$, $\alpha$, $\eta$ and $g$ ($e\simeq 0.3$). Following Ref.\ \cite{Haber:2017oqb}, we have displayed the $\eta$-$H$ and $g$-$H$ slices of this space in Fig.\ \ref{fig:Heta}. These are at $g=3.5$ and $\eta=-0.5$ respectively, values that later calculations will use, and are shown for $\alpha=0$ and $\alpha=0.3$. For reference, $\alpha=0.3$ at $T_c/\mu_q\sim0.1$, corresponds to  $m_s/\mu_q\sim0.5$. From Fig.\ \ref{fig:Heta} we can see that there are triple points where three phases meet. One can obtain their coordinates in terms of $g$, $\alpha$ and $\eta$ by equating the appropriate critical fields. There are two in the left panel of Fig.\ \ref{fig:Heta} and one in the right. The one unique to the $\eta$-$H$ plane is
\begin{equation}
    \left(\eta\,,\,\frac{H}{\mu^2/\sqrt{\lambda}}\right)=\left(\frac{7\alpha-6}{6+2\alpha},\frac{\sqrt{(3g^2+4e^2)(3g^2+e^2)\alpha}}{6ge} \right)\,,
    \label{3PtEta2}
\end{equation}
where the  2SC, CFL and dSC phases meet. This triple point is absorbed into the horizontal axis ($H=0$) when $\alpha=0$ due to dSC being disfavoured in the massless limit. The second point in the left panel is a transition between 2SC, CFL and the NOR phases, whose coordinates are
\begin{equation}
    \left(\eta,\,\frac{H}{\mu^2/\sqrt{\lambda}}\right)=\left(\frac{1}{2} - \frac{3g^2+4e^2}{2\left(3g^2+e^2\right)}\frac{3(3-2\alpha)}{2(6-\alpha)},\,\sqrt{\frac{3g^2+e^2}{2e^2}}\sqrt{1-\frac{\alpha}{6}} \right).   
    \label{3PtEta1}
\end{equation}
This is the same triple point as the one in the $g$-$H$ plane, given by
\begin{equation}
    \left(g\,,\,\frac{H}{\mu^2/\sqrt{\lambda}}\right)=\left( 2e\sqrt{\frac{2}{3}}\sqrt{\frac{1-\frac{11\alpha}{12}+\eta(1-\frac{\alpha}{6})}{1+\frac{4\alpha}{3}-8\eta\left(1-\frac{\alpha}{6}\right)}},\,\frac{3}{\sqrt{2}} \sqrt{\frac{1-\frac{5\alpha}{6} + \frac{\alpha^2}{9}}{1+\frac{4\alpha}{3}-8\eta\left(1-\frac{\alpha}{6}\right)}}\right).  
    \label{HgTriple2}
\end{equation}
Unlike the first triple point, this point also occurs in the massless limit. 

Including the correction terms already changes the phase structure at the homogeneous level. The most relevant remark is that CFL tends to be disfavoured over 2SC when the strange quark mass is treated as non-zero, which is in accordance with many other calculations. In both panels in Fig.\ \ref{fig:Heta} it can be appreciated that the 2SC phase encroaches on the massless CFL region as we move to a non-zero\footnote{In fact, as $\alpha\rightarrow0.6$, the CFL phase disappears from the right panel, with the triple point merging with the vertical axis. However, $\alpha\approx 0.6$ is comparatively large and our approximation might be less trustworthy at these values. Therefore, it is mainly a qualitative observation that the CFL phase may be completely disfavoured for certain parameters at high strange quark mass.} $\alpha$. Also observable in both panels is the slight favouring of the NOR phase over the 2SC phase in the presence of a strange quark mass. As mentioned previously, the dSC phase only emerges in the left panel when we consider these corrections and only for a small range of $\eta$ for the values of $\alpha$ shown. In future calculations we set $g=3.5$ and $\eta=-0.5$. This is outside the homogeneous dSC region and therefore it doesn't affect the calculation. Furthermore, at these values of $g$ and $\eta$ there is no NOR/CFL transition, thus with the focus on 2SC flux tubes we need only consider transitions to the NOR phase from 2SC. This applies also in the massless case and is mentioned in Ref.\ \cite{Haber:2017oqb}.

The 2SC phase becomes the ground state below certain $\eta$ (dependent on $g$). In Fig.\ \ref{fig:Heta} this occurs when $\eta\lesssim-0.85$, where at $H=0$ 2SC remains the preferred phase. This would suggest that as we decrease $\eta$, the 2SC flux tube region would also increase, which is interesting to note. However, at the weak-coupling value $\eta=-0.5$, the ground state is the CFL phase. All things considered, as we increase the magnetic field we progress through the phases in the order CFL, 2SC, NOR at our weak-coupling value, extrapolated to strong coupling $g=3.5$. Hence, taking the homogeneous phase structure to roughly resemble the inhomogeneous phase structure, we only consider transition between the 2SC phase and the NOR and CFL phases for the purposes of our flux tube calculations.

For comparison with astrophysical data, we give an estimation for the strength of the magnetic field at $T_c/\mu_q=0.5$ and $\mu_q=400\,\mathrm{MeV}$, which are typical values for the interior of compact stars. Taking the critical magnetic field for the NOR/2SC transition, $H_c\simeq14\,\mu^2/\sqrt{\lambda}$ roughly translates to $H_c\simeq 1.1\times 10^{19}\, {\rm G}$.

\subsection{Upper critical field $H_{c2}$}
\label{Project1:subsec:Hc2}

Before we turn to the flux tubes themselves, it is useful to compute their upper 
critical field $H_{c2}$. In the standard scenario of a single condensate, this is the maximum magnetic field which can sustain a non-zero condensate, under the assumption of a second-order transition to the normal phase. It is therefore the  critical field below which an array of flux tubes is expected. In our multi-component system the situation is more complicated, and we have to calculate different critical fields $H_{c2}$ depending on which condensates melt. Having $H_c$ and $H_{c2}$ at hand, we can then determine the parameter regime where the colour superconductor is of type II, and in particular where we expect 2SC flux tubes. 

The calculation of $H_{c2}$ is a generalisation to non-zero strange quark mass of the analogous calculation done in Ref.\ \cite{Haber:2017oqb}. That calculation, in turn, was a generalisation of the standard single-component calculation which can be found in many textbooks e.g. in Ref.\,\cite{tinkham2004introduction}. In a single-component superconductor, one linearises the GL equations for a small condensate. The equation for the condensate then 
has the form of the Schr{\"o}dinger equation for the harmonic  oscillator, from which one reads off the maximal possible magnetic field $H_{c2}$, corresponding to the ground state energy. We know from the previous subsection that for strong coupling, as we decrease the magnetic field within the NOR phase, we encounter the 2SC phase. Consequently, for the corresponding critical field $H_{c2}$ we only have to take into account a single condensate, i.e., this case is analogous to the textbook scenario and leads to the simple generalisation of Eq.\,(40) in Ref.\ \cite{Haber:2017oqb},
\begin{equation} \label{Hc22SCNOR}
\mbox{2SC/NOR:} \qquad H_{c2} = \frac{3(\mu^2-m_3^2)}{e} = \frac{3\mu^2}{e}\left(1-\frac{\alpha}{12}\right) \, .
\end{equation}
This is reminiscent of the result derived for a conventional superconductor in Sec.\,(\ref{Background:subsec:Hc2}).  In this standard scenario all three critical magnetic fields $H_c$, $H_{c1}$, and $H_{c2}$ intersect at a single point (as a function of a model parameter, usually the GL parameter $\kappa$). Therefore, this intercept defines the transition between type-I and type-II behaviour (usually at $\kappa=1/\sqrt{2}$). 
Here, by equating $H_{c2}$ with $H_c$ for the 2SC/NOR transition in Eq.\,\eqref{Hcs} we find for this transition point 
\begin{equation} \label{type2SCNOR}
\frac{T_c}{\mu_q} = \frac{\sqrt{7\zeta(3)}}{12\sqrt{3}\pi^2} \sqrt{g^2+\frac{e^2}{3}} +{\cal O}\left(\frac{m_s^4}{\mu_q^4}\right) \, ,
\end{equation}
where the weak-coupling expression of $\lambda$ in Eq.\,\eqref{weak} has been used. It thus turns out that $T_c/\mu_q$ is a natural parameter to distinguish between type-I and type-II behaviour -- with large $T_c/\mu_q$ corresponding to type II. Interestingly, we see that there is no mass correction to the transition point within the order of our approximation. We shall see later that indeed $H_{c1}$ intersects $H_c$ and $H_{c2}$ at the same $T_c/\mu_q$. The reason is that in the vicinity of this point the system effectively behaves as a single-component system. Additional condensates can be induced  in the cores of 2SC flux tubes -- and our main results concern such unconventional flux tubes -- but we will see that this is not the case close to the point \eqref{type2SCNOR}. 

The transition from the homogeneous 2SC phase, where $\phi_3$ is non-zero, to an inhomogeneous phase is slightly more complicated. Assuming a second-order transition, we linearise the GL equations in $\phi_1$ and $\phi_2$. In the massless limit, the resulting two (decoupled) equations yield the same critical magnetic field \cite{Haber:2017oqb}. In other words, as we approach the flux tube phase by decreasing $H$, both $\phi_1$ and $\phi_2$ become non-zero simultaneously (and continuously). This is different for non-zero $m_s$, in which case the two relevant equations are 
\begin{subequations}
\begin{eqnarray}
\left[\left(\nabla+i\tilde{q}_3
\tilde{\bm{A}}_3\right)^2+(\mu^2-m_1^2)+2h|\phi_3|^2\right]\phi_1 &\simeq& 0 \, , \\[2ex]
\left[\left(\nabla-i\tilde{q}_3
\tilde{\bm{ A}}_3\right)^2+(\mu^2-m_2^2)+2h|\phi_3|^2\right]\phi_2 &\simeq& 0 \, ,
\end{eqnarray}
\end{subequations}
where we have set $\tilde{\bm{ A}}_8=0$ since $\tilde{B}_8=0$ in the 2SC phase and where $\phi_3 = \rho_3/\sqrt{2}$ is the condensate in the homogeneous 2SC phase \eqref{rho32SC}. With the usual arguments and using the 2SC relation between $\tilde{B}_3$ and $H$ from Eq.\,\eqref{B32SC}, we obtain two different critical fields,
\begin{subequations} \label{Hc212}
\begin{eqnarray}
H_{c2}^{(1)} &=\frac{2\mu^2(3g^2+e^2)}{3eg^2}\left(1+\eta-\frac{4+\eta}{12}\alpha\right)  \, , \label{Hc2one}
\\[2ex]
H_{c2}^{(2)} &=\frac{2\mu^2(3g^2+e^2)}{3eg^2}\left(1+\eta-\frac{7+\eta}{12}\alpha\right) \, . 
\end{eqnarray}
\end{subequations} 
The most relevant case for us is the one where both $H_{c2}^{(1)}$  and $H_{c2}^{(2)}$ are positive (a formally negative value indicates that the critical field does not exist, indicating that the homogeneous phase persists down to $H=0$). This is the case for $\eta=-1/2$ and all reasonable, i.e., not too large, values of $\alpha$. In this scenario, there is a transition at  $H_{c2}^{(1)}$ from the homogeneous 2SC phase to a phase where both $\phi_1$ and $\phi_3$ are non-zero, which is an inhomogeneous version of the dSC phase (see Sec.\,(\ref{Project1:subsec:fSC})). Then, as we reach the ``would-be'' $H_{c2}^{(2)}$ by further decreasing $H$, the approximation by which this critical field was computed is no longer valid, and thus the value for $H_{c2}^{(2)}$ becomes irrelevant. Nevertheless, it can be expected that there will be some transition from an inhomogeneous dSC phase to an inhomogeneous CFL phase. The existence of an intermediate inhomogeneous dSC phase due to  the non-zero strange quark mass is an interesting new observation, but it is beyond the scope of this project to construct this phase explicitly. 

We may again compute the transition point between type-I and type-II behaviour. For $\eta=-1/2$ (where there is no homogeneous dSC phase), we equate $H_{c2}^{(1)}$ with $H_c$ for the 2SC/CFL transition from Eq.\,\eqref{Hcs}. Dropping terms quadratic in $\alpha$ we find
\begin{equation}
\frac{T_c}{\mu_q} \simeq  c\left(1-\frac{\alpha}{4}\right) \, , \qquad 
c\equiv \frac{\sqrt{7\zeta(3)}}{12\sqrt{2}\pi^2}\frac{g\sqrt{3g^2+4e^2}}{\sqrt{3g^2+e^2}} \,.
\end{equation}
Since $\alpha$ depends on $T_c/\mu_q$ this is an implicit equation for $T_c/\mu_q$. To lowest non-trivial order in $m_s^2/\mu_q^2$ the solution is 
\begin{equation} \label{type2SCCFL}
\frac{T_c}{\mu_q} = c\left(1+\frac{m_s^2}{8\mu_q^2}\ln c\right) +{\cal O}\left(\frac{m_s^4}{\mu_q^4}\right) \, .
\end{equation}
Therefore, this transition point between type-I and type-II behaviour {\it does} receive a correction quadratic in $m_s$ (linear in $\alpha$), in contrast to the transition point \eqref{type2SCNOR}. The detailed phase structure around this point is expected to be complicated. This is due to the intermediate inhomogeneous dSC phase, as just discussed, but even without mass correction this transition point is affected in a non-trivial way by the multi-component nature of the system \cite{Haber:2017kth,Haber:2017oqb}. Most importantly, if the lower boundary of the flux tube region $H_{c1}$ is computed in the usual way, i.e., assuming a second-order transition, it turns out that the three critical fields no longer intersect in a single point, and the situation becomes more complicated due to a first-order entrance into the flux tube phase. Here we do not have to deal with these complications, since the precise location of the transition point \eqref{type2SCCFL} and the phase transitions in its vicinity are not relevant for the 2SC flux tubes. 

\subsection{Flux tubes and lower critical field $H_{c1}$}
\label{Project1:subsec:Hc1}

Having identified the parameter range where type-II behaviour with respect to 2SC flux tubes is expected, we can now turn to the explicit construction of these flux tubes. Just like in Sec.\,(\ref{Background:subsec:Hc1}), we will restrict ourselves to the calculation of an isolated, straight flux tube, such that we can employ cylindrical symmetry and our calculation becomes effectively one-dimensional in the radial direction. This is sufficient to compute the critical field $H_{c1}$, which is defined as the  field at which it becomes favourable to place a single flux tube in the system, indicating a second-order transition to a phase containing an array of flux tubes. Since the distance between the flux tubes goes to infinity as $H_{c1}$ is approached from above, the interaction between flux tubes plays no role. As explained in the introduction, our main goal is to determine the fate of the 2SC domain walls in the presence of a non-zero strange quark mass. Therefore, we focus exclusively on 2SC flux tubes, i.e., configurations which asymptote to the 2SC phase far away from the centre of the flux tube. 

We will find that the method for deriving $H_{c1}$ in Sec.\,(\ref{Background:subsec:Hc1}) cannot be applied here and we must proceed numerically. In order to compute the profiles of the condensates and the gauge fields we need to derive their equations 
of motion and bring them into a form convenient for the numerical evaluation. We work in cylindrical coordinates $(r,\theta,z)$, where, as above, the $z$-axis is aligned with the external magnetic field $\bm{ H}$
and thus with the flux tube. We use similar ansatz and boundary conditions used for the asymptotics in Sec.\,(\ref{Background:subsec:Hc1}). We introduce dimensionless condensates $f_i$ ($i=1,2,3$), which only depend on the radial distance to the centre of the flux tube, 
\begin{equation}
    \rho_i(\bm{ r}) = f_i(r)\rho_{\rm 2SC} \, , 
\end{equation}
where we have denoted the condensate of the homogeneous 2SC phase \eqref{rho32SC} by $\rho_{\rm 2SC}$. Since we are interested in 2SC flux tubes, we impose the boundary conditions $f_3(\infty)=1$ and $f_1(\infty)=f_2(\infty)=0$. As in ordinary single-component flux tubes, we allow for a non-zero winding number $n\in \mathbb{Z}$, such that the phases of the condensates are 
\begin{equation} \label{windings}
    \psi_1(\bm{ r})=\psi_2(\bm{ r})=0 \, , \qquad \psi_3(\bm{ r})=n\theta \, .
\end{equation}
Here we have set the winding numbers for the ``non-2SC'' condensates $f_1$ and $f_2$ to zero. In principle, we might include configurations where these windings are non-zero. (The baryon circulation around the flux tube vanishes for arbitrary 
choices of the winding numbers as long as $f_1(\infty)=f_2(\infty)=0$.) In such configurations, $f_1$ and/or $f_2$ would have to vanish far away from the flux tube {\it and} in the centre of the flux tube, i.e., at best they would be non-vanishing in an intermediate domain. These configurations do not play a role in the massless limit \cite{Haber:2017oqb} and there is no obvious reason why they should become important if a strange quark mass is taken into account. Therefore, we shall work with Eq.\,\eqref{windings}. As a consequence, the boundary condition for the 2SC condensate in the core is $f_3(0)=0$, while $f_1(0)$ and $f_2(0)$ can be non-zero and must be determined dynamically. 

As we have seen in Sec.\,(\ref{Project1:sec:setup}), after the rotation of the gauge fields, $\tilde{\bm{ A}}$ decouples from the condensates. Therefore, we are left with two non-trivial gauge fields, for which we introduce the dimensionless versions $\tilde{a}_3$ and $\tilde{a}_8$ via
\begin{equation} \label{A3A8tt}
 \tilde{\bm{ A}}_3(\bm{ r}) = \left[\frac{H\cos\vartheta_1\sin\vartheta_2}{2}r+\frac{\tilde{a}_3(r)}{r}\right]\hat{\bm{e}}_\theta \, ,
 \qquad \tilde{\bm{ A}}_8(\bm{ r}) = \frac{\tilde{a}_8(r)}{r}\hat{\bm{e}}_\theta \, ,
\end{equation}
with the boundary conditions $\tilde{a}_3(0) = \tilde{a}_8(0)=0$. This yields the magnetic fields 
\begin{equation} \label{B3B8}
\tilde{\bm{ B}}_3 = \left(H\cos\vartheta_1\sin\vartheta_2+\lambda\rho_{\rm 2SC}^2\frac{\tilde{a}_3'}{R}\right)\hat{\bm{e}}_z \, , \qquad \tilde{\bm{ B}}_8 =\lambda\rho_{\rm 2SC}^2\frac{\tilde{a}_8'}{R}
\hat{\bm{e}}_z \, ,
\end{equation}
where prime denotes derivative with respect to $R$, which is the dimensionless radial coordinate
\begin{equation} \label{R}
R=\frac{r}{\xi_3} \, , \qquad \xi_3 \equiv \frac{1}{\sqrt{\lambda}\rho_{\rm 2SC}} \, . 
\end{equation}
Here, $\xi_3$ is the coherence length; we have added a subscript 3 to indicate that in a 2SC flux tube it is the $ud$ condensate $f_3$ whose asymptotic behaviour is characterised by the coherence length. Following Ref.\ \cite{Haber:2017oqb}, we have separated a term in $\tilde{\bm{ A}}_3$ for convenience, which gives rise to the non-zero field $\tilde{\bm{ B}}_3$ in the 2SC phase, i.e., far away from the flux tube. Therefore, the dimensionless gauge fields do not create any additional magnetic fields at infinity, $\tilde{a}_3'(\infty)=\tilde{a}'_8(\infty)=0$ (alternatively, this effect could have been implemented in the boundary condition for $\tilde{a}_3$). The behaviour of $\tilde{B}_3$ is another qualitative difference of the 2SC flux tube to a textbook flux tube (besides potentially induced additional condensates and the existence of two gauge fields). In a standard flux tube, the magnetic field is expelled in the superconducting phase far away from the flux tube and penetrates through the normal-conducting centre. Here, we have three magnetic fields: $\tilde{B}$, which fully penetrates the superconductor and thus is irrelevant for the calculation of the flux tube profiles; $\tilde{B}_8$, which behaves analogously to the ordinary magnetic field in an ordinary flux tube; and $\tilde{B}_3$, which is non-zero far away from the flux tube {\it and} is affected non-trivially by the flux tube profile. As a consequence, since $\tilde{B}_3$ depends on the external field $H$, the flux tube profiles and flux tube energies also depend on $H$, which poses a technical complication. We should keep in mind, however, that it is the ordinary magnetic field $\bm{ B}$ that dictates the formation of magnetic defects. A flux tube configuration, in which the condensation energy is necessarily reduced, may become favoured if this energy cost is overcompensated by admitting magnetic $\bm{ B}$-flux in the system (this is the meaning of the $-\bm{ B}\cdot\bm{ H}$ term in the Gibbs free energy). It is therefore useful for the interpretation of our results to compute the unrotated field $\bm{ B}$ from the profiles. Undoing the rotation \eqref{rotatetwice} and using Eqs.\,\eqref{tildeB} and \eqref{B3B8}, we find 
\begin{equation} \label{BoverH}
\frac{B}{H} = \cos^2\vartheta_1\left[1 +\frac{e\sin\vartheta_1}{4\Xi}\left( \frac{\tilde{a}_8'}{R} +\frac{3g}{\sqrt{3g^2+4e^2}}\frac{\tilde{a}_3'}{R}\right)\right] \, ,
\end{equation}
where we have introduced the  dimensionless magnetic field 
\begin{equation} \label{Xi}
\Xi \equiv \frac{\tilde{q}_3 H\cos\vartheta_1\sin\vartheta_2}{2\lambda\rho_{\rm 2SC}^2} = \frac{3eg^2}{4\sqrt{\lambda}(3g^2+e^2)}\left(1-\frac{m_3^2}{\mu^2}\right)^{-1}\frac{H}{\mu^2/\sqrt{\lambda}} \, .
\end{equation} 
We can now express the potential \eqref{U0} in terms of the dimensionless condensates and gauge fields,
\begin{equation}\label{U01}
    \begin{split}
        \Omega_0  =& \Omega_{\rm 2SC} + \frac{\lambda\rho_{\rm 2SC}^4}{2}\bigg[f_1'^2+f_2'^2+f_3'^2
        \\[2ex]
        &+f_1^2\left(\frac{f_1^2}{2}-\frac{\mu^2-m_1^2}{\mu^2-m_3^2}\right) +f_2^2\left(\frac{f_2^2}{2}-\frac{\mu^2-m_2^2}{\mu^2-m_3^2}\right)
        +\frac{(1-f_3^2)^2}{2} 
        \\[2ex]
    &+\frac{({\cal N}_1+\Xi R^2)^2f_1^2+({\cal N}_2-\Xi R^2)^2f_2^2+{\cal N}_3^2f_3^2}{R^2} -\eta(f_1^2f_2^2+f_1^2f_3^2+f_2^2f_3^2)\bigg] \, .
    \end{split}
\end{equation}
Here we have denoted the potential of the homogeneous 2SC phase 
by 
\begin{equation}
    \Omega_{\rm 2SC}=-\frac{(\mu^2-m_3^2)^2}{4\lambda}\, , 
\end{equation}
and we have abbreviated 
\begin{equation}
    {\cal N}_1 \equiv \tilde{q}_3\tilde{a}_3+\tilde{q}_{81}\tilde{a}_8 \, ,\qquad 
    {\cal N}_2 \equiv -\tilde{q}_3\tilde{a}_3+\tilde{q}_{82}\tilde{a}_8 \, ,\qquad 
    {\cal N}_3 \equiv n-\tilde{q}_{83}\tilde{a}_8 \, .  
\end{equation}
Inserting Eq.\,\eqref{U01} into the Gibbs free energy density \eqref{gibbs} and using the expressions for the magnetic fields from Eq.\,\eqref{B3B8}, we derive the equations of motion for the gauge fields,
\begin{subequations} \label{eqsa}
    \begin{align}
        \tilde{a}_3''-\frac{\tilde{a}_3'}{R}= \frac{\tilde{q}_3}{\lambda}[({\cal N}_1+\Xi R^2)f_1^2-({\cal N}_2-\Xi R^2)f_2^2] \, ,\label{a3t}  \\[2ex]
        \tilde{a}_8''-\frac{\tilde{a}_8'}{R}= \frac{1}{\lambda}[\tilde{q}_{81}({\cal N}_1+\Xi R^2)f_1^2+\tilde{q}_{82}({\cal N}_2-\Xi R^2)f_2^2-\tilde{q}_{83}{\cal N}_3f_3^2] \, , \label{a8t}
    \end{align}
\end{subequations}
and for the condensates, 
\begin{subequations} \label{eqsf}
    \begin{align}
        0= f_1''+\frac{f_1'}{R}+f_1\left[\frac{\mu^2-m_1^2}{\mu^2-m_3^2}-f_1^2-\frac{({\cal N}_1+\Xi R^2)^2}{R^2} +\eta(f_2^2+f_3^2)\right] \, , 
        \\[2ex]
        0= f_2''+\frac{f_2'}{R}+f_2\left[\frac{\mu^2-m_2^2}{\mu^2-m_3^2}-f_2^2-\frac{({\cal N}_2-\Xi R^2)^2}{R^2} +\eta(f_1^2+f_3^2)\right] \, , \label{eqf2} 
        \\[2ex]
        0= f_3''+\frac{f_3'}{R}+f_3\left[1-f_3^2-\frac{{\cal N}_3^2}{R^2} +\eta(f_1^2+f_2^2)\right] \, . \label{eqf3}
    \end{align}
\end{subequations}
By taking the limit $R\to \infty$ of Eq.\,\eqref{a8t}  we conclude
\begin{equation} \label{a8infty}
\tilde{a}_8(\infty) = \frac{n}{\tilde{q}_{83}} \, ,
\end{equation}
while no condition for $\tilde{a}_3(\infty)$ can be derived, hence this value has to be determined dynamically. Due to the boundary value \eqref{a8infty}, the baryon circulation around the flux tube vanishes, as in a standard magnetic flux tube. We discuss the asymptotic behaviour far away from the centre of the flux tube in Appendix \ref{app:Asymptotic}.

The Gibbs free energy density can be written as  
\begin{equation} \label{Gflux}
    \mathcal{G} 
    =\Omega_{\rm 2SC}-\frac{H^2\cos^2\vartheta_1}{2}+\frac{L}{V} \left({\cal F}_{\rm{VORT}}-\tilde{\Phi}_8H\sin\vartheta_1\right) \, , 
\end{equation}
where $L$ is the size of the system in the $z$-direction, and 
\begin{equation}
    \tilde{\Phi}_8 = \oint d\bm{ s}\cdot\tilde{\bm{ A}}_8 = \frac{2\pi n}{\tilde{q}_{83}} \, ,  
\end{equation}
with a closed integration contour encircling the flux tube at infinity, is the magnetic $\tilde{\bm{ B}}_8$-flux through the flux tube (like in Eq.\,\eqref{TotalFlux}). Employing  partial integration and the equations of motion \eqref{eqsf}, the free energy of a single flux tube per unit length is  
${\cal F}_{\rm{VORT}} = \pi\rho_{\rm 2SC}^2{\cal I}$ with 
\begin{equation}
    {\cal I} \equiv \int_0^\infty dR\,R\left[\frac{\lambda(\tilde{a}_3'^2+\tilde{a}_8'^2)}{R^2}-\frac{f_1^4}{2}-\frac{f_2^4}{2}+\frac{1-f_3^4}{2}+\eta
    (f_1^2f_2^2+f_1^2f_3^2+f_2^2f_3^2)\right] \, 
    \label{FreeEnergyIntegral}.
\end{equation}
 Written in this form, the free energy does not have any explicit dependence on the mass correction and is identical to the one in Ref.\ \cite{Haber:2017oqb} (of course, the dependence on the mass enters implicitly through the equations
of motion). 
The critical field $H_{c1}$ is defined as the field above which $\mathcal{G}$ is lowered by the addition of a flux tube, i.e., by the point at which the term in parentheses in Eq.\,\eqref{Gflux} is zero. This condition can be written as 
\begin{equation} \label{Xictube}
    \Xi_{c1} =\frac{g^2 {\cal I}(\Xi_{c1},n)}{8\lambda n}  \, ,
\end{equation}
where $\Xi_{c1}$ is the dimensionless version of $H_{c1}$  via Eq.\,\eqref{Xi}. 
In the ordinary textbook scenario, presented in Sec.\,(\ref{Background:subsec:Hc1}), the free energy of a flux tube does not depend on the external magnetic field, and thus \eqref{Xictube} would be an explicit expression for the (dimensionless) critical magnetic field. The free energy of a 2SC flux tube, however, {\it does} depend on the external magnetic field. Therefore, Eq.\,\eqref{Xictube}
is an {\it implicit} equation for $\Xi_{c1}$, which has to be solved numerically. 

As we shall see in the next section, $H_{c1}$ may intersect with $H_{c2}^{(1)}$, which indicates the second-order transition from the homogeneous 2SC phase to an inhomogeneous phase where the condensate $f_1$ is switched on. Since $H_{c1}$ becomes meaningless beyond this intercept, it is useful to compute this point explicitly. To this end, we write $H_{c2}^{(1)}$ from Eq.\,\eqref{Hc2one} (setting $\eta=-1/2$) in terms of the dimensionless field $\Xi$ \eqref{Xi}, which yields to linear order in $\alpha$ 
\begin{equation} \label{Xic2}
    \Xi_{c2}^{(1)} \simeq  \frac{1}{4}\left(1-\frac{\alpha}{2}\right)\, .
\end{equation}
Solving this equation simultaneously with Eq.\,\eqref{Xictube}, i.e., setting $\Xi_{c2}^{(1)} =\Xi_{c1}$, gives the intersection point. In the practical calculation, this is best done by inserting Eq.\,\eqref{Xic2} into Eq.\,\eqref{Xictube} and solving the resulting equation for $T_c/\mu_q$ (for given strange quark mass and winding number).
This calculation is relevant for the phase diagrams in Fig.\ \ref{fig:phasediagram}. 

\section{Numerical results and discussion}
\label{Project1:sec:results}

We are now prepared to compute the flux tube profiles and the resulting critical fields $H_{c1}$, which we will put together with $H_c$ and $H_{c2}$ from the previous section. As in the calculations of the critical fields $H_c$ and $H_{c2}$ we use the bosonic masses \eqref{m123}, keep terms linear in $\alpha$ in Eqs.\ \eqref{eqsf}, and, as discussed in Sec.\,(\ref{Project1:subsec:para}), we set $\eta=-1/2$ and $g=3.5$ for all following results. Then we solve the coupled second-order differential equations \eqref{eqsa} and \eqref{eqsf} numerically without further approximations using the successive over-relaxation method, which has been used before in similar contexts \cite{Haber:2017kth,Haber:2017oqb,Haber:2018tqw,Haber:2018yyd,Fraga:2018cvr,Schmitt:2020tac}. Details on this numerical method are given in Appendix \ref{app:Numerical}. Each numerical run yields a flux tube profile for given $m_s/\mu_q$, $T_c/\mu_q$ (from which $\alpha$ and $\lambda$ are obtained), dimensionless external magnetic field $\Xi$, and winding number $n$. Since we are interested in the critical field $\Xi_{c1}$, we need to solve Eq.\,\eqref{Xictube}, for which we employ the bisection method. This requires solving the differential equations about 10 -- 20 times until a reasonable accuracy for $\Xi_{c1}$ is reached. This whole procedure thus yields a critical field $H_{c1}$ for given $m_s/\mu_q$, $T_c/\mu_q$, and $n$. As argued in the introduction and as we shall see below, solutions with high winding numbers are expected to play an important role. Therefore, 
in principle, the procedure has to be repeated for all $n$ to find the preferred flux tube configuration for each point in the parameter space spanned by $m_s/\mu_q$ and $T_c/\mu_q$. In practice, we have performed the calculation for the lowest few $n$, and for selected parameter sets for much larger $n$ to check our conclusions.
An additional complication arises because in certain parameter regions there is more than one solution to the set of differential equations. The single-component flux tube with $f_1\equiv f_2\equiv 0$ always exists. Configurations where the condensate $f_2$ is induced in the core of the flux tube, but not $f_1$, turn out to be preferred over the single-component configuration whenever they exist and we shall discuss them in detail. Configurations where both $f_1$ and $f_2$ are induced in the core do exist as well, but only in a parameter region where $H_{c2}$ indicates that the ground state is a more complicated flux tube array. We shall therefore not discuss these three-component configurations.

\subsection{Flux tube properties}
\label{Project1:subsec:tubes}

\begin{figure} 
\begin{center}
\includegraphics[width=\textwidth]{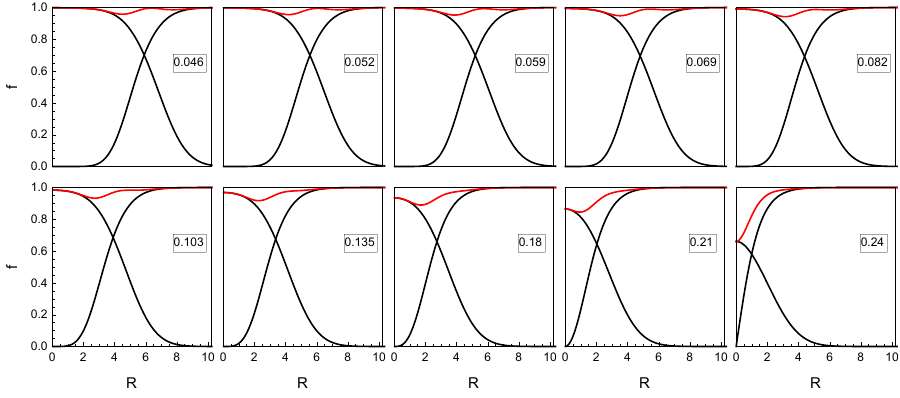}

\vspace{0.5cm}
\includegraphics[width=\textwidth]{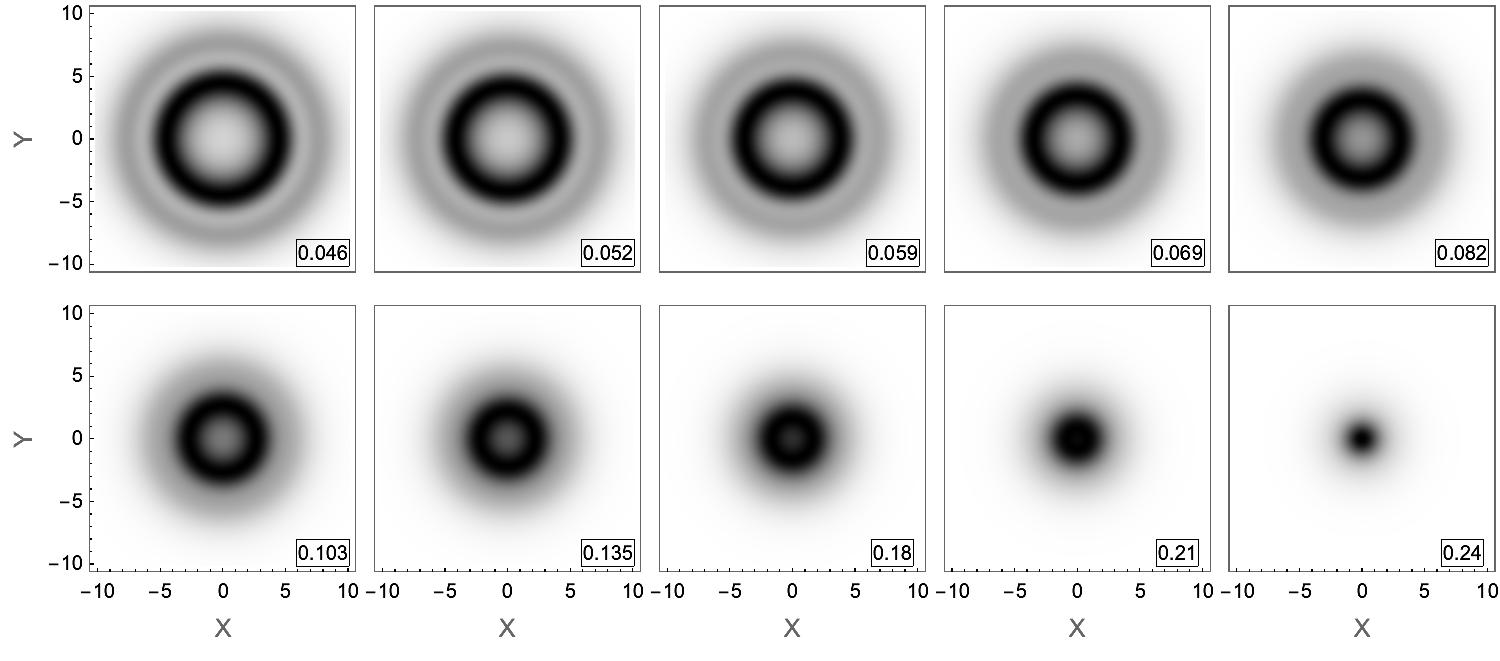}
\caption{{\it Upper panels:} Flux tube profiles of the dimensionless condensates $f_2$ and $f_3$ (black) and $\sqrt{f_2^2+f_3^2}$ (red) as  functions of the dimensionless radial coordinate $R$. The condensate $f_3$ is the usual 2SC condensate of $ud$ Cooper pairs and asymptotes to 1 for large $R$, while the $us$ condensate $f_2$ is induced in the core of the flux tube. For all plots $g=3.5$ and $T_c=0.0856\mu_q$, while $m_s/\mu_q$ assumes the values given in each panel, i.e., it increases from upper left to lower right. The masses are chosen such that the 
preferred configurations are flux tubes with winding numbers $n=10,\ldots, 1$ from upper left to lower right (and each plot shows the preferred configuration). For compact star conditions, $R=10$ translates to about $r\simeq 7.7\, {\rm fm}$. {\it Lower panels:} Ratio of the induced magnetic field over the 
external magnetic field, $B/H$, in the $X$-$Y$ plane perpendicular to the flux tube for the 10 configurations of the upper panels ($X$ and $Y$ in the same dimensionless units as $R$). The scale of the shading is adjusted for each plot separately, from black (maximal) to white (minimal). The magnetic field enters the superconductor in ring-like structures for small strange quark mass (large winding number) and turns into the conventional flux tube behaviour for large mass (small winding number). This structure is reflected in the red curves of the upper panels. 
}\label{fig:profiles}
\end{center}

\vspace{-0.8cm}
\end{figure}

\begin{figure} 
\begin{center}
\includegraphics[width=0.5\textwidth]{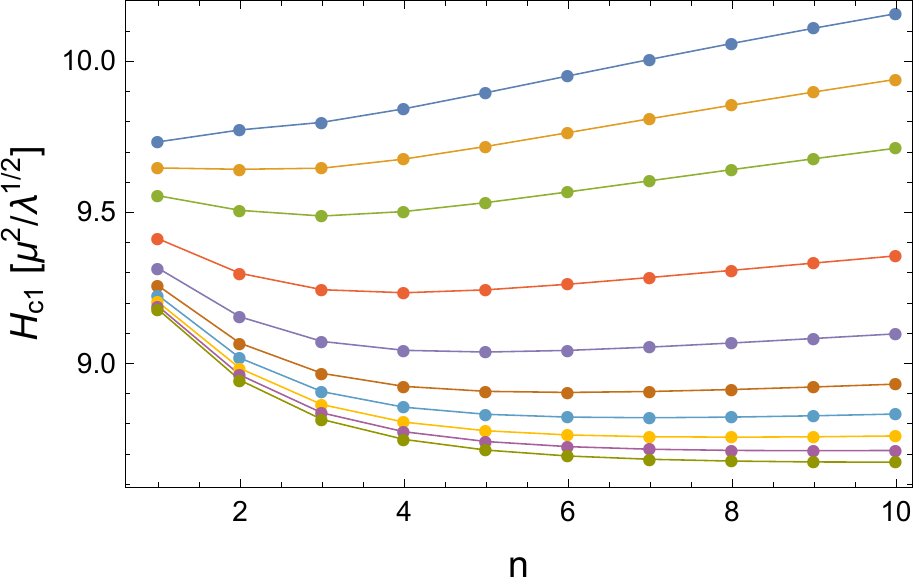}
\caption{Critical magnetic fields $H_{c1}$ as a function of the winding number $n$ (continuous lines inserted to guide the eye) for the same parameters as in Fig.\ \ref{fig:profiles}. Each line corresponds to a different strange quark mass, which increases from bottom to top (the values are given in the caption of Fig.\ \ref{fig:profiles}). The configuration with lowest $H_{c1}$ goes from $n=10$ (bottom) successively to $n=1$ (top). 
} \label{fig:winding}
\end{center}
\end{figure}

We start by discussing individual flux tube profiles and the associated magnetic fields. We do so by choosing a fixed $T_c/\mu_q$ such that for vanishing strange quark mass there is a magnetic field at which it is energetically favourable to put a domain wall in the system. The values of $T_c/\mu_q$ for which this is the case are known from Ref.\ \cite{Haber:2017oqb} (see Fig. 5 in that reference). The domain wall interpolates between the 2SC$_{\rm ud}$ and 2SC$_{\rm us}$ 
phases, i.e., on one side, and far away from it, we have $f_3=1$, $f_2=0$, and on the other side $f_2=1$, $f_3=0$. At non-zero $m_s$ the free energies of 2SC$_{\rm ud}$ and 2SC$_{\rm us}$ are obviously no longer equal, as we have seen explicitly in Sec.\,(\ref{Project1:sec:hom}), and thus the domain wall configuration becomes unstable. In Fig.\ \ref{fig:profiles} we have chosen 10 different non-zero values of $m_s/\mu_q$, such that the configuration with lowest $H_{c1}$ (to which we will refer as the ``energetically preferred'' or simply the ``preferred'' configuration)  has winding number $n=10, \ldots, 1$ as $m_s/\mu_q$ is increased. The profiles and magnetic fields are plotted at the corresponding minimal $H_{c1}$. 
The critical fields as a function of the winding number for the parameters of Fig.\ \ref{fig:profiles} are shown in Fig.\ \ref{fig:winding}, which proves the successive decrease in the preferred winding from low to high strange quark mass.

Let us first discuss the profiles themselves in Fig.\ \ref{fig:profiles}. Comparing with Fig.\,\ref{fig:Fluxtube}, we already see the similarities and differences from the single-component case. For the smallest masses shown here the profiles of the condensates are reminiscent of a domain wall profile: The second condensate, which is induced in the core, assumes essentially the value of the homogeneous 2SC condensate. Of course, in contrast to a domain wall, the flux tubes have a finite radius, which decreases as the winding number decreases (with increasing strange quark mass). For a quantitative discussion of the size of the flux tubes we use the various characteristic length scales obtained from the asymptotic 
behaviour of the flux tube profile, see Appendix \ref{app:Asymptotic}. The penetration depth $\ell$ usually is characteristic for the decay of the magnetic field away from the centre of the flux tube. In our case this scale corresponds to the decay of the $\tilde{B}_8$ field (in Fig.\ \ref{fig:profiles} we have restricted ourselves to plotting the ordinary magnetic field $B$). The coherence length  $\xi_3$ introduced in Eq.\,\eqref{R} indicates the scale on which the $ud$ condensate $f_3$ recovers its homogeneous value and corresponds to the usual coherence length of a single-component flux tube. Finally, and in contrast to the standard single-component scenario, our calculation in Appendix \ref{app:Asymptotic} shows that there is a third scale $\xi_2$, characteristic for the induced $us$ condensate $f_2$. Using Eqs.\ \eqref{lpen}, \eqref{lcoh}, and \eqref{lind} we can write these three length scales in physical units as 
\begin{subequations}\allowdisplaybreaks
\label{ellxi}
\begin{align}
    \ell &\simeq 0.94 \left[1+\frac{1}{48}\left(\frac{m_s}{\mu_q}\right)^2\ln\frac{\mu_q}{T_c}\right] \left(\frac{\mu_q}{400\, {\rm MeV}}\right)^{-1}\, {\rm fm} \, , \label{ell1} 
    \\[2ex]
    \xi_3 &\simeq \frac{0.066}{T_c/\mu_q}   \left[1+\frac{1}{48}\left(\frac{m_s}{\mu_q}\right)^2\ln\frac{\mu_q}{T_c}\right]\left(\frac{\mu_q}{400\, {\rm MeV}}\right)^{-1}\, {\rm fm} \, , \label{xi31}  
    \\[2ex]
    \label{lind0}
    \xi_2 &\simeq 1.3 \left(\frac{H}{\mu^2/\sqrt{\lambda}}\frac{T_c}{\mu_q}\right)^{-1/2}\left(\frac{\mu_q}{400\, {\rm MeV}}\right)^{-1} {\rm fm} \, , 
\end{align}
\end{subequations}
where we have used Eq.\,\eqref{weak} with $T=0$, the 2SC condensate \eqref{rho32SC}, and $g=3.5$. The length scale $\xi_2$ on which the induced condensate varies depends on the external magnetic field $H$, for which the dimensionless value (in units of $\mu^2/\sqrt{\lambda}$) can be inserted into Eq.\,\eqref{lind0}.  
For the parameters used in Fig.\ \ref{fig:profiles} the mass terms in Eqs.\ \eqref{ell1} and \eqref{xi31} are negligibly small and setting  $\mu_q=400\, {\rm MeV}$ we find $\ell\simeq 0.94\, {\rm fm}$, $\xi_3 \simeq 0.77 \, {\rm fm}$, and $\xi_2\simeq (1.4-1.5)\, {\rm fm}$, while the maximal value of the radial coordinate in this figure $R=10$ corresponds to $r\simeq 7.7\,{\rm fm}$. These values, characterising the radial extent of the core of the flux tube do not depend on the winding number $n$. In the derivation in Appendix \ref{app:Asymptotic} we have worked with a fixed and finite $n$ while sending the radial coordinate to infinity. However, from the full numerical profiles we see that the core of the flux tube clearly grows with increasing $n$. For instance, taking the point where the two condensates have the same value as a rough measure for the size of the core, we see that for the largest winding shown here, $n=10$, corresponding to the smallest mass, $m_s=0.046\mu_q$, this size is $r\simeq 4.6\, {\rm fm}$. This is about six times larger than the coherence length $\xi_3$, but it also shows that it only takes a relatively small strange quark mass to reduce the size of the flux tubes from infinity (domain wall) to a few fm.

The plots in the upper panels also show the profile of the function $\sqrt{f_2^2+f_3^2}$, which is the radial coordinate of the space spanned by $(f_2,f_3)$, i.e., a domain wall can be understood as a rotation in this space from the horizontal to the vertical axis. This function illustrates the depletion of the ``combined'' condensate in a cylindrical layer at non-zero radius for large winding numbers and in the centre of the tube for low winding numbers. This depletion reflects the structure of the magnetic field, which is shown in the lower panels. We recall that the 2SC phase (just like the CFL phase) admits a fraction of the external field $H$ even in the homogeneous phase, and this fraction is close to one for $g\gg e$. Therefore, the energetic benefit of a 2SC flux tube is to admit additional magnetic flux on top of the flux already present. As a consequence of our choice $g=3.5$, the ratio $B/H$ is  larger than 99\% already in the homogeneous 2SC phase, as one can see for instance from Eq.\,\eqref{BoverH}. The magnetic field profiles in the figure, showing $B/H$ between its minimal homogeneous 2SC value (white) and the maximal value (black) turn out to be in the range $0.9975 \lesssim B/H \lesssim 0.9995$. Therefore, the magnetic field variations are unlikely to have any physical impact in the strong coupling regime. One might also wonder about the numerical accuracy needed to resolve such tiny variations of $B/H$. However, from Eq.\,\eqref{BoverH} it is obvious that it is the small mixing angle $\vartheta_1$ ($\sin\vartheta_1\simeq 0.05$ for $g=3.5$)  that maps the variations into a tiny interval and thus the required numerical accuracy is not as high as one might think.

The magnetic profiles show interesting features. We see that the flux tubes with higher winding (small $m_s$) have their excess magnetic field concentrated in ring-like structures. This effect has been observed in the literature before with the help of a two-component abelian Higgs model \cite{Chernodub:2010sg} and a one-component model with non-standard coupling between the condensate and the abelian gauge field \cite{Bazeia:2018hyv} (see also Refs.\ \cite{Bazeia:2018eta,Bazeia:2018fhg} for similar structures of magnetic monopoles in a non-abelian model). In these studies, the so-called Bogomolny limit was considered for simplicity, which corresponds to the transition point between type-I and type-II behaviour. Here we observe the effect in a numerical evaluation of the general GL equations, and in particular we can point out regions in the phase diagram where the exotic ring-like structures are the preferred configuration (for a systematic discussion of the phase diagram see next subsection). The ring-like structure is  easy to understand: In the domain wall ($m_s=0$) the excess magnetic field is concentrated in the wall for symmetry reasons. As the strange quark mass is increased, the wall ``bends'' to form a flux tube with finite radius, and it is obvious for continuity reasons that for very small masses the maximum of the magnetic field is still sitting in the transition region. Then, as the sequence in Fig.\ \ref{fig:profiles} shows, the magnetic rings gradually turn into ordinary flux tubes as $m_s$ is increased (and the winding number decreases). Furthermore, we observe a double-ring structure, clearly visible in the multi-winding flux tubes. We have found that the double ring does not always occur. For instance, for the smaller value $T_c=0.08\mu_q$ the double ring is replaced by a single ring.

\begin{figure} 
\begin{center}
\hbox{\includegraphics[width=0.5\textwidth]{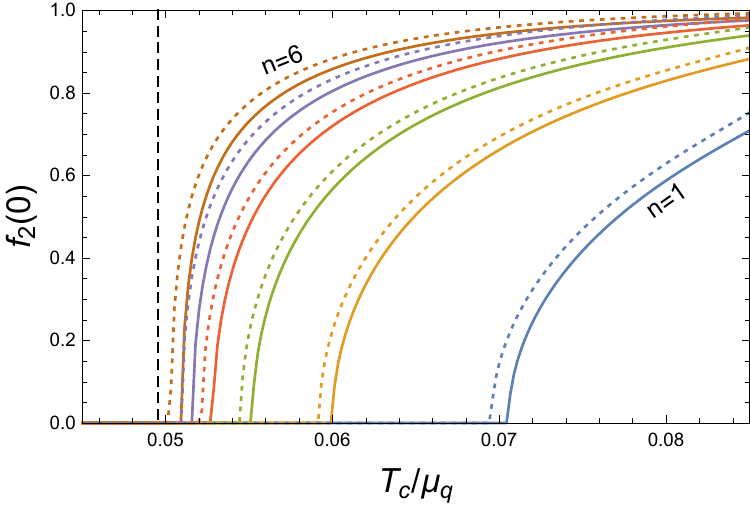}\includegraphics[width=0.5\textwidth]{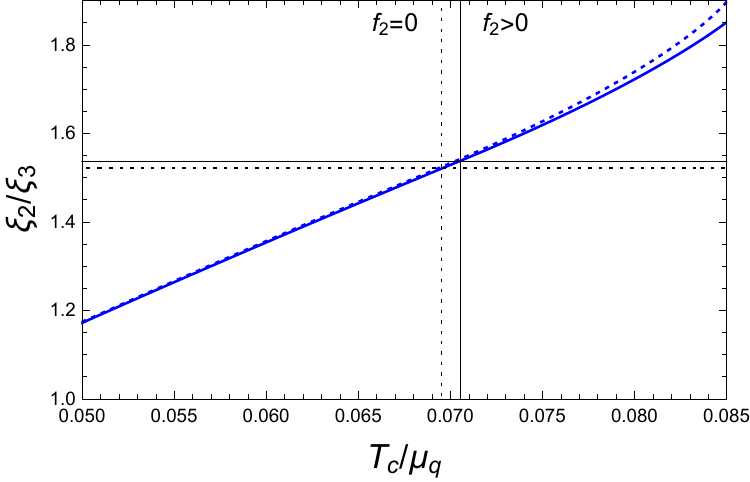}}
\caption{{\it Left panel:} Value of the dimensionless $us$ condensate $f_2(R=0)$ in the centre of the 2SC flux tube as a function of $T_c/\mu_q$ at the critical field $H_{c1}$ (which depends on $T_c/\mu_q$ and $n$) for strange quark mass $m_s=0.15\mu_q$ (solid lines)  and $m_s=0$ (dotted lines), each for winding numbers $n=1,\ldots, 6$, from right to left. The standard, single-component flux tube solution exists throughout the parameter space. The vertical dashed line marks the transition point between type-I and type-II behaviour. {\it Right panel:} Ratio of length scales for depleted condensate $f_3$ and induced condensate $f_2$ according to Eqs.\ \eqref{ellxi} for the $n=1$ curves of the left panel, with the blue curves corresponding to the massless (dotted) and massive (solid) cases. The vertical lines indicate the transition from zero to non-zero induced condensate, while the horizontal lines mark the corresponding values of the ratio $\xi_2/\xi_3$ at this transition.}  \label{fig:f2}
\end{center}
\end{figure}

All profiles in Fig.\ \ref{fig:profiles} are ``unconventional'' in the sense that they contain two condensates. As already mentioned above, these two-component solutions are not found everywhere in  the parameter space. In the left panel of Fig.\ \ref{fig:f2} we show the value of the induced condensate in the core $f_2(0)$ as a function of the parameter $T_c/\mu_q$ for winding numbers $n=1,\ldots,6$ and for a non-zero strange quark mass, compared to the massless case. We see that as $T_c/\mu_q$ decreases, the induced condensate becomes smaller until it continuously goes to zero at a point that depends on the winding number. Interestingly, as the winding number is increased, this point seems to converge to the point \eqref{type2SCNOR} that distinguishes type-I from type-II superconductivity. In particular, the behaviour of $f_2(0)$ becomes more and more
step-like for larger windings, i.e., $f_2(0)$ is close to one until it sharply decreases near the type-I/type-II transition point. However, flux tubes with higher winding are energetically disfavoured in the vicinity of the 
type-I/type-II transition point (as we shall see below), such that this interesting behaviour does not seem to be physically relevant. 

The right panel of Fig.\ \ref{fig:f2} is useful to understand why the induced condensate only appears in a certain parameter regime. In this plot we show the ratio of $\xi_2$ and $\xi_3$ given in Eqs.\ \eqref{ellxi}. We see that the induced condensate vanishes if this ratio
becomes too small. For a possible interpretation note that the coherence length 
is connected to the size of the Cooper pair. If we view $\xi_3$ as a typical size of the Cooper pair (also for $us$ pairing), then the right panel of Fig.\ \ref{fig:f2} suggests that the length scale for the variation of the induced condensate, which is determined by the external magnetic field, must be sufficiently large (at least by a factor of about 1.5) compared to the size of a Cooper pair. Otherwise the condensate ``does not fit'' into the core of the flux tube.

\subsection{Phase structure}
\label{pt1sec:phase}

We have seen that there are parameter choices for $(g, T_c/\mu_q, m_s/\mu_q)$ where multi-winding 2SC flux tubes with a ring-like structure of the magnetic field are energetically preferred. In this section we 
investigate the parameter space more systematically. We should keep in mind that in QCD the parameters, $g$, $T_c$, and $m_s$ are uniquely given by $\mu_q$ (at $T=0$). Therefore, as we vary the quark chemical potential, the system will move along a unique, but unknown, curve in this three-dimensional parameter space. For instance, at asymptotically large $\mu_q$, we start from small $g\ll 1$, exponentially suppressed $T_c$ and negligibly small $m_s$ (compared to $\mu_q$). Since we do not know the values of $m_s$ and $T_c$ at more moderate densities we keep our parameters as general as possible. In this sense, keeping $g=3.5$ fixed for our results is a simplification, in an even more general calculation one might also vary $g$.

\begin{figure} 
\begin{center}
\hbox{\includegraphics[width=0.5\textwidth]{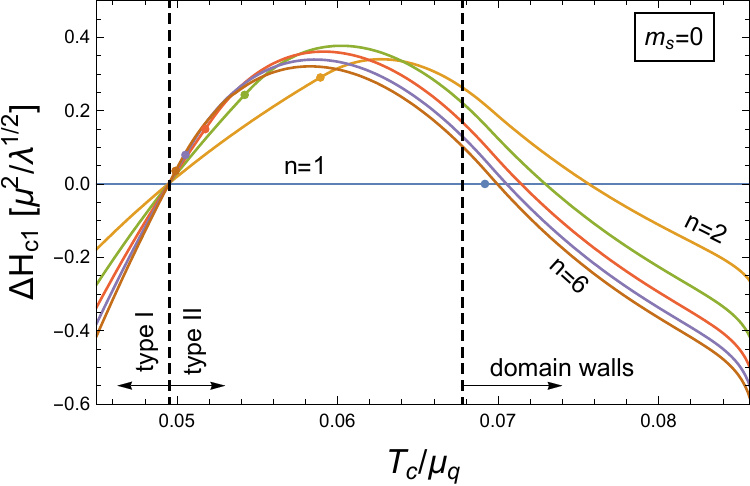}\includegraphics[width=0.5\textwidth]{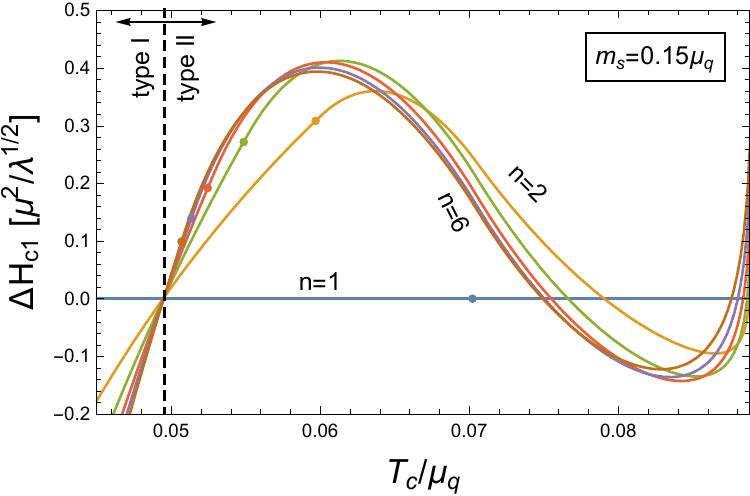}}
\caption{Critical fields for the emergence of flux tubes with winding $n$ relative to the critical field of the $n=1$ flux tube, $\Delta H_{c1} = H_{c1}(n)-H_{c1}(1)$, as a function of $T_c/\mu_q$ for the massless case (left panel) and $m_s=0.15\mu_q$ (right panel). In the massless case, domain walls ($n=\infty$) are the preferred configuration for $T_c/\mu_q \gtrsim 0.068$, as indicated by the vertical dashed line. The other vertical dashed line marks the change from type-I to type-II behaviour. The dots on the curves indicate the points above which the flux tubes have a core with induced condensate $f_2$ (see also Fig.\ \ref{fig:f2}).
}\label{fig:DeltaHc1}
\end{center}
\end{figure}

In Fig.\ \ref{fig:DeltaHc1} we compare the critical magnetic fields of flux tubes with different winding numbers for two different values of $m_s$ as a function of $T_c/\mu_q$. We plot the difference of the critical fields to the critical field of the $n=1$ configuration because the different curves would be barely distinguishable had we plotted the critical fields themselves. It is instructive to start with the massless case (left panel). For $T_c/\mu_q\simeq 0.05$ we observe the standard behaviour at the type-I/type-II transition: In the type-I regime, the critical field $H_{c1}$ can be lowered by increasing the winding number, and as $n\to\infty$ one expects $H_{c1}(n)$ to converge to $H_c$, the critical field at which the NOR and 2SC phases coexist. Just above the critical value $T_c/\mu_q\simeq 0.05$ the order of the different $H_{c1}(n)$ is exactly reversed, and $H_{c1}$ is given by the flux tube with lowest winding, $n=1$, the higher-winding flux tubes are disfavoured.
It is a good check for our numerics that all curves intersect at the 
same point, and this point is given by Eq.\,\eqref{type2SCNOR}, whose derivation is completely independent of the numerical evaluation. This conventional behaviour in the vicinity of the type-I/type-II transition point is expected because the second condensate plays no role here, at least for the lowest winding numbers, as we have seen in Fig.\ \ref{fig:f2}.

Now, as $T_c/\mu_q$ is increased, the unconventional 2SC behaviour becomes apparent. For any winding $n>1$ there is a critical $T_c/\mu_q$ at which this higher winding becomes favoured over $n=1$. Similar to the type-I regime, there is a region in which $H_{c1}(n)$ decreases monotonically with $n$ until, for $n\to \infty$, it is expected to intersect $H_{c1}(1)$ at the critical point where domain walls set in (we have only plotted the curves for a few winding numbers, obviously the numerics become more challenging for large $n$). This point, $T_c/\mu_q \simeq 0.068$, is taken from  Ref.\ \cite{Haber:2017oqb}, where it was computed explicitly by using a planar instead of a cylindrical geometry  (see Figs.\ 4 and 5 in that reference). Here we have marked the onset of domain walls by a vertical dashed line.  

In the right panel we first note that the behaviour in the vicinity of the type-I/type-II transition in the presence of a strange quark mass is qualitatively the same as in the massless case. The transition point itself, as we already know from Eq.\,\eqref{type2SCNOR}, is even exactly the same, at least up to lowest non-trivial order  in $m_s$. Now, however, the behaviour for larger $T_c/\mu_q$, where the second condensate does play a role, is qualitatively different. For the chosen value of $m_s/\mu_q$, we find that for windings $n\le 6$ there is a regime in which flux tubes with winding $n$ are preferred, while flux tubes with $n > 6$ are never preferred. We see that the lowest $H_{c1}$ corresponds to flux tubes with winding numbers 1,6,5,4,3,2, as $T_c/\mu_q$ is increased. These structures can be viewed as ``remnants'' of the domain wall.

\begin{figure} 
\begin{center}
\hbox{\includegraphics[width=0.5\textwidth]{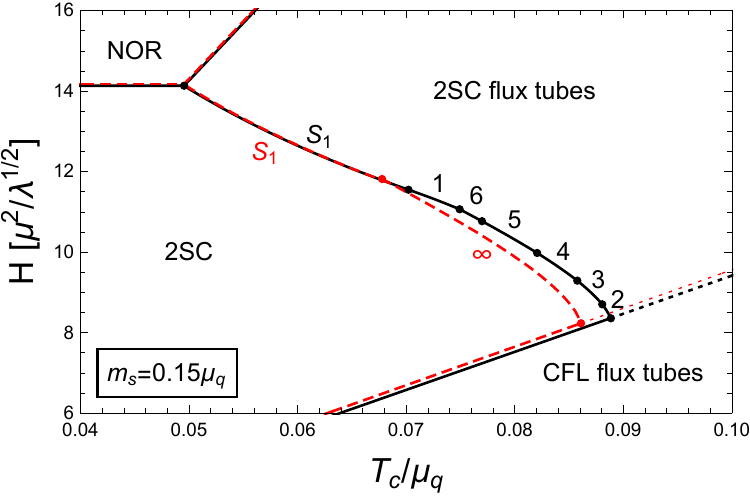}
\includegraphics[width=0.5\textwidth]{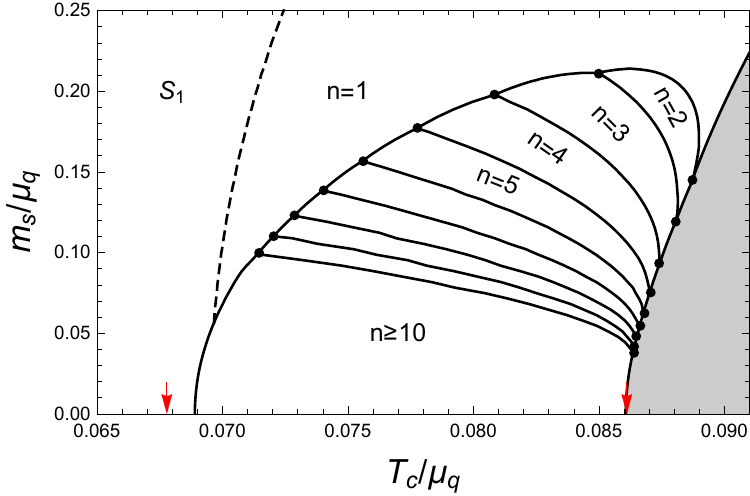}}
\caption{{\it Left panel:} Phase diagram in the plane of external field $H$ and $T_c/\mu_q$ for $m_s=0.15\mu_q$ (black solid lines) compared to the massless result from Ref.\ \cite{Haber:2017oqb} (red dashed lines, see Fig.\,\ref{fig:ChSBtoCSCmuB}). The numbers at the critical field for the emergence of 2SC flux tubes with $us$ core indicate the winding number $n$, with $n=\infty$ indicating the domain wall in the massless case. The critical lines for the emergence of 2SC flux tubes without a $us$ core (which always have winding $n=1$) are marked with $S_1$. {\it Right panel:} Phase diagram in the $m_s/\mu_q$-$T_c/\mu_q$ plane indicating the preferred flux tube configurations. In the shaded region dSC and CFL flux tubes appear, and the red arrows indicate the range where, at $m_s=0$, domain walls are preferred. This phase diagram shows one of the main results of the project, namely that remnants of the domain wall in the form of multi-winding flux tubes survive up to $m_s\simeq 0.21\mu_q$.  
}\label{fig:phasediagram}
\end{center}
\end{figure}

With the help of the results of this panel we can construct the phase diagram in the $H$-$T_c/\mu_q$ plane for this particular mass $m_s=0.15\mu_q$. To this end, we need to bring together the critical fields $H_{c1}$ from Fig.\ \ref{fig:DeltaHc1} with the critical fields $H_c$ from Sec.\,(\ref{Project1:subsec:Hc}) and the critical fields $H_{c2}$ from Sec.\,(\ref{Project1:subsec:Hc2}). The result is shown in the left panel of Fig.\ \ref{fig:phasediagram}. This panel contains the first-order transition between NOR and 2SC phases \eqref{Hcs}, the second-order transitions from the NOR to the ``2SC flux tube'' phase \eqref{Hc22SCNOR} and from the 2SC to the ``CFL flux tube'' phase \eqref{Hc2one}, and the second-order transition from the 2SC to the ``2SC flux tube'' phase just discussed, including the different segments corresponding to different winding numbers. We have included the phase transitions of the massless case for comparison, including the segment where domain walls are formed. The single-component flux tube configuration, i.e., the one without a $us$ condensate in the core, is denoted by $S_1$, following the notation of Ref.\ \cite{Haber:2017oqb}. This configuration always has winding number 1. The region labelled by ``CFL flux tubes'' is bounded from above by $H_{c2}^{(1)}$, which actually indicates a transition to dSC flux tubes (see discussion below Eqs.\ \eqref{Hc212}). For the given parameters, the ``would-be'' critical magnetic field $H_{c2}^{(2)}$ is very close to $H_{c2}^{(1)}$, and thus we have, slightly inaccurately, termed the entire region ``CFL flux tubes'', although we know that there is at least a thin slice where the inhomogeneous phase is not made of CFL flux tubes. 

Since our calculation only allows us to determine the critical fields for the entrance into the flux tube phases, we can only speculate about the structure of these inhomogeneous phases away from the second-order lines. It is obvious that the structure will be more complicated than in a standard single-component superconductor. This, firstly, concerns the transition between the phases labelled by ``CFL flux tubes''  and ``2SC flux tubes''. We have continued the $H_{c2}$ transition curve into the flux tube regime as a thin dotted line, but of course this curve should not be taken too seriously because it was calculated under the assumption of a homogeneous 2SC phase on one side of the transition. Secondly, and more closely related to our main results, the different winding numbers that occur upon entering the 2SC flux tube phase also suggest a complicated structure of the flux tube arrays. It is conceivable that there are transitions between ``pure'' arrays of a single winding number or that there are arrays composed of flux tubes with different winding numbers.  

We may now ask up to which values of the strange quark mass the multi-winding flux tubes survive. More precisely, at which value of $m_s$ do only $n=1$ flux tubes form at the entire phase transition curve $H_{c1}$ between the homogeneous 2SC phase and the ``2SC flux tube'' phase? This question can be answered by repeating the calculation needed for the left panel of Fig.\ \ref{fig:phasediagram} for different values of $m_s$ and determining the sequence of the preferred configurations as a  function of $T_c/\mu_q$ in each case. The result is shown in the right panel of Fig.\ \ref{fig:phasediagram}. In this panel, the $m_s/\mu_q$-$T_c/\mu_q$ plane is divided into different regions, each region labelled by the preferred flux tube configuration. One can check that a slice through this plot at $m_s=0.15\mu_q$ reproduces exactly the sequence of phases shown in the left panel. In particular, we have indicated the transition between the standard flux tubes $S_1$ and the two-component flux tubes with winding number $n=1$ by a dashed curve. We have restricted our calculation to winding numbers $n\le 10$, but it is obvious how the pattern continues: As we move towards $m_s=0$, say at fixed $T_c/\mu_q$, the winding number of the preferred configuration increases successively and more rapidly. Eventually, the winding number and thickness of the flux tube go to infinity, which corresponds to the domain wall, whose range is indicated with (red) arrows. Therefore, the lines that bound the region of multi-winding configurations
at small $m_s$ from both sides -- here calculated from $n=10$ -- will slightly change if higher windings are taken into account and they are expected to converge to the arrows at $m_s=0$. In the shaded region, dSC and CFL flux tubes start to become relevant, its boundary is given by the condition $H_{c1}=H_{c2}^{(1)}$ (see also the discussion around Eq.\,\eqref{Xic2}).

The main conclusion of this phase diagram is that the maximal strange quark mass for which remnants of the domain wall (i.e., 2SC flux tubes with winding number larger than one) exist, is $m_s\simeq 0.21\mu_q$. This is on the lower end of the range expected for compact stars. Therefore, at least within our approximations, we conclude that it is conceivable, but unlikely that 2SC flux tubes with exotic structures as shown in Fig.\ \ref{fig:profiles} play an important role in the astrophysical setting. However, our phase diagram shows that two-component flux tubes (with $n=1$) do survive for larger values of $m_s$. Since they have a $us$ condensate in their core, they can also be considered as remnants of the domain wall, albeit with  conventional magnetic field structure. We have checked that if the phase diagram in the right panel of Fig.\ \ref{fig:phasediagram} is continued to about $m_s=0.5\mu_q$, there is still a sizeable range (of roughly the same size as for $m_s=0.25\mu_q$), starting at $T_c/\mu_q\simeq 0.083$, where 2SC flux tubes with $us$ core are preferred. Consequently, they may well have to be taken into account for the physics of compact stars.

In the discussion of the applicability of our results to compact stars, we should also return to the actual magnitude of the magnetic fields considered here. Using the conversion to physical units given in the Introduction and the expressions for $\mu$ and $\lambda$ in Eq.\,\eqref{weak}, we find that the magnetic field in Gauss can be computed from 
\begin{equation}
H = 1.6\times 10^{19} \frac{H}{\mu^2/\sqrt{\lambda}}\ \frac{T_c}{\mu_q} \left(1-\frac{T}{T_c}\right)\left(\frac{\mu_q}{400\, {\rm MeV}}\right)^2 \rm{G} \, .
\end{equation}
Consequently, setting $T=0$, the triple point at $T_c/\mu_q \simeq 0.05$ in the left panel of Fig.\ \ref{fig:phasediagram}, where NOR, 2SC, and 2SC flux tube phases meet, corresponds to about $H\simeq 1.1\times 10^{19}\, {\rm G}$, while the critical fields at $T_c/\mu_q=0.08$ are $H_{c1}\simeq 1.3 \times 10^{19}\, {\rm G}$ and $H_{c2}\simeq 2.9\times 10^{19}\, {\rm G}$ (note that $H$ 
in physical units also varies along the {\it horizontal} direction of this plot since $\mu^2/\lambda^{1/2}$ depends on $T_c/\mu_q$). Magnetic fields of this magnitude are getting close to the limit for the stability of the star, and they are several orders of magnitude larger than observed at the surface. It is thus highly speculative, although possible, that these huge magnetic fields are reached in the interior of the star. However, exotic flux tubes as discussed here may be formed dynamically, for instance if the star cools through the superconducting phase transition at a roughly constant magnetic field \cite{Alford:2010qf}. And, if  colour-magnetic flux tubes exist in compact stars, they  help to sustain possible ellipticities of the star, resulting in detectable gravitational waves \cite{Glampedakis:2012qp}. 
Therefore, even though the equilibrium situation studied here may never be reached in an astrophysical setting, it is important to understand the various possible -- non-standard -- flux tube configurations.

\chapter{Superconducting baryon crystal induced via the chiral anomaly}
\label{chpt:Project2}

\begin{figure}[h]
    \centering
    \includegraphics[width=0.65\textwidth]{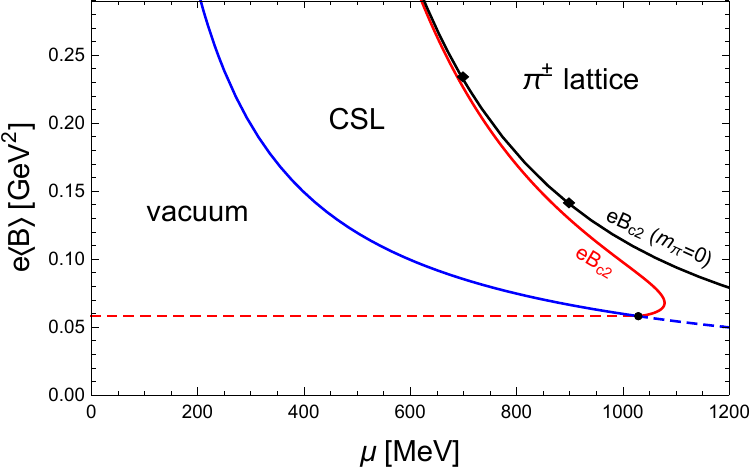}
    \caption{Phase diagram in the $\mu_B$-$e\langle B\rangle$ ($\mu\equiv\mu_B$ in this plot and $\langle B\rangle$ is the spatial average of the magnetic field) plane for a physical pion mass (red and blue curves) and in the chiral limit (black curve; in this case the vacuum only exists on the two axes). The solid blue curve marks the transition between the vacuum and the CSL phase and the solid red curve indicates the instability of the CSL phase towards charged pion condensation. This part of the thesis constructs the $\pi^{\pm}$ lattice and presents plots of the resulting lattice at the points indicated by the diamonds in Fig.\,\ref{fig:BaryonNumberDensity}. The dashed curves are transitions between metastable phases. A first-order transition between the vacuum and the $\pi^\pm$ lattice is expected somewhere between the dashed lines. Baryons in the form of nuclear matter are expected to compete with and possibly replace the phases at sufficiently large $\mu_B$, but are omitted here for simplicity (see left panel of Fig.\,\ref{fig:BaryonOnsetmuB}). To put the scale of the magnetic field into context, note that within natural Heaviside-Lorentz units $e\langle B\rangle = 0.1\, {\rm GeV}^2\simeq 5.18 \, m_\pi^2 $ corresponds to $e\langle B\rangle \simeq 5.12\times 10^{18}\, {\rm G}$. }
    \label{fig:eBmu}
\end{figure}

The lightest baryons are the nucleons with mass $m_N= 939\,\rm{MeV}$. In both planes involving $\mu_B$
 reviewed in Sec.\,(\ref{Background:subsec:muBPlane}), there is an expected baryon onset at $\mu_B\sim m_N$. Below the deconfinement transition and baryon onset we can expect the QCD vacuum and mesons. Mesons can emerge via thermal excitation at finite $T$ and meson condensation at finite $\mu_I$. Naively, one might expect no mesons when $T=\mu_I=0$.
This expectation was shown to be false in Ref.\,\cite{Son:2007ny}, where the authors demonstrated that the chiral anomaly can affect the phase structure at strong magnetic fields. In particular, it was shown that a neutral pion domain wall was locally stable above $eB=3 m_{\pi}^2$. This was expanded upon in Ref.\,\cite{Brauner:2016pko}, where this phase was generalised to a chiral soliton lattice (CSL) phase which the QCD vacuum was unstable to above the critical magnetic field $B_{\rm{CSL}}$. In addition, they showed there was a further instability to fluctuations in the charged pion field at a higher, second critical magnetic field, which implies that the charged pions condense. The nature of the instability is reminiscent of the instability of a conventional superconductor between the vortex lattice and normal phase (see Sec.\,(\ref{Background:subsec:Hc2})) at $B_{c2}$. The construction of this phase was beyond the scope of Ref.\,\cite{Brauner:2016pko}, and thus became the aim of the work presented in this section. This was later published in Ref.\,\cite{Evans:2022hwr} and the resulting phase structure is shown in Figure \,\ref{fig:eBmu}.

To this aim, we used a method inspired by the approach detailed in Sec.\,(\ref{Background:subsec:Hc2}) within the ChPT framework used in Ref.\,\cite{Brauner:2016pko}. Questions may be raised as to the applicability of this method given the differences between our scenario and the conventional case.  We will find that only minor adaptation is necessary to account for additional degrees of freedom, and it leads to analytical solutions in the chiral limit. We have already mentioned ChPT in passing and how it is an effective theory of QCD at low energies. The ChPT Lagrangian is a power expansion in some parameter (e.g. momentum), built by writing all terms that obey the symmetries of QCD at low energy up to a certain order. The notable symmetry for ChPT is the approximate chiral symmetry which when explicitly broken leads to massive pseudo-Nambu-Goldstone bosons \cite{Nambu:1960tm,Goldstone:1961eq}. Restraining to the two lightest quark flavours ($u$ and $d$), these bosons are the pions. Classed as mesons, there are four possible combinations of quark and anti-quark, two of which form electrically neutral particles and two of which have charge $\pm e$. They can be grouped into the neutral, $\pi^0$, and charged pions, $\pi^{\pm}$, respectively. The mass difference between them is small ($\sim 5\,\mathrm{MeV}$) and so we ascribe them the same pion mass, $m_{\pi}\simeq 140\,\mathrm{MeV}$ from here on.

Chiral Perturbation Theory does not include anomalies. An anomaly refers to a symmetry that does not survive the quantisation process i.e. a symmetry preserved in the classical field theory, but not the corresponding quantum field theory. Terms accounting for anomalies can be added to incorporate the missing physics. This can be achieved via a so-called Wess-Zumino-Witten (WZW) term \cite{Wess:1971yu,Witten:1983tw}. One such anomaly is the chiral anomaly which should be accounted for when considering pions and electromagnetism, as it describes the observed decay of neutral pions into two photons \cite{PhysRev.177.2426, Bell:1969ts}. Otherwise, neutral pions do not interact with electromagnetic fields and therefore $B$ as they are electrically neutral.
The consideration of the axial chiral anomaly by incorporating the WZW term is in fact key to the emergence of the CSL and its instability to charged pion condensation in Ref.\ \cite{Brauner:2016pko}. As explained originally in Refs.\,\cite{PhysRev.177.2426, Bell:1969ts}, the axial chiral anomaly is responsible for the decay of neutral pions into two photons via the electromagnetic interaction. It may then come as less of a surprise to note that by incorporating a Wess-Zumino-Witten term to account for the anomaly within ChPT, the neutral pions which are not electrically charged couple to the electromagnetic field. This term also provides the coupling to $\mu_B$ \cite{Son:2007ny}, shedding some light on why mesons appear in the $\mu_B$-$B$ plane at $T(=\mu_I)=0$ of the QCD phase diagram. Knowing this makes the rather unusual non-zero, inhomogeneous baryon number density, $n_B$, in the CSL phase where there are no baryons less unexpected. We will see that further effects from the anomaly on the baryon number density emerge when we keep the charged pion terms dropped from the WZW term in Ref.\,\cite{Brauner:2016pko}. 

For wider context, pionic superconductivity in ChPT has been discussed previously in the context of non-zero $\mu_I$, where the situation is much the same as conventional superconductivity - charged pions condense below some critical field and are expected to be a type-II superconductor \cite{Adhikari:2015wva} with a vortex lattice phase \cite{Adhikari:2018fwm,Adhikari:2018tvz}. 
The competition of this phase with a phase analogous to the CSL has been investigated in Ref.\,\cite{Gronli:2022cri} (see also Ref.\,\cite{Adhikari:2015pva}), where a WZW term couples neutral pions to $\mu_I$ instead of $\mu_B$. Pions aside, the scenario where an inhomogeneous superconducting phase occurs above rather than below a critical field is shared with the $\rho^{\pm}$ meson condensation of vacuum superconductivity \cite{Chernodub:2010qx,Chernodub:2011gs} discussed in Sec.\,(\ref{Background:subsec:BTPlane}) in the context of the $B$-$T$ plane at $\mu_B=0$. It bears similarity to the phase proposed, exhibiting some interplay between the charged and neutral varieties of the meson. A distinguishing feature of our proposed phase is that the superconducting transition occurs from the CSL phase which is an inhomogeneous phase with finite magnetisation. Moreover, the CSL phase is periodic, which would suggest a possible interplay between the CSL and superconducting vortex lattice, forming a three-dimensional structure. Even in the chiral limit where the neutral pion phase below $B_{c2}$ is homogeneous, we observe periodicity in the baryon number density.

This chapter begins with Sec.\,(\ref{Project2:sec:setup}), where we construct our Lagrangian including the WZW term and recapitulate results from Ref.\ \cite{Brauner:2016pko}. Our application of the expansion near $B_{c2}$ and analysis of lattice structures adapted from Sec.(\,\ref{Background:subsec:Hc2}) is applied in Sec.\,(\ref{Project2:subsec:exp0}). Solving the subsequent equations in the chiral limit and discussion of our results is done in Sec.\,(\ref{Project2:subsec:chiral}). The work draws from Sec.\,(\ref{Background:subsec:GL}) and Sec.\,(\ref{Background:subsec:Hc2}) which should be referred to if more detail is required. As with Chapter \ref{chpt:Project1}, the summary and discussion is reserved for the final chapter, Chapter \ref{chpt:Outlook}.

\section{Setup}
\label{Project2:sec:setup}
In this section, we construct our Lagrangian and derive the WZW term. We also define our choice of parameterisation for the chiral fields. Additionally, we re-derive the charged pion instability and other results from Ref.\ \cite{Brauner:2016pko}.  

\subsection{Lagrangian}
\label{Project2:subsec:Lagrangian}
Our starting point is the Lagrangian containing an electromagnetic part, a chiral part, and a WZW part,
\begin{equation}\label{Lfull}
    {\cal L} = {\cal L}_{\rm em} + {\cal L}_\Sigma+ \mathcal{L}_{\text{WZW}} \, .
\end{equation} 
The electromagnetic part is the same as defined in Eq.\,\eqref{Lem}. To leading order in the momentum expansion, the two-flavour chiral Lagrangian from Chiral Perturbation Theory \cite{Gasser:1983yg,Scherer:2002tk} is
\begin{equation}
    \mathcal{L}_{\Sigma} = \frac{f_\pi^2}{4}\mathrm{Tr}\left[\left(\nabla_\mu \Sigma\right)^{\dagger}\nabla^{\mu}\Sigma\right] + \frac{m_\pi^2 f_\pi^2}{4}\mathrm{Tr}\left[ \Sigma + \Sigma^{\dagger} \right]\,,
    \label{ChiralLagrangian}
\end{equation}
where $\Sigma$ is the chiral $SU(2)$ field, $f_{\pi}$ is the pion decay constant and the covariant derivative is  
\begin{equation}
    \nabla^{\mu}\Sigma = \partial^{\mu}\Sigma - i\left[ \mathcal{A}^{\mu},\Sigma \right]\,,
    \label{CovariantDerivative}
\end{equation}
with general gauge field $\mathcal{A}^{\mu}$, which can be written in the form
\begin{equation}
    \mathcal{A}_{\mu}= \mathsf{I} A_{\mu}^B +\frac{\tau_3}{2}A_{\mu}^I + e Q A^{\mu}\,.
\end{equation}
Here, $\mathsf{I}$ is the $2\times 2$ identity matrix, $A^B_{\mu}$ and $A^I_{\mu}$ are auxiliary gauge fields for the baryon and isospin charge respectively, $Q$ is the generator of the electromagnetic gauge group, and $\tau_a$ denotes the Pauli matrices ($a=1,2,3$ under usual convention). We are only interested in $\mu_B$ so we drop the $A_{\mu}^I$ term and set $A_{\mu}^B=(\mu_B,0,0,0)$. As it stands, the $A_{\mu}^B$ term drops out due to the commutator. In the same way, the diagonal part of $Q$ vanishes leaving only the part proportional to $\tau_3$, which is clear when it is decomposed like so;
\begin{equation}
    Q = \mathrm{diag}(2/3,-1/3)=\frac{1}{6}\mathsf{I} + \frac{\tau_{3}}{2}\,.
\end{equation}
This just leaves $\mathcal{A}^{\mu}=e\tau_3A^{\mu}/2$. The disappearance of $A_{\mu}^B$ can seem rather concerning as we are interested in the $\mu_B$-$B$ plane. Reassured, it will re-emerge as part of the WZW term discussed later, though this does emphasise the significance of said term.

A common parameterisation of the chiral field is
\begin{equation} \label{Sigma1}
    \Sigma = e^{i\phi_a\tau_a}  = \frac{\sigma+i\pi_a\tau_a}{f_\pi} \,,
\end{equation}
with
\begin{equation}
    \frac{\sigma}{f_\pi} = \cos\phi \, , \qquad \frac{\pi_a}{f_\pi} = \frac{\phi_a}{\phi}\sin\phi \, , 
\end{equation}
where $\phi^2\equiv\phi_1^2+\phi_2^2+\phi_3^2$. Note, $\pi_a$ are the pion fields and $\sigma$ is a spurious degree of freedom obeying the constraint $f_\pi^2 = \sigma^2+\pi_a\pi_a$. The charged pions can be summarised into the complex scalar field
\begin{equation}
    \varphi \equiv \frac{1}{\sqrt{2}}\left(\pi_1 +i\pi_2\right)\,,
\end{equation}
leaving $\pi_3\equiv \pi_0$ as the neutral pion field. With this parameterisation, the Lagrangian can be written in the form
\begin{equation}
    \begin{split}
    \mathcal{L}_{\Sigma}= \frac{1}{2} \partial_{\mu} \pi_0 \partial^{\mu} \pi_0 + \mathcal{D}_{\mu} \varphi \left(\mathcal{D}^{\mu} \varphi\right)^* + \frac{1}{2} \partial_{\mu} \sigma \partial^{\mu} \sigma  + m_{\pi}^2 f_{\pi} \sigma\,,   
    \end{split}
    \label{PiLagrangian}
\end{equation}
with the covariant derivative $\mathcal{D}_{\mu}\equiv \partial_{\mu} + ieA_{\mu}$. While this clearly separates both neutral and charged pions, we will adopt the parameterisation used in Appendix A of Ref.\ \cite{Brauner:2016pko} as it is more convenient for our purposes. It separates the third component as follows,
\begin{equation}
\label{SigmaU}
    \Sigma = e^{i\alpha\tau_3} U \,, 
\end{equation} 
where
\begin{equation} \label{Usig}
    U = \frac{\sigma_0+i(\pi_1'\tau_1+\pi_2'\tau_2)}{f_\pi} \,.
\end{equation}
Comparing both sides of Eq.\,\eqref{SigmaU}, the fields $\sigma_0$, $\pi_1'$ and $\pi_2'$ can be related to $\sigma$, $\pi_1$ and $\pi_2$ through
\begin{equation}
    \begin{array}{rl}
        \sigma = &\sigma_0\cos\alpha\\
        \pi_0= &\sigma_0\sin\alpha
    \end{array} \,\, , \qquad 
    \begin{array}{rl}
        \pi_1 = &\pi_1'\cos\alpha + \pi_2'\sin\alpha  \\
        \pi_2 =& -\pi_1'\sin\alpha + \pi_2'\cos\alpha 
    \end{array} \, \,,
    \label{alphaTrans}
\end{equation}
which amounts to going from cartesian to polar coordinates in the $(\sigma,\pi_0)$ sector and applying a rotation by $\alpha$ in the $(\pi_1,\pi_2)$ sector. We define a new complex field 
\begin{equation} \label{phiprime}
\varphi' = \frac{1}{\sqrt{2}}(\pi_1'+i\pi_2') \, ,
\end{equation}
which obeys the transformation $\varphi = e^{-i\alpha} \varphi'$. In the new coordinates, $\sigma_0$ is the (new) spurious degree of freedom, the constraint becomes $f_\pi^2 = \sigma_0^2+\pi_1'^2+\pi_2'^2 = \sigma_0^2+2|\varphi'|^2$, and the Lagrangian \eqref{PiLagrangian} is
\begin{equation} \label{TransLagrangian}
    {\cal L}_\Sigma = \frac{\sigma_0^2}{2}\partial_\mu\alpha\partial^\mu\alpha+D_\mu \varphi'(D^\mu\varphi')^*+\frac{1}{2}\partial_\mu\sigma_0\partial^\mu\sigma_0 +m_\pi^2f_\pi\sigma_0\cos\alpha \,,
\end{equation}
with the covariant derivative $D^\mu \equiv\partial^\mu+i(eA^\mu-\partial^\mu\alpha)$. The primary benefit of this re-parameterisation is that $\nabla\alpha$ turns out to be constant in our proposed phase in the chiral limit, which facilitates the calculation. We wish to express everything in terms of $\alpha$ (related to the neutral pion field in the absence of charged pions by $\pi_0 = f_\pi\sin\alpha$) and the complex field $\varphi'$ (related to the original charged pion fields by a rotation by the angle $\alpha$) only. This is achieved by using the constraint to eliminate $\sigma_0$ in Eq.\,\eqref{TransLagrangian}, leaving 
\begin{equation}  \label{LSig2}
    {\cal L}_\Sigma =
        D_\mu \varphi(D^\mu\varphi)^* + \frac{\partial_\mu|\varphi|^2\partial^\mu|\varphi|^2}{2(f_\pi^2-2|\varphi|^2)}  + \frac{f_\pi^2-2|\varphi|^2}{2}\partial_\mu\alpha\partial^\mu\alpha +m_\pi^2f_\pi\sqrt{f_\pi^2-2|\varphi|^2}\cos\alpha \,, 
\end{equation}
where we set $\varphi'\rightarrow\varphi$ from now on for brevity. The above can be verified by directly inserting Eq.\,\eqref{SigmaU} into Eq.\,\eqref{ChiralLagrangian}.

We adopt the same form for the WZW part as Refs.\ \cite{Son:2007ny,Brauner:2016pko}, namely
\begin{equation}
    \mathcal{L}_{\text{WZW}} = \left( A^B_{\mu} - \frac{e}{2}A_{\mu}\right)j^{\mu}_B\,,
    \label{ActionWZW}
\end{equation}
with the Goldstone-Wilczek baryon current \cite{Goldstone:1981kk,Son:2007ny,Brauner:2016pko}
\begin{equation}
    \begin{split}
        j^{\mu}_{B} 
        &= -\frac{\epsilon^{\mu\nu\rho\lambda}}{24\pi^2}\text{Tr}\left[(\Sigma\nabla_{\nu}\Sigma^{\dagger})( \Sigma\nabla_{\rho}\Sigma^{\dagger})( \Sigma\nabla_{\lambda}\Sigma^{\dagger}) +\frac{3ie}{4}F_{\nu\rho}\tau_3\left( \Sigma\nabla_{\lambda}\Sigma^{\dagger} +\nabla_{\lambda}\Sigma^{\dagger}\Sigma\right)\right]
        \\[2ex]
        &=-\frac{\epsilon^{\mu\nu\rho\lambda}}{4\pi^2}\partial_{\nu}\alpha\left(\frac{e}{2}F_{\rho\lambda}+ \frac{\partial_{\rho}j_{\lambda}}{ef_{\pi}^2}
        \right)\,,
    \end{split}
    \label{BCurrent}
\end{equation}
where   
\begin{equation}
    j^{\mu} = ie\left(\varphi^*\partial^{\mu}\varphi - \varphi\partial^{\mu}\varphi^*\right) -2e\left(eA^{\mu}-\partial^{\mu}\alpha\right)|\varphi|^2\,,
    \label{QCurrent}
\end{equation}
is the non-anomalous contribution to the charged current which we have identified from the equations of motion (discussed in Sec.\,(\ref{Project2:subsec:EoM})). Details on the evaluation of the traces leading to the second line of Eq.\,\eqref{BCurrent} can be found in Appendix \ref{app:Anomalous}. Interestingly, we see that besides the electromagnetic term, there is a vorticity contribution. These two terms are reminiscent of the CME and chiral vortical effect, which are generated by the chiral anomaly and which have been studied extensively mostly in the context of HICs, see for instance Ref.\ \cite{Landsteiner:2016led} for an introductory review. A similar vorticity term  gives rise to the CSL phase in a rotating system even without magnetic field \cite{Huang:2017pqe}, in which case also inhomogeneous pion-condensed phases have been predicted \cite{Eto:2021gyy}. Here we do not impose any rotation on the system, but we shall see that the dynamically created charged current does render the vorticity term non-zero, which has a direct impact on the baryon number. Note that we have introduced an additional term proportional to $\partial^{\mu}\alpha$ in Eq.\,\eqref{QCurrent} that does not appear in the derivation of $j_{B}^{\mu}$ from Eq.\,\eqref{BCurrent}. This does not change our Lagrangian or the equations of motion to follow. Though the field $\alpha$ is itself physical, this inclusion can effectively be thought of as a gauge transformation. In the end, it will amount to a shift to the rest frame of the current in $z$-direction (i.e. perpendicular to the vortex lattice).

From the explicit result for the baryon current \eqref{BCurrent} we  immediately conclude its conservation,
\begin{equation} \label{jB0}
\partial_\mu j^\mu_B = 0 \,,
\end{equation}
as it should be. In summary, our total Lagrangian is given by Eqs.\ \eqref{Lem}, \eqref{LSig2}, and \eqref{ActionWZW} with the baryon current \eqref{BCurrent}.

\subsection{Equations of motion and free energy}\label{Project2:subsec:EoM}
Here, we perform the same procedure as Sec.\,(\ref{Background:subsec:GL}). The equations of motion for $\varphi^*$, $\alpha$ and $A_{\mu}$ become, respectively, ($A_{\mu}^{B}$ is not treated as a dynamical field)
\begin{subequations}\allowdisplaybreaks
    \begin{align}
        \begin{split}
            0&= \left[D_{\mu}D^{\mu} +\partial_{\mu}\alpha\partial^{\mu}\alpha +\frac{\partial_{\mu}\partial^{\mu}|\varphi|^2}{f_{\pi}^2-2|\varphi|^2} +\frac{\partial_{\mu}|\varphi|^2\partial^{\mu}|\varphi|^2}{\left(f_{\pi}^2-2|\varphi|^2\right)^{2}}\right.
            \\[2ex]
            &\left.+\frac{m_{\pi}^2 f_{\pi}\cos{\alpha}}{\sqrt{f_{\pi}^2 -2|\varphi|^2}}
            +\frac{ie\epsilon^{\mu\nu\rho\lambda}}{8\pi^2f_{\pi}^2}\partial_{\nu}\alpha\, F_{\rho\lambda}D_{\mu}\right]\varphi\,,
            \label{phiEOMFull}
        \end{split}
        \\[2ex]
        \begin{split}
            0&= \partial_\mu\left[(f_{\pi}^2 -2|\varphi|^2)\partial^{\mu}\alpha\right] +m_{\pi}^2f_{\pi}\sqrt{f_{\pi}^2 -2|\varphi|^2}\sin{\alpha} 
            \\[2ex]
            &-\frac{e\epsilon^{\mu\nu\rho\lambda}}{16\pi^2}F_{\mu\nu}\left( \frac{e}{2}F_{\rho\lambda}+\frac{\partial_{\rho}j_{\lambda}}{ef_{\pi}^2}\right)\,,
        \label{alphaEOMFull}
        \end{split}
        \\[2ex]
           0 &= -\partial_{\nu}F^{\nu\mu}+j^{\mu} +\frac{e}{2}j^{\mu}_B
        -\frac{e^2\epsilon^{\mu\nu\rho\lambda}}{16\pi^2}\partial_{\nu}\alpha \, F_{\rho\lambda}\left(1 -\frac{2|\varphi|^2}{f_{\pi}^2}\right)
        \label{gaugeEOMFull}
        \,.
    \end{align}
\end{subequations}
From the Maxwell equation \eqref{gaugeEOMFull} we see that the charged current receives anomalous contributions. Making use of the 
baryon number conservation \eqref{jB0} we find that the total electric charge conservation reads
\begin{equation}
    0 = \partial_\mu j^\mu+\frac{e^2\epsilon^{\mu\nu\rho\lambda}}{8\pi^2f_\pi^2}\partial_\mu|\varphi|^2\partial_\nu\alpha\,F_{\rho\lambda} \, .
\end{equation}
This relation has been used to simplify the equation of motion for $\alpha$ \eqref{alphaEOMFull}.

We will only be interested in the static limit, and we assume the system to be locally charge neutral, which can be achieved for instance by adding a gas of electrons or positrons. This is similar to the standard GL treatment of an electronic superconductor presented in Sec.\,(\ref{Background:subsec:Hc2}), where the negative charge of the electron Cooper pairs is cancelled by the surrounding lattice of ions. As a consequence, the electric field vanishes and we may ignore Gauss' law, i.e.\ the temporal component of Eq.\,\eqref{gaugeEOMFull}. For the remaining equations we can therefore ignore all time derivatives and set $A_0=0$. (Note that even for $\partial_t\varphi=\partial_t\alpha=A_0=0$ there is an anomalous electric charge contribution in Gauss' law which can only be ignored under the assumption of a neutralising lepton gas.) As a result, all anomalous contributions to the equations of motion vanish and we arrive at
\begin{subequations}
    \begin{align}
        0&=\left[K +\frac{\nabla^2|\varphi|^2}{f_{\pi}^2-2|\varphi|^2} +\frac{\left(\nabla|\varphi|^2\right)^2}{\left(f_{\pi}^2-2|\varphi|^2\right)^{2}}+m_\pi^2\cos\alpha\left(1-\frac{f_\pi}{\sqrt{f_\pi^2-2|\varphi|^2}}\right) \right]\varphi\,, 
        \label{phiEOM}
        \\[2ex]
        0&=\nabla\cdot\left[(f_\pi^2-2|\varphi|^2)\nabla\alpha\right] -m_{\pi}^2f_\pi\sqrt{f_\pi^2-2|\varphi|^2}\sin{\alpha}\,,
        \label{alphaEOM}
        \\[2ex]
        \bm{j}&=\nabla\times \bm{B} = -ie\left( \varphi^*\nabla\varphi - \varphi\nabla\varphi^* \right)
        -2e\left(e\bm{A} +\nabla\alpha\right)|\varphi|^2\,,
        \label{gaugeEOM}
    \end{align}
    \label{EOMs}%
\end{subequations}
where $\bm{B} = \nabla\times \bm{A}$ is the magnetic field, $\bm{j}$ is the electromagnetic three current defined by $j^{\mu}=(j^0,\bm{j})$ and we have defined the operator
\begin{equation} \label{Ddef}
    K \equiv \nabla^2 - 2i(e\bm{A}+\nabla\alpha)\cdot\nabla-i\nabla\cdot(e\bm{A}+\nabla\alpha)-(e\bm{A}+\nabla\alpha)^2+(\nabla\alpha)^2-m_\pi^2\cos\alpha \, .
\end{equation}
Like in Sec.\,(\ref{Background:subsec:GL}), we are ignoring fluctuations to the thermodynamics and the grand canonical potential density is simply given by $\Omega=-{\cal L}$. Implementing our assumptions of a static system and vanishing electric field we obtain 
\begin{equation}
    \begin{split}
        \Omega(\bm{x}) &= \frac{B^2}{2}+|\left[\nabla -i\left(e\bm{ A}+\nabla\alpha\right)\right]\varphi|^2 +\frac{\left(\nabla|\varphi|^2\right)^2}{2\left(f_{\pi}^2-2|\varphi|^2\right)} +\frac{f_{\pi}^2 -2|\varphi|^2}{2}\left(\nabla\alpha\right)^2 
        \\[2ex]&-m_{\pi}^2f_{\pi}\sqrt{f_{\pi}^2-2|\varphi|^2}\cos{\alpha}  -\mu_B n_B(\bm{x})\, , 
    \end{split}
   \label{ThermoPotential}
\end{equation}
where 
\begin{equation}
   n_{B}(\bm{x})= j_B^0 = \frac{\nabla\alpha}{4\pi^2}\cdot\left(e\bm{B}+\frac{\nabla\times\bm{j}}{ef_\pi^2}\right) \,,
   \label{nblocal}
\end{equation}
is the (local) baryon density which follows from the baryon current \eqref{BCurrent} with the convention $\epsilon^{0123}=+1$.
With the help of the equation of motion \eqref{phiEOM} we can write the free energy (density)  as  
\begin{equation}
    \begin{split}
    \mathcal{F}&=\frac{1}{V}\int d^3\bm{x}\, \Omega(\bm{x}) \,, 
    \\[2ex]
    &=\frac{1}{V}\int d^3\bm{x}\bigg[\frac{B^2}{2}+ \frac{f_{\pi}^2}{2}\left(\nabla\alpha\right)^2 -\frac{m_{\pi}^2 f_{\pi}\cos{\alpha}}{\sqrt{f_{\pi}^2-2|\varphi|^2}}\left(f_{\pi}^2 -|\varphi|^2\right)-\frac{f_{\pi}^2}{2}\frac{(\nabla|\varphi|^2)^2}{\left(f_{\pi}^2-2|\varphi|^2\right)^2} 
     -\frac{e\mu_B}{4\pi^2}\nabla \alpha \cdot \bm{B} \bigg]
     \\[2ex]
    &
    +\frac{1}{V}\int_{S}d\bm{S}\cdot\bigg\{\varphi^*\left[\nabla -i\left(e\bm{A}+\nabla\alpha\right) +\frac{\nabla|\varphi|^2}{f_{\pi}^2-2|\varphi|^2}\right]\varphi 
    +\frac{\mu_B(\nabla\alpha\times \bm{j})}{4\pi^2ef_{\pi}^2}   \bigg\}\,,
    \end{split}
    \label{FreeEnergySimplified}
\end{equation}
where $V$ is the volume of the system and we have separated the surface terms that we can drop in our evaluation later.

\subsection{Instability at the critical magnetic field}
\label{Project2:subsec:CSL}
Here, we will recapitulate some results from Ref.\ \cite{Brauner:2016pko}. Importantly, we review the instability analysis which derives the critical magnetic field originally labelled $B_{\text{BEC}}$ above which charged pion condensation will occur. This is done in the same manner as in Sec.\,(\ref{Background:subsec:Hc2}) which will be useful to refer to for comparison. 

In the absence of charged pions, $\varphi=0$, the potential \eqref{ThermoPotential} reduces to
\begin{equation}
    \Omega_{\varphi=0} = \frac{B^2}{2}+\frac{f_{\pi}^2}{2}\left(\nabla\alpha\right)^2 -m_{\pi}^2f_{\pi}^2\cos{\alpha} -\frac{e\mu_B}{4\pi^2}\nabla \alpha \cdot \bm{B}\, ,
    \label{CSLThermoPotential}
\end{equation}
while the equation of motion for $\alpha$ \eqref{alphaEOM} is
\begin{equation}
   \nabla^2\alpha = m_{\pi}^2\sin{\alpha}\,.
\end{equation}
The above yields the first integral of motion,
\begin{equation}
    \frac{1}{2}\left( \nabla \alpha\right)^2 +m_{\pi}^2\cos{\alpha} = a^2\,,
    \label{alphaIOM}
\end{equation}
where $a$ is a constant. Without loss of generality, $\bm{B}$ (which is homogeneous) is chosen to align with the $z$-direction and as a result, $\nabla \alpha=\partial_{z}\alpha \hat{\bm{e}}_{z}$. The solution to the differential equation \eqref{alphaIOM} is then
\begin{equation}
    \partial_{z}\alpha = \frac{2m_{\pi}}{k} \sqrt{1 -k^2 \mathrm{sn}^2(z,k^2)}\,,
    \label{alphaMassiveSol}
\end{equation}
where $k^2 = 2 m_{\pi}^2/(a^2 + m_{\pi}^2)$ and $\mathrm{sn}$ denotes one of the Jacobi Elliptic functions. Requiring the free energy be minimal with respect to $k$, 
\begin{equation}
     \mathcal{F}_{\rm{CSL}} =\frac{1}{V}\int d^3\bm{x}\,\Omega_{\varphi=0} = -2m_{\pi}^2f_{\pi}^2\left(\frac{1}{k^2} -1\right)\,,
\end{equation}
which is preferred over the vacuum provided $B$ is greater than
\begin{equation}
    B_{\mathrm{CSL}} \equiv \frac{16\pi m_{\pi}f_{\pi}^2}{e \mu_B} \,.
\end{equation}
This critical magnetic field is the blue curve in Fig.\ \ref{fig:eBmu}.

This is the CSL phase. From Eq.\,\eqref{nblocal} one can appreciate that this is a spatially inhomogeneous phase with a non-zero baryon number density. Ideally we would proceed with these results. However, due to the relatively complicated form of the solution, we have only concerned ourselves with the results in the chiral (i.e. massless) limit. In this limit, the solution that minimises the potential \eqref{CSLThermoPotential} is
\begin{equation}
    \nabla \alpha = \frac{e \mu_B}{4\pi^2 f_{\pi}^2} \bm{B}\,.
    \label{GradAlpha}
\end{equation}
We see that the neutral pion phase has become homogeneous and exists all the way down to $B=0$, unlike the massive result. Note that it is the baryon number density which is homogeneous and not the neutral pion condensate. From Eq.\eqref{alphaTrans}, we see that $\sigma$ and $\pi_0$ condensates are periodic in $z$ i.e.\ along the direction of the magnetic field. The minimised free energy is then
\begin{equation}
    {\cal F}_{\rm CSL} = \frac{\langle B\rangle^2}{2}-\frac{1}{2} \left(\frac{e \mu_B \langle B\rangle}{4\pi^2 f_{\pi}}\right)^2\,.
    \label{FreeEnergyAlpha}
\end{equation}
The magnetic field is uniform in this phase. Nevertheless, we have replaced it by its spatial average, just like before using Eq.\,\eqref{avdef}. Here we simply have $\langle B\rangle=B$, but in the form \eqref{FreeEnergyAlpha} the free energy density can be compared more easily with our main results, where $B$ is no longer uniform. The baryon density in the chiral limit is uniform (in contrast to the general 
 density \eqref{nblocal}) and is given by 
\begin{equation} \label{nBCSL}
    n_B^{\rm CSL} = -\frac{\partial \Omega_{\varphi=0}}{\partial \mu_B} = \frac{e^2\mu_B\langle B\rangle^2}{16\pi^4 f_\pi^2} \, ,
\end{equation}
Note that we have retained the label ``CSL'' in Eqs.\ \eqref{FreeEnergyAlpha} and \eqref{nBCSL}. This was done for notational convenience though it is strictly speaking an abuse of the term given that there is no lattice in the direction of the magnetic field in the chiral limit.

The stability of the CSL phase can be probed by considering the fluctuations in the field $\varphi$. To this end, we go back to the equation of motion \eqref{phiEOMFull}. Setting $A_0=\partial_t\alpha=0$ but keeping the time dependence of $\varphi$ and linearising this equation yields
\begin{align}
     0\simeq
     \left(\partial_{t}^2-K -\frac{ie\nabla\alpha\cdot\bm{B}}{4\pi^2f_{\pi}^2}\,\partial_{t}\right)\varphi 
     \, .
     \label{LOInstability}
\end{align}
To proceed we align the $z$-axis with the magnetic field, $\bm{B}=B\hat{\bm{e}}_z$, such that we can choose $e\bm{A}+\nabla\alpha = eBx\hat{\bm{e}}_y$. Moreover, we employ the ansatz $\varphi(t,\bm{x})= e^{-i\omega t}e^{ik_y y}f(x,z)$ to obtain 
\begin{align}
     0= \left[-\omega^2 -\partial_x^2 - \partial_z^2+e^2B^2\left(x-\frac{k_y}{eB}\right)^2     -(\nabla\alpha)^2 +m_{\pi}^2\cos{\alpha}-\frac{e\nabla\alpha\cdot\bm{B}}{4\pi^2f_{\pi}^2}\, \omega\right]f(x,z)
     \, .\hspace{0.5cm} \label{ompartial}
\end{align}
Returning to the chiral limit, we abbreviate $\nabla\alpha = c\,\hat{\bm{e}}_z$ with 
\begin{equation} \label{cdef}
c \equiv \frac{e \mu_B B}{4\pi^2 f_{\pi}^2}  \, ,
\end{equation}
and further simplify the ansatz by writing $f(x,z)=e^{ik_{z}z} g(x)$ to give
\begin{equation}
\left[(\omega+\mu_*)^2 - k_z^2 -m_*^2\right]g(x) = \left[-\partial_x^2+e^2B^2\left(x-\frac{k_y}{eB}\right)^2\right]g(x) \, .
\end{equation}
Written in this form, this equation is identical to the standard GL scenario from $\varphi^4$ theory (see Eq.\,\eqref{omegapsi}), having identified an effective chemical potential  and an effective mass by 
\begin{equation}
\mu_* = \frac{c^2}{2\mu_B} \, , \qquad m_*^2 = \mu_*^2 - c^2   \, .
\end{equation}
Therefore, following exactly the same arguments as in Sec.\,(\ref{Background:subsec:Hc2}), the dispersion relation of the $\varphi$ field in the (massless) CSL phase is
\begin{equation}\label{wstar}
\omega = \sqrt{(2\ell+1)eB+m_*^2+k_z^2}-\mu_* \,, 
\end{equation}
and we encounter an instability of the $\ell=k_z=0$ mode for $eB< \mu_*^2-m_*^2=c^2$.
However, crucially, $\mu_*$ and $m_*$ depend on the magnetic field themselves. As a consequence, this condition translates into an instability for magnetic fields {\it larger} than the critical field 
\begin{equation}
    B_{c2}= \frac{16\pi^4f_{\pi}^4}{e\mu_B^2}\,.
    \label{Bc2}
\end{equation}
This is in contrast to the scenario of an ordinary type-II superconductor where the instability towards a superconducting flux tube lattice occurs upon {\it decreasing} the magnetic field. 

This critical magnetic field is equivalent to the aforementioned $B_{\rm{BEC}}$ in Ref.\ \cite{Brauner:2016pko} (where $e$ was set to 1). In this reference, the critical field was also computed for the case of a non-zero pion mass (we have used this result in the phase diagram of Fig.\ \ref{fig:eBmu}). The details of the massive case are not relevant for the following since we shall construct the superconducting state only in the chiral limit. Our derivation deviates in one detail from that of Ref.\ \cite{Brauner:2016pko}: The anomalous contribution in Eq.\,\eqref{phiEOMFull} generates the term linear in $\omega$ in Eq.\,\eqref{ompartial}, which we then have absorbed in the effective chemical potential $\mu_*$. It is possible to discard this term on the ground of a consistent power counting scheme. As argued in Ref.\ \cite{Brauner:2021sci}, in addition to the usual power counting in chiral perturbation theory in terms of the momentum scale $p\ll 4\pi f_\pi$, namely $\partial_\mu, m_\pi,A_\mu \sim {\cal O}(p)$, the baryon chemical potential should be counted as $A_\mu^B \sim {\cal O}(p^{-1})$. This ensures that the contribution $A_\mu^B j_B^\mu \sim {\cal O}(p^2)$ in the WZW Lagrangian \eqref{ActionWZW} is consistent with our chiral Lagrangian ${\cal L}_{\Sigma} \sim {\cal O}(p^2)$. In contrast, the second WZW contribution $eA_\mu j_B^\mu \sim {\cal O}(p^4)$ is of higher order. This is the term that gives rise to the effective chemical potential $\mu_*$. If $\mu_*$ is set to zero we reproduce the dispersion relation of Ref.\ \cite{Brauner:2016pko} exactly. However, since we also include the electromagnetic contribution ${\cal L}_{\rm em}\sim {\cal O}(p^4)$, which is crucial for our main results, our expansion is not consistent with respect to this scheme even in the absence of the WZW term.  Therefore, we have included all terms from ${\cal L}_{\rm WZW}$ \eqref{ActionWZW}, resulting in a formally higher-order term in the equation of motion \eqref{phiEOMFull}. An alternative power counting with respect to the electromagnetic field, namely $A_\mu \sim {\cal O}(p^0)$, $e \sim {\cal O}(p)$ \cite{Gronli:2022cri}, ensures consistency of the electromagnetic and chiral parts of the Lagrangian. Then, all our terms {\it are} consistently of order $p^2$ if we omit $eA_\mu j_B^\mu \sim {\cal O}(p^4)$ in the WZW Lagrangian, which is of higher order also within this alternative scheme. 

For the location of the instability the term linear in $\omega$ has no consequence because the critical magnetic field is given by $\mu_*^2-m_*^2$, which is identical to $c^2$ irrespective of whether $\mu_*$ is set to zero or not. Interestingly, however, the nature of the instability is affected: from Eq.\,\eqref{wstar} we see that $\omega$ turns negative at the critical magnetic field, whereas, if we set $\mu_*=0$ in that equation, $\omega$ turns {\it imaginary} at the critical magnetic field. 
These two cases are sometimes referred to as ``energetic'' and ``dynamical'' instabilities. Only a dynamical instability indicates a time scale on which the unstable modes grow, whereas an energetic instability can be turned dynamical if the system is allowed to exchange momentum with an external system, see for instance Refs.\ \cite{Haber:2015exa,Andersson:2019ezz}. We thus see that only in the presence of $\mu_*$, the nature of the instability is the same as in the GL treatment of a standard superconductor. In that case, as one can check with the help of Eq.\,\eqref{wstar}, $\omega$ can only become complex in a regime which is already energetically unstable. For the following, this aspect of the instability as well as the difference in the dispersion of the charged pions in the CSL phase is irrelevant. In other words, if we count powers of the momentum scale according to $A_\mu^B \sim {\cal O}(p^{-1}), A_\mu \sim {\cal O}(p^0), e \sim {\cal O}(p)$ all our main results in the subsequent sections follow consistently from an order $p^2$ Lagrangian.

\section{Flux tube lattice}
\label{Project2:sec:Lattice}

This section contains the main calculation. Here, we adapt and apply the original method used in Ref.\ \cite{Abrikosov:1956sx} to our system, find the free energy in the chiral limit, and determine the preferred configuration of the lattice followed by a discussion of the result(s). More so than the previous subsection, this section assumes some familiarity with the content of Sec.\,(\ref{Background:subsec:Hc2}).

\subsection{Expansion at the critical magnetic field} 
\label{Project2:subsec:exp0}

The instability discussed in Sec.\,(\ref{Project2:subsec:CSL}) indicates that there is a phase that includes charged pion condensation and which has lower free energy than the CSL phase for magnetic fields above the critical field $B_{c2}$. In this section, we construct such a phase by applying an expansion in the parameter $\epsilon \sim \sqrt{B-B_{c2}}$, exploiting the analogy with the standard type-II superconductor of Sec.\,(\ref{Background:subsec:Hc2}). We shall present the expansion for the general case, including the pion mass, but restrict ourselves to the chiral limit in the solution of the resulting equations in Sec.\,(\ref{Project2:subsec:CSL}). In contrast to ordinary gauged $\varphi^4$ theory, we now have the additional scalar field $\alpha$, which we also have to expand,
\begin{equation} \label{expansion}
    \varphi  = \varphi_0 +\delta\varphi +\ldots \,,  \qquad \bm{A} = \bm{A}_0 +\delta\bm{A} + \ldots  \, , \qquad \alpha = \alpha_0 + \delta\alpha + \ldots \, . 
\end{equation}
As in the conventional calculation, respectively, $\varphi_0$ and $\delta\varphi$  are of order $\epsilon$ and $\epsilon^3$, and $\bm{A}_0$ and $\delta\bm{A}$ are of order $\epsilon^0$ and $\epsilon^2$. Our expansion for $\alpha$ follows that of the gauge field with $\alpha_0$ of order $\epsilon^0$ and $\delta\alpha$ of order $\epsilon^2$. The motivation behind this being that in the chiral limit $\alpha$ is directly proportional to the magnetic field (as can be seen from Eq.\,\eqref{GradAlpha} below $B_{c2}$). We insert these expansions into the equations of motion \eqref{EOMs} to obtain the following order-by-order equations. From the equation of motion for $\varphi^*$ \eqref{phiEOM} we obtain the 
$\epsilon^1$ and $\epsilon^3$ contributions
\begin{subequations}
    \begin{align}
         K_0\varphi_0 &=0 \,, 
        \label{phi0EoMEx} 
        \\[2ex]
        \begin{split}
            K_0\delta\varphi &= \Bigg[2i(e\delta\bm{A}+\nabla\delta\alpha)\cdot\nabla+i\nabla\cdot(e\delta\bm{A}+\nabla\delta\alpha)+2(e\bm{A}_0+\nabla\alpha_0)\cdot(e\delta\bm{A}+\nabla\delta\alpha)
            \\[2ex]
            &-(\nabla^2\alpha_0+2\nabla\alpha_0\cdot\nabla)\delta\alpha-\frac{(\nabla^2-m_\pi^2\cos\alpha_0)|\varphi_0|^2}{f_\pi^2}\Bigg]\varphi_0 \,,
        \label{deltaphiEoMEx}
        \end{split}
    \end{align}
\end{subequations} 
where $K_0$ is the lowest-order contribution to the operator $K$, i.e. Eq.\,\eqref{Ddef} with $\bm{A}$ and $\alpha$ replaced by $\bm{A}_0$ and $\alpha_0$. From the equation of motion for $\bm{A}$, we derive the $\epsilon^0$ and $\epsilon^2$ contributions
\begin{subequations}
    \begin{align}
        \nabla\times\bm{B}_0 &= 0 \, , \label{A0EoMEx}
        \\[2ex]
        \nabla\times\delta\bm{B} &= -ie\left(\varphi_0^*\nabla\varphi_0 -\varphi_0\nabla\varphi_0^*\right) -2e\left(e\bm{A}_0 +\nabla\alpha_0\right)|\varphi_0|^2 \, , \label{deltaAEoMEx}
    \end{align}
\end{subequations}
where we have once again defined $\bm{B}_0 = \nabla\times \bm{A}_0$, $\delta\bm{B} = \nabla\times\delta\bm{A}$, such that we can denote the magnetic field up to ${\cal O}(\epsilon^2)$ by $\bm{B} \simeq \bm{B}_0  + \delta\bm{B}$. The lowest-order magnetic field corresponds to the critical magnetic field, $B_0 = B_{c2}$ and we satisfy Eq.\,\eqref{A0EoMEx} trivially by a constant $\bm{B}_0$. 
Finally, the equation of motion for $\alpha$ \eqref{alphaEOM} yields the following $\epsilon^0$ and $\epsilon^2$ equations,
\begin{subequations}
    \begin{align}
        \nabla^2\alpha_0 &= m_\pi^2 \sin\alpha_0 \, , \label{alpha0EoMEx} 
        \\[2ex]
        f_\pi^2(\nabla^2-m_\pi^2\cos\alpha_0)\delta\alpha &= (\nabla^2\alpha_0+2\nabla\alpha_0\cdot\nabla) |\varphi_0|^2 \, , \label{deltaalphaEoMEx}
    \end{align}
\end{subequations}
where the first equation has already been used to simplify the second. 

We keep terms up to fourth order in the free energy density \eqref{FreeEnergySimplified}. By dropping the surface terms, it can be brought into the form   
\begin{align} \label{FreeEnergyEx}
    {\cal F} \simeq \frac{1}{V}\int d^3\bm{x}\, \left\{\frac{B^2}{2}+\frac{f_\pi^2}{2}[(\nabla\alpha)^2-2m_\pi^2\cos\alpha] -\lambda_*|\varphi_0|^4 -\frac{e\mu_B}{4\pi^2}\nabla\alpha\cdot\bm{B}
    \right\} \,.
\end{align}
It is more convenient to hold off expanding $\alpha$ and $\bm{A}$ in the above expression until after solving the expanded equations of motion. Furthermore, we have introduced $\lambda_{*}$ defined by
\begin{equation}\label{lamstar}
    \lambda_*\langle|\varphi_0|^4\rangle\equiv \frac{\langle 
    (\nabla|\varphi_0|^2)^2\rangle +m_\pi^2\langle\cos\alpha_0 |\varphi_0|^4\rangle}{2f_\pi^2\langle|\varphi_0|^4\rangle}\langle|\varphi_0|^4\rangle \, .
\end{equation}
Written in this way, the free energy density would then resemble the one of $\varphi^4$ theory, see Eq.\,\eqref{F}, which will be helpful for the remaining evaluations. In particular, $\lambda_*$ plays the role of an effective self-coupling of the complex field.

As in the $\varphi^4$ calculation of Sec.\,(\ref{Background:subsec:Hc2}) we use the higher-order equations \eqref{deltaphiEoMEx} and \eqref{deltaAEoMEx} to derive an identity that will later be needed to evaluate the free energy. We multiply Eq.\,\eqref{deltaphiEoMEx} by $\varphi^*$ and Eq.\,\eqref{deltaAEoMEx} by $\delta\bm{A}+\nabla\delta\alpha/e$, and combine the resulting equations to obtain 
\begin{equation}
    \begin{split}
            \varphi_0^*K_0\delta\varphi 
            =&\, i\nabla\cdot\left[(e\delta\bm{A}+\nabla\delta\alpha)|\varphi_0|^2\right]-\left(\delta\bm{A}+\nabla\delta\alpha/e\right)\cdot(\nabla\times\delta\bm{B})
            \\[2ex]
            &-\frac{|\varphi_0|^2(\nabla^2-m_\pi^2\cos\alpha_0)|\varphi_0|^2}{f_\pi^2}-|\varphi_0|^2(\nabla^2\alpha_0+2\nabla\alpha_0\cdot\nabla)\delta\alpha\, .
    \end{split}
\end{equation}
Integrating over the volume on both sides and dropping the surface terms gives
\begin{equation}
    \begin{split}
        0= \int d^3\bm{x}\,\left[2\lambda_*|\varphi_0|^4-(\delta\bm{A}+\nabla\delta\alpha/e)\cdot(\nabla\times\delta\bm{B})\right. 
        \\[2ex]
        \left.-|\varphi_0|^2(\nabla^2\alpha_0+2\nabla\alpha_0\cdot\nabla)\delta\alpha\right] \, .
    \end{split}
     \label{intAdB}
\end{equation}
This is the analogue to Eq.\,\eqref{appBorthogonal}. The extra term due to the scalar field $\alpha$ will play an important role below. 

\subsection{Solution in the chiral limit}
\label{Project2:subsec:chiral}

We now solve the equations of motion and compute the free energy density in the chiral limit, $m_\pi=0$. According to our expansion, the lowest-order terms of $\alpha$ and $\bm{A}$ correspond to their CSL values at the critical field $B_{c2}$. Aligning the magnetic field with the $z$-direction, we can thus write 
\begin{equation}\label{alpha0}
\alpha_0 = \frac{z}{\xi} \, , \qquad e\bm{A}_0+\nabla\alpha_0 = eB_{c2} x\hat{\bm{e}}_y \, , 
\end{equation}
where $\xi^{-1}$ is the constant $c$ from Eq.\,\eqref{cdef} evaluated at 
$B=B_{c2}$, 
\begin{equation}\label{xidef}
    \xi = \frac{1}{\sqrt{eB_{c2}}} = \frac{\mu_B}{4\pi^2f_\pi^2} \, , 
\end{equation}
where we have used the explicit form of $B_{c2}$ \eqref{Bc2}. Our gauge choice for $\bm{A}_0$ implies $\bm{B}_0 = B_{c2} \hat{\bm{e}}_z$. Moreover, by assigning a $z$-component to $\bm{A}_0$ that absorbs $\nabla \alpha_0$ we ensure that there is no charged current in the $z$-direction, $j_z=0$, which is a convenient choice for the calculation. 

With Eq.\,\eqref{alpha0} we can solve the 
equation of motion for $\varphi_0$ \eqref{phi0EoMEx} in exactly the same way as for the standard superconductor. 
Consequently, following Sec.\,(\ref{Background:subsec:Hc2}), we have 
\begin{equation} \label{phi0xy1}
    \varphi_0(x,y) =\sum_{n=-\infty}^{\infty}C_{n}e^{in qy}\psi_n(x)\,, \qquad \psi_n(x) = e^{-\frac{(x-x_n)^2}{2\xi^2}} \, , 
\end{equation}
with complex coefficients $C_n$, the wave number $k_y=nq$, and $x_n \equiv nq\xi^2$. In particular, $\varphi_0$ does not depend on $z$ (which would be different in the presence of a pion mass because in that case $\nabla\alpha_0$ depends on $z$). The coherence length $\xi$, which characterises the variation of the condensate in the $x$-$y$ plane, is the same length scale as in $\alpha_0=z/\xi$. 

Next, we determine $\delta\bm{A}$ from Eq.\,\eqref{deltaAEoMEx}. 
Again, we can follow exactly the same arguments as in Sec.\,(\ref{Background:subsec:Hc2}). We can choose a gauge in which  $\delta\bm{A} = \delta A_y \hat{\bm{e}}_y$ such that $\delta \bm{B} = \delta B \hat{\bm{e}}_z$ with $\delta B = \partial_x \delta A_y$, and find
\begin{equation}
    \delta A_y = \left(\langle B\rangle -B_{c2} +e\langle |\varphi_0|^2\rangle\right) x - e\int dx\,|\varphi_0|^2\,.
    \label{deltaA}
\end{equation}
As a consequence, the magnetic field varies in the $x$-$y$ plane
and its $z$-component is 
\begin{equation}
    B =B_{c2} + \delta B =  \langle B\rangle + e(\langle|\varphi_0|^2\rangle-|\varphi_0|^2) \, ,
\end{equation}
where $\langle B\rangle$ will act as our external thermodynamic variable.

In the chiral limit, the ${\cal O}(\epsilon^2)$ equation of motion \eqref{deltaalphaEoMEx} becomes $\nabla^2\delta\alpha=0$. Assuming $\delta\alpha$ to be independent of $x$ and $y$, we pick the ansatz
\begin{equation}
    \delta\alpha = \delta c\,z\,.
\end{equation}
where $\delta c$ is a constant to be determined. We further assume we can fix this constant such that the total scalar field $\alpha$ is identical to its CSL value \eqref{GradAlpha} at the magnetic field $\langle B\rangle$, 
\begin{equation}
    \alpha_0+\delta\alpha = \frac{e\mu_B\langle B\rangle}{4\pi^2f_\pi^2} z \, .
\end{equation}
This condition is satisfied by  
\begin{equation}
    \delta\alpha = e(\langle B\rangle - B_{c2})\xi z \, .
\end{equation}
One can check our choice of $\delta c$ is correct by keeping it general throughout the calculation and then minimising the final free energy with respect to it at the end. The $\delta c$ that minimises the free energy is then found to be the one assumed above.
The higher-order correction $\delta\varphi$ can now in principle be calculated from Eq.\,\eqref{deltaphiEoMEx}. However, we shall not need the explicit result. We already extracted information from that equation in the derivation of the relation \eqref{intAdB}. To make use of this relation we first compute
\begin{equation} \label{dAdda}
    (\delta\bm{A}+\nabla\delta\alpha/e)\cdot(\nabla\times\delta\bm{B}) = -e\left(\langle B\rangle - B_{c2} +e \langle|\varphi_0|^2\rangle\right) |\varphi_0|^2 +e^2 |\varphi_0|^4 +\mbox{total derivatives} \, ,
\end{equation}
and 
\begin{equation} \label{phiDelta}
    |\varphi_0|^2(\nabla^2\alpha_0+2\nabla\alpha_0\cdot\nabla)\delta\alpha = 2e|\varphi_0|^2(\langle B\rangle - B_{c2}) \, . 
\end{equation}
Interestingly, the expression \eqref{phiDelta}, when added to Eq.\,\eqref{dAdda}, effectively flips the sign of $\langle B\rangle-B_{c2}$, and Eq.\,\eqref{intAdB} can be brought into the form
\begin{equation}
    e\langle|\varphi_0|^2\rangle = \frac{\langle B\rangle-B_{c2}}{\left(2\kappa^2-1\right)\beta +1}\,,
    \label{phi0BKbeta}
\end{equation}
with the effective GL parameter
\begin{equation} \label{kappaeff}
    \kappa \equiv \frac{\sqrt{\lambda_*}}{e} = \frac{1}{\sqrt{2}\,ef_\pi\xi} \, , 
\end{equation}
where we have used the definition of $\lambda_*$ \eqref{lamstar} and the identity
\begin{equation} \label{phi22}
    \langle \nabla|\varphi_0|^2\rangle^2 = \frac{\langle|\varphi_0|^4\rangle}{\xi^2} \, , 
\end{equation}
which we prove in Appendix \ref{app:Compute}. Moreover, $\beta$ is the same parameter as introduced by Abrikosov in the standard GL scenario. The left-hand side of Eq.\,\eqref{phi0BKbeta} is obviously positive, and thus the right-hand side must be positive too. For the denominator, we find with Eqs.\,\eqref{Bc2}, \eqref{xidef}, and \eqref{kappaeff}, that $2\kappa^2-1>0$ for all $\mu_B \lesssim 12\, {\rm GeV}$ and thus for all relevant $\mu_B$. Therefore, our result is only valid for $\langle B\rangle - B_{c2}\ge 0$. This reflects the fact that our charged pion superconductor occurs for magnetic fields larger than the critical field. Hence the sign flip in front of $\langle B\rangle - B_{c2}$ due to Eqs.\,\eqref{dAdda} and \eqref{phiDelta} was crucial. In the absence of the scalar field $\alpha$, the contribution \eqref{phiDelta} is absent and the numerator on the right-hand side of Eq.\,\eqref{phi0BKbeta} becomes $B_{c2}-\langle B\rangle $, which is positive in the standard scenario. 

We may also use Eq.\,\eqref{phi0BKbeta} to determine the coefficients $C_n$ in the charged pion condensate \eqref{phi0xy1}. As explained in Sec.\,(\ref{Background:subsec:Hc2}), we consider periodic solutions where $C_n = C$ for even $n$ and $C_n=iC$ for odd $n$. Then, comparing Eq.\,\eqref{N=2AvgVarPhi} with Eq.\,\eqref{phi0BKbeta} we read off
\begin{equation} \label{Csq}
    |C|^2 = \frac{\sqrt{R}}{e}\frac{\langle B\rangle-B_{c2}}{\left(2\kappa^2-1\right)\beta +1} \, , 
\end{equation}
where $R=q^2\xi^2/\pi$ determines the lattice structure of the solution. As a consistency check, this result shows that $|\varphi_0|^2$ is of order $\epsilon^2\sim \langle B\rangle-B_{c2}$, in accordance with our expansion \eqref{expansion}.  

Our results can now be inserted into the free energy density \eqref{FreeEnergyEx}. The magnetic energy and the $|\varphi_0|^4$ contribution have exactly the same form as in the $\varphi^4$ model, and we can use the result \eqref{FreeEnergyFinal} for these terms. The remaining terms simply reproduce the free energy density of the CSL phase \eqref{FreeEnergyAlpha}, such that we obtain 
\begin{equation}
    {\cal F} =  {\cal F}_{\rm CSL} - \frac{1}{2}\frac{\left(\langle B\rangle-B_{c2}\right)^2}{\left(2\kappa^2-1\right)\beta +1}\,.
    \label{FreeEnergyFinal}
\end{equation}
This is one of the chapter's main results since it shows that the free energy of the inhomogeneous charged pion superconductor is indeed lower than that of the (massless) CSL state for $\langle B\rangle >B_{c2}$. Let us discuss this result and the properties of our flux tube lattice in more detail. 

\begin{figure}
    \centering
   \hbox{\includegraphics[width=0.49\textwidth]{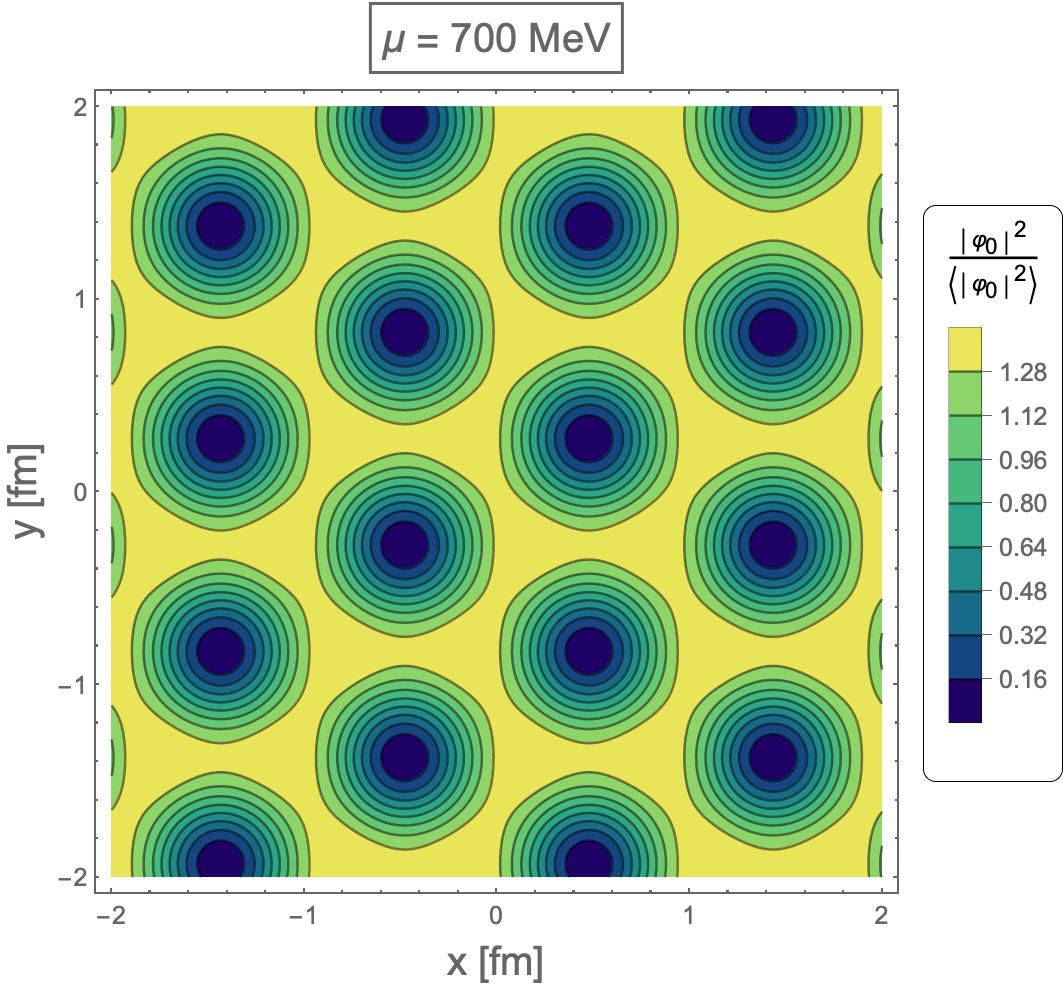} \hspace{0.2cm}\includegraphics[width=0.49\textwidth]{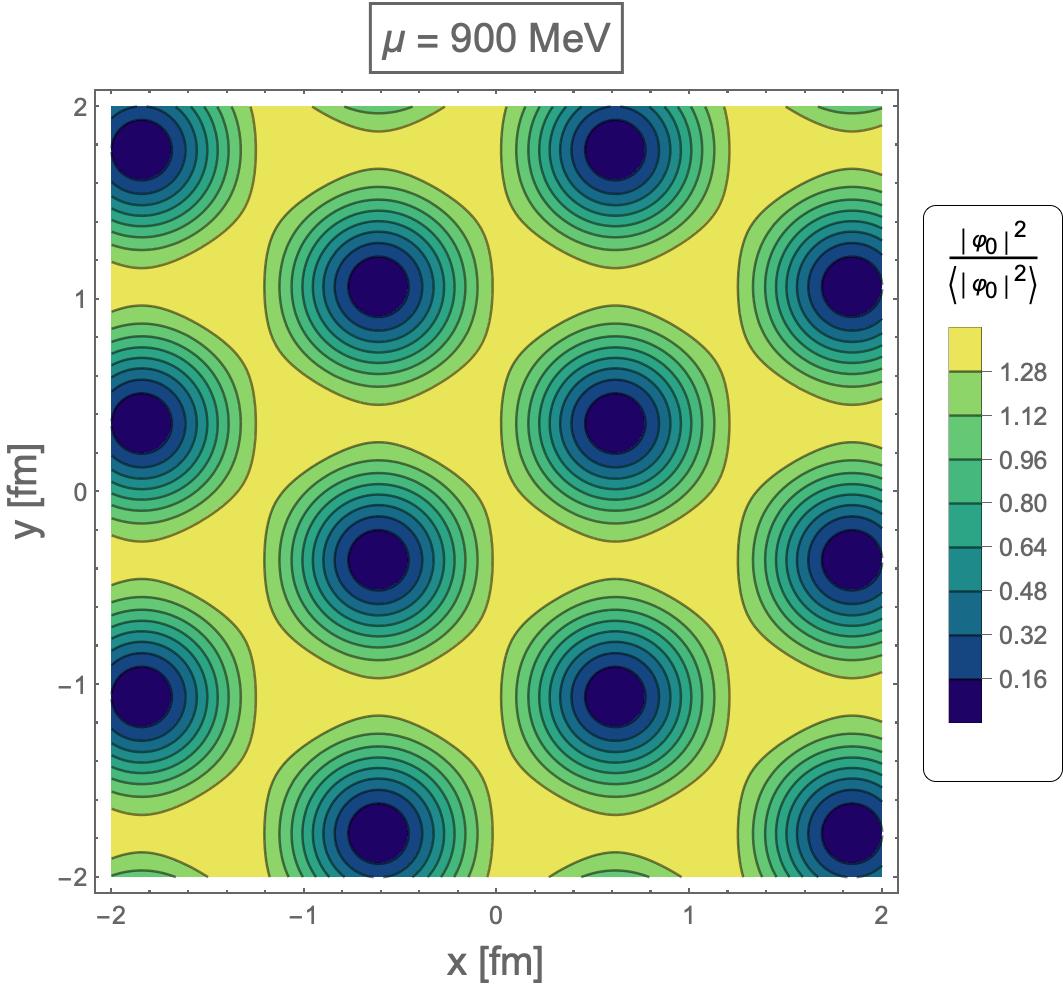}}
   \hbox{\includegraphics[width=0.49\textwidth]{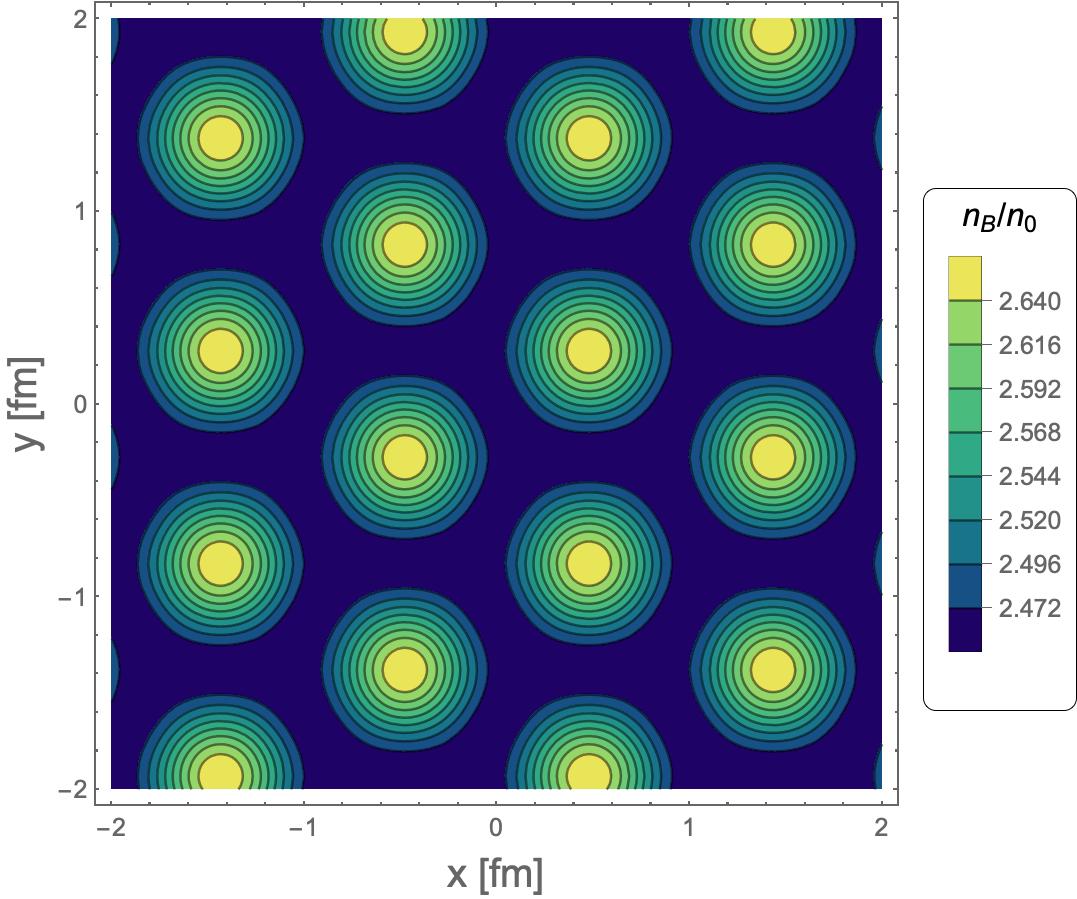} \hspace{0.2cm}\includegraphics[width=0.49\textwidth]{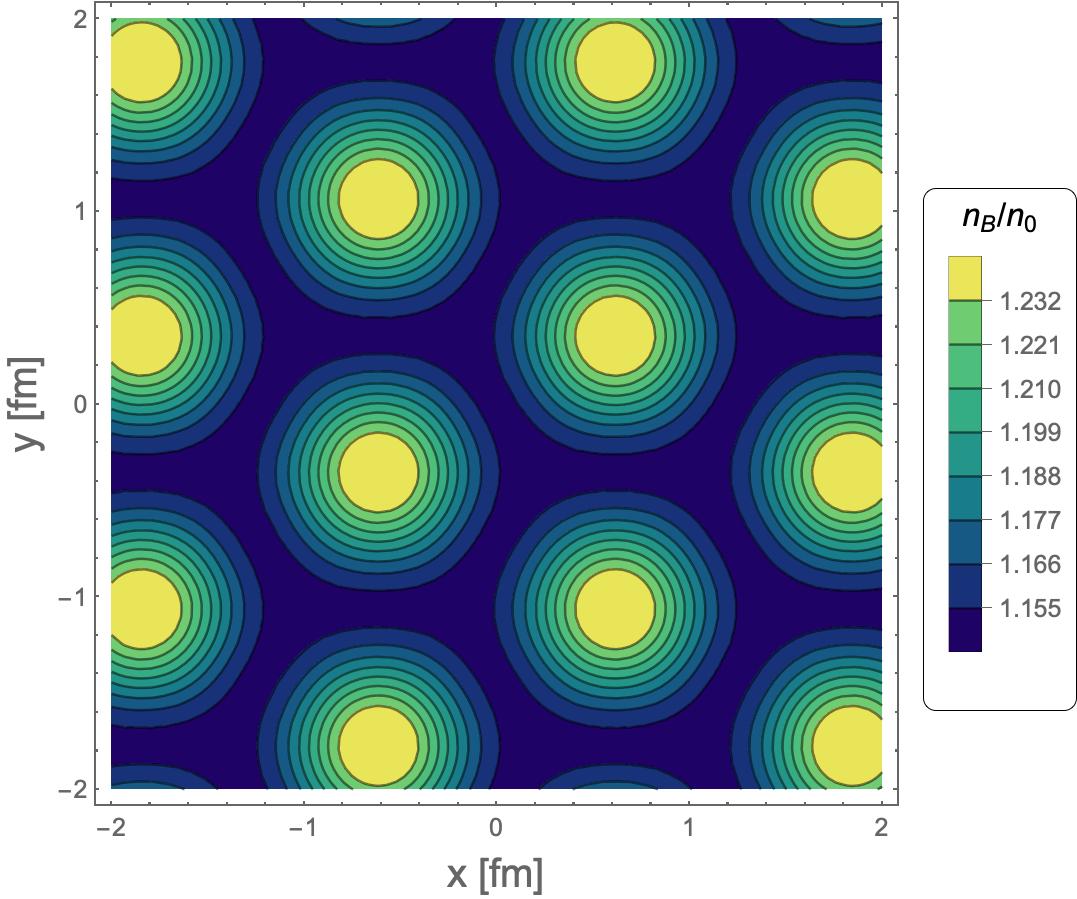}}
    \caption{{\it Upper panels:} Modulus of the charged pion condensate (squared) normalised to its spatial average in the plane perpendicular to the magnetic field. The plots show the energetically preferred hexagonal lattice structure just above the critical magnetic field, in both panels $\langle B\rangle = 1.01 B_{c2}$, where $eB_{c2}\simeq 0.2319\, {\rm GeV}^2$ for a baryon chemical potential $\mu_B=700\, {\rm MeV}$ (left panel) and $eB_{c2}\simeq 0.1403\, {\rm GeV}^2$ for $\mu_B=900\, {\rm MeV}$ (right panel). The distance between the minima turns out to be about $1.10\, {\rm fm}$ (left) and $1.42\, {\rm fm}$ (right). {\it Lower panels:} Baryon number density in units of the nuclear saturation density for the same $\mu_B$ and $\langle B\rangle$ as the corresponding upper panels. We have used the numerical values $f_\pi=92.4\, {\rm MeV}$ and $n_0=0.15\, {\rm fm}^{-3}$. 
    }  \label{fig:BaryonNumberDensity}
\end{figure}

The inhomogeneous state we have constructed is preferred over the CSL state above the black curve in Fig.\ \ref{fig:eBmu}. The transition at this critical curve is continuous in the sense that the charged pion condensate goes to zero as the curve is approached from above. The charged pion condensate gives rise to a crystalline structure in the plane perpendicular to the external magnetic field, while there is no variation of any physical observable in the direction parallel to the magnetic field. The magnetic field itself varies in the $x$-$y$ plane as well, just like in an ordinary type-II superconductor. It is identical to the external field $\langle B\rangle$ at the points where the condensate vanishes and is expelled in the regions with non-vanishing condensate. The result is a flux tube lattice, whose structure is determined by the parameter $\beta$. Since the free energy is minimised by the minimal $\beta$ the preferred lattice structure is given by the minimum of the function \eqref{betaa}. There is no difference in this function to the case of an ordinary type-II superconductor, and thus we find the same result, i.e. the free energy is minimised by a hexagonal flux tube lattice, $R=\sqrt{3}$. Using the expression \eqref{phi0xy1} and the coefficients \eqref{Csq} we  plot the condensate for two different points in the $\mu_B$-$\langle B\rangle$ plane in the upper panels of Fig.\ \ref{fig:BaryonNumberDensity}. The oscillation in the amplitude becomes larger as one moves away from the critical field and our expansion in $\epsilon$ becomes less applicable. We have thus chosen two points very close to the critical field, where we can trust our expansion,  $(\mu_B,\langle B\rangle) = (\mu_B,1.01 \,B_{c2}(\mu_B))$, for two different values of $\mu_B$. These points are marked by diamonds in the phase diagram of Fig.\ \ref{fig:eBmu}.

We see that the lattice spacing increases with $\mu_B$. This is obvious since the characteristic length scale is the coherence length $\xi \propto \mu_B$. The lattice spacing also becomes larger as one moves away from the critical field $B_{c2}$. This is identical to an ordinary type-II superconductor, where for $\langle B\rangle \to 0$ (or $H\to H_{c1}$) the spacing becomes infinite, indicating a transition to a Meissner state where the magnetic field is completely expelled. Here, such a state is not possible: a homogeneous charged pion condensate would expel the magnetic field completely. However, then there is no anomalous coupling to the neutral pions and in turn there is no effective potential that makes the charged pions condense. Therefore, we do not expect our lattice to be continuously connected to a Meissner state, at least not in the absence of an isospin chemical potential. It is therefore not obvious - even if we keep using chiral perturbation theory for such large magnetic fields and chemical potentials - how the flux tube lattice evolves far beyond the critical field $B_{c2}$. 

Our crystalline state is not only a lattice for the charged pion condensate and thus the magnetic field, but also for baryon number due to the WZW term. The local baryon number \eqref{nblocal} receives contributions both from $\nabla\alpha \cdot\bm{B}$ and $\nabla\alpha\cdot\nabla\times \bm{j}$. With the help of Eqs.\ \eqref{iphi0} and the definition of the (non-anomalous) charge current \eqref{QCurrent} we have $\bm{j} = -e(\hat{\bm{e}}_x\partial_y - \hat{\bm{e}}_y\partial_x)|\varphi_0|^2$, which implies 
\begin{equation}
    \nabla\times \bm{j} \simeq e\nabla^2|\varphi_0|^2 \hat{\bm{e}}_z \,,
\end{equation}
to order $\epsilon^2$. Consequently, the baryon density \eqref{nblocal} becomes
\begin{equation} \label{nBxy}
    n_B(x,y) \simeq n_B^{\rm CSL} + \frac{e^3\mu_B\langle B\rangle}{16\pi^4 f_\pi^2}\left[\langle|\varphi_0|^2\rangle-|\varphi_0|^2+\frac{\nabla^2|\varphi_0|^2}{e^2f_\pi^2}\right] \, , 
\end{equation}
to the same order\footnote{Note that we cannot consistently write the non-anomalous charge current or baryon number density to fourth order as is perhaps desired since we don't have an expression for $\delta\varphi$.} with the uniform CSL density $n_B^{\rm CSL}$ \eqref{nBCSL}, around which the baryon density oscillates, $\langle n_B \rangle = n_B^{\rm CSL}$. We plot the result in the lower panels of Fig.\ \ref{fig:BaryonNumberDensity} for the same points in the $\mu_B$-$\langle B\rangle$ plane as the upper two panels. We see that going above the critical field by 1\% leads to a periodic oscillation in the baryon number by about 8\%. The largest effect comes from the last term in Eq.\,\eqref{nBxy}, which originates from the vorticity contribution $\nabla\times \bm{j}$. This can be seen numerically or by estimating the ratio $[\nabla^2 |\varphi_0|^2/(e^2 f_\pi^2)]/|\varphi_0|^2\sim 2\kappa^2 \gg 1$, where we have simply replaced the derivative by the inverse coherence length. We have checked numerically that this ratio is a good estimate for the relative importance of the two terms.   

By comparing the upper with the lower panels, Fig.\ \ref{fig:BaryonNumberDensity} also demonstrates that the baryon density is enhanced where the charged pion condensate is depleted. As a consequence, we obtain a ``baryon crystal'' where baryon number is maximised at the triangular points of the lattice, just like the magnetic field. This two-dimensional lattice translates to ``baryon tubes'' in three dimensions since our system is translationally invariant in the direction of the magnetic field. This is due to our approximation of a vanishing pion mass. In the physical case, we expect a three-dimensional crystal with baryon number oscillating in all three dimensions, as expected by the CSL solution \eqref{alphaMassiveSol}. As the phase diagram in Fig.\ \ref{fig:eBmu} suggests, the chiral limit becomes a good approximation for very large magnetic fields, $e\langle B\rangle \gg m_\pi^2$. Therefore, one might expect a change in the structure of the phase from tube-like at ultra-large magnetic fields to bubble-like at more moderate fields. Of course, this is under the assumption of the validity of our chiral approach. At large magnetic fields, effects of chiral restoration and deconfinement will set in at some point, while at large chemical potentials the usual baryons have to be taken into account even before chiral restoration and deconfinement are expected. 
\chapter{Conclusion}
\label{chpt:Outlook}

\section{Summary}
\label{Outlook:summary}

This thesis has presented work on superconducting phases in the $\mu_B$-$B$ plane at $T=0$ of the QCD phase diagram, demonstrating the potential richness of this comparatively unexplored parameter space to the neighbouring $\mu_B$-$T$ at $B=0$ and $B$-$T$ at $\mu_B=0$ planes. We began by reviewing these planes to introduce some common phases, transitions and phenomena in the  QCD phase diagram and gain insight into the phase structure near $T=0$. By collecting the limited literature on the phase structure, we then conjectured and discussed what the full $\mu_B$-$B$ plane at $T=0$ might look like. Due to the lack of experimental input and first-principle calculations on the phase structure, even a schematic diagram is difficult to draw with confidence. Thus, we gave some speculation and posed some questions regarding the construction of this plane. These included the competition between the MC and IMC effects, the possibility of a quark-hadron continuity, and what phases to expect at magnetic fields $eB>0.3\rm{GeV^2}$. What can be said is that at low and high $\mu_B$, there are interesting inhomogeneous phases to be found, such as the CSL phase and CSC phases respectively. Some of these phases were explored in greater detail in the remaining parts of the thesis, which presented results from the published works Ref.\,\cite{Evans:2020uui,Evans:2022hwr}.

After briefly discussing the GL theory of type-II superconductors which reviewed many of the methods and procedures we went on to use, we turned our attention to the high $\mu_B$ region of this diagram. Here, quark matter is expected to be colour-superconducting and can be either type-I or type-II. Starting from a GL potential, we studied dense, colour-superconducting quark matter in the presence of a non-zero strange quark mass and external magnetic field. Aiming to improve upon previous results obtained in a massless regime \cite{Haber:2017oqb}, we included the effects of the strange quark mass up to second order in the free energy. With the main objective of determining whether the domain walls persist as the preferred configuration of magnetic defects in type-II 2SC matter, we focused primarily on the 2SC phase. We have shown that this is not the case, finding that instead, multi-winding 2SC$_{ud}$ flux tubes with a second, 2SC$_{us}$ condensate in the core are the favoured magnetic defects. As the mass decreases, increasing the winding number is preferred with an expanding region of high-winding phases becoming denser in the phase diagram as we approach $m_s\rightarrow0$ and $n\rightarrow\infty$. In this limit, the configuration tends towards the domain wall result. At preferred winding as low as $n=3$ we observe the magnetic flux forms ring-like structures, ``remnants'' of the domain wall configuration. In higher winding configurations, a second ring can be seen to appear. 

We then jumped to the opposite side of the plane where $\mu_B$ was not large enough to sustain baryons let alone CSC. In this chapter, we constructed a new inhomogeneous phase in the $\mu_B$-$B$ plane of the zero-temperature QCD phase diagram. It comprises charged and neutral pions where the charged pions form a vortex lattice akin to that of conventional type-II superconductors. They condense due to their interaction with the neutral pions which are coupled to the magnetic field and baryon chemical potential via the chiral anomaly. We have shown that above $B_{c2}$ this phase is preferred over the CSL phase in the chiral limit and that the favoured configuration of the vortex lattice is a hexagonal array. Just like an ordinary superconductor, the magnetic flux penetrates through these vortices where the charged pion condensate is depleted. Because of the anomalous coupling of the pions, this is reflected in the local, non-zero, inhomogeneous baryon number density, which is also largest in the centre of vortices. Thus, the three-dimensional structure of this phase is an array of ``baryon tubes''. We have pointed out that the main contribution to the spatial variation of the baryon number comes from the vorticity term arising from the anomaly. Our results are valid within ChPT and the chiral limit, where a WZW term was incorporated to account for the chiral anomaly. To construct this phase, we employed an expansion near the critical magnetic field like was used originally for type-II superconductors within GL theory.

\section{Discussion}
\label{Outlook:discussion}

In the context of the $\mu_B$-$B$ plane, our charged pion vortex lattice phase may be relevant when discussing the nuclear matter phase. As discussed, the transition to nuclear matter is expected at $\mu_B=923\,\rm{MeV}$. The transition line then behaves non-monotonically as we increase $B$, disfavouring nuclear matter initially, then favouring nuclear matter above $eB\sim 0.1\,\rm{GeV^2}$. This is near the $\mu_B$ and $B$ where we expect CSL and our charged pion lattice, and we may therefore expect competition between these inhomogeneous pion phases with nuclear matter. This is further compounded with the anomalous magnetic moment of nucleons possibly making this transition occur at lower $\mu_B$ than expected from Ref.\,\cite{Haber:2014ula}. However, if magnetic catalysis does indeed kick in at above $eB\sim 0.1\,\rm{GeV^2}$, then there could be a region where the vortex lattice exists uncontested at high magnetic fields. Of course, the calculation presented is only valid in the chiral limit, which is less physical, but at $eB\gg m_{\pi}^2$ we expect $m_{\pi}=0$ to be a good approximation. Furthermore, chirally restored phases may be favoured at similar values of the magnetic field. This effect however is comparatively slight in comparison to the wider area gained by the vortex lattice and CSL at higher $eB$. Furthermore, our results are found within ChPT, which make them a prediction of low-energy QCD and should be trusted more than the effective model predictions. On the other hand, the further we go in $B$ or $\mu_B$, the less reliable the leading order momentum expansion of ChPT theory becomes, which should be kept in mind when extrapolating to higher $B$ and $\mu_B$ .

The colour-superconducting results expand upon Ref.\,\cite{Haber:2017oqb}, improving the results of on the phase structure at lower $\mu_q$ by incorporating corrections for the strange quark mass which becomes appreciably large at lower densities.
The structure of the phase diagrams is not greatly affected when compared to the results of Ref.\,\cite{Haber:2017oqb} and therefore much of what we can learn from our results about the $\mu_B$-$B$ plane at $T=0$ can already be inferred from the previous work (see Sec.\,(\ref{Background:subsec:muBPlane}) for more on what Ref.\,\cite{Haber:2017oqb} implies for the phase diagram). Our results do imply that the 2SC becomes more favourable when quarks are not massless as expected. As a result, if we take our results to be valid at a $\mu_B$ realistic for NS then this would support the NJL model calculations that show a 2SC phase beyond the chiral transition to a greater extent than Ref.\,\cite{Haber:2017oqb}. Of course the NJL models calculations study CSC in this plane are typically $SU(2)$ which do not take CFL into account, which our results do not show diminishing significantly, thus not greatly reducing the possibility of a quark-hadron continuity. These are speculations however, as our results use extrapolated values from a weak-coupling regime. Perhaps a more substantial result is the existence of an inhomogeneous dSC phase at (the very least) large $\mu_B$ showing that there is a similar phase structure to the one found from the melting pattern of CSC condensates in the neighbouring $\mu$-$T$ plane at $B=0$.

The $\mu_B$-$B$ plane aside, both projects contribute to the study of unconventional superconductors. Chapter \ref{chpt:Project1} concerned a system where multiple superconducting condensates could be present as well as coupling to multiple gauge fields. Despite the specificity of our calculations in Chapter \ref{chpt:Project1}, ring-like magnetic flux structures are found in more general, yet largely unrelated models \cite{Chernodub:2010sg,Bazeia:2018hyv}. Thus, our results may be of interest to the wider discussion of multi-component superconductors. Though the presence of additional condensates and gauge fields did enrich the phase structure, with phase transitions in between different superconducting phases, the type-I and type-II regimes did not deviate too much from what was expected in single-component superconductors. In Chapter \ref{chpt:Project2} the consequence of the coupling of a real scalar field to the magnetic field was investigated. In contrast to the multi-component superconductor, this setup inverted the type-II vortex lattice to normal phase transition. While not unheard of \cite{Chernodub:2010qx, Chernodub:2011gs}, superconductivity being enhanced by the magnetic field rather than destroyed is a very interesting phenomenon. Furthermore, the presence of a CSL phase which is magnetised below the transition raises some interesting questions regarding the type-I and type-II regimes. The complete absence of a homogeneous phase is also a peculiar consequence.

From the perspective of NS, the work of Chapter \ref{chpt:Project2} is not relevant since $\mu_q\sim400\,\rm{MeV}$ is already beyond the region of validity for ChPT.
On the other hand, application to NSs was a motivation for the content of Chapter \ref{chpt:Project1}. A key question is whether we expect these multi-winding flux tubes to affect observable. Around $m_s\sim0.2\mu_q$, which is near the lower end of the range we expect in NS environments, multi-winding flux tubes begin becoming disfavoured, above which single winding flux tubes are preferred. Thus, it is improbable they are prominent features at equilibrium in NS. However, there are non-standard single winding flux tubes that have a second condensate in the core, which seem to survive up to $m_s/\mu_q\sim 0.5$. While the second condensate is not large enough to visibly effect the magnetic flux, it does indicate that NS may have non-standard flux tube configurations if there is a type-II colour-superconducting core. On the other hand, as discussed at the end of the chapter, the external magnetic field required for these exotic structures is getting near to the largest magnetic field sustainable by a NS and several orders of magnitude larger than observed at the surface. However, this does not mean that these multi-winding flux tubes cannot appear dynamically by cooling through the superconducting transition for instance. Therefore, they may still have a role to play in NS physics and leave signatures in observational tools \cite{Glampedakis:2012qp}. Detection of these defects would be important for the QCD phase diagram, especially the $\mu_B$-$B$ plane where compact stars offer the only experimental insight into the phase structure in that plane at present. 

\section{Outlook}
\label{Outlook:out}

Until we have better experimental data, more can be done theoretically. Specific to our colour-superconducting project, there are a few improvements and tasks to undertake. A natural next step would be to determine the effect of the mass term on the type-II CFL phase and compare with the results in \cite{Haber:2017oqb}. There are no features akin to the domain walls of type-II massless 2SC in the massless type-II CFL Flux Tube phase, so the effect is expected to be less substantial. Upon an initial inspection this extension seems straightforward. Tied to the semi-massive CFL phase structure is the appearance of the briefly discussed inhomogeneous dSC phase. This could also be explored numerically within this framework. Results from these two regimes could rectify the ambiguous transition line between the ``2SC Flux Tubes'' and (rather carelessly labelled) ``CFL Flux tubes'' regions in Fig.\ \ref{fig:phasediagram} and improve our diagram further. Another straightforward extension would be a greater exploration of the parameter space as we have restricted ourselves to fixed coupling parameters for our results, extrapolating from the weak-coupling regime. Restricting ourselves to weak-coupling values and extrapolating to large coupling constant is not desirable and has been done for simplicity.

A more considerable extension is the inclusion of temperature, which may add some relevance to our results to the larger $\mu_B$-$B$-$T$ phase structure as well as NS mergers and HICs where $T\approx 0$ is no longer a good approximation. Future colliders \cite{Ablyazimov:2017rf,Schmitt:2016pre,Kekelidze:2017ual, Hachiya:2020bjg, Xiaohong:2018weu} are seeking to produce data at higher $\mu_B$ and lower $T$. Thus, this would in some sense make our predictions more accessible to future experiments.  Furthermore, we have used a general critical temperature for all phases, where in reality they would not all be expected to transition at the same exact temperature. Not only would incorporating multiple critical values via a temperature dependence be more physical, but also improve our approximation, as GL is technically only valid near the transition temperature. This does perhaps highlight the ambiguity around which critical temperature we have expanded the free energy. On the topic of our framework, GL is purely bosonic in nature. A more rigorous analysis would seek to include the effect of the fermionic constituents. This could prove to be significant as previous studies have shown that magnetic fields of the magnitudes we have considered can affect the underlying structure of the Cooper pairs \cite{Ferrer:2005vd,Ferrer:2006vw,Noronha:2007wg,Fukushima:2007fc,Yu:2012jn,}. A fermionic approach would also account for any additional effect from the unpaired fermions in each of the 2SC phases. Furthermore, the explicitly broken flavour symmetry means the off-diagonal elements of the order-parameter matrix that we have neglected could become important. Such an approach would be difficult (hence the decision to remain in a bosonic framework) but conceivable. 

Improvements to our framework and inclusion of the temperature is also something that can be improved in our project in Chapter \ref{chpt:Project2}. Finite $T$ extensions have been applied to the CSL system in Refs.\ \cite{Brauner:2021sci,Brauner:2023ort}, and thus could be incorporated in future work. This once again adds relevance to the wider $\mu_B$-$B$-$T$ structure and heavy-ion colliders. As already touched upon, the current results are already near the limits of validity for ChPT theory. The possibility of the competition with baryonic phases was also discussed earlier. Thus, further exploration beyond $B_{c2}$ would warrant a different approach including baryonic degrees of freedom. For example, we could adopt a model similar to Ref.\,\cite{Haber:2014ula} which is valid for larger $\mu_B$ but smaller $B$. Such extensions would allow us to explore the competition between the pionic and baryonic phases and the lattice structure beyond $B_{c2}$. Moving deeper into the flux tube phase usually reduces the density of vortices, admitting less magnetic flux. Since the transition is inverted in this plane, it is unclear how additional flux would penetrate the system as we move away from $B_{c2}$, especially if the expectation is a lower density of vortices. Of particular interest would be the conjectured phase transition between the vacuum and the inhomogeneous charged pion phase without an intermediate transition to the CSL phase. Of course, our approximations are only valid near $B_{c2}$, therefore, we would most likely need to adopt a numerical approach.  

The clear next step is the inclusion of finite mass. The instability is still present in the massive phase, which would indicate that an inhomogeneous charged pion phase similar to the one discussed here also exists above $B_{c2}$ for physical masses. One can envisage a phase where the superconducting flux tube lattice interplays with the CSL periodicity perpendicular to it to form a rich three-dimensional structure. In anticipation of this, we endeavoured to keep our equations general throughout the relevant chapter, only moving to the chiral limit when necessary for analytical evaluation. Initial calculations (not presented here) with a single domain wall, i.e.\ the simplest massive CSL, are promising, showing that we can proceed largely analytically. Likely, the full massive case would eventually require numerical methods to fully solve. However, resigning to numerical methods could have the advantage of allowing us to explore deeper into the parameter space as already mentioned.


Finally, we turn to broader tasks. In a compact star, the quark matter would most likely occur in the core where there would be an interface with nuclear matter. One can then ask; what happens to these 2SC flux tubes when they reach the boundary of the system? It would be interesting to see how the magnetic field behaves at the interface with nuclear matter, where it is also expected to be affected by proton superconductivity. In general, these flux tubes could be significant to the overall structure of the magnetic field within the compact star. Even if they would only appear dynamically during the evolution of the star and not as a permanent, equilibrium feature of the core, they may have a lasting effect. Thus far we have discussed two of the extremes related to NSs; density and magnetic field. Neutron Stars are also known to rotate, the pulsars discussed in our observational review first and foremost among them. If CFL is truly the ground state of matter in the core of a NS, rotation could have a significant effect on its structure. We have already discussed its colour-superconducting properties, but not remarked much on its superfluid behaviour. Rotation would induce superfluid vortices within the CFL phase also, separate from the colour-magnetic defects already present. Some work on classifying these vortices was done in Refs.\,\cite{Haber:2017oqb,Haber:2018yyd}, but there is an unaddressed question how these vortices interact. Presumably, we would have a superfluid and colour-superconducting vortex lattice. Is one preferred over the other? Do they coexist or interplay in any way? Does this influence the phase structure or properties of compact stars? Resolving the vortex lattice of coupled neutron superfluid and proton superconductors in an external magnetic field has been explored \cite{Haber:2017kth,Wood:2020czv,Wood:2022smd}, which could be used as guidance in this undertaking. 

Continuing on the properties of vortex lattices, the CSL is known to occur due to other non-zero external parameters in the QCD phase diagram \cite{Brauner:2019aid}. The question is then if the vortex lattice phase due to the chiral anomaly appears due to the same external parameters also. Rotational frequency $\nu$ is once again another alternative external parameter that has recently seen investigation in Ref.\,\cite{Yamamoto:2021oys}, motivated by the possible detection of the chiral vortical effect in HICs. Indeed, a charged pion instability is also possible in the $\mu_B$-$\mu_I$-$\nu$ plane \cite{Huang:2017pqe}. The anomalous terms are not derived from a WZW term, making direct adaptation of our work less straightforward, but also more interesting as the origin of the instability changes. A plane where a WZW term does cause an instability is the $\mu_I$-$B$ plane \cite{Brauner:2019aid}, which shows promise for a direct adaptation. It should be noted that as mentioned in the introduction of Chapter \ref{chpt:Project2}, charged pions already condense due to $\mu_I$ without the presence of the anomaly term. At the ranges of magnetic field explored in this project, they are expected to be in the type-II regime \cite{Adhikari:2015wva}, though the homogeneous phase can also be present. Both are stable below certain critical magnetic fields, analogous to the situation in conventional superconductivity. Competition of these phases with the CSL phase was investigated in Ref.\ \cite{Gronli:2022cri}, with the CSL phase favoured at large magnetic fields and the vortex lattice at lower magnetic fields. It would therefore be interesting to apply the same calculation in this plane. One could even look at the three-dimensional phase diagram $\mu_B$-$\mu_I$-$B$ and see how the $\mu_B$-$B$ and $\mu_I$-$B$ planes connect. Investigations into the $\mu_B$-$\mu_I$ plane at $B=0$ have been done (e.g. see \cite{Mao:2020xvc}) which aid in this pursuit. Preliminary calculations do suggest similar structure to the $\mu_B$-$B$ plane, though questions regarding the exact form of the WZW term and appropriate power-counting schemes must be addressed before proceeding.



\appendix

\chapter{Details of lattice configuration calculations}

\label{app:Lattice}
Here we elucidate some details of the lattice structure calculations to complement the main text, in particular the lattice calculation of Sec.\,(\ref{Background:subsec:Hc2}) of Chapter \ref{chpt:Background} which is then used in Sec\,(\ref{Project2:subsec:chiral}) of Chapter \ref{chpt:Project2}.  

\section{Infinite sums of integrals with periodicity $N$}
\label{app:LatticeN}
To derive the results \eqref{N=2VarphiAvgs} by setting $C_n=C_{n+N}$ in Eqs.\,\eqref{GenAvgs}, we first look at Eq.\,\eqref{AvgVarphi} and re-write it as
\begin{equation}
        \langle |\varphi_0|^2 \rangle 
        = \frac{1}{N q} \sum_{n=-\infty}^{\infty}  |C_n|^2  \int_{0} ^{N q} dx \, e^{-\left( x-nq \right)^2 } 
        =
        \frac{1}{N q} \lim_{M\rightarrow\infty }\sum_{n=-M}^{M-1} |C_n|^2  \int_{-nq} ^{(N-n)q }\, dz\, e^{-z^2 }\,,
\end{equation}
such that the integral limits now depend on $n$. We have also expressed the sum within a  limit. Let $M$ be divisible by $N$. We will then expand the sum and collect similar terms through the imposed periodicity condition. We find that the associated integral limits between $C_n$ and $C_{n+N}$ are spaced $Nq$ apart and thus can be combined into the same integral. Through this, we can combine the integrals for each  $C_n$ from $n=0$ to $n=N-1$ into a single integral. The remaining terms can then be expressed as a sum. This is shown below;
\begin{equation}
 \begin{split}
        \sum_{n=-M}^{M-1} |C_n|^2  \int_{-nq} ^{(N-n)q } &=
        \bigg(
        |C_{-M}|^2\int_{Mq}^{(M+N)q} +|C_{-M+N}|^2\int^{Mq}_{(M-N)q}  
        \\&
        +|C_{-M+2N}|^2\int_{(M-2N)q}^{(M-N)q} +\ldots +|C_{M-2N}|^2\int_{-(M-2N)q}^{-(M-3N)q} 
        \\&
        +|C_{M-N}|^2\int_{-(M-N)q}^{-(M-2N)q}\bigg) +\bigg(
        |C_{-M+1}|^2\int_{(M-1)q}^{(M+N-1)q} 
        \\&
         +|C_{-M+N+1}|^2\int_{(M+N-1)q}^{(M+2N-1)q} +
        \ldots +|C_{M-2N+1}|^2\int_{-(M-2N+1)q}^{-(M-3N+1)q}
        \\&
        +|C_{M-N+1}|^2\int_{-(M-N+1)q}^{-(M-2N+1)q}\bigg)
        +\bigg(|C_{-M+2}|^2\int^{(M+N-2)q}_{(M-2)q} 
        \\&
        +\ldots 
        +|C_{M-N+2}|^2\int_{-(M-N+2)q}^{-(M-2N+2)q}\bigg)
        + \ldots +
        \bigg(|C_{-M+N-1}|^2\int_{(M-N+1)q}^{(M+1)q}
        \\&
        +\ldots +|C_{M-1}|^2\int_{-(M-1)q}^{-(M-N-1)q}\bigg)
        \\&=        
        |C_{-M}|^2\int_{-(M-N)q}^{(M+N)q} +|C_{-M+1}|^2\int_{-(M-N+1)q}^{(M+N-1)q} 
        \\&
        +|C_{-M+2}|^2\int^{(M+N-2)q}_{-(M-N+2)q} + \ldots +
        |C_{-M+N-1}|^2\int_{-(M-1)q}^{(M+1)q}
        \\&=\sum_{n=0}^{N-1}|C_{-M+n}|^2 \int_{-(M-N+n)q}^{(M+N-n)q}\,,
    \end{split}    
\end{equation}
where we have used the shorthand $\int^{b}_a \equiv \int^{b}_{a} dz\, e^{-z^2} $. Then in the limit $M\rightarrow \infty$ we can evaluate the Gaussian integral to obtain
\begin{equation}
    \langle|\varphi_0|^2\rangle =\frac{\sqrt{\pi}}{N q}\sum_{n=0}^{N-1}|C_{n}|^2\,, 
\end{equation}
where $C_{-M+n}=C_{n}$ since $M$ is divisible by $N$. This is the result \eqref{NAvgVarPhi}. 

The same method can be applied for $\langle |\varphi_0|^4\rangle$. Similarly, we re-write Eq.\,\eqref{AvgVarphi4} as
\begin{equation}
    \begin{split}
        \langle|\varphi_0|^4\rangle
        =& \frac{1}{Nq}\sum_{n_3} \sum_{n_2} e^{-\frac{\left(n_2^2 +n_3^2\right)q^2}{2}} \sum_{n_1} C^*_{n_1} C_{n_1+n_2} C_{n_1+n_3} C^*_{n_1+n_2+n_3}\int_0^{Nq}  dx\,  e^{- 2\left( x-\left(n_1+\frac{n_2+n_3}{2}\right) q \right)^2}
        \\
        =& \frac{1}{Nq}\sum_{n_3} \sum_{n_2} e^{-\frac{\left(n_2^2 +n_3^2\right)q^2}{2}} \lim_{M\rightarrow\infty} \sum_{n_1=-M/2}^{M/2} \tilde{C}^{n_1}_{n_2,n_3} \int_{-(n_1+\tilde{n})q}^{\left[N-(n_1+\tilde{n})\right]q} dz\,  e^{- 2z^2} \,,
    \end{split}
\end{equation}
where, for brevity, $\tilde{n}=(n_2+n_3)/2$ and $\tilde{C}^{n_1}_{n_2,n_3}=C^*_{n_1} C_{n_1+n_2} C_{n_1+n_3} C^*_{n_1+n_2+n_3}$. Note $\tilde{C}^{n_1+N}_{n_2,n_3}=\tilde{C}^{n_1}_{n_2,n_3}$ since the indices retain the periodicity from the original indices $n$, $m$, $p$ and $l$. Then,
\begin{equation}
    \begin{split}
        \sum_{n_1=-M/2}^{M/2} \tilde{C}^{n_1}_{n_2,n_3} \int_{-(n_1+\tilde{n})q}^{\left[N-(n_1+\tilde{n})\right]q}   &
        = 
        \bigg(
        \tilde{C}^{-M}_{n_2,n_3}\int_{(M-\tilde{n})q}^{(M+N-\tilde{n})q} 
        \\&
        +\tilde{C}^{-M+N}_{n_3,n_3}\int^{(M-\tilde{n})q}_{(M-N-\tilde{n})q} 
        + \tilde{C}^{-M+2N}_{n_2,n_3}\int_{(M-2N-\tilde{n})q}^{(M-N-\tilde{n})q}
        \\&
         +\ldots +\tilde{C}^{M-2N}_{n_2,n_3}\int_{-(M-2N+\tilde{n})q}^{-(M-3N+\tilde{n})q} 
        +\tilde{C}^{M-N}_{n_2,n_3}\int_{-(M-N+\tilde{n})q}^{-(M-2N+\tilde{n})q}\bigg)
        \\&
        +\bigg(
        \tilde{C}^{-M+1}_{n_2,n_3}\int_{(M-1-\tilde{n})q}^{(M+N-1-\tilde{n})q} +\tilde{C}^{-M+N+1}_{n_2,n_3}\int_{(M+N-1-\tilde{n})q}^{(M+2N-1-\tilde{n})q} 
        \\&
        +\ldots +\tilde{C}^{M-2N+1}_{n_2,n_3}\int_{-(M-2N+1+\tilde{n})q}^{-(M-3N+1+\tilde{n})q} +\tilde{C}^{M-N+1}_{n_2,n_3}\int_{-(M-N+1+\tilde{n})q}^{-(M-2N+1+\tilde{n})q}\bigg)
        \\&
        +\bigg(\tilde{C}^{-M+2}_{n_2,n_3}\int^{(M+N-2-\tilde{n})q}_{(M-2-\tilde{n})q} +\ldots 
        +\tilde{C}^{M-N+2}_{n_2,n_3}\int_{-(M-N+2+\tilde{n})q}^{-(M-2N+2+\tilde{n})q}\bigg)
        \\&
        + \ldots +
        \bigg(\tilde{C}^{-M+N-1}_{n_2,n_3}\int_{(M-N+1-\tilde{n})q}^{(M+1-\tilde{n})q}
        +\ldots +\tilde{C}^{M-1}_{n_2,n_3}\int_{-(M-1+\tilde{n})q}^{-(M-N-1+\tilde{n})q}\bigg)
        \\&=\sum_{n_1=0}^{N-1}\tilde{C}^{-M+n_1}_{n_2,n_3} \int_{-(M-N+n_1+\tilde{n})q}^{(M+N-n-\tilde{n})q}\,.
    \end{split}
\end{equation}
Taking the limit once more,
\begin{equation}
    \langle|\varphi_0|^4\rangle =\frac{1}{N q}\sqrt{\frac{\pi}{2}}\sum_{n_3} \sum_{n_2} e^{-\frac{\left(n_2^2 +n_3^2\right)q^2}{2}}\sum_{n=0}^{N-1}C^*_{n_1} C_{n_1+n_2} C_{n_1+n_3} C^*_{n_1+n_2+n_3}\,,
\end{equation}
which is the result \eqref{NAvgVarPhi4}.

\section{Translational symmetry}
\label{app:LatticeSymmetry}

Relation \eqref{TriangCond} is more easily deduced in Fourier space. We begin with general periodicity and
\begin{equation}
    |\varphi_0|^2\equiv\omega (x,y) = \sum_{n=-\infty}^{\infty}\sum_{m=-\infty}^{\infty} C_n C_m^* e^{i(n-m)qy}e^{-\left[ x- \frac{(n+m)q}{2}\right]^2 - \frac{(n-m)^2 q}{2}}\,,
    \label{PositionSpace}
\end{equation}
where we have used the de-dimensionalisation \eqref{dimless}. Its Fourier series is then
\begin{equation}
   \omega (x,y) = \sum_{r=-\infty}^{\infty}\sum_{s=-\infty}^{\infty} \tilde{\omega}(r,s) v_{rs}(x,y)\,,
   \label{FourierSeries}
\end{equation}
where $\tilde{\omega}(r,s)$ are the Fourier coefficients and $v_{rs}(x,y)$ are the basis functions given by
\begin{equation}
    v_{rs}(x,y) = e^{\frac{i2\pi r}{L_y}y}e^{\frac{i2\pi s}{L_x}x}= e^{\frac{iqr}{N_y}y} e^{\frac{i2\pi s}{N_x q}x}\,,
\end{equation}
where we have used the general expressions $L_x=N_x q$ and $L_y= 2\pi N_y /q$. The basis functions obey the orthogonality relation
\begin{equation}
    \frac{1}{L_x L_y}\int_0^{L_x}\int_0^{L_y} dx dy \, v^*_{\tilde{r}\tilde{s}}(x,y)v_{rs}(x,y)  = \delta_{\tilde{r},r}\delta_{\tilde{s},s}\,,
\end{equation}
and are used in the inverse transformation
\begin{equation}
    \tilde{\omega}(r,s) = \frac{1}{L_x L_y}\int_0^{L_x}\int_0^{L_y} dxdy\,  v^*_{\tilde{r}s}(x,y) \omega(x,y)\,.
\end{equation}
Inserting Eq.\,\eqref{PositionSpace} into the above,
\begin{equation}
    \tilde{\omega}(r,s) = \frac{e^{-\frac{r^2}{4}}e^{-\frac{\pi^2 s^2}{N_x^2 q^2} }}{N_x q} \sum_{n}C_n C_{n-r}^* e^{-\frac{i\pi s}{N_x}(2n-r)} \int_0^{N_x q} dx \, 
    e^{- \left[ x- \left(n -\frac{r}{2}\right)q \frac{(n+m)q}{2} -\frac{i\pi s}{N_x q}\right]^2} \,,
\end{equation}
where the y-integration,
\begin{equation}
    \frac{q}{2\pi N_y}\int_0^{\frac{2\pi N_y}{q}} dy\, e^{-i\left[\frac{r}{N_y} -(n -m)\right]}  = \delta_{n-m,\,r}  \,,
    \label{Kronecker}
\end{equation}
and the Kronecker delta has been applied. Note that the Kronecker delta technically imposes $r/N_y=n-m$, such that $r$ must be divisible by $N_y$ because $n$ and $m$ are integers. In position space, specifying $N_y$ was unnecessary as it vanished. 
However, since $r$ is also any integer, we are forced to set $N_y=1$. We have also (re-)completed the square in the exponent like so;
\begin{equation}
    \begin{split}
        &\left[ x- \frac{(2n-r)q}{2}\right]^2 +\frac{r^2}{2}q +\frac{i2\pi s}{N_x q}x
        \\
        &=\left[ x- \left(n -\frac{r}{2}\right)q -\frac{i\pi s}{N_x q}\right]^2 + \frac{r^2}{4} 
        +\frac{i\pi s}{N_x}(2n-r) +\frac{\pi^2 s^2}{N_x^2 q^2}\,.
    \end{split}
\end{equation}
With the periodicity relation $C_n=C_{n+N_x}$, the integral with finite limits of the infinite sum is equivalent to the finite sum of integral with infinite limits in the same way as was shown in Appendix \ref{app:LatticeN} with some added technicalities. First of all we notice that 
\begin{equation}
    C_{n+N_x} C_{n+N_x-r}^* e^{-\frac{i\pi s}{N_x}(2(n+N_x)-r)}= C_n C_{n-r}^* e^{-\frac{i\pi s}{N_x}(2n-r)} \,,
\end{equation}
such that we can collect similar terms spaced $N_x q$ apart. Then, under the same assumptions of Appendix \ref{app:LatticeN}, we obtain
\begin{equation}
    \begin{split}
        &  \sum_{n}C_n C_{n-r}^* e^{-\frac{i\pi s}{N_x}(2n-r)}
        \int_0^{N_x q} dx\, e^{- \left[ x- \left(n -\frac{r}{2}\right)q -\frac{i\pi s}{N_x q}\right]^2}
        \\
        &=\lim_{M\rightarrow\infty}\sum_{n=0}^{N_x-1}C_{-M+n}C^*_{-M+n-r} e^{-\frac{i\pi s}{N_x}(2n-r)} dz\, \int_{-(M-N_x+\bar{n})q-\frac{i\pi s}{N_x}}^{(M+N_x-\bar{n})q-\frac{i\pi s}{N_x}} e^{-z^2} \,,
    \end{split}
\end{equation}
where $\bar{n}=n-r/2$ and $z=x- \left(n -r/2\right)q -i\pi s/(N_x q)$. We state the (standard) result
\begin{equation}
    \lim_{L\rightarrow\infty}\int_{-L +ik }^{L+ik} du\, e^{-u^2}  = \sqrt{\pi}\,,
\end{equation}
where $u$ is real. This is derived by considering an integral of $e^{-w^2}$ with respect to the complex variable $w=u+iv$ over a closed rectangular contour in the upper half of the complex plane $u$-$v$. In the limit, integrals along the imaginary axis disappear as they involve factors of $e^{-L^2}$. Using this we can complete the integral in the limit as before and find
\begin{equation}
    \tilde{\omega}(r,s)=\frac{\sqrt{\pi}}{N_x q} e^{-\frac{r^2}{4}}e^{-\frac{\pi^2 s^2}{N_x^2q^2}} \sum_{n=0}^{N_x-1} C_{n}C^*_{n-r}  e^{-\frac{i\pi s}{N_x}(2n-r)}\,.
\end{equation}

Specialising to the $N_x=2$ case,
\begin{equation}
    \tilde{\omega}(r,s)=\frac{\sqrt{\pi}}{2 q} e^{-\frac{r^2}{4}}e^{-\frac{\pi^2 s^2}{4q^2}}  \left[  C_{0}C^*_{-r} + C_{1}C^*_{1-r}(-1)^s \right] i^{rs} \,,
    \label{FourierN=2}
\end{equation}
where we simplified $e^{-i\pi s(2n-r)/2}=\left(-1\right)^{ns} i^{rs}$. We wish to use this Fourier representation to obtain relation \eqref{TriangCond} by investigating the translational symmetry $\omega(x,y)=\omega(x+N_1L_x/2,y+N_2L_y/2)$ where $N_1$, $N_2$ are integers. Putting this into Eq.\,\eqref{FourierSeries}, we see that
\begin{equation}
    v_{rs}(x+N_1q,y+N_2\pi/q) = v_{rs}(x,y)e^{i\pi N_1 s}e^{i\pi N_2 r}\,,
\end{equation}
and
\begin{equation}
    \left[  C_{0}C^*_{-r} + C_{1}C^*_{1-r}(-1)^s \right] \left( 1- e^{i\pi N_1 s}e^{i\pi N_2 r}\right)=0\,,
\end{equation}
must be satisfied. Recalling that $N_x=2$ case is equivalent to saying that all odd $n$ coefficients are equal and all even $n$ coefficients are equal, we look at all four cases of even and/or odd $r$ and $s$. The case when $r$ and $s$ are both even is automatically fulfilled, leaving the three conditions
\begin{subequations}
    \begin{align}
        |C_0|=|C_1| \, \quad &\text{or} \quad N_1=\text{odd number}\,, \label{Cond1} 
        \\[2ex] 
        \quad C_0 C_1^*+ C_0^* C_1=0 \quad &\text{or} \quad N_2=\text{even number}\,, \label{Cond2} 
        \\[2ex]
        \quad C_0 C_1^*-C_0^* C_1=0 \quad &\text{or} \quad N_1+N_2=\text{even number}\,. \label{Cond3} 
    \end{align}
\end{subequations}
First of all, imposing no conditions on $C_n$ is not possible as $N_1+N_2$ needing to be even does not work if $N_1$ is always odd and $N_2$ is always even. Second, imposing no conditions on $N_1$ and $N_2$ means one of the coefficients must vanish and since $|C_0|=|C_1|$ from condition \eqref{Cond1}, then this gives us the trivial solution. In fact, from either condition \eqref{Cond2} or \eqref{Cond3} we can show that $|C_0|=|C_1|$ if we choose the condition on the coefficient, thus not requiring $N_1$ to be odd if we choose the left-hand option of these conditions. This means we cannot choose both these conditions on the $C_n$ without obtaining the trivial solution. Our choices are therefore,
\begin{subequations}
    \begin{align}
        |C_0|=|C_1| \, \quad &\text{and} \qquad N_1,N_2=\text{even number}\,, \quad  \label{Choice1} 
        \\[2ex] 
         C_0=(-1)^k i C_1 \quad &\text{and} \qquad N_1+N_2=\text{even number}\,, \label{Choice2} \quad \text{or}
        \\[2ex]
         C_0=(-1)^k C_1 \quad &\text{and} \qquad N_2=\text{even number}\,, \label{Choice3} 
    \end{align}
\end{subequations}
where we have used polar representation $C_0=|C_0|e^{i\theta}$ and  $C_1=|C_1|e^{i\phi}$ to simplify the coefficient conditions in \eqref{Cond2} and \eqref{Cond3} and $k$ is an integer. We can calculate $\beta$ for all choices by putting them into Eqs.\,\eqref{N=2VarphiAvgs} and find
\begin{subequations}
    \begin{align}
        \beta_a &= \frac{q}{\sqrt{2\pi}}\left(f_0^2 +2f_0f_1 +\cos\left[2\left(\theta -\phi\right)\right]f_1^2\right)\,, \label{beta1} 
        \\[2ex] 
        \beta_b &= \frac{q}{\sqrt{2\pi}}\left(f_0^2 +2f_0f_1 -f_1^2\right)\,, \label{beta2} 
        \\[2ex]
        \beta_c &= \frac{q}{\sqrt{2\pi}}\left(f_0^2 +2f_0f_1 +f_1^2\right)\,, \label{beta3} 
    \end{align}
\end{subequations}
where we have labelled each $\beta$ with $a$,$b$ and $c$ corresponding to choices \eqref{Choice1}, \eqref{Choice2} and \eqref{Choice3} respectively and used polar notation for $C_0$ and $C_1$ once more. The smallest $\beta$ minimises the free energy and since $f_0$ and $f_1$ are always positive and real, $\beta_b<\beta_c$. Furthermore, $\beta_b<\beta_a$ except when $\theta-\phi=(2k+1)\pi/2$ for which $\beta_a=\beta_b$. This choice of $\theta$ and $\phi$ actually yields $ C_0=(-1)^k i C_1$ in condition \eqref{Choice1} and as such is a special case of choice \eqref{Choice2}. In fact,  $\beta_b=\beta_a$ with $\theta-\phi=k\pi$ which recovers $ C_0=(-1)^k C_1$ also. This might be expected since choice \eqref{Choice1} is contained within the other two choices. Therefore, we see that the translational symmetry of \eqref{Choice2} where $C_0=\pm i C_1$, minimises the free energy. The symmetry can be expressed as $|\varphi_0\left[x +(2M_x +Z)q, y+(2M_y+Z)\pi/q\right]|^2= |\varphi_0\left(x, y\right)|^2$ where $M_x$, $M_y$ and $Z$ are all integers. In other words, we have invariance under translations $(x+N_1q, y+N_2\pi/q)$ when $N_1$ and $N_2$ are even or $N_1$ and $N_2$ are odd (which is equivalent to saying $N_1 +N_2$ is even).


\chapter{Asymptotics for large distances}

\label{app:Asymptotic}
In this appendix, we discuss the asymptotic behaviour of the condensates and the 
gauge fields from Chapter \ref{chpt:Project1} far away from the centre of the flux tube, in particular the non-standard behaviour of the
condensate induced in the core, i.e., the condensate that vanishes at infinite radial distance. 

We restrict ourselves to the situation of two non-vanishing condensates, the $ud$ condensate $f_3$, which assumes its homogeneous value at large distances, and the induced $us$ condensate $f_2$. The third condensate, $f_1$, is omitted because it never plays a role in the preferred 2SC flux tube configurations. This is a more involved discussion of the asymptotics Sec.\,(\ref{Background:subsec:Hc1}) which is a useful warm-up for this appendix. As in the analogous calculation of CFL flux tubes \cite{Haber:2017oqb,Haber:2018tqw}, we use the 
following ansatz for the gauge fields and the condensates,
\begin{subequations} \label{asymp}
\begin{align}
    &\tilde{a}_3(R) = \tilde{a}_3(\infty)+R\tilde{v}_3(R) \, , \qquad  \tilde{a}_8(R) =\frac{n}{\tilde{q}_{83}}+R\tilde{v}_8(R) \, , \\
    &f_2(R) = u_2(R) \, , \qquad f_3(R) = 1+u_3(R) \, ,
\end{align}
\end{subequations}
where we have taken into account the boundary conditions at $R=\infty$ for $\tilde{a}_8$, $f_2$, and $f_3$, while the boundary value for $\tilde{a}_3$ is a dynamical quantity. We have introduced the functions $\tilde{v}_3$, $\tilde{v}_8$, $u_2$, $u_3$, which all approach 0 as $R\to\infty$. We assume (and this will be confirmed a posteriori) that they fall off exponentially at large $R$. 

With this ansatz we find that the equation of motion for $f_2$ \eqref{eqf2}
becomes
\begin{equation} \label{u2approx}
    0\simeq u_2''+\frac{u_2'}{R}+u_2(-\Xi^2R^2+\ldots) \, ,
\end{equation}
where the dots stand for terms that are suppressed exponentially or by powers of $R$ compared to the $\Xi^2R^2$ term. This term thus plays a crucial role in the asymptotic behaviour, and we find the asymptotic solution 
\begin{equation} \label{u2asymp}
    u_2(R) \simeq \frac{d_2}{R}e^{-\Xi R^2/2} \, , 
\end{equation}
where $d_2$ is an integration constant which can only be computed from solving the 
full differential equations numerically. The solution \eqref{u2asymp} fulfils 
Eq.\,\eqref{u2approx} if we neglect terms that are suppressed by powers of $R$
compared to the leading-order terms. Note that here and in the rest of this appendix we treat the winding number $n$ as fixed and finite. A different analysis, carefully taking into account the two limits $R\to\infty$ and $n\to \infty$ would be needed for the transition from flux tube to domain wall solutions, see Ref.\ \cite{Penin:2020cxj} for a recent study in the abelian Higgs model. 

Next, we consider the equation of motion for $\tilde{a}_3$ \eqref{a3t}, which becomes
\begin{equation}
    R\tilde{v}_3''+\tilde{v}_3'-\frac{\tilde{v}_3}{R} \simeq \frac{\tilde{q}_3d_2^2}{\lambda R^2}(\Xi^2 R^2+\ldots )e^{-\Xi R^2}\, .
\end{equation}
Again only keeping the leading-order contributions we find that the asymptotic behaviour of the gauge field $\tilde{a}_3$ is given by
\begin{equation}
    \tilde{v}_3 \simeq  \frac{\tilde{q}_3d_2^2}{4\lambda \Xi R^3}e^{-\Xi R^2} \, . 
\end{equation}
We see that the condensate $f_2$ and the deviation of the gauge field $\tilde{a}_3$ from $\tilde{a}_3(\infty)$ are both suppressed by $\exp(-{\rm const}\times R^2)$. This is in contrast to the behaviour of a single-component flux tube, where the 
suppression is $\exp(-{\rm const}\times R)$ as in Eq.\,\eqref{assymbf}. This dependence, as we shall now see, is assumed by the functions $\tilde{a}_8$ and $f_3$. Anticipating the milder suppression of $\tilde{v}_8$ and $u_3$, i.e., neglecting terms of order $\exp(-{\rm const}\times R^2)$, we can approximate the remaining equations of motion \eqref{a8t} and \eqref{eqf3} by 
\begin{subequations}
\begin{align}
    0\simeq R^2 \tilde{v}_8''+R\tilde{v}_8'-\tilde{v}_8\left(1+\frac{\tilde{q}_{83}^2}{\lambda}R^2\right)  \, , \\
    0\simeq R^2u_3''+Ru_3'-2u_3R^2 \, .
\end{align}
\end{subequations}
These equations are identical to the standard single-component case discussed in Sec.\,(\ref{Background:subsec:Hc1}), and their solution can be written in terms of modified Bessel functions of the second kind. Here, we simply quote the leading asymptotic behaviour,
\begin{equation}
    \tilde{v}_8(R) \simeq \frac{\tilde{c}_8}{\sqrt{R}} e^{-\tilde{q}_{83}R/\sqrt{\lambda}}\, , 
    \qquad u_3(R) = \frac{d_3}{\sqrt{R}} e^{-\sqrt{2}R} \, ,
\end{equation}
where $\tilde{c}_8$ and $d_3$ are constants that can be determined from the full numerical solution. Recalling the definition of the dimensionless radial coordinate $R$ \eqref{R}, we see that this result indicates the usual behaviour: The 
length scale for the exponential decay of the rotated magnetic field $\tilde{B}_8$ is given by 
the penetration depth,
\begin{equation} \label{lpen}
    \frac{\tilde{q}_{83}R}{\sqrt{\lambda}} = \frac{r}{\ell} \, , \qquad \ell = \frac{1}{\tilde{q}_{83}\rho_{\rm 2SC}} \, , 
\end{equation}
and the length scale on which the condensate approaches its homogeneous value is the coherence length, already introduced in Eq.\,\eqref{R},
\begin{equation} \label{lcoh}
    R = \frac{r}{\xi_3}  \, , \qquad \xi_3= \frac{1}{\sqrt{\lambda}\rho_{\rm 2SC}} \, . 
\end{equation}
These can be compared to Eqs.\,\eqref{Lambda} and \eqref{xi} respectively. In contrast, the induced condensate decays on a different length scale, given by the external magnetic field, 
\begin{equation} \label{lind}
    \Xi^{1/2}R = \frac{r}{\xi_2}  \, , \qquad \xi_2 = 
    \left(\frac{\tilde{q}_3H\cos\vartheta_1\sin\vartheta_2}{2}\right)^{-1/2} = \left[\frac{3eg^2 H}{4(3g^2+e^2)}\right]^{-1/2} \, ,
\end{equation}
where we have used the definition of the dimensionless magnetic field $\Xi$ \eqref{Xi}. We have added the subscript 2 to indicate that it is the condensate $f_2$ that varies on the length scale $\xi_2$. 

\chapter{Numerical methods}

\label{app:Numerical}
Here, we discuss the methods used to numerically solve the differential equations \eqref{eqsa} and \eqref{eqsf} in Chapter \ref{chpt:Project1}. One can view our numerical approach as being inspired by the following idea: A solution $f(x)$ can be thought of as a ``root'' of a differential equation $F(f)=0$. Motivated by this, one might then adopt a numerical solution-finding method for a differential equation in analogy to a root-finding method for an ordinary equation.

Consider a differential equation
\begin{equation}
    F\left(\frac{d^{m}f}{d x^m},\frac{d^{m-1}f}{d x^{m-1}},\dots,\frac{d^{2}f}{d x^2},\frac{d f}{d x},f(x)\right)=0\,,
    \label{FDiffrntlEqn}
\end{equation}
which we need to solve numerically. Applying the finite, central difference approximation
\begin{equation}
    \frac{d f}{d x}\Bigr\rvert_{x=x_i} \approx \frac{f_{i+1}-f_{i-1}}{2 h_x}\,,
    \label{CentDiffApprox2}
\end{equation}
expresses all derivatives in terms of $f_i \equiv f(x_i)$ where $h_x=x_{i+1}-x_i$ with $i=\{I,\dots,0\}$. Here, the $x$-axis has been discretised into $I$ points, in turn discretising the solution $f(x)$. Imagining a plot of $f(x)$, this forms an $I\times I$ grid of points. The differential equation \eqref{FDiffrntlEqn} is then approximated as
\begin{equation}
   F_i(f_{I},f_{I-1},...,f_{i+1},f_{i},f_{i-1},...f_1,f_0)\approx0\,,
   \label{DiscreteF}
\end{equation}
where $F_i \equiv F(f_i)$ is the differential equation evaluated at the solution at point $x=x_i$. Now, instead of a differential equation with solution $f(x)$, we have a system of $I$ linear equations individually labelled by $i$ with roots $f_i$. In principle, each equation could depend on $f_{i}$ at each point $i$ (as implied above) though, with $I$ sufficiently large to ensure $h_x$ is small and our approximation \eqref{CentDiffApprox2} is accurate enough, in most circumstances this will certainly not be the case (for $F_i$ to depend on both $f_{I}$ and $f_{0}$ i.e.\ both boundary values simultaneously, would require $F$ to be a differential equation of an unreasonably high order).   

For each $F_i$, we wish to find the corresponding root $f_i$. The idea is now to apply the Newton-Raphson, root-finding  method in some form. For the simple case of a function $y(x)$ of a single variable, the procedure uses the iterative equation, 
\begin{equation}
    x^{k+1}=x^{k}-\frac{y(x^{k})}{y'(x^{k})}\,,
    \label{NewRaph}
\end{equation}
where $y'(x)$ is the derivative with respect to $x$ and $x^{k}$ is the $k$-th iteration approximation to the root. Since we seek the roots of a system of linear, multi-variable equations, the above must be generalised if we are to use it for our purposes. In particular, the derivative of $F_{i}$ with respect to the solution at every point (labelled $j=\{I,\dots, 0\}$) must be taken into account. Therefore, the analogous term to $y'(x)$ in Eq.\,\eqref{NewRaph} is the Jacobian matrix 

\begin{equation}
    \mathsf{J}=
    \begin{pmatrix}
        \frac{\partial F_0}{\partial f_0}    & \dots  & \frac{\partial F_0}{\partial f_j} & \dots   & \frac{\partial F_0}{\partial f_I}  \\
        \vdots &\ddots & \vdots & & \vdots \\
        \frac{\partial F_i}{\partial f_0} & \dots &  \frac{\partial F_j}{\partial f_j} & \dots &  \frac{\partial F_i}{\partial f_I} \\
        \vdots & & \vdots & \ddots & \vdots \\
        \frac{\partial F_I}{\partial f_0} & \dots & \frac{\partial F_I}{\partial f_j} & \dots & \frac{\partial F_I}{\partial f_I} 
    \end{pmatrix}
    \,.
\end{equation}
The Newton-Raphson method for our system of differential equations is then
\begin{equation}
    f^{k+1}_{i}=f^{k}_i-\mathsf{J}^{-1}\mathbf{F}\,,
    \label{MatrixOverRelax}
\end{equation}
where $\mathbf{F}=(F_{0}\quad F_{1}\quad \dots \quad  F_{j} \quad \dots \quad F_{I})^{\mathsf{T}}$ is the vector containing all equations in our system. Going through all the points (here, from $i=I$ to $i=0$) yields a sequence of coordinates ($x_i$, $f_i$) which approximates the solution $f(x)$. Repeating the procedure with these newly obtained coordinates then produces an improved approximation, with further repetition giving better and better approximations (barring any complications e.g.\ divergences or issues with stationary points).

For our routine to ``over-relax'' to the solution, we multiply the difference between two iterations $n$ (that is the term involving the ratio of the differential equation and the Jacobian) with the parameter $w$. The purpose of this parameter is to speed up convergence. It can be shown that the optimal value lies in the range $1<w<2$. Typically, as one approaches the maximum of this range the faster the convergence but above some value (which can be $w<2$) the solution can diverge. This value depends on the problem and thus it is a question of choosing a $w$ close enough to it for faster computation without causing divergences. This can be included in \eqref{MatrixOverRelax} like so;

\begin{equation}
   f^{k+1}_{i}=f^{k}_i-w(\mathsf{J}^{-1})_{ij}F_{j}\,,
   \label{IndexOverRelaxW}
\end{equation}
where it has been stated in index form.

To improve convergence further, one can use the Gauss-Seidel method. An equation $F_{i}$ (unless it is trivial) is dependent on more than one value of the solution at any point, which may include a point for which we have already calculated the new $(k+1)$-th approximation. Assuming we are going from $i=I$ to $i=0$ (large $x$ to small $x$), we have already calculated the next iteration of guesses $k+1$ for $i+1,\dots,I$ (except when $i=I$). Therefore, we can use these new values to improve the approximation for our current point $i$. For example, if $F$ depended on the adjacent values $f_{i\pm1}$ as well as $f_i$ itself, in the Gauss-Seidel method we would use $F_{i}(f^{k+1}_{i+1},f_{i}^{k},f_{i-1}^{k})$ to calculate $f_{i}^{k+1}$ using \eqref{IndexOverRelaxW} instead of $F_{i}(f^{k}_{i+1},f_{i}^{k},f_{i-1}^{k})$. All this together is classified as a successive over-relaxation method (SOR).

The method detailed above was used to numerically solve the Eqs.\ \eqref{eqsa} and \eqref{eqsf} simultaneously in order to evaluate the free energy integral \eqref{FreeEnergyIntegral}. In our formulation, the central difference approximation was applied twice, such that the second derivative used points $i\pm1$ and $i\pm2$ 
like so
\begin{equation}
  \frac{d^2 f}{d x^2} \approx \frac{f_{i+2}-2f_i + f_{i-2}}{4h_x^2}\,.
\end{equation}
It should be remarked that the approximation can be applied which only requires points $i\pm1$ for the second derivatives
\begin{equation}
       \frac{d^2 f}{d x^2}  \approx \frac{f_{i+1}- 2f_i + f_{i-1}}{h_x^2}\,.
\end{equation}
The errors for both are second-order and brief testing shows no significant difference in the accuracy or speed of the computation between the two. In similar, future projects, more consideration may be given to this. Initial guess curves were required for the first iteration in addition to boundary conditions for the beginning and end points of each curve for each iteration. We began each run from $R=\infty$ to $R=0$ so to speak as we are interested in how the value $f_2(0)$ would vary with certain parameters. As a result, the start points of $a_3(R)$ were determined dynamically due to $a_3(\infty)$ being without an analytically derived boundary condition. Each iteration produced a set of solutions which were then used as the base curves for the next (recalling that the differential equations are not independent of each other). In the procedure itself, the non-diagonal elements of the Jacobian were ignored to make our calculations simpler and less computationally taxing (this was tested in Ref.\ \cite{Haber:2017oqb} which used the same method and the authors concluded its inclusion was negligible to the results). 

This routine was then incorporated into a second-root finding method, this time using the bisection method, to numerically solve the Eq.\,\eqref{Xictube}. The routine would get called 10-20 times in the second root-finding procedure, which was employed for multiple $T_c/\mu_q$ points and multiple values of $m_s/\mu_q$.
\chapter{Anomalous baryon current}

\label{app:Anomalous}
In this appendix, we provide some details of the derivation leading to the result of the Goldstone-Wilczek baryon current in Eq.\,\eqref{BCurrent} in Chapter \ref{chpt:Project2}. 
The starting point is the first line of Eq.\,\eqref{BCurrent}. Inserting the definition of the covariant derivative \eqref{CovariantDerivative} into this expression gives
\begin{equation}\label{jmuB1}
    \begin{split}
        j^{\mu}_{B} 
    = -\frac{\epsilon^{\mu\nu\rho\lambda}}{24\pi^2}\text{Tr}\left[-\Sigma\partial_{\nu}\Sigma^{\dagger}\partial_{\rho}\Sigma\partial_{\lambda}\Sigma^{\dagger} + \frac{3ie}{2}A_{\nu}\tau_3\left(\partial_{\rho}\Sigma^{\dagger}\partial_{\lambda}\Sigma -\partial_{\rho}\Sigma\partial_{\lambda}\Sigma^{\dagger} \right)\right.
    \\[2ex]
    \left.+\frac{3ie}{4}F_{\nu\rho}\tau_3\left( \Sigma\partial_{\lambda}\Sigma^{\dagger} +\partial_{\lambda}\Sigma^{\dagger}\Sigma\right)\right]\,.
    \end{split}
\end{equation} 
Next, we use the parameterisation given by Eqs.\,\eqref{SigmaU} and \eqref{Usig}, $\Sigma = \Sigma_0U$ with $\Sigma_0=e^{i\alpha\tau_3}$.  One easily confirms $\partial_\nu\Sigma_0 = i\partial_\nu \alpha\, \tau_3 \Sigma_0$ and $[\Sigma_0,\tau_3]=[\Sigma_0^\dag,\tau_3]=0$. Moreover, one can check explicitly that $\tau_3U=U^\dag\tau_3$, and, since $U$ is unitary, we have $\partial_\mu U^\dag U =  - U^\dag\partial_\mu U$.
With the help of these relations the traces in Eq.\,\eqref{jmuB1} become 
\begin{subequations}
\begin{align}
\begin{split}
    \frac{\epsilon^{\mu\nu\rho\lambda}}{24\pi^2} \text{Tr}[\Sigma\partial_\lambda\Sigma^\dag\partial_\nu\Sigma\partial_\rho\Sigma^\dag]=&\frac{\epsilon^{\mu\nu\rho\lambda}}{24\pi^2}\text{Tr}[U\partial_\lambda U^\dag\partial_\nu U\partial_\rho U^\dag] 
    \\[2ex]
    &-\frac{i\epsilon^{\mu\nu\rho\lambda}}{8\pi^2}\partial_\nu\alpha\,\text{Tr}[\tau_3 \partial_\rho U\partial_\lambda U^\dag] \, ,
\end{split}
\\[2ex]
-\frac{ie\,\epsilon^{\mu\nu\rho\lambda}}{16\pi^2} A_\lambda \text{Tr}[\tau_3(\partial_\nu\Sigma^\dag\partial_\rho\Sigma-\partial_\nu\Sigma\partial_\rho\Sigma^\dag)]=&-\frac{e\,\epsilon^{\mu\nu\rho\lambda}}{8\pi^2}A_\lambda \partial_\nu \alpha \,\text{Tr}[U\partial_\rho U] \, , 
\\[2ex]
-\frac{ie\,\epsilon^{\mu\nu\rho\lambda}}{32\pi^2}F_{\nu\rho} \text{Tr}[\tau_3(\Sigma\partial_\lambda\Sigma^\dag+\partial_\lambda\Sigma^\dag\Sigma)] =&-\frac{e\,\epsilon^{\mu\nu\rho\lambda}}{32\pi^2}F_{\rho\lambda}\partial_\nu\alpha\,\text{Tr}[1+U^2] \, .
\end{align}
\end{subequations}
These terms can be combined to the compact result
\begin{equation} 
    j_B^\mu=\frac{\epsilon^{\mu\nu\rho\lambda}}{24\pi^2}\text{Tr}[U\partial_\lambda U^\dag\partial_\nu U\partial_\rho U^\dag] +\partial_\nu G^{\mu\nu} \,, 
\label{appAjBtop}
\end{equation}
where
\begin{equation}
    G^{\mu\nu} = -\frac{\alpha \, \epsilon^{\mu\nu\rho\lambda}}{32\pi^2}\left(4i\,\text{Tr}[\tau_3\partial_\rho U\partial_\lambda U^\dag]+4eA_\lambda\text{Tr}[U\partial_\rho U]+e F_{\rho\lambda}\text{Tr}[1+U^2]\right) \, .
    \label{appAGmunu}
\end{equation}
Finally, by evaluating the traces we arrive at
\begin{equation}
     j_{B}^{\mu}=-\frac{\epsilon^{\mu\nu\rho\lambda}}{4\pi^2}\partial_{\nu}\alpha\left\{\frac{e}{2}F_{\rho\lambda}+\frac{1}{f_{\pi}^2}\partial_{\rho}\left[ i\left(\varphi^*\partial_{\lambda}\varphi -\varphi\partial_{\lambda}\varphi^* \right) -2eA_{\lambda}|\varphi|^2\right] \right\}\,.
\end{equation}
(The complex scalar field $\varphi$ is the rotated field of Eq.\,\eqref{phiprime}, but, as in the main part, we have dropped the prime for notational convenience.) One can now replace $eA_{\lambda}\rightarrow eA_{\lambda}-\partial_{\lambda}\alpha$ without changing the result and thus we arrive at the second line of Eq.\,\eqref{BCurrent}.

\chapter{Computing $\langle (\nabla |\varphi_0|^2)^2\rangle$}

\label{app:Compute}
This appendix relates to Chapter \ref{chpt:Project2}. Here, we prove the identity \eqref{phi22}, which is needed for the effective coupling $\lambda_*$ in the calculation of the pion superconductor. We work with the dimensionless quantities \eqref{dimless}, such that with the form of the condensate \eqref{phi0xy1} we find   
\begin{align}
    \begin{split}
        (\nabla |\varphi_0|^2)^2 
        =& (\partial_x|\varphi_0|^2)^2+(\partial_y|\varphi_0|^2)^2 
        \\[2ex]
        =& \sum_{l,p,m,n}  C_l^*C_p C_m^*C_n e^{i(p-l+n-m)qy} \psi_l(x)\psi_p(x)\psi_m(x)\psi_n (x)
        \\[2ex]
        &\times \left\{[2x-(m+n)q][2x-(l+p)q]-q^2(m-n)(l-p)\right\} \,,
    \end{split}
\end{align}
where all indices run form $-\infty$ to $\infty$ once more and should be assumed from here on. Following Sec.\,(\ref{Background:subsec:Hc2}), we do the $y$-integration and switch to the indices defined in Eq.\,\eqref{n123def}, leaving
\begin{align}
    \begin{split}
        \langle(\nabla |\varphi_0|^2)^2\rangle
        =&\frac{1}{N q}\sum_{n_1,n_2,n_3} C_{n_1+n_2+n_3}^* C_{n_1+n_3} C_{n_1}^* C_{n_1+n_2} e^{-\frac{n_2^2q^2 + n_3^2q^2}{2} } \int_{0}^{N q} dx\, e^{- 2\left[ x-\frac{(2n_1+n_2+n_3) q}{2} \right]^2} 
        \\[2ex]
        &\times\left\{4\left[ x-\frac{(2n_1+n_2+n_3) q}{2} \right]^2 +\left(n_2^2-n_3^2\right)q^2\right\}\,.
    \end{split}
\end{align}
Once again, we assume the general periodicity $C_n=C_{n+N}$ and express things in terms of an infinite limit to evaluate the $x$-integral, following a similar procedure to Appendix \ref{app:LatticeN}. Using partial integration on the first term in the curly brackets, we evaluate the Gaussian integral(s) to obtain
\begin{equation}
    \begin{split}
        \langle\left(\nabla|\varphi_0|^2\right)^2\rangle 
        = \frac{1}{Nq}\sqrt{\frac{\pi}{2}} \sum_{n_2,n_3} e^{-\frac{\left(n_2^2 + n_3^2\right)q^2}{2} }\sum_{n_1=0}^{N-1}  C_{n_1+n_2+n_3}^* C_{n_1+n_3} C_{n_1}^* C_{n_1+n_2} \left[1 +\left(n_2^2-n_3^2\right)q^2\right]\,.
    \end{split}
\end{equation}
It can be appreciated that the second term in the square brackets is zero by renaming the summation indices  $n_2\leftrightarrow n_3$ suitably. Therefore, we are left with 
\begin{equation} \label{dPhiPhi}
    \langle(\nabla|\varphi_0|^2)^2\rangle = \langle|\varphi_0|^4\rangle \,,
\end{equation}
where we identified the general result Eq.\,\eqref{NAvgVarPhi4}. Since in the notation of this appendix the gradient denotes derivatives with respect to the dimensionless coordinates, we obtain Eq.\,\eqref{phi22} after reinstating the coherence length $\xi$.

\backmatter
\printbibliography
\end{document}